\documentclass[11pt]{article}
\usepackage{axodraw,epsfig,graphics}
\usepackage[margin=0.4cm]{caption}
\renewcommand{\theequation}{\arabic{section}.\arabic{equation}}

\def\signofmetric{1}

\if0\signofmetric
\def\BDpos{}
\def\BDneg{-}
\def\BDplus{+}
\def\BDminus{-}
\def\thetasigmamuthetadagger{\theta\sigma^\mu\theta^\dagger}
\def\thetasigmamuloweredthetadagger{\theta\sigma_\mu\theta^\dagger}
\fi

\if1\signofmetric
\def\BDpos{-}
\def\BDneg{}
\def\BDplus{-}
\def\BDminus{+}
\def\thetasigmamuthetadagger{\theta^\dagger\sigmabar^\mu\theta}
\def\thetasigmamuloweredthetadagger{\theta^\dagger\sigmabar_\mu\theta}
\fi

\if2\signofmetric
\def\BDpos{\oplus}
\def\BDneg{\ominus}
\def\BDplus{\oplus}
\def\BDminus{\ominus}
\def\thetasigmamuthetadagger{\theta\sigma^\mu\theta^\dagger}
\def\thetasigmamuloweredthetadagger{\theta\sigma_\mu\theta^\dagger}
\fi

\if3\signofmetric
\def\BDpos{\ominus}
\def\BDneg{\oplus}
\def\BDplus{\ominus}
\def\BDminus{\oplus}
\def\thetasigmamuthetadagger{\theta^\dagger\sigmabar^\mu\theta}
\def\thetasigmamuloweredthetadagger{\theta^\dagger\sigmabar_\mu\theta}
\fi

\newcommand{\dagg}[1]{#1^\dagger}

\newcommand{\thdthd}{\theta^\dagger\hspace{-1pt}\theta^\dagger}
\newcommand{\nablasubmu}{\nabla\hspace{-2pt}{}_\mu}

\def\beq{\begin{eqnarray}}
\def\eeq{\end{eqnarray}}
\def\bea{\begin{eqnarray*}}
\def\eea{\end{eqnarray*}}
\def\Baryon{{\rm B}}
\def\Lepton{{\rm L}}
\def\sbar{\overline}
\def\stilde{\widetilde}
\def\sst{\scriptscriptstyle}
\def\vac{|0\rangle}
\def\antivac{\langle 0|}
\def\G{\stilde G}
\def\Wmess{W_{\rm mess}}
\def\NI{\stilde N_1}
\def\nmess{N_5}
\def\lagr{{\cal L}}
\def\drbar{\overline{\rm DR}}
\def\msbar{\overline{\rm MS}}
\def\conj{{{\rm c.c.}}}
\def\Et{{\slashchar{E}_T}}
\def\Etot{{\slashchar{E}}}
\def\MPlanck{M_{\rm P}}
\def\cbeta{c_{\beta}}
\def\sbeta{s_{\beta}}
\def\cW{c_{W}}
\def\sW{s_{W}}
\def\deltaeps{\delta}
\def\sigmabar{\overline\sigma}

\def\half{{1\over 2}}
\def\FX{F}
\def\Branching{{\rm Br}}
\def\Splus{S_+}
\def\Sminus{S_-}
\def\mAMSB{F_\phi}
\def\Dcon{\overline D}

\def\centeron#1#2{{\setbox0=\hbox{#1}\setbox1=\hbox{#2}\ifdim
\wd1>\wd0\kern.5\wd1\kern-.5\wd0\fi
\copy0\kern-.5\wd0\kern-.5\wd1\copy1\ifdim\wd0>\wd1
\kern.5\wd0\kern-.5\wd1\fi}}
\def\ltap{\;\centeron{\raise.35ex\hbox{$<$}}{\lower.65ex\hbox{$\sim$}}\;}
\def\gtap{\;\centeron{\raise.35ex\hbox{$>$}}{\lower.65ex\hbox{$\sim$}}\;}
\def\gsim{\mathrel{\gtap}}
\def\lsim{\mathrel{\ltap}}

\def\slashchar#1{\setbox0=\hbox{$#1$}           
   \dimen0=\wd0                                 
   \setbox1=\hbox{/} \dimen1=\wd1               
   \ifdim\dimen0>\dimen1                        
      \rlap{\hbox to \dimen0{\hfil/\hfil}}      
      #1                                        
   \else                                        
      \rlap{\hbox to \dimen1{\hfil$#1$\hfil}}   
      /                                         
   \fi}                                        %

\begin{document}
\setcounter{footnote}{1}
\begin{flushright}
hep-ph/9709356\\
version 7, January 2016
\end{flushright}

\begin{center}
{\Large\bf A Supersymmetry Primer}\\

\vspace{0.14in}

{\sc Stephen P.~Martin} \\
Department of Physics, Northern Illinois University, DeKalb IL 60115 
\end{center}

\begin{center}
\begin{minipage}[]{0.86\linewidth}
I provide a pedagogical introduction to supersymmetry. The level of 
discussion is aimed at readers who are familiar with the Standard Model 
and quantum field theory, but who have had little or no prior exposure to 
supersymmetry. Topics covered include: motivations for supersymmetry, the 
construction of supersymmetric Lagrangians, superspace and superfields, 
soft supersymmetry-breaking interactions, the Minimal Supersymmetric 
Standard Model (MSSM), $R$-parity and its consequences, the origins of 
supersymmetry breaking, the mass spectrum of the MSSM, decays of 
supersymmetric particles, experimental signals for supersymmetry, and 
some extensions of the minimal framework.
\end{minipage}
\end{center}

\tableofcontents
\newpage
\setlength{\baselineskip}{1.05\baselineskip}
\begin{quotation}
\noindent
\mbox{``We are, I think, in the right Road of Improvement, for we are making Experiments."}\\
--Benjamin Franklin
\end{quotation}

\section{Introduction}\label{sec:intro}
\setcounter{equation}{0}
\setcounter{figure}{0}
\setcounter{table}{0}
\setcounter{footnote}{1}

The Standard Model of high-energy physics, augmented by neutrino masses,
provides a remarkably successful description of presently known phenomena.
The experimental frontier has advanced into the TeV range with no
unambiguous hints of additional structure. Still, it seems clear that the
Standard Model is a work in progress and will have to be extended to
describe physics at higher energies. Certainly, a new framework will be
required at the reduced Planck scale
$\MPlanck = (8 \pi G_{\rm Newton})^{-1/2} = 2.4 \times 10^{18}$ GeV, 
where quantum gravitational effects become important. Based only on a
proper respect for the power of Nature to surprise us, it seems nearly as
obvious that new physics exists in the 16 orders of magnitude in energy
between the presently explored territory near the electroweak scale,
$M_W$, and the Planck scale. 

The mere fact that the ratio $\MPlanck/M_W$ is so huge is already a
powerful clue to the character of physics beyond the Standard Model,
because of the infamous ``hierarchy problem" \cite{hierarchyproblem}. This
is not really a difficulty with the Standard Model itself, but rather a
disturbing sensitivity of the Higgs potential to new physics in almost any
imaginable extension of the Standard Model. The electrically neutral part
of the Standard Model Higgs field is a complex scalar $H$ with a classical
potential
\beq
V = m_H^2 |H|^2 + {\lambda} |H|^4\> .
\label{higgspotential}
\eeq
The Standard Model requires a non-vanishing vacuum expectation value (VEV)
for $H$ at the minimum of the potential. This occurs if $\lambda > 0$
and $m_H^2 < 0$, resulting in $\langle H \rangle =
\sqrt{-m_H^2/2\lambda}$. 
We know experimentally that $\langle H \rangle$ is approximately 174 GeV 
from measurements of the properties of
the weak interactions. 
The 2012 discovery 
\cite{Higgsdiscovery}-\cite{Higgsmass} 
of the Higgs boson with a mass
near 125 GeV implies that, assuming the Standard Model is correct as an 
effective field theory, $\lambda = 0.126$ and $m_H^2 = -(\mbox{92.9 GeV})^2$. 
(These are running $\msbar$ parameters evaluated at a renormalization scale 
equal to the top-quark mass, and include the effects of 2-loop corrections.)
The problem is that $m_H^2$ receives enormous quantum
corrections from the virtual effects of every particle or other phenomenon
that couples, directly or indirectly, to the Higgs field. 

\begin{figure}[b]
\begin{center}
\begin{picture}(108,63)(-54,-27)
\SetWidth{0.9}
\DashLine(-55,0)(-22,0){4}
\DashLine(55,0)(22,0){4}
\CArc(0,0)(22,0,360)
\Text(-50,7)[c]{$H$}
\Text(0,32)[c]{$f$}
\Text(0,-34.5)[c]{(a)}
\end{picture}
\hspace{1.4cm}
\begin{picture}(108,63)(-54,-27)
\SetWidth{0.9}
\Text(0,37)[c]{$S$}
\Text(-41,-5)[c]{$H$}
\DashLine(-46,-12)(46,-12){4}
\SetWidth{1.5}
\DashCArc(0,8)(20,-90,270){5}
\Text(0,-34.5)[c]{(b)}
\end{picture}
\end{center}
\vspace{-0.3cm}
\caption{One-loop quantum corrections to the Higgs squared mass parameter
$m_H^2$, due to (a) a Dirac fermion $f$, and (b) a scalar $S$.
\label{fig:higgscorr1}}
\end{figure}
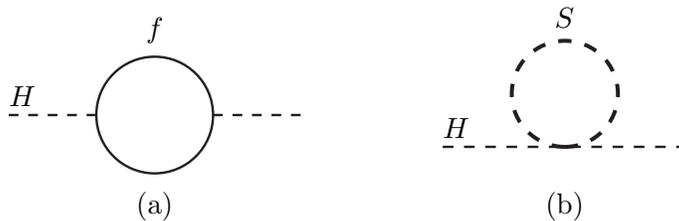

For example, in Figure \ref{fig:higgscorr1}a we have a correction to $m_H^2$
from a loop containing a Dirac fermion $f$ with mass $m_f$.%
If the Higgs
field couples to $f$ with a term in the Lagrangian $-\lambda_f H \sbar f
f$, then the Feynman diagram in Figure \ref{fig:higgscorr1}a yields a
correction
\beq
\Delta m_H^2 \>=\>  
-{|\lambda_f|^2\over 8 \pi^2} \Lambda_{\rm UV}^2 + \ldots .
\label{quaddiv1}
\eeq
Here $\Lambda_{\rm UV}$ is an ultraviolet momentum cutoff used to regulate 
the loop integral; it should be interpreted as at least the energy scale 
at which new physics enters to alter the high-energy behavior of the 
theory. The ellipses represent terms proportional to $m_f^2$, which 
grow at most logarithmically 
with $\Lambda_{\rm UV}$ (and actually differ for the real and imaginary 
parts of $H$). Each of the leptons and quarks of the Standard Model can 
play the role of $f$; for quarks, eq.~(\ref{quaddiv1}) should be 
multiplied by 3 to account for color. The largest correction comes when 
$f$ is the top quark with $\lambda_f\approx 0.94$. The problem is that if 
$\Lambda_{\rm UV}$ is of order $\MPlanck$, say, then this quantum 
correction to $m_H^2$ is some 30 orders of magnitude larger than the 
required value of $m_H^2 \approx -(92.9$ GeV$)^2$. This is only directly a 
problem for corrections to the Higgs scalar boson squared mass, because 
quantum corrections to fermion and gauge boson masses do not have the 
direct quadratic sensitivity to $\Lambda_{\rm UV}$ found in 
eq.~(\ref{quaddiv1}).  However, the quarks and leptons and the electroweak 
gauge bosons $Z^0$, $W^\pm$ of the Standard Model all obtain masses from 
$\langle H \rangle$, so that the entire mass spectrum of the Standard 
Model is directly or indirectly sensitive to the cutoff 
$\Lambda_{\rm UV}$.

One could imagine that the solution is to simply pick a $\Lambda_{\rm UV}$
that is not too large. But then one still must concoct some new physics at
the scale $\Lambda_{\rm UV}$ that not only alters the propagators in the
loop, but actually cuts off the loop integral. This is not easy to do in a
theory whose Lagrangian does not contain more than two derivatives, and
higher-derivative theories generally suffer from a failure of either
unitarity or causality \cite{EliezerWoodard}. In string theories, loop
integrals are nevertheless cut off at high Euclidean momentum $p$ by
factors $e^{-p^2/\Lambda^2_{\rm UV}}$.  However, then $\Lambda_{\rm UV}$
is a string scale that is usually\footnote{Some attacks on the
hierarchy problem, not reviewed here, 
are based on the proposition that the ultimate cutoff
scale is actually close to the electroweak scale, rather 
than the apparent Planck scale.} thought to
be not very far below $\MPlanck$. 

Furthermore, there are contributions similar to eq.~(\ref{quaddiv1}) from the virtual 
effects of any heavy particles that might exist, and these involve
the masses of the heavy particles (or other high physical mass scales), 
not just the cutoff. It cannot be overemphasized that merely choosing 
a regulator with no quadratic divergences does {\em not} address the hierarchy problem. 
The problem is not really the quadratic divergences, but rather the quadratic sensitivity 
to high mass scales. The latter are correlated with quadratic divergences for some, 
but not all, choices of ultraviolet regulator. The absence of quadratic divergences 
is a necessary, but not sufficient, criterion for avoiding the hierarchy problem.

For example, suppose there exists a heavy complex scalar particle $S$ with
mass $m_S$ that couples to the Higgs with a Lagrangian term $ -\lambda_S
|H|^2 |S|^2$. Then the Feynman diagram in Figure~\ref{fig:higgscorr1}b
gives a correction
\beq
\Delta m_H^2 \>=\> {\lambda_S\over 16 \pi^2}
\left [\Lambda_{\rm UV}^2 - 2 m_S^2
\> {\rm ln}(\Lambda_{\rm UV}/m_S) + \ldots
\right ].
\label{quaddiv2}
\eeq
If one rejects the possibility of a physical interpretation of
$\Lambda_{\rm UV}$ and uses dimensional regularization on the loop
integral instead of a momentum cutoff, then there will be no $\Lambda_{\rm
UV}^2$ piece. However, even then the term proportional to $m_S^2$ cannot
be eliminated without the physically unjustifiable tuning of a
counter-term specifically for that purpose. This illustrates that 
$m_H^2$ is sensitive to the
masses of the {\it heaviest} particles that $H$ couples to; if $m_S$ is
very large, its effects on the Standard Model do not decouple, but instead
make it difficult to understand why $m_H^2$ is so small. 

\begin{figure}
\begin{center}
\begin{picture}(102,52)(-49,-17)
\SetWidth{0.9}
\DashLine(-52,-22)(-22,-22){4}
\DashLine(52,-22)(22,-22){4}
\DashLine(-22,-22)(22,-22){4}
\CArc(0,12)(22,0,180)
\Line(-22,12)(22,12)
\Photon(-22,-22)(-22,12){2}{4.5}
\Photon(22,-22)(22,12){-2}{4.5}
\Text(0,28)[]{$F$}
\Text(-48,-15)[]{$H$}
\end{picture}
\hspace{1.6cm}
\begin{picture}(98,52)(-49,-17)
\SetWidth{0.9}
\DashLine(-45,-22)(0,-22){4}
\DashLine(45,-22)(0,-22){4}
\CArc(0,12)(22,0,180)
\Line(-22,12)(22,12) 
\Photon(0,-22)(-22,12){2}{4.5}
\Photon(0,-22)(22,12){-2}{4.5}
\Text(0,28)[]{$F$}
\Text(-41,-15)[]{$H$}
\end{picture}
\vspace{-0.25cm}
\end{center}
\caption{Two-loop corrections to the Higgs squared mass parameter
involving a heavy fermion $F$ that couples only indirectly to the Standard
Model Higgs through gauge interactions. \label{fig:higgscorr2}}
\end{figure}
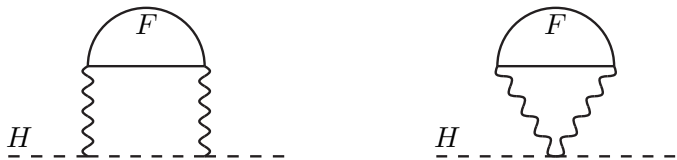

This problem arises even if there is no direct coupling between the
Standard Model Higgs boson and the unknown heavy particles. For example,
suppose there exists a heavy fermion $F$ that, unlike the quarks and
leptons of the Standard Model, has vectorlike quantum numbers and
therefore gets a large mass $m_F$ without coupling to the Higgs field. [In
other words, an arbitrarily large mass term of the form $m_F \overline F
F$ is not forbidden by any symmetry, including weak isospin $SU(2)_L$.] In
that case, no diagram like Figure~\ref{fig:higgscorr1}a exists for $F$.
Nevertheless there will be a correction to $m_H^2$ as long as $F$ shares
some gauge interactions with the Standard Model Higgs field; these may be
the familiar electroweak interactions, or some unknown gauge forces that
are broken at a very high energy scale inaccessible to experiment. In
either case, the two-loop Feynman diagrams in Figure~\ref{fig:higgscorr2}
yield a correction%
\beq
\Delta m_H^2 \>=\> C_H T_F \left ( {g^2 \over 16 \pi^2} \right )^2
\left [ a \Lambda_{\rm UV}^2 + 24 m_F^2 \>{\rm ln} (\Lambda_{\rm UV}/m_F)
+ \ldots \right ],
\label{quaddiv3}
\eeq
where $C_H$ and $T_F$ are group theory factors\footnote{Specifically, 
$C_H$ is the quadratic Casimir invariant of $H$, and $T_F$ is the Dynkin 
index of $F$ in a normalization such that $T_F=1$ for a Dirac fermion (or 
two Weyl fermions) in a fundamental representation of $SU(n)$.} of order 
1, and $g$ is the appropriate gauge coupling. The coefficient $a$ depends 
on the method used to cut off the momentum integrals. It does not arise at 
all if one uses dimensional regularization, but the $m_F^2$ contribution 
is always present with the given coefficient. The numerical factor 
$(g^2/16 \pi^2)^2$ may be quite small (of order $10^{-5}$ for electroweak 
interactions), but the important point is that these contributions to 
$\Delta m_H^2$ are sensitive both to the largest masses and to the 
physical ultraviolet cutoff in the theory, presumably of order $\MPlanck$. The 
``natural" squared mass of a fundamental Higgs scalar, including quantum 
corrections, therefore seems to be more like $\MPlanck^2$ than the 
experimental value. Even very indirect contributions from 
Feynman diagrams with three or more loops can give unacceptably large 
contributions to $\Delta m_H^2$. The argument above applies not just for 
heavy particles, but for arbitrary high-scale physical phenomena such as 
condensates or additional compactified dimensions.

It could be that the Higgs boson field is not fundamental, but rather is the result of
a composite field or collective phenomenon. Such ideas are certainly still worth
exploring, although they typically present difficulties in their simplest
forms. In particular, so far the 125 GeV Higgs boson does appear to have properties
consistent with a fundamental scalar field. Or, it could be that the ultimate 
ultraviolet cutoff scale, and therefore the mass scales of all presently undiscovered particles and condensates, are much lower than the Planck scale. 
But, if the Higgs boson is a fundamental particle, and there really
is physics far above the electroweak scale, then we have two remaining
options: either we must make the rather bizarre assumption that {\it none} of the
high-mass particles or condensates couple (even
indirectly or extremely weakly) to the Higgs scalar field, or else some
striking cancellation is needed between the various contributions to
$\Delta m_H^2$. 

The systematic cancellation of the dangerous contributions to $\Delta
m_H^2$ can only be brought about by the type of conspiracy that is better
known to physicists as a symmetry. Comparing eqs.~(\ref{quaddiv1}) and
(\ref{quaddiv2}) strongly suggests that the new symmetry ought to relate
fermions and bosons, because of the relative minus sign between fermion
loop and boson loop contributions to $\Delta m_H^2$. (Note that
$\lambda_S$ must be positive if the scalar potential is to be bounded from
below.) If each of the quarks and leptons of the Standard Model is
accompanied by two complex scalars with $\lambda_S = |\lambda_f|^2$, then
the $\Lambda_{\rm UV}^2$ contributions of Figures~\ref{fig:higgscorr1}a
and \ref{fig:higgscorr1}b will neatly cancel \cite{quadscancel}. Clearly,
more restrictions on the theory will be necessary to ensure that this
success persists to higher orders, so that, for example, the contributions
in Figure~\ref{fig:higgscorr2} and eq.~(\ref{quaddiv3}) from a very heavy
fermion are canceled by the two-loop effects of some very heavy bosons.
Fortunately, the cancellation of all such contributions to scalar masses
is not only possible, but is actually unavoidable, once we merely assume
that there exists a symmetry relating fermions and bosons, called a 
{\it supersymmetry}. 

A supersymmetry transformation turns a bosonic state into a fermionic
state, and vice versa. The operator $Q$ that generates such
transformations must be an anti-commuting spinor, with
\beq
Q |{\rm Boson}\rangle = |{\rm Fermion }\rangle, \qquad\qquad
Q |{\rm Fermion}\rangle = |{\rm Boson }\rangle .
\eeq
Spinors are intrinsically complex objects, so $Q^\dagger$ (the hermitian
conjugate of $Q$) is also a symmetry generator. Because $Q$ and
$Q^\dagger$ are fermionic operators, they carry spin angular momentum 1/2,
so it is clear that supersymmetry must be a spacetime symmetry. The
possible forms for such symmetries in an interacting quantum field theory
are highly restricted by the Haag-Lopuszanski-Sohnius extension \cite{HLS} of the
Coleman-Mandula theorem \cite{ColemanMandula}. For realistic theories that, like the
Standard Model, have chiral fermions (i.e., fermions whose left- and
right-handed pieces transform differently under the gauge group) and thus
the possibility of parity-violating interactions, this theorem implies
that the generators $Q$ and $Q^\dagger$ must satisfy an algebra of
anticommutation and commutation relations with the schematic form
\beq
&&\{ Q, Q^\dagger \} = P^\mu , \label{susyalgone}
\\
&&\{ Q,Q \} = \{ Q^\dagger , Q^\dagger \} = 0 , \label{susyalgtwo}
\\
&&[ P^\mu , Q  ] = [P^\mu, Q^\dagger ] = 0 ,\label{susyalgthree}
\eeq
where $P^\mu$ is the four-momentum generator of spacetime translations.
Here we have ruthlessly suppressed the spinor indices on $Q$ and
$Q^\dagger$; after developing some notation we will, in section
\ref{subsec:susylagr.freeWZ}, derive the precise version of
eqs.~(\ref{susyalgone})-(\ref{susyalgthree}) with indices restored. In the
meantime, we simply note that the appearance of $P^\mu$ on the right-hand
side of eq.~(\ref{susyalgone}) is unsurprising, because it transforms under
Lorentz boosts and rotations as a spin-1 object while $Q$ and $Q^\dagger$
on the left-hand side each transform as spin-1/2 objects. 

The single-particle states of a supersymmetric theory fall into
irreducible representations of the supersymmetry algebra, called {\it
supermultiplets}. Each supermultiplet contains both fermion and boson
states, which are commonly known as {\it superpartners} of each other. By
definition, if $|\Omega\rangle$ and $|\Omega^\prime \rangle$ are members
of the same supermultiplet, then $|\Omega^\prime\rangle$ is proportional
to some combination of $Q$ and $\dagg{Q}$ operators acting on
$|\Omega\rangle $, up to a spacetime translation or rotation. The
squared-mass operator $\BDpos P^2$ commutes with the operators $Q$,
$\dagg{Q}$, and with all spacetime rotation and translation operators, so
it follows immediately that particles inhabiting the same irreducible
supermultiplet must have equal eigenvalues of $\BDpos P^2$, and therefore
equal masses. 

The supersymmetry generators $Q,Q^\dagger$ also commute with the
generators of gauge transformations. Therefore particles in the same
supermultiplet must also be in the same representation of the gauge group,
and so must have the same electric charges, weak isospin, and color
degrees of freedom. 

Each supermultiplet contains an equal number of fermion and boson degrees
of freedom. To prove this, consider the operator $(-1)^{2s}$ where $s$ is
the spin angular momentum. By the spin-statistics theorem, this operator
has eigenvalue $+1$ acting on a bosonic state and eigenvalue $-1$ acting
on a fermionic state. Any fermionic operator will turn a bosonic state
into a fermionic state and vice versa. Therefore $(-1)^{2s}$ must
anticommute with every fermionic operator in the theory, and in particular
with $Q$ and $Q^\dagger$. Now, within a given supermultiplet, consider the
subspace of states $| i \rangle$ with the same eigenvalue $p^\mu$ of the
four-momentum operator $P^\mu$. In view of eq.~(\ref{susyalgthree}), any
combination of $Q$ or $Q^\dagger$ acting on $|i\rangle$ must give another
state $|i^\prime\rangle$ with the same four-momentum eigenvalue. Therefore
one has a completeness relation $\sum_i |i\rangle\langle i | = 1$ within
this subspace of states. Now one can take a trace over all such states of
the operator $(-1)^{2s} P^\mu$ (including each spin helicity state
separately): 
\beq
\sum_i \langle i | (-1)^{2s} P^\mu | i \rangle
&=&
\sum_i \langle i | (-1)^{2s} Q Q^\dagger|i\rangle
+\sum_i\langle i | (-1)^{2s} Q^\dagger Q | i \rangle
\nonumber\\
&=&
\sum_i \langle i | (-1)^{2s} Q Q^\dagger | i \rangle
+ \sum_i \sum_j \langle i | (-1)^{2s} Q^\dagger |j \rangle \langle j | Q
| i \rangle\qquad{}
\nonumber\\
&=&
\sum_i \langle i | (-1)^{2s} Q Q^\dagger | i \rangle +
\sum_j \langle j | Q (-1)^{2s}  Q^\dagger | j \rangle
\nonumber\\
&=&\sum_i \langle i | (-1)^{2s} Q Q^\dagger | i \rangle -
\sum_j \langle j |  (-1)^{2s} Q Q^\dagger | j \rangle
\nonumber \\
&=& 0.
\eeq
The first equality follows from the supersymmetry algebra relation
eq.~(\ref{susyalgone}); the second and third from use of the completeness
relation; and the fourth from the fact that $(-1)^{2s}$ must anticommute
with $Q$. Now $\sum_i \langle i | (-1)^{2s} P^\mu | i \rangle = \, p^\mu$
Tr[$(-1)^{2s}$] is just proportional to the number of bosonic degrees of
freedom $n_B$ minus the number of fermionic degrees of freedom $n_F$ in
the trace, so that
\beq
n_B= n_F
\label{nbnf}
\eeq
must hold for a given $p^\mu\not= 0$ in each supermultiplet.

The simplest possibility for a supermultiplet consistent with
eq.~(\ref{nbnf}) has a single Weyl fermion (with two spin helicity states,
so $n_F=2$) and two real scalars (each with $n_B=1$). It is natural to
assemble the two real scalar degrees of freedom into a complex scalar
field; as we will see below this provides for convenient formulations of
the supersymmetry algebra, Feynman rules, supersymmetry-violating effects,
etc. This combination of a two-component Weyl fermion and a complex scalar
field is called a {\it chiral} or {\it matter} or {\it scalar}
supermultiplet. 

The next-simplest possibility for a supermultiplet contains a spin-1
vector boson. If the theory is to be renormalizable, this must be a gauge
boson that is massless, at least before the gauge symmetry is
spontaneously broken. A massless spin-1 boson has two helicity states, so
the number of bosonic degrees of freedom is $n_B=2$. Its superpartner is
therefore a massless spin-1/2 Weyl fermion, again with two helicity
states, so $n_F=2$. (If one tried to use a massless spin-3/2 fermion
instead, the theory would not be renormalizable.) Gauge bosons must
transform as the adjoint representation of the gauge group, so their
fermionic partners, called {\it gauginos}, must also. Because the adjoint
representation of a gauge group is always its own conjugate, the gaugino
fermions must have the same gauge transformation properties for
left-handed and for right-handed components. Such a combination of
spin-1/2 gauginos and spin-1 gauge bosons is called a {\it gauge} or {\it
vector} supermultiplet. 

If we include gravity, then the spin-2 graviton (with 2 helicity states,
so $n_B=2$) has a spin-$3/2$ superpartner called the gravitino. The
gravitino would be massless if supersymmetry were unbroken, and so it has
$n_F=2$ helicity states. 

There are other possible combinations of particles with spins that can 
satisfy eq.~(\ref{nbnf}). However, these are always reducible to 
combinations\footnote{For example, if a gauge symmetry were to 
spontaneously break without breaking supersymmetry, then a massless vector 
supermultiplet would ``eat'' a chiral supermultiplet, resulting in a 
massive vector supermultiplet with physical degrees of freedom consisting 
of a massive vector ($n_B=3$), a massive Dirac fermion formed from the 
gaugino and the chiral fermion ($n_F=4$), and a real scalar ($n_B=1$).} of 
chiral and gauge supermultiplets if they have renormalizable interactions, 
except in certain theories with ``extended" supersymmetry. Theories with 
extended supersymmetry have more than one distinct copy of the 
supersymmetry generators $Q,\dagg{Q}$. Such models are mathematically 
interesting, but evidently do not have any phenomenological prospects. The 
reason is that extended supersymmetry in four-dimensional field theories 
cannot allow for chiral fermions or parity violation as observed in the 
Standard Model. So we will not discuss such possibilities further, 
although extended supersymmetry in higher-dimensional field theories might 
describe the real world if the extra dimensions are compactified in an 
appropriate way, and extended supersymmetry in four dimensions provides 
interesting toy models and calculation tools. 
The ordinary, non-extended, phenomenologically 
viable type of supersymmetric model is sometimes called $N=1$ 
supersymmetry, with $N$ referring to the number of supersymmetries (the 
number of distinct copies of $Q, \dagg{Q}$).

In a supersymmetric extension of the Standard Model 
\cite{FayetHsnu}-\cite{Rparity}, 
each of the known fundamental 
particles is therefore in either a chiral or gauge supermultiplet, and 
must have a superpartner with spin differing by 1/2 unit. The first step 
in understanding the exciting phenomenological consequences of this 
prediction is to decide exactly how the known particles fit into 
supermultiplets, and to give them appropriate names. A crucial observation 
here is that only chiral supermultiplets can contain fermions whose 
left-handed parts transform differently under the gauge group than their 
right-handed parts. All of the Standard Model fermions (the known quarks 
and leptons) have this property, so they must be members of chiral 
supermultiplets. The bosonic partners of the quarks and leptons therefore must
be spin-0, and not spin-1 vector bosons.\footnote{In particular, one cannot attempt to make a 
spin-1/2 neutrino be the superpartner of the spin-1 photon; the neutrino 
is in a doublet, and the photon is neutral, under weak isospin.} 

The names 
for the spin-0 partners of the quarks and leptons are constructed by 
prepending an ``s", for scalar. So, generically they are called {\it 
squarks} and {\it sleptons} (short for ``scalar quark" and ``scalar 
lepton"), or sometimes {\it sfermions}. 
The left-handed and right-handed pieces of the quarks and 
leptons are separate two-component Weyl fermions with different gauge 
transformation properties in the Standard Model, so each must have its own 
complex scalar partner. The symbols for the squarks and sleptons are the 
same as for the corresponding fermion, but with a tilde 
($\phantom{.}\stilde{\phantom{.}}\phantom{.}$) 
used to denote the superpartner of a Standard 
Model particle. For example, the superpartners of the left-handed and 
right-handed parts of the electron Dirac field are called left- and 
right-handed selectrons, and are denoted $\stilde e_L$ and $\stilde e_R$. 
It is important to keep in mind that the ``handedness" here does not refer 
to the helicity of the selectrons (they are spin-0 particles) but to that 
of their superpartners. A similar nomenclature applies for smuons and 
staus: $\stilde \mu_L$, $\stilde\mu_R$, $\stilde\tau_L$, $\stilde \tau_R$. 
The Standard Model neutrinos (neglecting their very small masses) are 
always left-handed, so the sneutrinos are denoted generically by 
$\stilde\nu$, with a possible subscript indicating which lepton flavor 
they carry: $\stilde\nu_e$, $\stilde\nu_\mu$, $\stilde\nu_\tau$. Finally, 
a complete list of the squarks is $\stilde q_L$, $\stilde q_R$ with 
$q=u,d,s,c,b,t$. The gauge interactions of each of these squark and 
slepton fields are the same as for the corresponding Standard Model 
fermions; for instance, the left-handed squarks $\stilde u_L$ and $\stilde 
d_L$ couple to the $W$ boson, while $\stilde u_R$ and $\stilde d_R$ do 
not.\setcounter{footnote}{1}

It seems clear that the Higgs scalar boson must reside in a chiral
supermultiplet, since it has spin 0. Actually, it turns out that just one
chiral supermultiplet is not enough. One reason for this is that
if there were only one Higgs chiral supermultiplet, the electroweak gauge
symmetry would suffer a gauge anomaly, and would be inconsistent as a
quantum theory. This is because the conditions for cancellation of gauge
anomalies include $ {\rm Tr}[T_3^2 Y] = {\rm Tr}[Y^3] = 0, $ where $T_3$
and $Y$ are the third component of weak isospin and the weak hypercharge,
respectively, in a normalization where the ordinary electric charge is
$Q_{\rm EM} = T_3 + Y$. The traces run over all of the left-handed Weyl
fermionic degrees of freedom in the theory. In the Standard Model, these
conditions are already satisfied, somewhat miraculously, by the known
quarks and leptons. Now, a fermionic partner of a Higgs chiral
supermultiplet must be a weak isodoublet with weak hypercharge $Y=1/2$ or
$Y=-1/2$. In either case alone, such a fermion will make a non-zero
contribution to the traces and spoil the anomaly cancellation. This can be
avoided if there are two Higgs supermultiplets, one with each of $Y=\pm
1/2$, so that the total contribution to the anomaly traces from the two
fermionic members of the Higgs chiral supermultiplets vanishes by
cancellation. As we will see in section \ref{subsec:mssm.superpotential},
both of these are also necessary for another completely different reason:
because of the structure of supersymmetric theories, only a $Y=1/2$ Higgs
chiral supermultiplet can have the Yukawa couplings necessary to give
masses to charge $+2/3$ up-type quarks (up, charm, top), and only a
$Y=-1/2$ Higgs can have the Yukawa couplings necessary to give masses to
charge $-1/3$ down-type quarks (down, strange, bottom) and to the charged
leptons. 

We will call the $SU(2)_L$-doublet complex scalar fields with
$Y=1/2$ and $Y=-1/2$ by the names $H_u$ and $H_d$,
respectively.\footnote{Other notations in the literature have
$H_1, H_2$ or $H,\sbar H$ instead of $H_u, H_d$. The notation used here
has the virtue of making it easy to remember which Higgs VEVs
gives masses to which type of quarks.} The weak isospin components of
$H_u$ with $T_3=(1/2$, $-1/2$) have electric charges $1$, $0$
respectively, and are denoted ($H_u^+$, $H_u^0$). Similarly, the
$SU(2)_L$-doublet complex scalar $H_d$ has $T_3=(1/2$, $-1/2$) components
($H_d^0$, $H_d^-$). The neutral scalar that corresponds to the physical
Standard Model Higgs boson is in a linear combination of $H_u^0$ and
$H_d^0$; we will discuss this further in section
\ref{subsec:MSSMspectrum.Higgs}.  The generic nomenclature for a spin-1/2
superpartner is to append ``-ino" to the name of the Standard Model
particle, so the fermionic partners of the Higgs scalars are called
higgsinos. They are denoted by $\stilde H_u$, $\stilde H_d$ for the
$SU(2)_L$-doublet left-handed Weyl spinor fields, with weak isospin
components $\stilde H_u^+$, $\stilde H_u^0$ and $\stilde H_d^0$, $\stilde
H_d^-$. 

\renewcommand{\arraystretch}{1.4}
\begin{table}[tb]
\begin{center}
\begin{tabular}{|c|c|c|c|c|}
\hline
\multicolumn{2}{|c|}{Names} 
& spin 0 & spin 1/2 & $SU(3)_C ,\, SU(2)_L ,\, U(1)_Y$
\\  \hline\hline
squarks, quarks & $Q$ & $({\stilde u}_L\>\>\>{\stilde d}_L )$&
 $(u_L\>\>\>d_L)$ & $(\>{\bf 3},\>{\bf 2}\>,\>{1\over 6})$
\\
($\times 3$ families) & $\sbar u$
&${\stilde u}^*_R$ & $u^\dagger_R$ & 
$(\>{\bf \overline 3},\> {\bf 1},\> -{2\over 3})$
\\ & $\sbar d$ &${\stilde d}^*_R$ & $d^\dagger_R$ & 
$(\>{\bf \overline 3},\> {\bf 1},\> {1\over 3})$
\\  \hline
sleptons, leptons & $L$ &$({\stilde \nu}\>\>{\stilde e}_L )$&
 $(\nu\>\>\>e_L)$ & $(\>{\bf 1},\>{\bf 2}\>,\>-{1\over 2})$
\\
($\times 3$ families) & $\sbar e$
&${\stilde e}^*_R$ & $e^\dagger_R$ & $(\>{\bf 1},\> {\bf 1},\>1)$
\\  \hline
Higgs, higgsinos &$H_u$ &$(H_u^+\>\>\>H_u^0 )$&
$(\stilde H_u^+ \>\>\> \stilde H_u^0)$& 
$(\>{\bf 1},\>{\bf 2}\>,\>+{1\over 2})$
\\ &$H_d$ & $(H_d^0 \>\>\> H_d^-)$ & $(\stilde H_d^0 \>\>\> \stilde H_d^-)$& 
$(\>{\bf 1},\>{\bf 2}\>,\>-{1\over 2})$
\\  \hline
\end{tabular}
\caption{Chiral supermultiplets in the Minimal Supersymmetric Standard Model.
The spin-$0$ fields are complex scalars, and the spin-$1/2$ fields are 
left-handed two-component Weyl fermions.\label{tab:chiral}}
\vspace{-0.6cm}
\end{center}
\end{table}
We have now found all of the chiral supermultiplets of a minimal
phenomenologically viable extension of the Standard Model. They are
summarized in Table \ref{tab:chiral}, 
classified according to their transformation
properties under the Standard Model gauge group $SU(3)_C\times SU(2)_L
\times U(1)_Y$, which combines $u_L,d_L$ and $\nu,e_L$ degrees of freedom
into $SU(2)_L$ doublets. Here we follow a standard convention, that all
chiral supermultiplets are defined in terms of left-handed Weyl spinors,
so that the {\it conjugates} of the right-handed quarks and leptons (and
their superpartners) appear in Table \ref{tab:chiral}. 
This protocol for defining chiral
supermultiplets turns out to be very useful for constructing
supersymmetric Lagrangians, as we will see in section \ref{sec:susylagr}.
It is also useful to have a symbol for each of the chiral supermultiplets
as a whole; these are indicated in the second column of 
Table \ref{tab:chiral}. Thus, for
example, $Q$ stands for the $SU(2)_L$-doublet chiral supermultiplet
containing $\stilde u_L,u_L$ (with weak isospin component $T_3=1/2$), and
$\stilde d_L, d_L$ (with $T_3=-1/2$), while $\sbar u$ stands for the
$SU(2)_L$-singlet supermultiplet containing $\stilde u_R^*, u_R^\dagger$.
There are three families for each of the quark and lepton supermultiplets,
Table \ref{tab:chiral} lists the first-family representatives. A family
index $i=1,2,3$ can be affixed to the chiral supermultiplet names ($Q_i$,
$\sbar u_i, \ldots$) when needed, for example 
$(\sbar e_1, \sbar e_2, \sbar e_3)=
(\sbar e, \sbar \mu, \sbar \tau)$. The bar on $\sbar u$, $\sbar d$, $\sbar
e$ fields is part of the name, and does not denote any kind of
conjugation. 

The Higgs chiral supermultiplet $H_d$
(containing $H_d^0$, $H_d^-$, $\stilde H_d^0$, $\stilde H_d^-$) has
exactly the same Standard Model gauge quantum numbers as the left-handed
sleptons and leptons $L_i$, for example ($\stilde \nu$, $\stilde e_L$, $\nu$,
$e_L$). Naively, one might therefore suppose that we could have been more
economical in our assignment by taking a neutrino and a Higgs scalar to be
superpartners, instead of putting them in separate supermultiplets. This
would amount to the proposal that the Higgs boson and a sneutrino should
be the same particle. This attempt played a key role in some of the first
attempts to connect supersymmetry to phenomenology \cite{FayetHsnu}, but
it is now known to not work. Even ignoring the anomaly cancellation
problem mentioned above, many insoluble phenomenological problems would
result, including lepton-number non-conservation and a mass for at least
one of the neutrinos in gross violation of experimental bounds. Therefore,
all of the superpartners of Standard Model particles are really new
particles, and cannot be identified with some other Standard Model state. 

\renewcommand{\arraystretch}{1.55}
\begin{table}[t]
\begin{center}
\begin{tabular}{|c|c|c|c|}
\hline
Names & spin 1/2 & spin 1 & $SU(3)_C, \> SU(2)_L,\> U(1)_Y$\\
\hline\hline
gluino, gluon &$ \stilde g$& $g$ & $(\>{\bf 8},\>{\bf 1}\>,\> 0)$
\\
\hline
winos, W bosons & $ \stilde W^\pm\>\>\> \stilde W^0 $&
 $W^\pm\>\>\> W^0$ & $(\>{\bf 1},\>{\bf 3}\>,\> 0)$
\\
\hline
bino, B boson &$\stilde B^0$&
 $B^0$ & $(\>{\bf 1},\>{\bf 1}\>,\> 0)$
\\
\hline
\end{tabular}
\caption{Gauge supermultiplets in
the Minimal Supersymmetric Standard Model.\label{tab:gauge}}
\vspace{-0.45cm}
\end{center}
\end{table}
The vector bosons of the Standard Model clearly must reside in gauge
supermultiplets. Their fermionic superpartners are generically referred to
as gauginos. The $SU(3)_C$ color gauge interactions of QCD are mediated by
the gluon, whose spin-1/2 color-octet supersymmetric partner is the
gluino. As usual, a tilde is used to denote the supersymmetric partner of
a Standard Model state, so the symbols for the gluon and gluino are $g$
and $\stilde g$ respectively. The electroweak gauge symmetry
$SU(2)_L\times U(1)_Y$ is associated with spin-1 gauge bosons $W^+, W^0,
W^-$ and $B^0$, with spin-1/2 superpartners $\stilde W^+, \stilde W^0,
\stilde W^-$ and $\stilde B^0$, called {\it winos} and {\it bino}. After
electroweak symmetry breaking, the $W^0$, $B^0$ gauge eigenstates mix to
give mass eigenstates $Z^0$ and $\gamma$. The corresponding gaugino
mixtures of $\stilde W^0$ and $\stilde B^0$ are called zino ($\stilde Z^0$) 
and photino ($\stilde \gamma$); if supersymmetry were unbroken, they would
be mass eigenstates with masses $m_Z$ and 0. Table \ref{tab:gauge} 
summarizes the gauge
supermultiplets of a minimal supersymmetric extension of the Standard
Model. 

The chiral and gauge supermultiplets in Tables \ref{tab:chiral} and 
\ref{tab:gauge} make up the
particle content of the Minimal Supersymmetric Standard Model (MSSM). The
most obvious and interesting feature of this theory is that none of the
superpartners of the Standard Model particles has been discovered as of
this writing. If supersymmetry were unbroken, then there would have to be
selectrons $\stilde e_L$ and $\stilde e_R$ with masses exactly equal to
$m_e = 0.511...$ MeV. A similar statement applies to each of the other
sleptons and squarks, and there would also have to be a massless gluino
and photino. These particles would have been extraordinarily easy to
detect long ago. Clearly, therefore, {\it supersymmetry is a broken
symmetry} in the vacuum state chosen by Nature. 

An important clue as to the nature of supersymmetry breaking can be
obtained by returning to the motivation provided by the hierarchy problem.
Supersymmetry forced us to introduce two complex scalar fields for each
Standard Model Dirac fermion, which is just what is needed to enable a
cancellation of the quadratically sensitive $(\Lambda_{\rm UV}^2)$ pieces
of eqs.~(\ref{quaddiv1}) and (\ref{quaddiv2}). This sort of cancellation
also requires that the associated dimensionless couplings should be
related (for example $\lambda_S = |\lambda_f|^2$). The necessary relationships
between couplings indeed occur in unbroken supersymmetry, as we will see
in section \ref{sec:susylagr}. In fact, unbroken supersymmetry guarantees
that quadratic divergences in scalar squared masses, and therefore the quadratic sensitivity to high mass scales, must vanish to all
orders in perturbation theory.\footnote{A simple way to understand this is
to recall that unbroken supersymmetry requires the degeneracy of scalar
and fermion masses. Radiative corrections to fermion masses are known to
diverge at most logarithmically in any renormalizable
field theory, so the same must be
true for scalar masses in unbroken supersymmetry.} Now, if broken
supersymmetry is still to provide a solution to the hierarchy problem even
in the presence of supersymmetry breaking, then the relationships between
dimensionless couplings that hold in an unbroken supersymmetric theory
must be maintained. Otherwise, there would be quadratically divergent
radiative corrections to the Higgs scalar masses of the form
\beq
\Delta m_H^2 = {1\over 8\pi^2} (\lambda_S - |\lambda_f|^2)
\Lambda_{\rm UV}^2 + \ldots .
\label{eq:royalewithcheese}
\eeq
We are therefore led to consider ``soft" supersymmetry breaking. This
means that the effective Lagrangian of the MSSM can be written in the form
\beq
\lagr = \lagr_{\rm SUSY} + \lagr_{\rm soft},
\eeq
where $\lagr_{\rm SUSY}$ contains all of the gauge and Yukawa interactions
and preserves supersymmetry invariance, and $\lagr_{\rm soft}$ violates
supersymmetry but contains only mass terms and coupling parameters with {\it
positive} mass dimension. Without further justification, soft
supersymmetry breaking might seem like a rather arbitrary requirement.
Fortunately, we will see in section \ref{sec:origins} that theoretical
models for supersymmetry breaking do indeed yield effective Lagrangians
with just such terms for $\lagr_{\rm soft}$. If the largest mass scale
associated with the soft terms is denoted $m_{\rm soft}$, then the
additional non-supersymmetric corrections to the Higgs scalar squared mass
must vanish in the $m_{\rm soft}\rightarrow 0$ limit, so by dimensional
analysis they cannot be proportional to $\Lambda_{\rm UV}^2$. More
generally, these models maintain the cancellation of quadratically
divergent terms in the radiative corrections of all scalar masses, to all
orders in perturbation theory. The corrections also cannot go like $\Delta
m_H^2 \sim m_{\rm soft} \Lambda_{\rm UV}$, because in general the loop
momentum integrals always diverge either quadratically or logarithmically,
not linearly, as $\Lambda_{\rm UV} \rightarrow \infty$. So they must be of
the form
\beq
\Delta m_{H}^2 =
m_{\rm soft}^2
\left [{\lambda\over 16 \pi^2}\> {\rm ln}(\Lambda_{\rm UV}/m_{\rm soft})
+ \ldots \right ].
\label{softy}
\eeq
Here $\lambda$ is schematic for various dimensionless couplings, and the
ellipses stand both for terms that are independent of $\Lambda_{\rm UV}$
and for higher loop corrections (which depend on $\Lambda_{\rm UV}$
through powers of logarithms). 

Because the mass splittings between the known Standard Model particles and
their superpartners are just determined by the parameters $m_{\rm soft}$
appearing in $\lagr_{\rm soft}$, eq.~(\ref{softy}) tells us that the
superpartner masses should not be too huge.\footnote{This is 
obviously fuzzy and subjective. Nevertheless, such subjective criteria can be useful,
at least on a personal level, for making choices about 
what research directions to pursue, given finite time and money.} 
Otherwise, we would lose our
successful cure for the hierarchy problem, since the $m_{\rm soft}^2$
corrections to the Higgs scalar squared mass parameter would be
unnaturally large compared to the square of the electroweak breaking scale
of 174 GeV. The top and bottom squarks and the winos and bino give
especially large contributions to $\Delta m_{H_u}^2$ and $\Delta
m_{H_d}^2$, but the gluino mass and all the other squark and slepton
masses also feed in indirectly, through radiative corrections to the top
and bottom squark masses. Furthermore, in most viable models of
supersymmetry breaking that are not unduly contrived, the superpartner
masses do not differ from each other by more than about an order of
magnitude. Using $\Lambda_{\rm UV} \sim \MPlanck$ and $\lambda \sim 1$ in
eq.~(\ref{softy}), one estimates that $m_{\rm soft}$, and therefore the masses
of at least the lightest few superpartners, should probably not be much 
greater than the TeV scale, in order for the MSSM scalar potential 
to provide a Higgs VEV resulting in $m_W,m_Z$ = 80.4, 91.2 GeV without 
miraculous cancellations. While this is a fuzzy criterion, it
is the best reason for the continued optimism among many theorists that
supersymmetry will be discovered at the CERN
Large Hadron Collider, and can be studied at a future $e^+ e^-$ linear
collider with sufficiently high energy. 

However, it should be noted that the hierarchy problem was {\it not} the 
historical motivation for the development of supersymmetry in the early 
1970's. The supersymmetry algebra and supersymmetric field theories were 
originally concocted independently in various disguises 
\cite{RNS}-\cite{Volkov} 
bearing little resemblance to the MSSM. It is quite impressive that a 
theory developed for quite different reasons, including purely aesthetic 
ones, was later found to provide a solution for the hierarchy problem.

One might also wonder whether there is any good reason why all of the
superpartners of the Standard Model particles should be heavy enough to
have avoided discovery so far. There is. All of the particles in the MSSM
that have been found so far, except the 125 GeV Higgs boson, 
have something in common; they would
necessarily be massless in the absence of electroweak symmetry breaking.
In particular, the masses of the $W^\pm, Z^0$ bosons and all quarks and
leptons are equal to dimensionless coupling constants times the Higgs VEV
$\sim 174 $ GeV, while the photon and gluon are required to be massless by
electromagnetic and QCD gauge invariance. Conversely, all of the
undiscovered particles in the MSSM have exactly the opposite property;
each of them can have a Lagrangian mass term in the absence of electroweak
symmetry breaking. For the squarks, sleptons, and Higgs scalars this
follows from a general property of complex scalar fields that a mass term
$m^2 |\phi|^2$ is always allowed by all gauge symmetries. For the
higgsinos and gauginos, it follows from the fact that they are fermions in
a real representation of the gauge group. So, from the point of view of
the MSSM, the discovery of the top quark in 1995 marked a quite natural
milestone; the already-discovered particles are precisely those that had
to be light, based on the principle of electroweak gauge symmetry. There
is a single exception: it has long been known that at least
one neutral Higgs scalar boson had to be lighter
than about 135 GeV if the minimal version of supersymmetry is correct, for
reasons to be discussed in section \ref{subsec:MSSMspectrum.Higgs}. 
The 125 GeV Higgs boson discovered in 2012 is presumably this particle, and the fact 
that it was not much heavier can be counted as a successful prediction of supersymmetry. 

An important feature of the MSSM is that the superpartners listed in
Tables \ref{tab:chiral} and \ref{tab:gauge} 
are not necessarily the mass eigenstates of the theory.
This is because after electroweak symmetry breaking and supersymmetry
breaking effects are included, there can be mixing between the electroweak
gauginos and the higgsinos, and within the various sets of squarks and
sleptons and Higgs scalars that have the same electric charge. The lone
exception is the gluino, which is a color octet fermion and therefore does
not have the appropriate quantum numbers to mix with any other particle.
The masses and mixings of the superpartners are obviously of paramount
importance to experimentalists. It is perhaps slightly less obvious that
these phenomenological issues are all quite directly related to one
central question that is also the focus of much of the theoretical work in
supersymmetry: ``How is supersymmetry broken?" The reason for this is that
most of what we do not already know about the MSSM has to do with
$\lagr_{\rm soft}$. The structure of supersymmetric Lagrangians allows
little arbitrariness, as we will see in section \ref{sec:susylagr}.
In fact, all of the dimensionless couplings and all but one mass term in
the supersymmetric part of the MSSM Lagrangian correspond directly to
parameters in the ordinary Standard Model that have already been measured
by experiment. For example, we will find out that the supersymmetric
coupling of a gluino to a squark and a quark is determined by the QCD
coupling constant $\alpha_S$. In contrast, the supersymmetry-breaking part
of the Lagrangian contains many unknown parameters and, apparently, a
considerable amount of arbitrariness. Each of the mass splittings between
Standard Model particles and their superpartners correspond to terms in
the MSSM Lagrangian that are purely supersymmetry-breaking in their origin
and effect. These soft supersymmetry-breaking terms can also introduce a
large number of mixing angles and CP-violating phases not found in the
Standard Model. Fortunately, as we will see in section
\ref{subsec:mssm.hints}, there is already strong evidence that the
supersymmetry-breaking terms in the MSSM are actually not arbitrary at
all. Furthermore, the additional parameters will be measured and
constrained as the superpartners are detected. From a theoretical
perspective, the challenge is to explain all of these parameters with a
predictive model for supersymmetry breaking. 

The rest of the discussion is organized as follows. Section
\ref{sec:notations} provides a list of important notations. In section
\ref{sec:susylagr}, we will learn how to construct Lagrangians for
supersymmetric field theories, while section \ref{sec:superfields} 
reprises the same subject, but using the more elegant superspace formalism.
Soft supersymmetry-breaking couplings are
described in section \ref{sec:soft}. In section \ref{sec:mssm}, we will
apply the preceding general results to the special case of the MSSM,
introduce the concept of $R$-parity, and explore the importance of the
structure of the soft terms. Section \ref{sec:origins} outlines some
considerations for understanding the origin of supersymmetry breaking, and
the consequences of various proposals. In section \ref{sec:MSSMspectrum},
we will study the mass and mixing angle patterns of the new particles
predicted by the MSSM. Their decay modes are considered in section
\ref{sec:decays}, and some of the qualitative features of experimental
signals for supersymmetry are reviewed in section \ref{sec:signals}.
Section \ref{sec:variations} describes some sample variations on the
standard MSSM picture. The discussion will be lacking in historical
accuracy or perspective; the reader is
encouraged to consult the many outstanding books
\cite{WessBaggerbook}-\cite{Shifman:2012zz}, 
review articles
\cite{HaberKanereview}-\cite{Bertolini:2013via} 
and the reprint volume
\cite{reprints}, which contain a much more consistent guide to the
original literature. 


\section{Interlude: Notations and Conventions}\label{sec:notations}
\setcounter{equation}{0}
\setcounter{figure}{0}
\setcounter{table}{0}
\setcounter{footnote}{1}

This section specifies my notations and conventions.  Four-vector indices
are represented by letters from the middle of the Greek alphabet
$\mu,\nu,\rho,\dots= 0,1,2,3$. The contravariant four-vector position and
momentum of a particle are
\beq
x^\mu = (t,\, \vec{x}), \qquad\qquad p^\mu = (E,\, \vec{p}) ,
\eeq
while the four-vector derivative is
\beq
\partial_\mu = ({\partial / \partial t},\, \vec{\nabla}) .
\eeq
The spacetime metric is 
\beq
\eta_{\mu\nu} = 
{\mbox{diag}}(\BDplus 1, \BDminus 1, \BDminus 1, \BDminus 1), 
\eeq
so that $p^2 = \BDpos m^2$ for an on-shell particle of mass $m$.

It is overwhelmingly convenient to employ two-component Weyl spinor
notation for fermions, rather than four-component Dirac or Majorana
spinors. The Lagrangian of the Standard Model (and any supersymmetric
extension of it) violates parity; each Dirac fermion has left-handed and
right-handed parts with completely different electroweak gauge
interactions. If one used four-component spinor notation instead, then
there would be clumsy left- and right-handed projection operators
\beq
P_L = (1 - \gamma_5)/2, \qquad\qquad
P_R = (1 + \gamma_5)/2
\eeq
all over the place. The two-component Weyl fermion notation has the
advantage of treating fermionic degrees of freedom with different gauge
quantum numbers separately from the start, as Nature intended for us to
do. But an even better reason for using two-component notation here is
that in supersymmetric models the minimal building blocks of matter are
chiral supermultiplets, each of which contains a single two-component Weyl
fermion. 

Because two-component fermion notation may be unfamiliar to some
readers, I now specify my conventions by showing how they correspond to
the four-component spinor language. A four-component Dirac fermion
$\Psi_{\sst D}$ with mass $M$ is described by the Lagrangian
\beq
\lagr_{\rm Dirac}
\,=\,  i \overline\Psi_{\sst D} \gamma^\mu \partial_\mu \Psi_{\sst D}
-  M \overline \Psi_{\sst D} \Psi_{\sst D}\> .
\label{diraclag} 
\eeq
For our purposes it is convenient to use the specific representation of
the 4$\times$4 gamma matrices given in $2\times$2 blocks by
\beq
\gamma^\mu = \pmatrix{ 0 & \sigma^\mu \cr
                       \sigmabar^\mu & 0\cr},
\qquad\qquad
\gamma_5 = \pmatrix{-1 & 0\cr 0 & 1\cr},
\eeq
where
\beq
&&\sigma^0 = \sigmabar^0 = \pmatrix{1&0\cr 0&1\cr},\qquad
\,\>\>\>\>\>\>\>\>\sigma^1 = -\sigmabar^1 = \pmatrix{0&1\cr 1&0\cr}, 
\nonumber\\
&&\sigma^2 = -\sigmabar^2 = \pmatrix{ 0&-i\cr i&0\cr},\qquad
\>\>\>\sigma^3 = -\sigmabar^3 = \pmatrix{1&0\cr 0&-1\cr}
\> .
\label{pauli}
\eeq
In this representation, a four-component Dirac spinor is written in terms of 2
two-component, complex,\footnote{For obscure reasons, 
in much of the specialized literature on supersymmetry 
a bar ($\overline{\psi}$) has been used to represent the 
conjugate of a two-component spinor, rather than a dagger ($\psi^\dagger$). 
Here, I maintain consistency with essentially all other 
quantum field theory textbooks by using the dagger notation for the 
conjugate of a two-component spinor.}  
anti-commuting objects $\xi_\alpha$ and
$(\chi^\dagger)^{\dot{\alpha}} \equiv \chi^{\dagger\dot\alpha}$, with two
distinct types of spinor indices $\alpha=1,2$ and $\dot{\alpha}=1,2$: 
\beq
\Psi_{\sst D} =
\pmatrix{\xi_\alpha\cr {\chi^{\dagger\dot{\alpha}}}\cr} .
\label{psidirac}
\eeq
It follows that
\beq
\overline\Psi_{\sst D}  =
\Psi_{\sst D}^\dagger \pmatrix{0 & 1\cr 1 & 0\cr} =
\pmatrix{\chi^\alpha &
                           \xi^\dagger_{\dot{\alpha}}\cr }
\> .
\label{psid}
\eeq
Undotted (dotted) indices from the beginning of the Greek alphabet are
used for the first (last) two components of a Dirac spinor.  The field
$\xi$ is called a ``left-handed Weyl spinor" and $\chi^\dagger$ is a
``right-handed Weyl spinor". The names fit, because
\beq
P_L \Psi_{\sst D} = \pmatrix{\xi_\alpha \cr 0\cr},\qquad\qquad
P_R \Psi_{\sst D} = \pmatrix{0\cr \chi^{\dagger\dot{\alpha}}\cr}
\> .
\eeq
The Hermitian conjugate of any left-handed Weyl spinor is a right-handed
Weyl spinor: 
\beq
\psi^{\dagger}_{\dot{\alpha}}
\equiv 
(\psi_\alpha)^\dagger = (\psi^\dagger)_{\dot{\alpha}}
\, ,
\eeq
and vice versa: 
\beq
( \psi^{\dagger\dot{\alpha}} )^\dagger =
\psi^\alpha. 
\eeq
Therefore, any particular fermionic degrees of freedom can be described
equally well using a left-handed Weyl spinor (with an undotted index) or
by a right-handed one (with a dotted index). By convention, all names of
fermion fields are chosen so that left-handed Weyl spinors do not carry
daggers and right-handed Weyl spinors do carry daggers, as in
eq.~(\ref{psidirac}).

The heights of the dotted and undotted spinor indices are important; for
example, comparing eqs.~(\ref{diraclag})-(\ref{psid}), we observe that the
matrices $(\sigma^\mu)_{\alpha\dot{\alpha}}$ and
$(\sigmabar^\mu)^{\dot{\alpha}\alpha}$ defined by eq.~(\ref{pauli}) carry
indices with the heights as indicated. The spinor indices are raised and
lowered using the antisymmetric symbol 
\beq
\epsilon^{12} = -\epsilon^{21} =
\epsilon_{21} = -\epsilon_{12} = 1, \qquad\qquad \epsilon_{11} = \epsilon_{22} =
\epsilon^{11} = \epsilon^{22} = 0,
\label{eq:defepstwo}
\eeq 
according to
\beq
\xi_\alpha = \epsilon_{\alpha\beta}
\xi^\beta,\qquad\qquad\!\!\!\!\!\!\!\!\!
\xi^\alpha = \epsilon^{\alpha\beta}
\xi_\beta,\qquad\qquad\!\!\!\!\!\!\!\!\!
\chi^\dagger_{\dot{\alpha}} = \epsilon_{\dot{\alpha}\dot{\beta}}
\chi^{\dagger\dot{\beta}},\qquad\qquad\!\!\!\!\!\!\!\!\!
\chi^{\dagger\dot{\alpha}} = \epsilon^{\dot{\alpha}\dot{\beta}}
\chi^\dagger_{\dot{\beta}}\>.\qquad{}
\eeq
This is consistent since
$\epsilon_{\alpha\beta} \epsilon^{\beta\gamma} =
\epsilon^{\gamma\beta}\epsilon_{\beta\alpha} = \delta_\alpha^\gamma$
and
$\epsilon_{\dot{\alpha}\dot{\beta}} \epsilon^{\dot{\beta}\dot{\gamma}} =
\epsilon^{\dot{\gamma}\dot{\beta}}\epsilon_{\dot{\beta}\dot{\alpha}} =
\delta_{\dot{\alpha}}^{\dot{\gamma}}$.

\vspace{0.1cm}

As a convention, repeated spinor indices contracted like 
\beq
{}^\alpha\hspace{0.03cm}{}_\alpha \qquad\quad \mbox{or} 
\qquad\quad 
{}_{\dot{\alpha}}\hspace{0.03cm}{}^{\dot{\alpha}}\, 
\eeq 
can be suppressed.
In particular,
\beq
\xi\chi \equiv \xi^\alpha\chi_\alpha = \xi^\alpha \epsilon_{\alpha\beta}
\chi^\beta = -\chi^\beta \epsilon_{\alpha\beta} \xi^\alpha =
\chi^\beta \epsilon_{\beta\alpha} \xi^\alpha = \chi^\beta \xi_\beta \equiv
\chi\xi
\label{xichi}
\eeq
with, conveniently, no minus sign in the end. [A minus sign appeared in
eq.~(\ref{xichi}) from exchanging the order of anti-commuting spinors, but
it disappeared due to the antisymmetry of the $\epsilon$ symbol.]
Likewise, $\xi^\dagger \chi^\dagger$ and $\chi^\dagger \xi^\dagger $ are
equivalent abbreviations for $\chi^\dagger_{\dot{\alpha}} \xi^{\dagger
\dot{\alpha}} = \xi^\dagger_{\dot{\alpha}} \chi^{\dagger \dot{\alpha}}$,
and in fact this is the complex conjugate of $\xi\chi$: 
\beq
(\xi\chi)^* = \chi^\dagger \xi^\dagger = \xi^\dagger \chi^\dagger.
\eeq
In a similar way, one can check that
\beq
(\chi^\dagger \sigmabar^\mu \xi)^* \,=\, 
\xi^\dagger \sigmabar^\mu \chi \,=\, -\chi \sigma^\mu \xi^\dagger
\,=\,
 -(\xi\sigma^\mu\chi^\dagger)^*
\label{yetanotheridentity}
\eeq
stands for $\xi^\dagger_{\dot{\alpha}}(\sigmabar^\mu)^{\dot{\alpha}\alpha}
\chi_\alpha$, etc. Note that when taking the complex conjugate of a spinor bilinear,
one reverses the order. The spinors here are assumed to be
classical fields; for quantum fields the complex conjugation operation in these 
equations would be replaced by Hermitian conjugation in the Hilbert
space operator sense. 

Some other identities that will be useful below include:
\beq
(\chi^\dagger \sigmabar^\nu \sigma^\mu \xi^\dagger)^* =
\xi\sigma^\mu \sigmabar^\nu \chi  \>=\>
\chi \sigma^\nu \sigmabar^\mu \xi \>=\>
(\xi^\dagger \sigmabar^\mu \sigma^\nu \chi^\dagger)^*,
\label{eq:dei}
\eeq
and the Fierz rearrangement identity:
\beq
\chi_{\alpha}\> (\xi\eta) &=&
- \xi_{\alpha}\> (\eta\chi) - \eta_\alpha\> (\chi\xi) ,
\label{fierce}
\eeq
and the reduction identities
\beq
&&
\sigma_{\alpha\dot{\alpha}}^\mu \,
\sigmabar_\mu^{\dot{\beta}\beta} 
\,\>=\>\, 
\BDpos 2 \delta_\alpha^\beta
\delta_{\dot{\alpha}}^{\dot{\beta}} ,
\label{eq:feif}
\\
&&
\sigma_{\alpha\dot{\alpha}}^\mu \,
\sigma_{\mu\beta\dot{\beta}} 
\,\>=\>\, 
\BDpos 2 \epsilon_{\alpha\beta}
\epsilon_{\dot{\alpha}\dot{\beta}} ,
\label{eq:rickettsisok}
\\
&&
\sigmabar^{\mu\dot\alpha\alpha} \,
\sigmabar_\mu^{\dot\beta\beta} 
\,\>=\>\, 
\BDpos 2 \epsilon^{\alpha\beta}
\epsilon^{\dot{\alpha}\dot{\beta}} ,
\label{eq:pagehousesux}
\\
&&
\bigl[ \sigma^\mu \sigmabar^\nu + \sigma^\nu \sigmabar^\mu
\bigr ]_\alpha{}^\beta  
\,=\,
\BDpos 2 \eta^{\mu\nu} \delta_\alpha^\beta 
,
\label{pauliidentA}
\\
&&
\bigl[ \sigmabar^\mu \sigma^\nu + \sigmabar^\nu \sigma^\mu \bigr
]^{\dot{\beta}}{}_{\dot{\alpha}}
\,=\,
\BDpos 2 \eta^{\mu\nu} \delta_{\dot{\alpha}}^{\dot{\beta}} ,
\label{pauliidentB}
\\
&&\sigmabar^\mu \sigma^\nu \sigmabar^\rho \>=\>
\BDpos \eta^{\mu\nu} \sigmabar^\rho 
\BDplus \eta^{\nu\rho} \sigmabar^\mu 
\BDminus \eta^{\mu\rho} \sigmabar^\nu 
\BDminus i \epsilon^{\mu\nu\rho\kappa} \sigmabar_\kappa ,
\label{eq:lloydhouserules}
\\
&&\sigma^\mu \sigmabar^\nu \sigma^\rho \>=\>
\BDpos \eta^{\mu\nu} \sigma^\rho 
\BDplus \eta^{\nu\rho} \sigma^\mu 
\BDminus \eta^{\mu\rho} \sigma^\nu 
\BDplus i \epsilon^{\mu\nu\rho\kappa} \sigma_\kappa ,
\label{eq:pagehouseisOK}
\eeq
where $\epsilon^{\mu\nu\rho\kappa}$ is the totally antisymmetric tensor
with $\epsilon^{0123}=+1$. 

With these conventions, the Dirac Lagrangian eq.~(\ref{diraclag}) can now
be rewritten: 
\beq
\lagr_{\rm Dirac}
\, = \,  i \xi^\dagger \sigmabar^\mu \partial_\mu \xi 
       + i \chi^\dagger \sigmabar^\mu \partial_\mu \chi 
       - M (\xi\chi + \xi^\dagger \chi^\dagger)
\eeq
where we have dropped a total derivative piece 
$-i\partial_\mu(\chi^\dagger \sigmabar^\mu \chi)$, 
which does not affect the action. 

A four-component Majorana spinor can be obtained from the Dirac spinor of
eq.~(\ref{psid}) by imposing the constraint $\chi = \xi$, so that
\beq
\Psi_{\rm M} = \pmatrix{\xi_\alpha \cr \xi^{\dagger\dot{\alpha}}\cr},
\qquad\qquad
\overline\Psi_{\rm M}
= \pmatrix{ \xi^\alpha &  \xi^\dagger_{\dot{\alpha}} \cr}.
\eeq
The four-component spinor form of the
Lagrangian for a Majorana fermion with mass $M$,
\beq
\lagr_{\rm Majorana} \,=\,  
{i\over 2}\overline\Psi_{\rm M} \gamma^\mu \partial_\mu \Psi_{\rm M}
- {1\over 2} M \overline\Psi_{\rm M} \Psi_{\rm M}
\eeq
can therefore be rewritten as
\beq
\lagr_{\rm Majorana} \,=\, 
i\xi^\dagger \sigmabar^\mu\partial_\mu \xi -
{1\over 2} M(\xi\xi + \xi^\dagger\xi^\dagger)
\eeq
in the more economical two-component Weyl spinor representation. Note that
even though $\xi_\alpha$ is anti-commuting, $\xi\xi$ and its complex
conjugate $\xi^\dagger\xi^\dagger$ do not vanish, because of the
suppressed $\epsilon$ symbol, see eq.~(\ref{xichi}). Explicitly, $\xi\xi =
\epsilon^{\alpha\beta} \xi_\beta \xi_\alpha = \xi_2\xi_1 - \xi_1 \xi_2 = 2
\xi_2 \xi_1$. 

More generally, any theory involving spin-1/2 fermions can always be
written in terms of a collection of left-handed Weyl spinors $\psi_i$
with
\beq
\lagr \,=\, i \psi^{\dagger i} \sigmabar^\mu \partial_\mu\psi_i
+ \ldots
\eeq
where the ellipses represent possible mass terms, gauge interactions, and
Yukawa interactions with scalar fields. Here the index $i$ runs over the
appropriate gauge and flavor indices of the fermions; it is raised or
lowered by Hermitian conjugation. Gauge interactions are obtained
by promoting the ordinary derivative to a gauge-covariant derivative: 
\beq
\lagr \,=\, i \psi^{\dagger i} \sigmabar^\mu \nabla_\mu\psi_i
+ \ldots
\eeq
with
\beq
\nabla_\mu\psi_i \,=\, \partial_\mu\psi_i \BDplus i g_a A^a_\mu {T^a_i}^j \psi_j,
\eeq
where $g_a$ is the gauge coupling corresponding to the Hermitian 
Lie algebra generator matrix $T^a$ with vector field $A^a_\mu$.

There is a different $\psi_i$ for the left-handed piece and for the
hermitian conjugate of the right-handed piece of a Dirac fermion. 
Given any expression involving bilinears of four-component
spinors
\beq
\Psi_i = \pmatrix{ \xi_i\cr\chi_i^\dagger\cr},
\eeq
labeled by a flavor or gauge-representation index $i$, one can
translate into two-component Weyl spinor language (or vice versa) using
the dictionary: 
\beq
&&\overline\Psi_i P_L \Psi_j = \chi_i\xi_j,\qquad\qquad\qquad
\overline\Psi_i P_R \Psi_j = \xi_i^\dagger \chi_j^\dagger,\qquad\>{}\\
&&\overline\Psi_i \gamma^\mu P_L \Psi_j = \xi_i^\dagger \sigmabar^\mu 
\xi_j
,\qquad\qquad
\overline\Psi_i \gamma^\mu P_R \Psi_j = \chi_i \sigma^\mu \chi^\dagger_j
\qquad\>\>\>{}
\eeq
etc. 

Let us now see how the Standard Model quarks and leptons are described in
this notation. The complete list of left-handed Weyl spinors can be given
names corresponding to the chiral supermultiplets in Table \ref{tab:chiral}: 
\beq
Q_i & = &
\pmatrix{u\cr d},\>\, 
\pmatrix{c\cr s},\>\, 
\pmatrix{t\cr b},
\\
\sbar u_i & = &
\>\>\sbar u ,\>\,\sbar c,\>\, \sbar t,
\\
\sbar d_i & = &
\>\>\sbar d ,\>\,\sbar s,\>\, \sbar b
\\
L_i & = &
\pmatrix{\nu_e\cr e},\>\, 
\pmatrix{\nu_\mu\cr \mu},\>\, 
\pmatrix{\nu_\tau\cr \tau},\>\, 
\\
\sbar e_i & = &
\>\>\sbar e ,\>\,\sbar \mu,\>\, \sbar \tau    .
\eeq
Here $i=1,2,3$ is a family index. The bars on these fields are part of the
names of the fields, and do {\it not} denote any kind of conjugation.
Rather, the unbarred fields are the left-handed pieces of a Dirac spinor,
while the barred fields are the names given to the conjugates of the
right-handed piece of a Dirac spinor. For example, $e$ is the same thing
as $e_L$ in Table \ref{tab:chiral}, 
and $\sbar e$ is the same as $e_R^\dagger$. Together
they form a Dirac spinor: 
\beq
\pmatrix{e\cr {\sbar e}^\dagger} \equiv \pmatrix{e_L \cr e_R} ,
\label{espinor}
\eeq
and similarly for all of the other quark and charged lepton Dirac
spinors. (The neutrinos of the Standard Model are not part of a Dirac
spinor, at least in the approximation that they are massless.) The fields
$Q_i$ and $L_i$ are weak isodoublets, which always go together when one is
constructing interactions invariant under the full Standard Model gauge
group $SU(3)_C\times SU(2)_L \times U(1)_Y$. Suppressing all color and
weak isospin indices, the kinetic and gauge part of the Standard Model
fermion Lagrangian density is then
\beq
\lagr \,=\,
 iQ^{\dagger i}\sigmabar^\mu \nabla_\mu Q_i
+ i\sbar u^{\dagger }_i\sigmabar^\mu \nabla_\mu \sbar u^i
+ i\sbar d^{\dagger }_i\sigmabar^\mu \nabla_\mu \sbar d^i
+ i L^{\dagger i}\sigmabar^\mu \nabla_\mu L_i
+ i \sbar e^{\dagger }_i\sigmabar^\mu \nabla_\mu \sbar e^i
\qquad{}
\eeq
with the family index $i$ summed over, and $\nabla_\mu$ the
appropriate Standard Model covariant derivative. For example,
\beq
\nabla_\mu \pmatrix{ \nu_e \cr e} &=&
\left [ \partial_\mu \BDplus i g W^a_\mu (\tau^a/2) 
                     \BDplus i g' Y_L B_\mu \right ]
\pmatrix{ \nu_e \cr e}
\\
\nabla_\mu \overline e &=& \left [ \partial_\mu 
\BDplus i g' Y_{\sbar e} B_\mu \right ] \sbar e
\eeq
with $\tau^a$ ($a=1,2,3$) equal to the Pauli matrices, $Y_L = -1/2$ and 
$Y_{\sbar e} = +1$. The gauge eigenstate weak bosons are related to
the mass eigenstates by 
\beq
W^\pm_\mu &=& (W_\mu^1 \mp i W_\mu^2)/\sqrt{2} ,
\\
\pmatrix{Z_\mu \cr A_\mu} &=& 
\pmatrix{\cos\theta_W & - \sin\theta_W \cr
         \sin\theta_W & \cos\theta_W \cr }
\pmatrix{W^3_\mu \cr B_\mu} .
\eeq
Similar expressions hold for the other quark and lepton gauge eigenstates,
with $Y_Q = 1/6$, $Y_{\sbar u} = -2/3$, and $Y_{\sbar d} = 1/3$. The
quarks also have a term in the covariant derivative corresponding to gluon
interactions proportional to $g_3$ (with $\alpha_S = g_3^2/4 \pi$) with
generators $T^a = \lambda^a/2$ for $Q$, and in the complex conjugate
representation $T^a = -(\lambda^a)^*/2$ for $\sbar u$ and $\sbar d$, where
$\lambda^a$ are the Gell-Mann matrices. 

For a more detailed discussion of the two-component fermion notation, including many worked examples in which it is employed to calculate cross-sections and decay rates in the Standard Model and in supersymmetry, see ref.~\cite{DHM}, or a more concise
account in \cite{Martin:2012us}.

\section{Supersymmetric Lagrangians}\label{sec:susylagr}
\renewcommand{\theequation}{\arabic{section}.\arabic{subsection}.\arabic{equation}}
\setcounter{equation}{0}
\setcounter{figure}{0}
\setcounter{table}{0}
\setcounter{footnote}{1}

In this section we will describe the construction of supersymmetric
Lagrangians. The goal is a recipe that will allow us
to write down the allowed interactions and mass terms of a general
supersymmetric theory, so that later we can apply the results to the
special case of the MSSM. In this section, we will not use the superfield
\cite{superfields} language, which is more elegant and efficient 
for many purposes, but requires a more specialized machinery and
might seem rather cabalistic at first. 
Section \ref{sec:superfields} will provide the 
superfield version of the same material.
We begin by considering the simplest example of a supersymmetric theory in
four dimensions. 

\subsection{The simplest supersymmetric model: a free chiral
supermultiplet}\label{subsec:susylagr.freeWZ}
\setcounter{footnote}{1}
\renewcommand{\theequation}{\arabic{section}.\arabic{subsection}.\arabic{equation}}
\setcounter{equation}{0}

The minimum fermion content of a field theory in four dimensions consists
of a single left-handed two-component Weyl fermion $\psi$. Since this is
an intrinsically complex object, it seems sensible to choose as its
superpartner a complex scalar field $\phi$. The simplest action we can
write down for these fields just consists of kinetic energy terms for
each: 
\beq
&&S = \int d^4x\>
\left (\lagr_{\rm scalar} + \lagr_{\rm fermion}\right ) ,
\label{Lwz} \\
&&\lagr_{\rm scalar} = 
  \BDpos \partial^\mu \phi^* \partial_\mu \phi ,
\qquad\qquad
\lagr_{\rm fermion} = 
  i \psi^\dagger \sigmabar^\mu \partial_\mu \psi .
\eeq
This is called the massless, non-interacting {\it Wess-Zumino model}
\cite{WessZumino}, and it corresponds to a single chiral supermultiplet as
discussed in the Introduction. 

A supersymmetry transformation should turn the scalar boson field $\phi$
into something involving the fermion field $\psi_\alpha$. The simplest
possibility for the transformation of the scalar field is
\beq
\deltaeps \phi = \epsilon \psi,\qquad\qquad
\deltaeps \phi^* = \epsilon^\dagger \psi^\dagger ,
\label{phitrans}
\eeq
where $\epsilon^\alpha$ is an infinitesimal, anti-commuting, two-component
Weyl fermion object that parameterizes the supersymmetry transformation. Until
section \ref{subsec:origins.gravitino}, we will be discussing global
supersymmetry, which means that $\epsilon^\alpha$ is a constant,
satisfying $\partial_\mu \epsilon^\alpha=0$. Since $\psi$ has dimensions
of [mass]$^{3/2}$ and $\phi$ has dimensions of [mass], it must be that
$\epsilon$ has dimensions of [mass]$^{-1/2}$. Using eq.~(\ref{phitrans}),
we find that the scalar part of the Lagrangian transforms as
\beq
\deltaeps
\lagr_{\rm scalar} \,=\, 
\BDpos \epsilon \partial^\mu \psi \> \partial_\mu \phi^*
\BDplus \epsilon^\dagger \partial^\mu \psi^\dagger \> \partial_\mu \phi .
\label{Lphitrans}
\eeq
We would like for this to be canceled by $\deltaeps\lagr_{\rm fermion}$,
at least up to a total derivative, so that the action will be invariant
under the supersymmetry transformation. Comparing eq.~(\ref{Lphitrans})
with $\lagr_{\rm fermion}$, we see that for this to have any chance of
happening, $\deltaeps \psi$ should be linear in $\epsilon^\dagger$ and in
$\phi$, and should contain one spacetime derivative. Up to a
multiplicative constant, there is only one possibility to try:
\beq
\deltaeps\psi_\alpha
\,=\,
- i (\sigma^\mu \epsilon^\dagger)_\alpha\> \partial_\mu \phi,
\qquad\qquad
\deltaeps\psi^\dagger_{\dot{\alpha}}
\,=\,
 i (\epsilon\sigma^\mu)_{\dot{\alpha}}\>   \partial_\mu \phi^* .
\label{psitrans}
\eeq
With this guess, one immediately obtains
\beq
\deltaeps \lagr_{\rm fermion} =
-\epsilon \sigma^\mu \sigmabar^\nu \partial_\nu \psi\> \partial_\mu \phi^*
+\psi^\dagger \sigmabar^\nu \sigma^\mu \epsilon^\dagger \>
\partial_\mu \partial_\nu \phi
\> .
\label{preLpsitrans}
\eeq
This can be simplified by employing the
Pauli matrix identities eqs.~(\ref{pauliidentA}), (\ref{pauliidentB}) 
and using the fact that partial derivatives commute
$(\partial_\mu\partial_\nu = \partial_\nu\partial_\mu)$. Equation
(\ref{preLpsitrans}) then becomes
\beq
\deltaeps \lagr_{\rm fermion} & = &
\BDneg \epsilon\partial^\mu\psi\> \partial_\mu\phi^*
\BDminus \epsilon^\dagger \partial^\mu\psi^\dagger\> \partial_\mu \phi
\nonumber\\ && 
- \partial_\mu \left (
\epsilon \sigma^\nu \sigmabar^\mu \psi \> \partial_\nu \phi^* 
\BDminus \epsilon \psi\> \partial^\mu \phi^*
\BDminus \epsilon^\dagger \psi^\dagger \> \partial^\mu \phi \right ).
\label{Lpsitrans}
\eeq
The first two terms here just cancel against $\deltaeps\lagr_{\rm
scalar}$, while the remaining contribution is a total derivative. So we
arrive at
\beq
\deltaeps S =
\int d^4x \>\>\, (\deltaeps \lagr_{\rm scalar} + \deltaeps
\lagr_{\rm fermion})
= 0,\>
\label{invar}
\eeq
justifying our guess of the numerical multiplicative factor made in
eq.~(\ref{psitrans}). 

We are not quite finished in showing that the theory described by
eq.~(\ref{Lwz}) is supersymmetric. We must also show that the
supersymmetry algebra closes; in other words, that the commutator of two
supersymmetry transformations parameterized by two different spinors
$\epsilon_1$ and $\epsilon_2$ is another symmetry of the theory. Using
eq.~(\ref{psitrans}) in eq.~(\ref{phitrans}), one finds
\beq
(\delta_{\epsilon_2} \delta_{\epsilon_1} -
\delta_{\epsilon_1} \delta_{\epsilon_2}) \phi 
\,\equiv\,
\delta_{\epsilon_2} (\delta_{\epsilon_1} \phi) -
\delta_{\epsilon_1} (\delta_{\epsilon_2} \phi) 
\,=\,
i (- \epsilon_1 \sigma^\mu \epsilon_2^\dagger 
   + \epsilon_2 \sigma^\mu \epsilon_1^\dagger)\> \partial_\mu \phi
. \label{coophi}
\eeq
This is a remarkable result; in words, we have found that the commutator
of two supersymmetry transformations gives us back the derivative of the
original field. In the Heisenberg picture of quantum mechanics
$\BDpos i\partial_\mu$ corresponds to the generator of
spacetime translations $P_\mu$, so eq.~(\ref{coophi}) implies the form of the
supersymmetry algebra that was foreshadowed in eq.~(\ref{susyalgone}) of
the Introduction. (We will make this statement more explicit before the
end of this section, and prove it again a different way in section \ref{sec:superfields}.)

All of this will be for nothing if we do not find the same result for the
fermion $\psi$. Using eq.~(\ref{phitrans}) in eq.~(\ref{psitrans}), we get
\beq
(\delta_{\epsilon_2} \delta_{\epsilon_1} -
\delta_{\epsilon_1} \delta_{\epsilon_2}) \psi_\alpha 
\,=\,
-i(\sigma^\mu\epsilon_1^\dagger)_\alpha\>  \epsilon_2 \partial_\mu\psi  
+i(\sigma^\mu\epsilon_2^\dagger)_\alpha \> \epsilon_1 \partial_\mu\psi
{}.
\eeq
This can be put into a more useful form by applying the Fierz identity
eq.~(\ref{fierce})
with $\chi = \sigma^\mu \epsilon_1^\dagger$,
$\xi = \epsilon_{2} $, $\eta = \partial_\mu \psi$, and again with
$\chi = \sigma^\mu \epsilon_2^\dagger$,
$\xi = \epsilon_{1} $, $\eta = \partial_\mu \psi$, followed
in each case by an application
of the identity eq.~(\ref{yetanotheridentity}). The result is
\beq
(\delta_{\epsilon_2} \delta_{\epsilon_1} -
\delta_{\epsilon_1} \delta_{\epsilon_2}) \psi_\alpha 
& = &
 i (-\epsilon_1 \sigma^\mu \epsilon_2^\dagger 
    +\epsilon_2 \sigma^\mu \epsilon_1^\dagger) \> \partial_\mu \psi_\alpha
+ i \epsilon_{1\alpha} \> \epsilon_2^\dagger \sigmabar^\mu \partial_\mu\psi
- i \epsilon_{2\alpha} \> \epsilon_1^\dagger \sigmabar^\mu 
\partial_\mu\psi
{}.\phantom{xxxxx}
\label{commpsi}
\eeq
The last two terms in (\ref{commpsi}) vanish on-shell; that is, if the
equation of motion $\sigmabar^\mu\partial_\mu \psi = 0$ following from the
action is enforced. The remaining piece is exactly the same spacetime
translation that we found for the scalar field.

The fact that the supersymmetry algebra only closes on-shell (when the
classical equations of motion are satisfied) might be somewhat worrisome,
since we would like the symmetry to hold even quantum mechanically. This
can be fixed by a trick. We invent a new complex scalar field $F$, which
does not have a kinetic term. Such fields are called {\it auxiliary}, and
they are really just book-keeping devices that allow the symmetry algebra
to close off-shell. The Lagrangian density for $F$ and its complex
conjugate is simply
\beq
\lagr_{\rm auxiliary} = F^* F \> .
\label{lagraux}
\eeq
The dimensions of $F$ are [mass]$^2$, unlike an ordinary scalar field,
which has dimensions of [mass]. Equation (\ref{lagraux}) implies the
not-very-exciting equations of motion $F=F^*=0$. However, we can use the
auxiliary fields to our advantage by including them in the supersymmetry
transformation rules. In view of eq.~(\ref{commpsi}), a plausible thing to
do is to make $F$ transform into a multiple of the equation of motion for
$\psi$: 
\beq
\deltaeps F = - i \epsilon^\dagger \sigmabar^\mu \partial_\mu \psi,
\qquad\qquad
\deltaeps F^* = i\partial_\mu \psi^\dagger \sigmabar^\mu \epsilon .
\label{Ftrans}
\eeq
Once again we have chosen the overall factor on the right-hand sides by
virtue of foresight. Now the auxiliary part of the Lagrangian density
transforms as
\beq
\delta \lagr_{\rm auxiliary} = 
-i \epsilon^\dagger \sigmabar^\mu \partial_\mu \psi \> F^* 
+i \partial_\mu \psi^\dagger \sigmabar^\mu \epsilon \> F ,
\eeq
which vanishes on-shell, but not for arbitrary off-shell field
configurations. Now, by adding an extra term to
the transformation law for $\psi$ and $\psi^\dagger$:
\beq
\delta \psi_\alpha =
- i (\sigma^\mu \epsilon^\dagger)_{\alpha}\> \partial_\mu\phi 
+ \epsilon_\alpha F,
\qquad\>\>
\delta \psi_{\dot{\alpha}}^\dagger =
i (\epsilon\sigma^\mu)_{\dot{\alpha}}\> \partial_\mu \phi^* 
+ \epsilon^\dagger_{\dot{\alpha}} F^* ,
\label{fermiontrans}
\eeq
one obtains an additional contribution to $\deltaeps \lagr_{\rm fermion}$,
which just cancels with $\deltaeps \lagr_{\rm auxiliary}$, up to a total
derivative term. So our ``modified" theory with $\lagr = \lagr_{\rm
scalar} +\lagr_{\rm fermion} + \lagr_{\rm auxiliary}$ is still invariant
under supersymmetry transformations. Proceeding as before, one now obtains
for each of the fields $X=\phi,\phi^*,\psi,\psi^\dagger,F,F^*$,
\beq
(\delta_{\epsilon_2} \delta_{\epsilon_1} -
\delta_{\epsilon_1} \delta_{\epsilon_2}) X &=&
i (-\epsilon_1 \sigma^\mu \epsilon_2^\dagger +
   \epsilon_2 \sigma^\mu \epsilon_1^\dagger) \> \partial_\mu X
\label{anytrans}
\eeq
using eqs.~(\ref{phitrans}), (\ref{Ftrans}), and (\ref{fermiontrans}), but
now without resorting to any equations of motion. So we have
succeeded in showing that supersymmetry is a valid symmetry of the
Lagrangian off-shell.

In retrospect, one can see why we needed to introduce the auxiliary field
$F$ in order to get the supersymmetry algebra to work off-shell. On-shell,
the complex scalar field $\phi$ has two real propagating degrees of
freedom, matching the two spin polarization states of $\psi$. Off-shell,
however, the Weyl fermion $\psi$ is a complex two-component object, so it
has four real degrees of freedom. (Going on-shell eliminates half of the
propagating degrees of freedom for $\psi$, because the Lagrangian is
linear in time derivatives, so that the canonical momenta can be
re-expressed in terms of the configuration variables without time
derivatives and are not independent phase space coordinates.) To make the
numbers of bosonic and fermionic degrees of freedom match off-shell as
well as on-shell, we had to introduce two more real scalar degrees of
freedom in the complex field $F$, which are eliminated when one goes
on-shell. This counting is summarized in Table \ref{table:WZdofcounting}. 
The auxiliary field formulation is especially useful when discussing
spontaneous supersymmetry breaking, as we will see in section
\ref{sec:origins}. 
\renewcommand{\arraystretch}{1.4}
\begin{table}[tb]
\begin{center}
\begin{tabular}{|c|c|c|c|}
\hline
 & $\phi$ & $\psi$ & $F$ \\
\hline
on-shell ($n_B=n_F=2$) & 2 & 2 & 0 \\
\hline
off-shell ($n_B=n_F=4$) & 2 & 4 & 2 \\
\hline
\end{tabular}
\caption{Counting of real degrees of freedom in the
Wess-Zumino model.\label{table:WZdofcounting}}
\vspace{-0.45cm}
\end{center}
\end{table}

Invariance of the action under a continuous symmetry transformation always implies
the existence of a conserved current, and supersymmetry is no exception.
The {\it supercurrent} $J^\mu_\alpha$ is an anti-commuting four-vector. It
also carries a spinor index, as befits the current associated with a
symmetry with fermionic generators \cite{ref:supercurrent}. By the usual
Noether procedure, one finds for the supercurrent (and its hermitian
conjugate) in terms of the variations of the fields
$X=\phi,\phi^*,\psi,\psi^\dagger,F,F^*$: 
\beq
\epsilon J^\mu + \epsilon^\dagger J^{\dagger\mu}
&\equiv & 
\sum_X \, \delta X\>{\delta\lagr\over \delta(\partial_\mu X)} 
- K^\mu ,
\label{Noether}
\eeq
where $K^\mu$ is an object whose divergence is the variation of the
Lagrangian density under the supersymmetry transformation, $\delta \lagr =
\partial_\mu K^\mu$. Note that $K^\mu$ is not unique; one can always
replace $K^\mu$ by $K^\mu + k^\mu$, where $k^\mu$ is any vector satisfying
$\partial_\mu k^\mu=0$, for example $k^\mu = \partial^\mu \partial_\nu
a^\nu - \partial_\nu\partial^\nu a^\mu$ for any four-vector $a^\mu$. 
A little work reveals that, up to
the ambiguity just mentioned,
\beq
J^\mu_\alpha = (\sigma^\nu\sigmabar^\mu\psi)_\alpha\> \partial_\nu \phi^*
, \qquad\qquad
J^{\dagger\mu}_{\dot{\alpha}}
=  (\psi^\dagger \sigmabar^\mu \sigma^\nu)_{\dot{\alpha}}
\> \partial_\nu \phi .
\label{WZsupercurrent}
\eeq
The supercurrent and its hermitian conjugate are separately conserved:
\beq
\partial_\mu J^\mu_\alpha = 0,\qquad\qquad
\partial_\mu J^{\dagger\mu}_{\dot{\alpha}} = 0 ,
\eeq
as can be verified by use of the equations of motion. From these currents
one constructs the conserved charges
\beq
Q_\alpha = {\sqrt{2}}\int d^3 \vec{x}\> J^0_\alpha,\qquad\qquad
Q^\dagger_{\dot{\alpha}} = {\sqrt{2}} \int d^3\vec{x} \> 
J^{\dagger 0}_{\dot{\alpha}} ,
\eeq
which are the generators of supersymmetry transformations. (The factor of
$\sqrt{2}$ normalization is included to agree with an arbitrary historical
convention.) As quantum mechanical operators, they satisfy
\beq
\left [ \epsilon Q + \epsilon^\dagger Q^\dagger , X \right ]
= -i{\sqrt{2}} \> \delta X
\label{interpolistheworstbandonearth}
\eeq
for any field $X$, up to terms that vanish on-shell. This
can be verified explicitly by using the canonical equal-time
commutation and anticommutation relations
\beq
&&[ \phi(\vec{x}), \pi(\vec{y}) ] \,=\,
[ \phi^*(\vec{x}), \pi^*(\vec{y}) ] \,=\, i \delta^{(3)}(\vec{x}-\vec{y}
\hspace{0.25mm}) , \qquad\>\>{}\\
&&
\{ 
\psi_\alpha (\vec{x}), 
\psi^\dagger_{\dot{\alpha}} (\vec{y}) \} \,=\,
(\sigma^0)_{{\alpha}\dot{\alpha}}\,\delta^{(3)}(\vec{x}-\vec{y}
\hspace{0.25mm} ), \qquad{}
\eeq
which follow from the free field theory Lagrangian eq.~(\ref{Lwz}). Here $\pi =
\partial_0 \phi^*$ and $\pi^* = \partial_0 \phi$ are the momenta conjugate
to $\phi$ and $\phi^*$ respectively.

Using eq.~(\ref{interpolistheworstbandonearth}), the content of
eq.~(\ref{anytrans}) can be expressed in terms of canonical commutators as
\beq
\Bigl [
\epsilon_2 Q + \epsilon_2^\dagger Q^\dagger,\,
\bigl [
\epsilon_1 Q + \epsilon_1^\dagger Q^\dagger
,\, X
\bigr ] \Bigr ]
-
\Bigl [
\epsilon_1 Q + \epsilon_1^\dagger Q^\dagger,\,
\bigl [
\epsilon_2 Q + \epsilon_2^\dagger Q^\dagger
,\, X
\bigr ] \Bigr ]
=\qquad\qquad &&
\nonumber
\\
2( \epsilon_1 \sigma^\mu \epsilon_2^\dagger
  -\epsilon_2 \sigma^\mu \epsilon_1^\dagger)\, i\partial_\mu X
,&&
\qquad\>{}
\label{epsalg}
\eeq
up to terms that vanish on-shell. The spacetime momentum operator is
$P^\mu = (H, \vec{P})$, where $H$ is the Hamiltonian and $\vec{P}$ is the
three-momentum operator, given in terms of the canonical fields by
\beq
H &=& 
\int d^3\vec{x} \left [
\pi^*\pi + 
(\vec{\nabla} \phi^*)
\cdot (\vec{\nabla} \phi)
+ i \psi^\dagger \vec{\sigma} \cdot \vec{\nabla} \psi
\right ] ,
\\
\vec P &=& 
-\int d^3\vec{x} \left (
\pi \vec{\nabla} \phi 
+\pi^* \vec{\nabla} \phi^* 
+ i \psi^\dagger \sigmabar^0 \vec{\nabla} \psi 
\right )
.
\eeq 
It generates spacetime translations on the fields $X$ according to
\beq
[P^\mu, X ] = \BDneg i \partial^\mu X.
\label{supergrass}
\eeq
Rearranging the terms in eq.~(\ref{epsalg}) using the Jacobi identity,
we therefore have
\beq
\Bigl [ \bigl [
\epsilon_2 Q + \epsilon_2^\dagger Q^\dagger,\,
\epsilon_1 Q + \epsilon_1^\dagger Q^\dagger \bigr ]
,\, X
\Bigr ]
&= & 2(\BDneg\epsilon_1 \sigma_\mu \epsilon_2^\dagger
       \BDplus\epsilon_2 \sigma_\mu \epsilon_1^\dagger )\,  [ P^\mu , X ],
\label{epsalg2}
\eeq
for any $X$, up to terms that vanish on-shell, so it must be that
\beq
 \bigl [
\epsilon_2 Q + \epsilon_2^\dagger Q^\dagger,\,
\epsilon_1 Q + \epsilon_1^\dagger Q^\dagger \bigr ]
&= & 2(\BDneg\epsilon_1 \sigma_\mu \epsilon_2^\dagger
       \BDplus\epsilon_2 \sigma_\mu \epsilon_1^\dagger)\,  P^\mu  .
\label{epsalg3}
\eeq
Now by expanding out eq.~(\ref{epsalg3}), one obtains the precise form of
the supersymmetry algebra relations
\beq
&&\{ Q_\alpha , Q^\dagger_{\dot{\alpha}} \} =
\BDpos 2\sigma^\mu_{\alpha\dot{\alpha}} P_\mu,
\label{nonschsusyalg1}\\
&&\{ Q_\alpha, Q_\beta\} = 0
, \qquad\qquad
\{ Q^\dagger_{\dot{\alpha}}, Q^\dagger_{\dot{\beta}} \} = 0
,
\phantom{XXX}
\label{nonschsusyalg2}
\eeq
as promised in the Introduction. [The commutator in eq.~(\ref{epsalg3})
turns into anticommutators in eqs.~(\ref{nonschsusyalg1}) and
(\ref{nonschsusyalg2}) when the anti-commuting
spinors $\epsilon_1$ and $\epsilon_2$ are extracted.] The results
\beq
[Q_\alpha, P^\mu ] = 0, \qquad\qquad [Q^\dagger_{\dot{\alpha}},
P^\mu] = 0
\eeq
follow immediately from eq.~(\ref{supergrass}) and the fact that the
supersymmetry transformations are global (independent of position in
spacetime). This demonstration of the supersymmetry algebra in terms of
the canonical generators $Q$ and $Q^\dagger$ requires the use of the
Hamiltonian equations of motion, but the symmetry itself is valid
off-shell at the level of the Lagrangian, as we have already shown. 

\subsection{Interactions of chiral supermultiplets}\label{subsec:susylagr.chiral}
\setcounter{footnote}{1}
\renewcommand{\theequation}{\arabic{section}.\arabic{subsection}.\arabic{equation}}
\setcounter{equation}{0}

In a realistic theory like the MSSM, there are many chiral
supermultiplets, with both gauge and non-gauge interactions. In this
subsection, our task is to construct the most general possible theory of
masses and non-gauge interactions for particles that live in chiral
supermultiplets. In the MSSM these are the quarks, squarks, leptons,
sleptons, Higgs scalars and higgsino fermions. We will find that the form
of the non-gauge couplings, including mass terms, is highly restricted by
the requirement that the action is invariant under supersymmetry
transformations. (Gauge interactions will be dealt with in the following
subsections.)

Our starting point is the Lagrangian density for a collection of free
chiral supermultiplets labeled by an index $i$, which runs over all gauge
and flavor degrees of freedom. Since we will want to construct an
interacting theory with supersymmetry closing off-shell, each
supermultiplet contains a complex scalar $\phi_i$ and a left-handed Weyl
fermion $\psi_i$ as physical degrees of freedom, plus a non-propagating 
complex auxiliary
field $F_i$. The results of the previous
subsection tell us that the free part of the Lagrangian is
\beq
\lagr_{\rm free} &=&
\BDpos \partial^\mu \phi^{*i} \partial_\mu \phi_i
+ i {\psi}^{\dagger i} \sigmabar^\mu \partial_\mu \psi_i
+ F^{*i} F_i ,
\label{lagrfree}
\eeq
where we sum over repeated indices $i$ (not to be confused with the
suppressed spinor indices), with the convention that fields $\phi_i$ and
$\psi_i$ always carry lowered indices, while their conjugates always carry
raised indices. It is invariant under the supersymmetry transformation
\beq
&
\delta \phi_i = \epsilon\psi_i ,
\qquad\>\>\>\>\>\qquad\qquad\qquad
\phantom{xxxi}
&
\delta \phi^{*i} = \epsilon^\dagger {\psi}^{\dagger i} ,
\label{phitran}
\\
&
\delta (\psi_i)_\alpha = 
- i (\sigma^\mu {\epsilon^\dagger})_{\alpha}\, \partial_\mu
\phi_i + \epsilon_\alpha F_i ,
\qquad
&
\delta ({\psi}^{\dagger i})_{\dot{\alpha}}=
 i (\epsilon\sigma^\mu)_{\dot{\alpha}}\, \partial_\mu
\phi^{*i} + \epsilon^\dagger_{\dot{\alpha}} F^{*i} ,
\phantom{xxxx}
\\
&
\delta F_i = - i \epsilon^\dagger \sigmabar^\mu\partial_\mu \psi_i ,
\qquad\qquad\qquad
\phantom{xxi}
&
\deltaeps F^{* i} = 
 i\partial_\mu {\psi}^{\dagger i} \sigmabar^\mu  \epsilon\> .
\phantom{xxx}
\label{eq:Ftran}
\eeq

We will now find the most general set of renormalizable interactions for
these fields that is consistent with supersymmetry. We do this working in
the field theory before integrating out the auxiliary fields. To begin,
note that in order to be renormalizable by power counting, each term must
have field content with total mass dimension $\leq 4$. So, the only
candidate terms are: 
\beq
\lagr_{\rm int} =
\left (-{1\over 2} W^{ij} \psi_i \psi_j + W^i F_i + x^{ij} F_i F_j \right ) 
+ \conj  - U,
\label{pretryint}
\eeq
where $W^{ij}$, $W^i$, $x^{ij}$, and $U$ are polynomials in the scalar
fields $\phi_i, \phi^{*i}$, with degrees $1$, $2$, $0$, and $4$,
respectively. [Terms of the form $F^{*i} F_j$ are already included in 
eq.~(\ref{lagrfree}), with the coefficient fixed by the transformation 
rules (\ref{phitran})-(\ref{eq:Ftran}).]

We must now require that $\lagr_{\rm int}$ is invariant under the
supersymmetry transformations, since $\lagr_{\rm free}$ was already
invariant by itself.  This immediately requires that the candidate term
$U(\phi_i, \phi^{*i})$ must vanish. If there were such a term, then under
a supersymmetry transformation eq.~(\ref{phitran}) it would transform into
another function of the scalar fields only, multiplied by $\epsilon\psi_i$
or ${\epsilon^\dagger}{\psi}^{\dagger i}$, and with no spacetime
derivatives or $F_i$, $F^{*i}$ fields. It is easy to see from
eqs.~(\ref{phitran})-(\ref{pretryint}) that nothing of this form can possibly
be canceled by the supersymmetry transformation of any other term in the
Lagrangian. Similarly, the dimensionless coupling $x^{ij}$ must be zero,
because its supersymmetry transformation likewise cannot possibly be
canceled by any other term. So, we are left with
\beq
\lagr_{\rm int} =
\left (-\half W^{ij} \psi_i \psi_j + W^i F_i \right ) + \conj
\label{tryint}
\eeq
as the only possibilities. At this point, we are not assuming that $W^{ij}$
and $W^i$ are related to each other in any way. However, soon we will 
find out that they {\it are} related, which is why we have chosen to use
the same letter for them. Notice that eq.~(\ref{xichi}) tells us that
$W^{ij}$ is symmetric under $i\leftrightarrow j$. 

It is easiest to divide the variation of $\lagr_{\rm int}$ into several
parts, which must cancel separately. First, we consider the part that
contains four spinors:
\beq
\delta \lagr_{\rm int} |_{\rm 4-spinor} = \left [
-{1\over 2} {\delta W^{ij} \over \delta \phi_k} 
  (\epsilon \psi_k)(\psi_i \psi_j)
-{1\over 2} {\delta W^{ij} \over \delta \phi^{*k}}
  ({\epsilon^\dagger}{\psi}^{\dagger k})(\psi_i \psi_j) \right ]
+ \conj
\label{deltafourferm}
\eeq
The term proportional to $(\epsilon \psi_k)(\psi_i\psi_j)$ cannot cancel
against any other term. Fortunately, however, the Fierz identity
eq.~(\ref{fierce}) implies
\beq
(\epsilon \psi_i) (\psi_j \psi_k) + (\epsilon \psi_j) (\psi_k \psi_i)
+ (\epsilon \psi_k) (\psi_i\psi_j) = 0 ,
\eeq
so this contribution to $\delta\lagr_{\rm int}$ vanishes identically if
and only if $\delta W^{ij}/\delta \phi_k$ is totally symmetric under
interchange of $i,j,k$. There is no such identity available for the term
proportional to $({\epsilon^\dagger } {\psi}^{\dagger k})(\psi_i\psi_j)$.
Since that term cannot cancel with any other, requiring it to be absent
just tells us that $W^{ij}$ cannot contain $\phi^{*k}$. In other words,
$W^{ij}$ is holomorphic (or complex analytic) in the complex fields $\phi_k$.

Combining what we have learned  so far, we can write
\beq
W^{ij} = M^{ij} + y^{ijk} \phi_k
\eeq
where $M^{ij}$ is a symmetric mass matrix for the fermion fields, and
$y^{ijk}$ is a Yukawa coupling of a scalar $\phi_k$ and two fermions
$\psi_i \psi_j$ that must be totally symmetric under interchange of
$i,j,k$. It is therefore possible, and it turns out to be convenient, to
write
\beq
W^{ij} = {\delta^2 \over \delta\phi_i\delta\phi_j} W
\label{expresswij}
\eeq
where we have introduced a useful object
\beq
W =
{1\over 2} M^{ij} \phi_i \phi_j + {1\over 6} y^{ijk} \phi_i \phi_j \phi_k,
\label{superpotential}
\eeq
called the {\it superpotential}.\index{superpotential} This is not a
scalar potential in the ordinary sense; in fact, it is not even real. It
is instead a holomorphic function of the scalar fields $\phi_i$ treated as
complex variables. 

Continuing on our vaunted quest, we next consider the parts of 
$\delta \lagr_{\rm int}$ that contain a spacetime derivative: 
\beq
\delta \lagr_{\rm int} |_\partial &=& \left (
 i W^{ij}\partial_\mu \phi_j \, \psi_i \sigma^\mu {\epsilon^\dagger}
+ i W^i\, \partial_\mu \psi_i\sigma^\mu {\epsilon^\dagger}
\right ) +\conj
\label{wijwi}
\eeq
Here we have used the identity eq.~(\ref{yetanotheridentity}) on the
second term, which came from $(\delta F_i)W^i$. Now we can use
eq.~(\ref{expresswij}) to observe that
\beq
W^{ij} \partial_\mu \phi_j =
\partial_\mu \left ( {\delta W\over \delta{\phi_i}}\right ) .
\label{parttwo}
\eeq
Therefore, eq.~(\ref{wijwi}) will be a total derivative if
\beq
W^i = {\delta W\over \delta \phi_i} 
= M^{ij}\phi_j + {1\over 2} y^{ijk} \phi_j \phi_k\> ,
\label{wiwiwi}
\eeq
which explains why we chose its name as we did. The remaining terms in
$\delta \lagr_{\rm int}$ are all linear in $F_i$ or $F^{*i}$, and it is
easy to show that they cancel, given the results for $W^i$ and $W^{ij}$
that we have already found. 

Actually, we can include a linear term in the superpotential without
disturbing the validity of the previous discussion at all: 
\beq
W = L^i \phi_i +
{1\over 2} M^{ij} \phi_i \phi_j + {1\over 6} y^{ijk} \phi_i \phi_j \phi_k
.
\label{superpotentialwithlinear}
\eeq
Here $L^i$ are parameters with dimensions of [mass]$^2$, which affect only
the scalar potential part of the Lagrangian. Such
linear terms are only allowed when $\phi_i$ is a gauge singlet, and there are
no such gauge singlet chiral supermultiplets in the MSSM with minimal
field content.  I will therefore omit this term from the remaining
discussion of this section.  However, this type of term does play an
important role in the discussion of spontaneous supersymmetry breaking, as
we will see in section \ref{subsec:origins.general}. 

To recap, we have found that the most general non-gauge interactions for
chiral supermultiplets are determined by a single holomorphic function of the
complex scalar fields, the superpotential $W$. The auxiliary fields $F_i$
and $F^{*i}$ can be eliminated using their classical equations of motion.
The part of $\lagr_{\rm free} + \lagr_{\rm int}$ that contains the
auxiliary fields is $ F_i F^{*i} + W^i F_{i} + W^{*}_i F^{*i}$, leading to
the equations of motion
\beq
F_i = -W_i^*,\qquad\qquad F^{*i} = -W^i \> .
\label{replaceF}
\eeq
Thus the auxiliary fields are expressible algebraically (without any
derivatives) in terms of the scalar fields. 

After making the replacement\footnote{Since $F_i$ and $F^{*i}$ appear only
quadratically in the action, the result of instead doing a functional
integral over them at the quantum level has precisely the same effect.}
eq.~(\ref{replaceF}) in $\lagr_{\rm free} + \lagr_{\rm int}$, we obtain
the Lagrangian density
\beq
\lagr = \BDpos\partial^\mu \phi^{*i} \partial_\mu \phi_i
+ i \psi^{\dagger i} \sigmabar^\mu \partial_\mu \psi_i
-\half \left (W^{ij} \psi_i \psi_j  + W^{*}_{ij} \psi^{\dagger i}
\psi^{\dagger j} \right )
- W^i W^{*}_i.
\label{noFlagr}
\eeq
Now that the non-propagating fields $F_i, F^{*i}$ have been eliminated, it
follows from eq.~(\ref{noFlagr}) that the scalar potential for the theory
is just given in terms of the superpotential by
\beq
V(\phi,\phi^*) = W^k W_k^* = F^{*k} F_k = 
\phantom{xxxxxxxxxxxxxxxxxxxxxxxxxxxxxxxxxxxx.}
&&
\nonumber
\\ 
M^*_{ik} M^{kj} \phi^{*i} \phi_{j}
+{1\over 2} M^{in} y_{jkn}^* \phi_i \phi^{*j} \phi^{*k}
+{1\over 2} M_{in}^{*} y^{jkn} \phi^{*i} \phi_j \phi_k
+{1\over 4} y^{ijn} y_{kln}^{*} \phi_i \phi_j \phi^{*k} \phi^{*l}
\, .
&&
\label{ordpot}
\eeq
This scalar potential is automatically bounded from below; in fact, since
it is a sum of squares of absolute values (of the $W^k$), it is always
non-negative. If we substitute the general form for the superpotential
eq.~(\ref{superpotential}) into eq.~(\ref{noFlagr}), we obtain for the
full Lagrangian density
\beq
\lagr &=&
\BDpos \partial^\mu \phi^{*i} \partial_\mu \phi_i - V(\phi,\phi^*)
+ i \psi^{\dagger i} \sigmabar^\mu \partial_\mu \psi_i
- \half M^{ij} \psi_i\psi_j - \half M_{ij}^{*} \psi^{\dagger i}
\psi^{\dagger j}
\nonumber\\
&& - \half y^{ijk} \phi_i \psi_j \psi_k - \half y_{ijk}^{*} \phi^{*i}
\psi^{\dagger j} \psi^{\dagger k}.
\label{lagrchiral}
\eeq

Now we can compare the masses of the fermions and scalars by looking at
the linearized equations of motion:
\beq
\partial^\mu\partial_\mu \phi_i &=& 
\BDneg 
M_{ik}^{*} M^{kj} \phi_j
+ \ldots,\\
i\sigmabar^\mu\partial_\mu\psi_i &=&  
M_{ij}^{*} \psi^{\dagger j}+\ldots,
\qquad\qquad
i\sigma^\mu\partial_\mu\psi^{\dagger i} \>=\> 
M^{ij} \psi_j +\ldots .\qquad\>\>\>{}
\label{linfermiontwo}
\eeq
One can eliminate $\psi$ in terms of $\psi^\dagger$ and vice versa in
eq.~(\ref{linfermiontwo}), obtaining [after use of the identities
eqs.~(\ref{pauliidentA}) and (\ref{pauliidentB})]:
\beq
\partial^\mu\partial_\mu \psi_i =   
\BDneg M_{ik}^{*} M^{kj} \psi_j
+ \ldots ,\qquad\qquad
\partial^\mu\partial_\mu \psi^{\dagger j} =
\BDneg \psi^{\dagger i} M_{ik}^{*} M^{kj}+\ldots
\> .
\eeq
Therefore, the fermions and the bosons satisfy the same wave equation with
exactly the same squared-mass matrix with real non-negative eigenvalues,
namely ${(M^2)_i}^j = M_{ik}^{*} M^{kj}$. It follows that diagonalizing
this matrix by redefining the fields with a unitary matrix gives a
collection of chiral supermultiplets, each of which contains a
mass-degenerate complex scalar and Weyl fermion, in agreement with the
general argument in the Introduction. 

\subsection{Lagrangians for gauge
supermultiplets}\label{subsec:susylagr.gauge}
\setcounter{footnote}{1}
\renewcommand{\theequation}{\arabic{section}.\arabic{subsection}.\arabic{equation}}
\setcounter{equation}{0}

The propagating degrees of freedom in a gauge supermultiplet are a
massless gauge boson field $A_\mu^a$ and a two-component Weyl fermion
gaugino $\lambda^a$. The index $a$ here runs over the adjoint
representation of the gauge group ($a=1,\ldots ,8$ for $SU(3)_C$ color
gluons and gluinos; $a=1,2,3$ for $SU(2)_L$ weak isospin; $a=1$ for
$U(1)_Y$ weak hypercharge). The gauge transformations of the vector
supermultiplet fields are
\beq
A^a_\mu &\rightarrow&  A^a_\mu
\BDminus \partial_\mu \Lambda^a + g f^{abc} A^b_\mu \Lambda^c ,
\label{Agaugetr}
\\
\lambda^a &\rightarrow& \lambda^a + g f^{abc} \lambda^b \Lambda^c
,
\label{lamgaugetr}
\eeq
where $\Lambda^a$ is an infinitesimal gauge transformation parameter, $g$
is the gauge coupling, and $f^{abc}$ are the totally antisymmetric
structure constants that define the gauge group. The special case of an
Abelian group is obtained by just setting $f^{abc}=0$;  the corresponding
gaugino is a gauge singlet in that case. The conventions are such that for
QED, $A^\mu = (V, \vec{A})$ where $V$ and $\vec{A}$ are the usual electric
potential and vector potential, with electric and magnetic fields given by
$\vec{E} = -\vec{\nabla} V - \partial_0 \vec{A}$ and $\vec{B} =
\vec{\nabla} \times \vec{A}$. 

The on-shell degrees of freedom for $A^a_\mu$ and $\lambda^a_\alpha$
amount to two bosonic and two fermionic helicity states (for each $a$), as
required by supersymmetry. However, off-shell $\lambda^a_\alpha$ consists
of two complex, or four real, fermionic degrees of freedom, while
$A^a_\mu$ only has three real bosonic degrees of freedom; one degree of
freedom is removed by the inhomogeneous gauge transformation
eq.~(\ref{Agaugetr}). So, we will need one real bosonic auxiliary field,
traditionally called $D^a$, in order for supersymmetry to be consistent
off-shell. This field also transforms as an adjoint of the gauge group
[i.e., like eq.~(\ref{lamgaugetr}) with $\lambda^a$ replaced by $D^a$] and
satisfies $(D^a)^* = D^a$. Like the chiral auxiliary fields $F_i$, the
gauge auxiliary field $D^a$ has dimensions of [mass]$^2$ and no kinetic
term, so it can be eliminated on-shell using its algebraic equation of
motion. The counting of degrees of freedom is summarized in Table
\ref{table:gaugedofcounting}.
\renewcommand{\arraystretch}{1.45}
\begin{table}[tb]
\begin{center}
\begin{tabular}{|c|c|c|c|}
\hline
 & $A_\mu$ & $\lambda$ & $D$ \\
\hline
on-shell ($n_B=n_F=2$) & 2 & 2 & 0 \\
\hline
off-shell ($n_B=n_F=4$) & 3 & 4 & 1 \\
\hline
\end{tabular}
\caption{Counting of real degrees of freedom for each gauge
supermultiplet. \label{table:gaugedofcounting}}
\vspace{-0.4cm}
\end{center}
\end{table}

Therefore, the Lagrangian density for a gauge supermultiplet ought to be
\beq
\lagr_{\rm gauge} = -{1\over 4} F_{\mu\nu}^a F^{\mu\nu a}
+ i \lambda^{\dagger a} \sigmabar^\mu \nabla_\mu \lambda^a
+ {1\over 2} D^a D^a ,
\label{lagrgauge}
\eeq
where
\beq
F^a_{\mu\nu} = \partial_\mu A^a_\nu - \partial_\nu A^a_\mu 
               \BDminus g f^{abc} A^b_\mu A^c_\nu
\label{eq:YMfs}
\eeq
is the usual Yang-Mills field strength, and
\beq
\nabla_\mu \lambda^a = \partial_\mu \lambda^a 
                  \BDminus g f^{abc} A^b_\mu \lambda^c
\label{ordtocovlambda}
\eeq
is the covariant derivative of the gaugino field. To check that
eq.~(\ref{lagrgauge}) is really supersymmetric, one must specify the
supersymmetry transformations of the fields. 
The forms of these follow from the requirements that they should be linear
in the infinitesimal parameters $\epsilon,\epsilon^\dagger$ with
dimensions of [mass]$^{-1/2}$, that $\delta A^a_\mu$ is real, and that
$\delta D^a$ should be real and proportional to the field equations for
the gaugino, in analogy with the role of the auxiliary field $F$ in the
chiral supermultiplet case. Thus one can guess, up to multiplicative
factors, that\footnote{The supersymmetry transformations
eqs.~(\ref{Atransf})-(\ref{Dtransf}) are non-linear for non-Abelian gauge
symmetries, since there are gauge fields in the covariant
derivatives acting on the gaugino fields and in the field strength $F_{\mu
\nu}^a$. By adding even more auxiliary fields besides $D^a$, one can make
the supersymmetry transformations linear in the fields; this is easiest to
do in superfield language (see sections \ref{subsec:vectorsuperfields}, 
\ref{subsec:superspacelagrabelian}, and \ref{subsec:superspacelagrnonabelian}). 
The version in this section, in which those extra auxiliary fields 
have been eliminated, is called ``Wess-Zumino gauge" \cite{WZgauge}.}
\beq
&& \delta A_\mu^a = 
- {1\over \sqrt{2}} \left (\epsilon^\dagger \sigmabar_\mu
\lambda^a + \lambda^{\dagger a} \sigmabar_\mu \epsilon \right )
,
\label{Atransf}
\\
&& \delta \lambda^a_\alpha =
\BDneg {i\over 2\sqrt{2}} (\sigma^\mu \sigmabar^\nu \epsilon)_\alpha
\> F^a_{\mu\nu} + {1\over \sqrt{2}} \epsilon_\alpha\> D^a 
,
\\
&& \delta D^a =  {i\over \sqrt{2}} \left (
-\epsilon^\dagger \sigmabar^\mu \nabla_\mu \lambda^a 
+\nabla_\mu \lambda^{\dagger a} \sigmabar^\mu \epsilon \right ) .
\label{Dtransf}
\eeq
The factors of $\sqrt{2}$ are chosen so that the action obtained by
integrating $\lagr_{\rm gauge}$ is indeed invariant, and the phase of
$\lambda^a$ is chosen for future convenience in treating the MSSM.

It is now a little bit tedious, but straightforward, to also check that
\beq
(\delta_{\epsilon_2} \delta_{\epsilon_1} -\delta_{\epsilon_1}
\delta_{\epsilon_2} ) X = 
i (-\epsilon_1\sigma^\mu \epsilon_2^\dagger
+\epsilon_2\sigma^\mu \epsilon_1^\dagger) \nabla_\mu X
\label{joeyramone}
\eeq
for $X$ equal to any of the gauge-covariant fields $F_{\mu\nu}^a$,
$\lambda^a$, $\lambda^{\dagger a}$, $D^a$, as well as for arbitrary
covariant derivatives acting on them. This ensures that the supersymmetry
algebra eqs.~(\ref{nonschsusyalg1})-(\ref{nonschsusyalg2}) is realized on
gauge-invariant combinations of fields in gauge supermultiplets, as they
were on the chiral supermultiplets [compare eq.~(\ref{anytrans})]. This
check requires the use of identities 
eqs.~(\ref{eq:dei}), (\ref{eq:feif}) and (\ref{eq:lloydhouserules}).
If we had not included the auxiliary field
$D^a$, then the supersymmetry algebra eq.~(\ref{joeyramone}) would hold
only after using the equations of motion for $\lambda^a$ and
$\lambda^{\dagger a}$. The auxiliary fields satisfies a trivial equation
of motion $D^a=0$, but this is modified if one couples the gauge
supermultiplets to chiral supermultiplets, as we now do. 

\subsection{Supersymmetric gauge
interactions}\label{subsec:susylagr.gaugeinter}
\setcounter{footnote}{1}
\renewcommand{\theequation}{\arabic{section}.\arabic{subsection}.\arabic{equation}}
\setcounter{equation}{0}

Now we are ready to consider a general Lagrangian density for a
supersymmetric theory with both chiral and gauge supermultiplets. Suppose
that the chiral supermultiplets transform under the gauge group in a
representation with hermitian matrices ${(T^a)_i}^j$ satisfying $[T^a,T^b]
=i f^{abc} T^c$. [For example, if the gauge group is $SU(2)$, then
$f^{abc} = \epsilon^{abc}$, and 
for a chiral supermultiplet transforming in the fundamental
representation the $T^a$ are $1/2$ times the Pauli
matrices.] Because supersymmetry and gauge transformations commute,
the scalar, fermion, and auxiliary fields must be in the same
representation of the gauge group, so
\beq
X_i &\rightarrow& X_i + ig \Lambda^a (T^a X)_i
\eeq
for $X_i = \phi_i,\psi_i,F_i$.  To have a gauge-invariant Lagrangian, we
now need to replace the ordinary derivatives $\partial_\mu \phi_i$,
$\partial_\mu \phi^{*i}$, and $\partial_\mu \psi_i$ 
in eq.~(\ref{lagrfree}) with covariant derivatives:
\beq
\nabla_\mu \phi_i &=& \partial_\mu \phi_i \BDplus i g A^a_\mu (T^a\phi)_i
\label{ordtocovphi}
\\
\nabla_\mu \phi^{*i} &=& \partial_\mu \phi^{*i} \BDminus i g A^a_\mu (\phi^* T^a)^i
\\
\nabla_\mu \psi_i &=& \partial_\mu \psi_i \BDplus i g A^a_\mu (T^a\psi)_i .
\label{ordtocovpsi}
\eeq
Naively, this simple procedure achieves the goal of coupling the vector
bosons in the gauge supermultiplet to the scalars and fermions in the
chiral supermultiplets. However, we also have to consider whether there
are any other interactions allowed by gauge invariance and involving the
gaugino and $D^a$ fields, which might have to be included to make a
supersymmetric Lagrangian. Since $A^a_\mu$ couples to $\phi_i$ and
$\psi_i$, it makes sense that $\lambda^a$ and $D^a$ should as well. 

In fact, there are three such possible interaction terms that are
renormalizable (of field mass dimension $\leq 4$), namely
\beq
(\phi^* T^a \psi)\lambda^a,\qquad
\lambda^{\dagger a} (\psi^\dagger T^a
\phi),
\qquad {\rm and} \qquad
(\phi^* T^a \phi) D^a .
\label{extrater}
\eeq
Now one can add them, with unknown dimensionless coupling coefficients, to
the Lagrangians for the chiral and gauge supermultiplets, and demand that
the whole mess be real and invariant under supersymmetry,
up to a total derivative. Not surprisingly, this is possible only if the
supersymmetry transformation laws for the matter fields are modified to
include gauge-covariant rather than ordinary derivatives. Also, it is
necessary to include one strategically chosen extra term in $\delta F_i$,
so: 
\beq
&&
\delta \phi_i = \epsilon\psi_i
\label{gphitran}\\
&&\delta \psi_{i\alpha} =
-i (\sigma^\mu \epsilon^\dagger)_{\alpha}\> \nabla_\mu\phi_i + \epsilon_\alpha F_i
\\
&&\deltaeps F_i = -i \epsilon^\dagger \sigmabar^\mu \nabla_\mu \psi_i
\> + \> \sqrt{2} g (T^a \phi)_i\> \epsilon^\dagger \lambda^{\dagger a} .
\eeq
After some algebra one can now fix the coefficients for the terms in
eq.~(\ref{extrater}), with the result that the full Lagrangian density
for a renormalizable supersymmetric theory is
\beq
\lagr & = & \lagr_{\rm chiral} + \lagr_{\rm gauge} 
\nonumber\\
        && - \sqrt{2} g 
(\phi^* T^a \psi)\lambda^a 
- \sqrt{2} g\lambda^{\dagger a} (\psi^\dagger T^a \phi)
+ g  (\phi^* T^a \phi) D^a .
\label{gensusylagr}
\eeq
Here $\lagr_{\rm chiral }$ means the chiral supermultiplet Lagrangian
found in section \ref{subsec:susylagr.chiral} [e.g., eq.~(\ref{noFlagr})
or (\ref{lagrchiral})], but with ordinary derivatives replaced everywhere
by gauge-covariant derivatives, and $\lagr_{\rm gauge}$ was given in
eq.~(\ref{lagrgauge}). To prove that eq.~(\ref{gensusylagr}) is invariant
under the supersymmetry transformations, one must use the identity
\beq
W^i (T^a \phi)_i = 0.
\label{wgaugeinvar}
\eeq 
This is precisely the condition that must be satisfied anyway in order for
the superpotential, and thus $\lagr_{\rm chiral}$, to be gauge invariant.

The second line in eq.~(\ref{gensusylagr}) consists of interactions whose
strengths are fixed to be gauge couplings by the requirements of
supersymmetry, even though they are not gauge interactions from the point
of view of an ordinary field theory. The first two terms are a direct
coupling of gauginos to matter fields; this can be thought of as the
``supersymmetrization" of the usual gauge boson couplings to matter fields.
The last term combines with the $D^a D^a/2$ term in $\lagr_{\rm gauge}$ to
provide an equation of motion
\beq
D^a = -g (\phi^* T^a \phi ).
\label{solveforD}
\eeq
Thus, like the auxiliary fields $F_i$ and $F^{*i}$, the $D^a$ are
expressible purely algebraically in terms of the scalar fields. Replacing
the auxiliary fields in eq.~(\ref{gensusylagr}) using
eq.~(\ref{solveforD}), one finds that the complete scalar potential is
(recall that $\lagr$ contains $-V$): 
\beq
V(\phi,\phi^*) = F^{*i} F_i + \half \sum_a D^a D^a = W_i^* W^i +
\half \sum_a g_a^2 (\phi^* T^a \phi)^2.
\label{fdpot}
\eeq
The two types of terms in this expression are called ``$F$-term" and
``$D$-term" contributions, respectively. In the second term in
eq.~(\ref{fdpot}), we have now written an explicit sum $\sum_a$ to cover
the case that the gauge group has several distinct factors with different
gauge couplings $g_a$. [For instance, in the MSSM the three factors
$SU(3)_C$, $SU(2)_L$ and $U(1)_Y$ have different gauge couplings $g_3$,
$g$ and $g^\prime$.] Since $V(\phi,\phi^*)$ is a sum of squares, it is
always greater than or equal to zero for every field configuration. It is
an interesting and unique feature of supersymmetric theories that the
scalar potential is completely determined by the {\it other} interactions
in the theory. The $F$-terms are fixed by Yukawa couplings and fermion
mass terms, and the $D$-terms are fixed by the gauge interactions.

By using Noether's procedure [see eq.~(\ref{Noether})], one finds the
conserved supercurrent
\beq
J_\alpha^\mu &\!\!\!=\!\!\!&
(\sigma^\nu\sigmabar^\mu \psi_i)_\alpha\, \nabla_\nu \phi^{*i}
+ i (\sigma^\mu \psi^{\dagger i})_\alpha\, W_i^*
\nonumber
\\ &&
\BDplus {1\over 2 \sqrt{2}}
(\sigma^\nu \sigmabar^\rho \sigma^\mu
\lambda^{\dagger a})_\alpha\, F^a_{\nu\rho}
+ {i\over {\sqrt{2}}} g_a \phi^* T^a \phi
\> (\sigma^\mu \lambda^{\dagger a})_\alpha , \>\>\>\>{}
\label{supercurrent}
\eeq
generalizing the expression given in eq.~(\ref{WZsupercurrent}) for the
Wess-Zumino model. This result will be useful when we discuss certain
aspects of spontaneous supersymmetry breaking in section
\ref{subsec:origins.gravitino}.

\subsection{Summary: How to build a supersymmetric
model}\label{subsec:susylagr.summary}
\setcounter{footnote}{1}
\renewcommand{\theequation}{\arabic{section}.\arabic{subsection}.\arabic{equation}}
\setcounter{equation}{0}

In a renormalizable supersymmetric field theory, the interactions and
masses of all particles are determined just by their gauge transformation
properties and by the superpotential $W$. By construction, we found that
$W$ had to be a holomorphic function of the complex scalar fields $\phi_i$,
which are always defined to transform under supersymmetry into left-handed
Weyl fermions. In an equivalent language, to be covered 
in section \ref{sec:superfields}, $W$ is
said to be a function of chiral {\it superfields} \cite{superfields}. A
superfield is a single object that contains as components all of the
bosonic, fermionic, and auxiliary fields within the corresponding
supermultiplet, for example $\Phi_i \supset (\phi_i,\psi_i,F_i)$. (This is
analogous to the way in which one often describes a weak isospin doublet
or a color triplet by a multicomponent field.) The gauge quantum numbers and
the mass dimension of a chiral superfield are the same as that of its
scalar component. In the superfield formulation, one writes instead of
eq.~(\ref{superpotentialwithlinear})
\beq
W &=& 
L^i \Phi_i +
{1\over 2}M^{ij} \Phi_i\Phi_j +{1\over 6} y^{ijk} \Phi_i \Phi_j \Phi_k 
,
\label{superpot}
\eeq
which implies exactly the same physics. The derivation of all of our
preceding results can be obtained somewhat more elegantly using superfield
methods, which have the advantage of making invariance under supersymmetry
transformations manifest by defining the Lagrangian in terms of integrals
over a ``superspace" with fermionic as well as ordinary commuting
coordinates. We have purposefully avoided this extra layer of notation so far, in
favor of the more pedestrian, but more familiar and accessible, component
field approach. The latter is at least more appropriate for making contact
with phenomenology in a universe with supersymmetry breaking.  
The specification of the superpotential
is really just a code for the terms that it implies in the Lagrangian, so the
reader may feel free to think of the superpotential either as a function
of the scalar fields $\phi_i$ or as the same function of the superfields
$\Phi_i$. 

Given the supermultiplet content of the theory, the form of the
superpotential is restricted by the requirement of gauge invariance [see
eq.~(\ref{wgaugeinvar})].  In any given theory, only a subset of the
parameters $L^i$, $M^{ij}$, and $y^{ijk}$ are allowed to be non-zero.  The
parameter $L^i$ is only allowed if $\Phi_i$ is a gauge singlet. (There are
no such chiral supermultiplets in the MSSM with the minimal field
content.) The entries of the mass matrix $M^{ij}$ can only be non-zero for
$i$ and $j$ such that the supermultiplets $\Phi_i$ and $\Phi_j$ transform
under the gauge group in representations that are conjugates of each
other. (In the MSSM there is only one such term, as we will see.)
Likewise, the Yukawa couplings $y^{ijk}$ can only be non-zero when
$\Phi_i$, $\Phi_j$, and $\Phi_k$ transform in representations that can
combine to form a singlet. 

\begin{figure}
\begin{center}
\begin{picture}(66,60)(0,0)
\SetWidth{0.85}
\ArrowLine(0,0)(33,12)  
\ArrowLine(66,0)(33,12)
\DashLine(33,52.5)(33,12){4}
\ArrowLine(33,32.2501)(33,32.25)
\Text(0.75,10.5)[c]{$j$}
\Text(66,10)[c]{$k$}
\Text(26,48)[c]{$i$}
\Text(33,-12)[c]{(a)}
\end{picture}
\hspace{1.8cm}
\begin{picture}(66,60)(0,0)
\SetWidth{0.85}
\ArrowLine(33,12)(0,0)  
\ArrowLine(33,12)(66,0)
\DashLine(33,52.5)(33,12){4}
\ArrowLine(33,32.25)(33,32.2501)
\Text(0.75,10.5)[c]{$j$}
\Text(66,10)[c]{$k$}
\Text(26,48)[c]{$i$}
\Text(33,-12)[c]{(b)}
\end{picture}
\hspace{1.8cm}
\begin{picture}(60,60)(0,0)
\SetWidth{0.85}
\DashLine(0,0)(30,30){4}
\DashLine(60,0)(30,30){4}   
\DashLine(0,60)(30,30){4} 
\DashLine(60,60)(30,30){4} 
\ArrowLine(16.5,16.5)(16.501,16.501)
\ArrowLine(43.5,16.5)(43.499,16.501)
\ArrowLine(16.5,43.5)(16.499,43.501)
\ArrowLine(43.5,43.5)(43.501,43.501)
\Text(-2,9)[c]{$i$}
\Text(64,10)[c]{$j$} 
\Text(-2,54)[c]{$k$}
\Text(63.5,54)[c]{$l$}
\Text(30,-12)[c]{(c)}
\end{picture}  
\end{center}
\caption{The dimensionless non-gauge interaction 
vertices in a supersymmetric theory: 
(a) scalar-fermion-fermion Yukawa
interaction $y^{ijk}$, 
(b) the complex conjugate interaction $y_{ijk}$, and
(c) quartic scalar interaction $y^{ijn}y^*_{kln}$.
\label{fig:dim0}}
\end{figure}
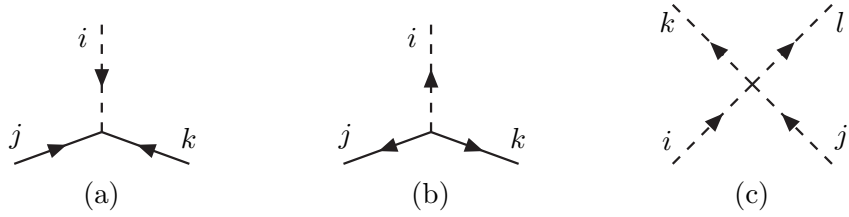

\begin{figure}
\begin{center}
\begin{picture}(66,62)(0,0)
\SetWidth{0.85}
\DashLine(33,52.5)(33,12){4}
\DashLine(0,0)(33,12){4}
\DashLine(66,0)(33,12){4}
\ArrowLine(33,32.25)(33,32.2501)
\ArrowLine(16.5,6)(16.5165,6.006)
\ArrowLine(49.5,6)(49.4835,6.006)
\Text(1.75,10.5)[c]{$j$}
\Text(65,10)[c]{$k$}
\Text(26,48)[c]{$i$}
\Text(33,-12)[c]{(a)}
\end{picture}
\hspace{0.93cm}
\begin{picture}(66,62)(0,0)
\SetWidth{0.85}
\DashLine(33,52.5)(33,12){4}
\DashLine(0,0)(33,12){4}
\DashLine(66,0)(33,12){4}
\ArrowLine(33,32.2501)(33,32.25)
\ArrowLine(16.5165,6.006)(16.5,6)
\ArrowLine(49.4835,6.006)(49.5,6)
\Text(1.75,10.5)[c]{$j$}
\Text(65,10)[c]{$k$}
\Text(26,48)[c]{$i$}
\Text(33,-12)[c]{(b)}
\end{picture}
\hspace{0.93cm}
\begin{picture}(72,62)(0,0)
\SetWidth{0.85}
\ArrowLine(0,12)(36,12)
\ArrowLine(72,12)(36,12)
\Line(33,9)(39,15)
\Line(39,9)(33,15)
\Text(0,19)[c]{$i$}
\Text(72,20)[c]{$j$}
\Text(36,-12)[c]{(c)}
\end{picture}
\hspace{0.93cm}
\begin{picture}(72,62)(0,0)
\SetWidth{0.85}
\ArrowLine(36,12)(0,12)
\ArrowLine(36,12)(72,12)
\Line(33,9)(39,15)
\Line(39,9)(33,15)
\Text(0,19)[c]{$i$}
\Text(72,20)[c]{$j$}
\Text(36,-12)[c]{(d)}
\end{picture}
\hspace{0.94cm}
\begin{picture}(72,62)(0,0)
\SetWidth{0.85}
\DashLine(0,12)(36,12){4}
\DashLine(72,12)(36,12){4}
\ArrowLine(17.99,12)(18,12)
\ArrowLine(55,12)(55.01,12)
\Line(33,9)(39,15)
\Line(39,9)(33,15)
\Text(72,20)[c]{$i$}
\Text(0,20)[c]{$j$}
\Text(36,-12)[c]{(e)}
\end{picture}
\end{center}
\caption{Supersymmetric dimensionful couplings: 
(a) (scalar)$^3$ interaction vertex $M^*_{in} y^{jkn}$ and 
(b) the conjugate interaction $M^{in} y^*_{jkn}$, 
(c) fermion mass term $M^{ij}$ and 
(d) conjugate fermion mass term $M^*_{ij}$,
and 
(e) scalar squared-mass term $M^*_{ik}M^{kj}$.
\label{fig:dim12}}
\end{figure}
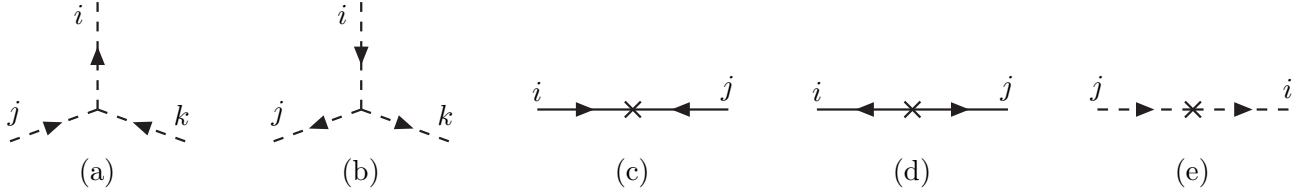

The interactions implied by the superpotential eq.~(\ref{superpot}) (with
$L^i=0$) were listed in
eqs.~(\ref{ordpot}), (\ref{lagrchiral}), 
and are shown\footnote{Here, the auxiliary fields have been
eliminated using their equations of motion (``integrated out"). 
One can instead give Feynman rules that include
the auxiliary fields, or directly in terms of superfields on superspace,
although this is usually less practical for phenomenological
applications.} in Figures~\ref{fig:dim0} and \ref{fig:dim12}.
Those in
Figure~\ref{fig:dim0} are all determined by the dimensionless parameters
$y^{ijk}$. The Yukawa interaction in Figure~\ref{fig:dim0}a corresponds to
the next-to-last term in eq.~(\ref{lagrchiral}).%
For each particular
Yukawa coupling of $\phi_i \psi_j \psi_k$ with strength $y^{ijk}$, there
must be equal couplings of $\phi_j \psi_i \psi_k$ and $\phi_k \psi_i
\psi_j$, since $y^{ijk}$ is completely symmetric under interchange of any
two of its indices as shown in section \ref{subsec:susylagr.chiral}.
The arrows on the fermion and scalar lines point in the direction for
propagation of $\phi$ and $\psi$ and opposite the direction of propagation
of $\phi^*$ and $\psi^\dagger$. Thus there is also a vertex corresponding
to the one in Figure~\ref{fig:dim0}a but with all arrows reversed,
corresponding to the complex conjugate [the last term in
eq.~(\ref{lagrchiral})]. It is shown in Figure \ref{fig:dim0}b. There is
also a dimensionless coupling for $\phi_i \phi_j \phi^{*k}\phi^{*l}$, with
strength $y^{ijn} y^*_{kln}$, as required by supersymmetry [see the last
term in eq.~(\ref{ordpot})]. The relationship between the Yukawa
interactions in Figures~\ref{fig:dim0}a,b and the scalar interaction of
Figure \ref{fig:dim0}c is exactly of the special type needed to cancel the
quadratic divergences in quantum corrections to scalar masses, as
discussed in the Introduction [compare Figure~\ref{fig:higgscorr1},
and eq.~(\ref{eq:royalewithcheese})]. 

Figure~\ref{fig:dim12} shows the only interactions corresponding to
renormalizable and supersymmetric vertices with coupling dimensions of
[mass] and [mass]$^2$.%
First, there are (scalar)$^3$ couplings in Figure
\ref{fig:dim12}a,b, which are entirely determined by the superpotential
mass parameters $M^{ij}$ and Yukawa couplings $y^{ijk}$, as indicated by
the second and third terms in eq.~(\ref{ordpot}). The propagators of the
fermions and scalars in the theory are constructed in the usual way using
the fermion mass $M^{ij}$ and scalar squared mass $M^*_{ik}M^{kj}$. The
fermion mass terms $M^{ij}$ and $M_{ij}$ each lead to a chirality-changing
insertion in the fermion propagator; note the directions of the arrows in
Figure~\ref{fig:dim12}c,d. There is no such arrow-reversal for a scalar
propagator in a theory with exact supersymmetry; as depicted in
Figure~\ref{fig:dim12}e, if one treats the scalar squared-mass term as an
insertion in the propagator, the arrow direction is preserved. 

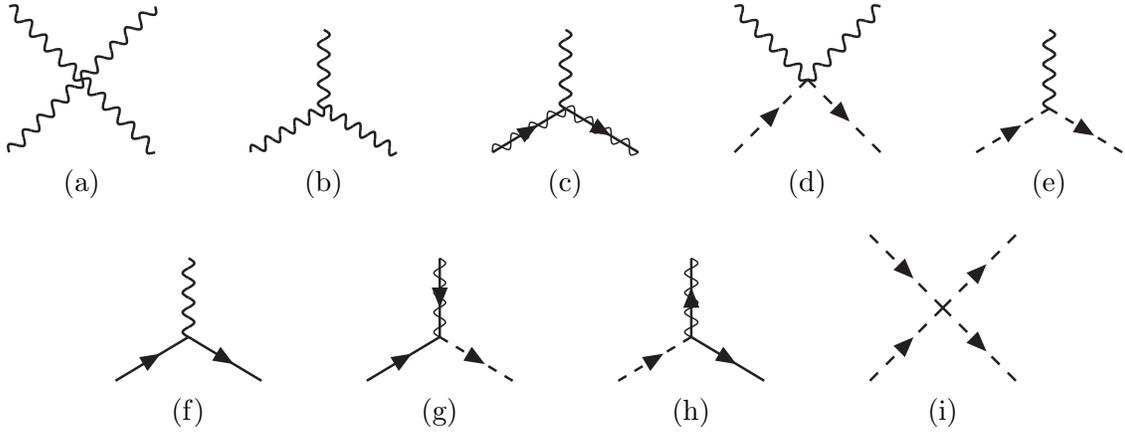
\begin{figure}
\begin{center}
\begin{picture}(50,66)(0,-7)
\SetScale{1.1}
\SetWidth{0.85}
\Photon(0,0)(25,25){2}{5}
\Photon(50,0)(25,25){2}{5}
\Photon(0,50)(25,25){2}{5}
\Photon(50,50)(25,25){2}{5}
\Text(27.5,-12.1)[c]{(a)}
\end{picture}
\hspace{1.2cm}
\begin{picture}(50,66)(0,-7)
\SetScale{1.1}
\SetWidth{0.85}
\Photon(0,0)(25,15){2}{5}
\Photon(50,0)(25,15){2}{5}
\Photon(25,42)(25,15){2}{4}
\Text(27.5,-12.1)[c]{(b)}
\end{picture}
\hspace{1.2cm}
\begin{picture}(50,66)(0,-7)
\SetScale{1.1}
\Photon(0,0)(25,15){2.1}{4}
\Photon(25,15)(50,0){2.1}{4}
\SetWidth{0.85}
\Photon(25,42)(25,15){2}{4}
\ArrowLine(0,0)(25,15)     
\ArrowLine(25,15)(50,0) 
\Text(27.5,-12.1)[c]{(c)}
\end{picture}
\hspace{1.2cm}
\begin{picture}(50,66)(0,-7)
\SetScale{1.1}
\SetWidth{0.85}
\DashLine(0,0)(25,25){3.8}
\DashLine(50,0)(25,25){3.8}
\Photon(0,50)(25,25){2}{5}
\Photon(50,50)(25,25){-2}{5}
\ArrowLine(13,13)(13.001,13.001)
\ArrowLine(36.999,13.001)(37,13)
\Text(27.5,-12.1)[c]{(d)}
\end{picture}
\hspace{1.2cm}
\begin{picture}(50,66)(0,-7)
\SetScale{1.1}
\SetWidth{0.85}
\DashLine(25,15)(0,0){3.6}  
\DashLine(50,0)(25,15){3.6}
\Photon(25,42)(25,15){2}{4}
\ArrowLine(12.5,7.5)(12.5125,7.5075)
\ArrowLine(37.5,7.5075)(37.5125,7.5)
\Text(27.5,-12.1)[c]{(e)}
\end{picture}

\vspace{1.0cm}

\begin{picture}(50,50)(0,0)
\SetScale{1.1}
\SetWidth{0.85}
\ArrowLine(0,0)(25,15)
\ArrowLine(25,15)(50,0)
\Photon(25,42)(25,15){2}{4} 
\Text(27.5,-12.1)[c]{(f)}
\end{picture}
\hspace{1.33cm}
\begin{picture}(50,50)(0,0)
\SetScale{1.1}
\Photon(25,42)(25,15){2.0}{4}   
\SetWidth{0.85}
\ArrowLine(0,0)(25,15)
\ArrowLine(25,42)(25,15)
\ArrowLine(37.5,7.5)(37.525,7.485)
\DashLine(50,0)(25,15){3.6}
\Text(27.5,-12.1)[c]{(g)}
\end{picture}
\hspace{1.33cm}
\begin{picture}(50,50)(0,0)
\SetScale{1.1}
\Photon(25,42)(25,15){2.25}{4}
\SetWidth{0.85}
\ArrowLine(12.5,7.5)(12.5125,7.5075)
\ArrowLine(25,15)(50,0)
\ArrowLine(25,15)(25,42)
\DashLine(25,15)(0,0){3.6}
\Text(27.5,-12.1)[c]{(h)}
\end{picture}
\hspace{1.33cm}
\begin{picture}(50,50)(0,0)
\SetScale{1.1}
\SetWidth{0.85}
\DashLine(0,0)(25,25){4}
\DashLine(50,0)(25,25){4}
\DashLine(0,50)(25,25){4}
\DashLine(50,50)(25,25){4}  
\ArrowLine(12.5,12.5)(12.501,12.501)
\ArrowLine(37.499,12.501)(37.5,12.5)
\ArrowLine(12.499,37.501)(12.5,37.5)
\ArrowLine(37.5,37.5)(37.501,37.501)
\Text(27.5,-12.1)[c]{(i)}
\end{picture}
\end{center}
\caption{Supersymmetric gauge interaction vertices.
\label{fig:gauge}}
\end{figure}

Figure~\ref{fig:gauge} shows the gauge
interactions in a supersymmetric theory. Figures \ref{fig:gauge}a,b,c
occur only when the gauge group is non-Abelian, for example for $SU(3)_C$
color and $SU(2)_L$ weak isospin in the MSSM.%
Figures \ref{fig:gauge}a and
\ref{fig:gauge}b are the interactions of gauge bosons, which derive from
the first term in eq.~(\ref{lagrgauge}). In the MSSM these are exactly the
same as the well-known QCD gluon and electroweak gauge boson vertices of
the Standard Model. (We do not show the interactions of ghost fields,
which are necessary only for consistent loop amplitudes.) Figures
\ref{fig:gauge}c,d,e,f are just the standard interactions between gauge
bosons and fermion and scalar fields that must occur in any gauge theory
because of the form of the covariant derivative; they come from
eqs.~(\ref{ordtocovlambda}) and (\ref{ordtocovphi})-(\ref{ordtocovpsi})
inserted in the kinetic part of the Lagrangian. Figure \ref{fig:gauge}c
shows the coupling of a gaugino to a gauge boson; the gaugino line in a
Feynman diagram is traditionally drawn as a solid fermion line
superimposed on a wavy line. In Figure~\ref{fig:gauge}g we have the
coupling of a gaugino to a chiral fermion and a complex scalar [the first
term in the second line of eq.~(\ref{gensusylagr})]. One can think of this
as the ``supersymmetrization" of Figure \ref{fig:gauge}e or
\ref{fig:gauge}f; any of these three vertices may be obtained from any
other (up to a factor of ${\sqrt{2}}$) by replacing two of the particles
by their supersymmetric partners. There is also an interaction in
Figure~\ref{fig:gauge}h which is just like Figure~\ref{fig:gauge}g but
with all arrows reversed, corresponding to the complex conjugate term in
the Lagrangian [the second term in the second line in
eq.~(\ref{gensusylagr})]. Finally in Figure~\ref{fig:gauge}i we have a
scalar quartic interaction vertex [the last term in eq.~(\ref{fdpot})],
which is also determined by the gauge coupling. 

The results of this section can be used as a recipe for constructing the
supersymmetric interactions for any model. In the case of the MSSM, we
already know the gauge group, particle content and the gauge
transformation properties, so it only remains to decide on the
superpotential. This we will do in section
\ref{subsec:mssm.superpotential}. However, first we will revisit the structure of
supersymmetric Lagrangians in section \ref{sec:superfields} using the manifestly
supersymmetric formalism of superspace and superfields, and then describe the
general form of soft supersymmetry breaking terms in section \ref{sec:soft}.

\section{Superspace and superfields}\label{sec:superfields}
\setcounter{figure}{0}
\setcounter{table}{0}
\setcounter{footnote}{2}
\renewcommand{\theequation}{\arabic{section}.\arabic{subsection}.\arabic{equation}}

In this section, the basic ideas of superspace and superfields are covered. These ideas
provide elegant tools for understanding the structure of supersymmetric theories, and
are essential for analyzing and 
communicating ideas about the formal structure of supersymmetry
in the most succinct ways. However, they are also not strictly necessary; 
the discussion given above shows how supersymmetry can be defined and studied
completely without the superspace and superfield notation. The reader who is mainly 
interested in phenomenological aspects of supersymmetric extensions 
of the Standard Model is encouraged to skip this section, especially
on a first reading. The other sections (mostly) do not depend on it.

\subsection{Supercoordinates, general superfields, and 
superspace differentiation and integration\label{subsec:supercoordinates}}
\setcounter{footnote}{2}
\setcounter{equation}{0}

Supersymmetry can be given a geometric interpretation using superspace,
a manifold obtained by adding four fermionic coordinates to the usual bosonic spacetime coordinates
$t,x,y,z$. Points in superspace are labeled by coordinates:
\beq
x^\mu,\> \theta^\alpha,\> \theta^\dagger_{\dot\alpha}.
\label{eq:supercoordinates}
\eeq
Here $\theta^\alpha$ and $\theta^\dagger_{\dot\alpha}$ 
are constant complex anti-commuting two-component spinors with dimension
[mass]$^{-1/2}$. In the superspace formulation, the component fields of 
a supermultiplet
are united into a single superfield, a function of these superspace coordinates. 
We will see below that infinitesimal translations in superspace coincide with the 
global supersymmetry transformations that we have already found in component field language. Superspace thus allows an elegant and 
manifestly invariant definition of supersymmetric field theories.

Differentiation and integration on spaces with anti-commuting coordinates are defined 
by analogy with ordinary commuting variables. Consider first, as a warm-up 
example, a single anti-commuting variable 
$\eta$ (carrying no spinor indices). Because $\eta^2=0$, 
a power series expansion in $\eta$ 
always terminates, and a general function is linear in $\eta$:
\beq
f(\eta) = f_0 + \eta f_1.
\label{eq:deffuneta}
\eeq
Here $f_0$ and $f_1$ may be functions of other commuting or 
anti-commuting variables, but not $\eta$. One of them will be  
anti-commuting (Grassmann-odd), and the other is commuting (Grassmann-even). 
Then define:
\beq
\frac{df}{d\eta} = f_1 .
\label{eq:defdereta}
\eeq
The differential operator $\frac{d}{d\eta}$ anticommutes 
with every Grassmann-odd object,
so that if $\eta'$ is distinct from $\eta$ but also anti-commuting, then 
\beq
\frac{d(\eta'\eta)}{d\eta} = -\frac{d(\eta\eta')}{d\eta} = -\eta'.
\eeq
To define an integration operation with respect to $\eta$, take
\beq
\int d\eta = 0, \qquad\quad
\int d\eta\> \eta = 1,
\label{eq:defderetabasis}
\eeq
and impose linearity.
This defines the Berezin integral \cite{Berezin} 
for Grassmann variables, and gives
\beq
\int d\eta \,f(\eta)  
\> = \> f_1 .
\label{eq:definteta}
\eeq
Comparing eqs.~(\ref{eq:defdereta}) and (\ref{eq:definteta}) shows
the peculiar fact that 
differentiation and integration are the same thing
for an anti-commuting variable. 
The definition eq.~(\ref{eq:defderetabasis}) is motivated by the fact that it implies
translation invariance,
\beq
\int d\eta \> f(\eta + \eta') = \int d\eta \> f(\eta),
\eeq
and the integration by parts formula
\beq
\int d\eta \> \frac{df}{d\eta} = 0,
\label{eq:funtheoeta}
\eeq
in analogy with the fundamental theorem of the calculus 
for ordinary commuting variables. 
The anti-commuting Dirac delta function has the defining property
\beq
\int d\eta\> \delta(\eta - \eta')\, f(\eta) &=& f(\eta'),
\eeq
which leads to
\beq
\delta (\eta - \eta') &=& \eta - \eta'.
\eeq

For superspace with coordinates $x^\mu, \theta^\alpha, \theta^\dagger_{\dot\alpha}$,
any superfield can be expanded in a power series in the anti-commuting variables, with components
that are functions of $x^\mu$. Since there are two 
independent components of $\theta^\alpha$ and 
likewise for $\theta^\dagger_{\dot\alpha}$, the expansion always 
terminates, with each term containing at 
most two $\theta$'s and two $\theta^\dagger$'s.
A general superfield is therefore:
\beq
S(x, \theta, \theta^\dagger) = 
a
+ \theta \xi 
+ \theta^\dagger\hspace{-1pt} \chi^\dagger 
+ \theta\theta b 
+ \thdthd c 
+ \thetasigmamuthetadagger v_\mu 
+ \thdthd  \theta \eta 
+ \theta \theta  \theta^\dagger\hspace{-1pt} \zeta^\dagger 
+ \theta\theta\thdthd d.
\label{eq:gensuperfield}
\eeq
To see that there are no other independent contributions, note the identities
\beq
\theta_\alpha \theta_\beta \>=\> \frac{1}{2} \epsilon_{\alpha\beta} \theta\theta
,
\qquad
\theta^\dagger_{\dot \alpha}\hspace{-1pt} \theta^\dagger_{\dot \beta} 
\>=\> \frac{1}{2} \epsilon_{\dot\beta\dot\alpha} 
\thdthd
,
\qquad
\theta_\alpha 
\theta^\dagger_{\dot\beta} 
\>=\> \frac{1}{2} 
\sigma^\mu_{\alpha\dot\beta}
(\thetasigmamuloweredthetadagger)
,
\eeq
derived from eqs.~(\ref{eq:defepstwo}) and (\ref{eq:feif}). 
These can be used to rewrite any term into the forms given
in eq.~(\ref{eq:gensuperfield}).
Some other identities involving the anti-commuting coordinates that are 
useful in checking results below are:
\beq
(\theta \xi)( \theta\chi) 
&=& -\frac{1}{2}(\theta\theta)( \xi\chi)
,
\qquad\qquad
(\theta^\dagger 
\hspace{-1pt} 
\xi^\dagger)
\, 
(\theta^\dagger 
\hspace{-1pt}
\chi^\dagger)
\>=\> 
-\frac{1}{2}(\thdthd ) (\xi^\dagger\hspace{-1pt}\chi^\dagger)
,
\\
(\theta \xi )(\theta^\dagger
\hspace{-1pt} 
\chi^\dagger )
&=&
\frac{1}{2} (\thetasigmamuthetadagger )
(\xi \sigma_\mu \chi^\dagger ),
\\
\theta^\dagger \sigmabar^\mu \theta &=& -\theta \sigma^\mu \theta^\dagger 
\>=\> 
(\theta^\dagger \sigmabar^\mu \theta)^*,
\\
\theta \sigma^\mu \sigmabar^\nu \theta &=& 
\BDpos\eta^{\mu\nu} \theta\theta ,
\qquad\qquad\qquad
\theta^\dagger\hspace{-1pt} \sigmabar^\mu \sigma^\nu \theta^\dagger \>=\> 
\BDpos\eta^{\mu\nu} \thdthd.
\eeq
These follow from identities already given in section 
\ref{sec:notations}.

The general superfield $S$ could be either commuting or anti-commuting, and could carry 
additional Lorentz vector or spinor indices. For simplicity, let us 
assume for the rest of this subsection that it is Grassmann-even and 
carries no other indices. Then, without further restrictions, the 
components of the general superfield $S$ are 8 bosonic fields $a,b,c,d$ 
and $v_\mu$, and 4 two-component fermionic fields 
$\xi,\chi^\dagger,\eta,\zeta^\dagger$. All of these are complex functions 
of $x^\mu$. The numbers of bosons and fermions do agree (8 complex, or 16 
real, degrees of freedom for each), but there are too many of them to 
match either the chiral or vector supermultiplets encountered in the 
previous section. This means that the general superfield is a reducible
representation of supersymmetry.
In sections \ref{chiralsuperfields} and \ref{subsec:vectorsuperfields} 
below, we will see how chiral and vector superfields are 
obtained by imposing constraints on the general case 
eq.~(\ref{eq:gensuperfield}).

Derivatives with respect to the anti-commuting coordinates are defined by
\beq
\frac{\partial\phantom{x}}{\partial\theta^\alpha} (\theta^\beta) 
\>=\> \delta^\beta_\alpha ,
\qquad
\frac{\partial\phantom{x}}{\partial\theta^\alpha} 
(\theta^\dagger_{\dot\beta}) \>=\> 0 ,
\qquad
\frac{\partial\phantom{x}}{\partial\theta^\dagger_{\dot\alpha}} 
(\theta^\dagger_{\dot\beta}) \>=\> 
\delta^{\dot\alpha}_{\dot\beta} ,
\qquad
\frac{\partial\phantom{x}}{\partial\theta^\dagger_{\dot\alpha}} 
(\theta^{\beta}) \>=\> 0 .
\eeq
Thus, for example, 
$\frac{\partial\phantom{x}}{\partial\theta^\alpha}(\psi\theta) = \psi_\alpha$
and
$\frac{\partial\phantom{x}}{\partial\theta_\alpha}(\psi\theta) = -\psi^\alpha$ 
for an anti-commuting spinor $\psi_\alpha$, and
$\frac{\partial\phantom{x}}{\partial\theta^\alpha}(\theta\theta) = 2 
\theta_\alpha$ and 
$\frac{\partial\phantom{x}}{\partial\theta_\alpha}(\theta\theta) = -2 
\theta^\alpha$.

To integrate over superspace, define
\beq
d^2\theta  
\>=\> -\frac{1}{4} d\theta^\alpha d \theta^\beta \epsilon_{\alpha\beta}
,
\qquad\quad
d^2\theta^\dagger \>=\> -\frac{1}{4} d\theta^\dagger_{\dot\alpha} 
d \theta^\dagger_{\dot\beta} 
\epsilon^{\dot\alpha\dot\beta}
,
\eeq
so that, using eq.~(\ref{eq:defderetabasis}),
\beq
\int d^2\theta \,\theta\theta = 1,\qquad\qquad
\int d^2\theta^\dagger \,\thdthd = 1.
\eeq
Integration of a general superfield therefore just picks out the 
relevant coefficients of $\theta\theta$ and/or 
$\thdthd$
in eq.~(\ref{eq:gensuperfield}):
\beq
\int d^2\theta\> S(x, \theta, \theta^\dagger) &=& 
b(x) 
+ \theta^\dagger\hspace{-1pt} \zeta^\dagger (x)
+ \thdthd d(x) ,
\\
\int d^2\theta^\dagger \> S(x, \theta, \theta^\dagger) &=& 
c(x) 
+ \theta \eta(x) 
+ \theta \theta d(x) ,
\\
\int d^2\theta d^2 \theta^\dagger \> S(x, \theta, \theta^\dagger) &=& d(x) .
\label{eq:heckuvajobtimmy}
\eeq
The Dirac delta functions with respect to integrations
$d^2\theta$ and $d^2\theta^\dagger$ are:
\beq
\delta^{(2)} (\theta - \theta') = (\theta - \theta')(\theta - \theta'),
\qquad\quad
\delta^{(2)} (\theta^\dagger - \theta^{\prime\dagger}) = 
(\theta^\dagger - \theta^{\prime\dagger})
(\theta^\dagger - \theta^{\prime\dagger}) ,
\label{eq:deltathetatheta}
\eeq
so that
\beq
\int d^2\theta \> \delta^{(2)}(\theta) \, S(x,\theta,\theta^\dagger) 
&=& S(x,0,\theta^\dagger) 
\>=\> 
a(x) + \theta^\dagger\hspace{-1pt} \chi^\dagger (x) + 
\thdthd c(x) ,
\phantom{xxxxx}
\\
\int d^2\theta^\dagger \> \delta^{(2)}(\theta^\dagger) \, 
S(x,\theta,\theta^\dagger) 
&=& 
S(x,\theta,0) 
\> = \>
a(x) + \theta \xi(x) + \theta\theta b(x) ,
\\
\int d^2\theta d^2\theta^\dagger \> 
\delta^{(2)}(\theta)\delta^{(2)}(\theta^\dagger) \, 
S(x,\theta,\theta^\dagger) 
&=& 
S(x,0,0) 
\>=\>
a(x).
\eeq
The integrals of total derivatives with respect to the fermionic 
coordinates vanish:
\beq
\int d^2\theta 
\frac{\partial\phantom{x}}{\partial\theta^{\alpha}} \mbox{(anything)} = 0
,
\qquad\quad
\int d^2\theta^\dagger 
\frac{\partial\phantom{x}}{\partial\theta^\dagger_{\dot\alpha}} 
\mbox{(anything)} = 0,
\label{eq:totalthetaderivsvanish}
\eeq
just as in eq.~(\ref{eq:funtheoeta}). This allows for integration by parts.

\subsection{Supersymmetry transformations the superspace 
way\label{subsec:superspacetransformations}}
\setcounter{footnote}{2}
\setcounter{equation}{0}

To formulate supersymmetry transformations in terms of superspace,
define the following differential operators that act on superfields:
\beq
\hat Q_\alpha &=& i \frac{\partial\phantom{x}}{\partial\theta^\alpha} 
- (\sigma^\mu \theta^\dagger)_\alpha \partial_\mu
,
\qquad\qquad
\hat Q^\alpha \>=\> -i\frac{\partial\phantom{x}}{\partial\theta_\alpha} 
+ (\theta^\dagger \sigmabar^\mu)^\alpha \partial_\mu
,
\label{eq:defQhat}
\\
\hat Q^{\dagger\dot\alpha} &=& 
i\frac{\partial\phantom{x}}{\partial\theta^\dagger_{\dot\alpha}} 
- (\sigmabar^\mu \theta)^{\dot\alpha} \partial_\mu
,
\qquad\qquad\>
\hat Q^{\dagger}_{\dot\alpha} \>=\> 
-i\frac{\partial\phantom{x}}{\partial\theta^{\dagger\dot\alpha}} 
+(\theta \sigma^\mu)_{\dot\alpha} \partial_\mu
.
\label{eq:defQdaggerhat}
\eeq
These obey the usual 
product rules for derivatives, but with a minus sign for anti-commuting through
a Grassmann-odd object. For example:
\beq
\hat Q_\alpha (S T) &=& (\hat Q_\alpha S) T + (-1)^S S (\hat Q_\alpha T)
\label{eq:grassmanproductrule}
\eeq
where $S$ and $T$ are any superfields, and $(-1)^S$ is equal to 
$-1$ if $S$ is Grassmann-odd, and $+1$ if $S$ is Grassmann-even.

Then the supersymmetry transformation parameterized by infinitesimal 
$\epsilon$, $\epsilon^\dagger$ for any superfield $S$ is given 
by\footnote{The factor of $\sqrt{2}$ is a convention, not universally 
chosen in the literature, but adopted here in order to avoid $\sqrt{2}$ 
factors in the supersymmetry transformations of section 
\ref{subsec:susylagr.freeWZ} while maintaining consistency.}
\beq
\sqrt{2}\, \delta_{\epsilon} S &=& 
-i
(\epsilon \hat Q + \epsilon^\dagger \hat Q^\dagger) S
\>\,=\>\,
\Bigl (\epsilon^\alpha \frac{\partial\phantom{x}}{\partial\theta^\alpha} 
    + \epsilon^\dagger_{\dot\alpha} 
\frac{\partial\phantom{x}}{\partial\theta^\dagger_{\dot\alpha}}
    + i \bigl [\epsilon \sigma^\mu \theta^\dagger 
       + \epsilon^\dagger \sigmabar^\mu \theta \bigr ] \partial_\mu \Bigr ) S
\label{eq:superspacedefsupertrans}
\\
&=&
S (x^\mu + i \epsilon \sigma^\mu \theta^\dagger 
+ i \epsilon^\dagger \sigmabar^\mu \theta 
, \> \theta\! +\! \epsilon
,\> \theta^\dagger\! +\! \epsilon^\dagger)
-  
S (x^\mu, \, \theta,\, \theta^\dagger)    ,
\label{eq:supertranslation}
\eeq
The last equality follows
by a Taylor expansion to first order in $\epsilon$ and $\epsilon^\dagger$.
Equation (\ref{eq:supertranslation}) shows that a supersymmetry 
transformation can be viewed as a translation in superspace, with:
\beq
\theta^\alpha &\rightarrow& \theta^\alpha + \epsilon^\alpha,
\\
\theta^\dagger_{\dot\alpha} &\rightarrow& \theta^\dagger_{\dot\alpha} 
+ \epsilon^\dagger_{\dot\alpha},
\\
x^\mu &\rightarrow& x^\mu + i \epsilon \sigma^\mu \theta^\dagger 
+ i \epsilon^\dagger \sigmabar^\mu \theta .
\eeq
Since $\hat Q$, $\hat Q^\dagger$ are linear differential operators, the product or 
linear combination of any superfields satisfying 
eq.~(\ref{eq:superspacedefsupertrans}) is again a superfield with the 
same transformation law.

It is instructive and useful to work out the supersymmetry 
transformations of all of the component fields of the general superfield 
eq.~(\ref{eq:gensuperfield}). They are:
\beq
\sqrt{2}\,
\delta_\epsilon a &=& \epsilon \xi 
+ \epsilon^\dagger\hspace{-1pt}\chi^\dagger ,
\label{eq:gensuperfieldtransa}
\\
\sqrt{2}\,
\delta_\epsilon \xi_{\alpha} &=& 2 \epsilon_\alpha b 
  \BDplus (\sigma^\mu \epsilon^\dagger)_\alpha (v_\mu \BDminus i \partial_\mu a)
,
\\
\sqrt{2}\,
\delta_\epsilon \chi^{\dagger{\dot\alpha}} &=& 
  2 \epsilon^{\dagger{\dot\alpha}} c 
  \BDminus (\sigmabar^\mu \epsilon)^{\dot \alpha} (v_\mu \BDplus i \partial_\mu a)
,
\\
\sqrt{2}\,
\delta_\epsilon b &=& \epsilon^\dagger\hspace{-1pt} \zeta^\dagger 
- \frac{i}{2} \epsilon^\dagger 
\sigmabar^\mu \partial_\mu \xi
,
\\
\sqrt{2}\,
\delta_\epsilon c &=& \epsilon \eta - \frac{i}{2} \epsilon 
\sigma^\mu \partial_\mu \chi^\dagger
,
\\
\sqrt{2}\,
\delta_\epsilon v^\mu &=& 
\epsilon \sigma^\mu \zeta^\dagger 
-\epsilon^\dagger \sigmabar^\mu \eta
-\frac{i}{2} \epsilon \sigma^\nu \sigmabar^\mu \partial_\nu \xi
+\frac{i}{2} \epsilon^\dagger \sigmabar^\nu \sigma^\mu \partial_\nu \chi^\dagger
,
\\
\sqrt{2}\,
\delta_\epsilon \eta_{\alpha} &=& 2 \epsilon_\alpha d 
- i (\sigma^\mu \epsilon^\dagger)_\alpha \partial_\mu c 
\BDplus \frac{i}{2} (\sigma^\nu \sigmabar^\mu \epsilon)_\alpha \partial_\mu v_\nu
,
\\
\sqrt{2}\,
\delta_\epsilon \zeta^{\dagger{\dot\alpha}} &=& 
2 \epsilon^{\dagger{\dot \alpha}} d 
- i (\sigmabar^\mu \epsilon)^{\dot\alpha} \partial_\mu b 
\BDminus \frac{i}{2} (\sigmabar^\nu \sigma^\mu \epsilon^\dagger )^{\dot \alpha} 
\partial_\mu v_\nu 
,
\\
\sqrt{2}\,
\delta_\epsilon d &=& 
-\frac{i}{2}
\epsilon^\dagger \sigmabar^\mu \partial_\mu \eta
-\frac{i}{2}
\epsilon \sigma^\mu \partial_\mu \zeta^\dagger .
\label{eq:gensuperfieldtransd}
\eeq
Note that since the terms on the right-hand sides all have exactly one $\epsilon$
or one $\epsilon^\dagger$, 
boson fields are always transformed into fermions and vice versa.

It is probably not obvious yet that the supersymmetry transformations as 
just defined coincide with those found in section \ref{sec:susylagr}. 
This will become clear below when we discuss the specific form of chiral 
and vector superfields and the Lagrangians that govern their dynamics. 
Meanwhile, however, we can compute the anticommutators of $\hat Q$, $\hat 
Q^\dagger$ from eqs.~(\ref{eq:defQhat}), (\ref{eq:defQdaggerhat}), with 
the results:
\beq
\Bigl \lbrace \hat Q_\alpha,\, \hat Q^\dagger_{\dot\beta} \Bigr \rbrace
&=& 
2 i \sigma^\mu_{\alpha\dot\beta}\partial_\mu
\>\,=\,\> \BDpos 2 \sigma^\mu_{\alpha\dot\beta} \hat P_\mu,
\label{eq:diffopSUSYalgebra1}
\\
\Bigl \lbrace \hat Q_\alpha,\, \hat Q_{\beta} \Bigr \rbrace &=& 0,\qquad\quad
\Bigl \lbrace \hat Q^\dagger_{\dot \alpha},\, \hat Q^\dagger_{\dot\beta} 
\Bigr \rbrace \>=\> 0.
\label{eq:diffopSUSYalgebra2}
\eeq
Here, the differential operator generating spacetime translations is   
\beq
\hat P_\mu = \BDpos i \partial_\mu.
\eeq
Eqs.~(\ref{eq:diffopSUSYalgebra1})-(\ref{eq:diffopSUSYalgebra2}) have the same 
form as the supersymmetry algebra given in 
eqs.~(\ref{nonschsusyalg1}), (\ref{nonschsusyalg2}). 

It is important to keep in 
mind the conceptual distinction between the unhatted objects 
$Q_\alpha, Q^\dagger_{\dot\alpha}, 
P^\mu$ appearing in section \ref{subsec:susylagr.freeWZ}, 
which are operators acting on the Hilbert space of quantum states,
and the corresponding hatted objects 
$\hat Q_\alpha, \hat Q^\dagger_{\dot\alpha},
\hat P^\mu$, which are differential operators acting on functions in 
superspace. For any superfield
quantum mechanical operator $X$ in the Heisenberg picture, 
the two kinds of operations are related by 
\beq
\bigl [X,\, \epsilon Q + \epsilon^\dagger\hspace{-1pt} Q^\dagger \bigr ]
&=&
(\epsilon \hat Q + \epsilon^\dagger\hspace{-1pt} \hat Q^\dagger) X ,
\\
\bigl [X ,\, P_\mu \bigr ] &=& \hat P_\mu X.
\eeq

\subsection{Chiral covariant derivatives\label{subsec:supercovariantderivatives}}
\setcounter{equation}{0}
\setcounter{footnote}{2}

To construct Lagrangians in superspace, we will later want to use 
derivatives with respect to the anti-commuting coordinates, just as 
ordinary Lagrangians are built using spacetime derivatives 
$\partial_\mu$. We will also use such derivatives to impose constraints 
on the general superfield in a way consistent with the supersymmetry 
transformations. However, $
\partial
/
\partial\theta^\alpha
$ is not appropriate for 
this purpose, because it is not supersymmetric covariant:
\beq
\delta_\epsilon \left (\frac{\partial S}{\partial \theta^\alpha} \right )
\>\not=\>
\frac{\partial\phantom{x}}{\partial\theta^\alpha} (\delta_\epsilon S),
\eeq
and similarly for 
$
\partial
/
\partial\theta^\dagger_{\dot\alpha}
$. This means
that derivatives of a superfield with respect to 
$\theta_\alpha$ or $\theta^\dagger_{\dot\alpha}$ are not superfields; they do not transform the right way.
To fix this, it is useful to define the chiral covariant derivatives:
\beq
D_\alpha &=& \frac{\partial\phantom{x}}{\partial\theta^\alpha} 
-i (\sigma^\mu \theta^\dagger)_\alpha \partial_\mu
,
\qquad\qquad
D^\alpha \>=\> -\frac{\partial\phantom{x}}{\partial\theta_\alpha} 
+i (\theta^\dagger \sigmabar^\mu)^\alpha \partial_\mu
.
\eeq
For a Grassmann-even superfield $S$, one can then define the anti-chiral covariant derivative to obey:
\beq
\Dcon_{\dot\alpha} S^* \equiv (D_\alpha S)^*,
\label{eq:defDcon}
\eeq
which implies
\beq
\Dcon^{\dot\alpha} &=& 
\frac{\partial\phantom{x}}{\partial\theta^\dagger_{\dot\alpha}} 
-i (\sigmabar^\mu \theta)^{\dot\alpha} \partial_\mu
,
\qquad\qquad\>
\Dcon_{\dot\alpha} \>=\> 
-\frac{\partial\phantom{x}}{\partial\theta^{\dagger\dot\alpha}} 
+i (\theta \sigma^\mu)_{\dot\alpha} \partial_\mu
.
\eeq
One may now check that
\beq
\Bigl \lbrace \hat Q_\alpha ,\, D_\beta \Bigr \rbrace \>=\>
\Bigl \lbrace \hat Q^\dagger_{\dot\alpha} ,\, D_\beta \Bigr \rbrace \>=\>
\Bigl \lbrace \hat Q_\alpha ,\, \Dcon_{\dot\beta} \Bigr \rbrace \>=\>
\Bigl \lbrace \hat Q^\dagger_{\dot\alpha} ,\, \Dcon_{\dot\beta} 
\Bigr \rbrace \>=\> 0 .
\label{eq:QDanticommute}
\eeq
Using the supersymmetry transformation definition of eq.~(\ref{eq:superspacedefsupertrans}), it follows that
\beq
\delta_\epsilon \left ( D_\alpha S \right )
= D_\alpha\left ( \delta_\epsilon  S \right ),
\qquad
\qquad
\delta_\epsilon \left ( \Dcon_{\dot\alpha} S \right )
= \Dcon_{\dot\alpha} \left ( \delta_\epsilon  S \right )
.
\label{eq:Dsupercon}
\eeq
Thus the derivatives $D_\alpha$ and $\Dcon_{\dot\alpha}$ are indeed 
supersymmetric covariant; acting on superfields, they return superfields. 
This crucial property makes them useful both for defining constraints on superfields in a 
covariant way, and for defining superspace Lagrangians involving 
anti-commuting spinor coordinate derivatives. These derivatives are linear differential operators, obeying product rules
exactly analogous to eq.~(\ref{eq:grassmanproductrule}).

The chiral and anti-chiral covariant derivatives also can be checked to 
satisfy the useful 
anticommutation identities:
\beq
\Bigl \lbrace D_\alpha,\, \Dcon_{\dot\beta} \Bigr \rbrace  &=& 
2 i \sigma^\mu_{\alpha\dot\beta}\partial_\mu,
\label{eq:diffopderivs}
\\
\Bigl \lbrace D_\alpha,\, D_{\beta} \Bigr \rbrace &=& 0,\qquad\quad
\Bigl \lbrace \Dcon_{\dot \alpha},\,
              \Dcon_{\dot\beta} \Bigr \rbrace \>=\> 0.
\label{eq:diffopderivs2}
\eeq
This has exactly the same form as the supersymmetry algebra in 
eqs.~(\ref{eq:diffopSUSYalgebra1}) and (\ref{eq:diffopSUSYalgebra2}),
but $D, \Dcon$ should not be 
confused with 
the differential operators for supersymmetry transformations,
$\hat Q, \hat Q^\dagger$. The operators $D, \Dcon$ do not represent a second supersymmetry.

The reader might be wondering why we use an overline notation for $\Dcon$, but a dagger
for $\hat Q^\dagger$. The reason is that the dagger and the overline 
denote different kinds of conjugation. The dagger
on $\hat Q$ represents Hermitian conjugation in the same sense that 
$\hat P = -i \partial_\mu$ is an Hermitian differential operator on an 
inner product space, but the overline on $\Dcon$ represents 
complex conjugation in the same sense that
$\partial_\mu$ is a real differential operator, with 
$(\partial_\mu \phi)^* = \partial_\mu \phi^*$. Recall that if we 
define the inner product on the space of functions of $x^\mu$ by:
\beq
\langle \psi | \phi \rangle = \int \!d^4x\> \psi^*(x) \phi(x),
\eeq
then, using integration by parts,
\beq 
\langle \psi| \hat P \phi \rangle = \left (\langle \phi| \hat P \psi \rangle\right )^*
\eeq
Similarly, the dagger on the differential operator $\hat Q^\dagger$ denotes Hermitian
conjugation with respect to the inner product defined by integration of 
complex superfields over superspace. To see this, define, for any 
two classical superfields
$S(x,\theta,\theta^\dagger)$ and $T(x,\theta,\theta^\dagger)$, the inner product:
\beq
\langle T | S \rangle = \int d^4x\int\! d^2\theta\! \int\! d^2\theta^\dagger\> T^* S.
\eeq
Now one finds, by integration by parts over superspace, that with the definitions
in eqs.~(\ref{eq:defQhat}) and (\ref{eq:defQdaggerhat}),
\beq
\langle T | \hat Q^\dagger_{\dot\alpha} S \rangle = 
\left ( \langle S |\hat Q_\alpha T \rangle \right )^* .
\label{eq:TSQcon}
\eeq
In contrast, the definition of $\Dcon$ in eq.~(\ref{eq:defDcon}) is analogous
to the equation $(\partial_\mu \phi)^* = \partial_\mu \phi^*$ for functions 
on ordinary spacetime; in that sense, $\partial_\mu$ is a real differential operator,
and similarly $\Dcon_{\dot\alpha}$ is the conjugate of $D_\alpha$. 
This is more than just notation; if we
defined $D^\dagger_{\dot\alpha}$ from $D_\alpha$ in a way analogous to 
eq.~(\ref{eq:TSQcon}), then one can check that 
it would not be equal to $\Dcon_{\dot\alpha}$ as defined above. 
Note that the dagger on the quantum field theory operator $Q^\dagger_{\alpha}$ 
(without the hat) represents yet another sort of Hermitian conjugation, 
in the quantum mechanics Hilbert space sense. 
 
It is useful to note that, using eq.~(\ref{eq:totalthetaderivsvanish}), 
\beq
\int d^2\theta \,D_\alpha \mbox{(anything)} \qquad\mbox{and}\qquad
\int d^2\theta^\dagger \,\Dcon_{\dot\alpha} \mbox{(anything)}
\eeq
are each total derivatives with respect to $x^\mu$. This enables integration by parts 
in superspace
of Lagrangian terms with respect to either $D_\alpha$ or $\Dcon_{\dot\alpha}$. Another useful fact is that
acting three consecutive times with either of $D_\alpha$ or $\Dcon_{\dot\alpha}$ always produces a vanishing result:
\beq
D_\alpha D_\beta D_\gamma \mbox{(anything)} = 0 \qquad\mbox{and}\qquad
\Dcon_{\dot\alpha} \Dcon_{\dot\beta} \Dcon_{\dot\gamma} \mbox{(anything)} = 0.
\label{eq:DDDeq0}
\eeq
This follows from eq.~(\ref{eq:diffopderivs2}), and is true essentially 
because the spinor indices on the anti-commuting derivatives can only have two values.

\subsection{Chiral superfields\label{chiralsuperfields}}
\setcounter{equation}{0}
\setcounter{footnote}{2}

To describe a chiral supermultiplet, 
consider the superfield $\Phi(x,\theta,\theta^\dagger)$ 
obtained by imposing the constraint
\beq
\Dcon_{\dot \alpha} \Phi &=& 0.
\label{eq:leftchiralsuperfieldconstraint}
\eeq
A field satisfying this constraint is said to be a chiral (or 
left-chiral) superfield, and its complex conjugate $\Phi^*$ is called 
anti-chiral (or right-chiral) and satisfies
\beq
D_{\alpha} \Phi^* &=& 0.
\label{eq:anti-chiralsuperfieldconstraint}
\eeq
These constraints are 
consistent with the transformation rule for general superfields 
because of eq.~(\ref{eq:Dsupercon}).

To solve the constraint eq.~(\ref{eq:leftchiralsuperfieldconstraint}) in 
general, it is convenient to define
\beq
y^\mu \equiv x^\mu \BDminus i \thetasigmamuthetadagger ,\>
\label{eq:defineycoord}
\eeq
and change coordinates on superspace to the set:
\beq
y^\mu,\>\theta^\alpha,\>\theta^\dagger_{\dot\alpha}.
\eeq
In terms of these variables, the chiral covariant derivatives have the 
representation:
\beq
D_\alpha &=& \frac{\partial\phantom{x}}{\partial\theta^\alpha} 
-2i (\sigma^\mu \theta^\dagger)_\alpha 
\frac{\partial\phantom{x}}{\partial y^\mu} ,
\qquad\qquad
D^\alpha \>=\> -\frac{\partial\phantom{x}}{\partial\theta_\alpha} 
+2i (\theta^\dagger \sigmabar^\mu)^\alpha 
\frac{\partial\phantom{x}}{\partial y^\mu} ,
\phantom{xxxx}
\label{eq:Dinyrep}
\\
\Dcon^{\dot\alpha} &=& 
\frac{\partial\phantom{x}}{\partial\theta^\dagger_{\dot\alpha}} , 
\qquad\qquad\qquad\qquad\qquad\>\>\>\>\>
\Dcon_{\dot\alpha} \>=\> 
-\frac{\partial\phantom{x}}{\partial\theta^{\dagger\dot\alpha}} . 
\label{eq:Ddaggerinyrep}
\eeq
Equation (\ref{eq:Ddaggerinyrep}) makes it clear that the chiral 
superfield 
constraint eq.~(\ref{eq:leftchiralsuperfieldconstraint}) is solved by any 
function of $y^\mu$ and $\theta$, as long as it is not a function of $\theta^\dagger$. 
Therefore, one can expand:
\beq
\Phi \,=\,
\phi(y) + \sqrt{2}\theta \psi(y) + \theta\theta F(y) ,
\label{eq:Phiinyrep}
\eeq
and similarly
\beq
\Phi^* \,=\, 
\phi^*(y^*) + \sqrt{2}\theta^\dagger\hspace{-1pt} \psi^\dagger(y^*) 
+ \thdthd F^*(y^*) .
\label{eq:Phistarinyrep}
\eeq
The factors of $\sqrt{2}$ are conventional, and
$y^{\mu *} = x^\mu \BDplus i \thetasigmamuthetadagger $.
The chiral covariant derivatives in terms of the coordinates 
$(y^*,\theta,\theta^\dagger)$ are also sometimes useful:
\beq
D_\alpha &=& \frac{\partial\phantom{x}}{\partial\theta^\alpha} 
,
\qquad\qquad
\qquad\qquad\>
\qquad\quad\>
D^\alpha \>=\> -\frac{\partial\phantom{x}}{\partial\theta_\alpha} 
,
\\
\Dcon^{\dot\alpha} &=& 
\frac{\partial\phantom{x}}{\partial\theta^\dagger_{\dot\alpha}} 
-2i (\sigmabar^\mu \theta)^{\dot\alpha} 
\frac{\partial}{\partial y^{\mu *}}
,
\qquad\qquad\>
\Dcon_{\dot\alpha} \>=\> 
-\frac{\partial\phantom{x}}{\partial\theta^{\dagger\dot\alpha}} 
+2i (\theta \sigma^\mu)_{\dot\alpha}
\frac{\partial}{\partial y^{\mu *}}
.
\eeq

According to eq.~(\ref{eq:Phiinyrep}),
the chiral superfield independent degrees of freedom are a complex scalar $\phi$, 
a two-component fermion $\psi$,
and an auxiliary field $F$, just as found in
subsection \ref{subsec:susylagr.freeWZ}. 
If $\Phi$ is a free fundamental chiral superfield, 
then assigning it dimension [mass]$^1$
gives the canonical mass dimensions to the component fields, because
$\theta$ and $\theta^\dagger$ have dimension [mass]$^{-1/2}$. 
Rewriting the chiral superfields in terms of the 
original coordinates $x, \theta, \theta^\dagger$, by expanding in a power series in the 
anti-commuting coordinates, gives 
\beq
\Phi \!\!&=&\!\!
\phi(x) 
\BDminus i \thetasigmamuthetadagger \partial_\mu \phi(x)
\BDminus \frac{1}{4} \theta\theta\thdthd \partial_\mu \partial^\mu 
\phi(x)
+ \sqrt{2}\theta \psi(x) 
\nonumber \\ &&
- \frac{i}{\sqrt{2}} \theta\theta 
\theta^\dagger \sigmabar^\mu \partial_\mu \psi(x)
+ \theta\theta F(x),
\label{eq:Phiinxrep}
\\
\Phi^*
\!\!&=&\!\!
\phi^*(x) 
\BDplus i \thetasigmamuthetadagger  \partial_\mu \phi^*(x)
\BDminus \frac{1}{4} \theta\theta\thdthd \partial_\mu \partial^\mu 
\phi^*(x)
+ \sqrt{2}\theta^\dagger\hspace{-1pt} \psi^\dagger(x) 
\nonumber \\ &&
- \frac{i}{\sqrt{2}} \thdthd \theta \sigma^\mu \partial_\mu 
\psi^\dagger(x)
+ \thdthd F^*(x).
\label{eq:Phistarinxrep}
\eeq
Depending on the situation, eqs.~(\ref{eq:Phiinyrep})-(\ref{eq:Phistarinyrep}) are sometimes 
a more convenient representation than 
eqs.~(\ref{eq:Phiinxrep})-(\ref{eq:Phistarinxrep}).

By comparing the general superfield case 
eq.~(\ref{eq:gensuperfield}) to eq.~(\ref{eq:Phiinxrep}), we see that 
the latter can be 
obtained from the former by identifying component fields:
\beq
&& a = \phi,
\qquad\quad
\xi_\alpha = \sqrt{2} \psi_\alpha,
\qquad\qquad
b = F,
\label{eq:specSPhione}
\\
&& 
\chi^{\dagger\dot\alpha}  =0,
\qquad\qquad
c = 0,
\qquad\qquad
v_\mu = \BDneg i \partial_\mu \phi,
\qquad\qquad
\eta_\alpha = 0,
\phantom{xxx}
\\ 
&&
\zeta^{\dagger\dot\alpha} = 
-\frac{i}{\sqrt{2}} (\sigmabar^\mu \partial_\mu \psi)^{\dot\alpha}  ,
\qquad\qquad
d = \BDneg \frac{1}{4} \partial_\mu \partial^\mu \phi.
\label{eq:specSPhitwo}
\eeq
It is now straightforward to obtain the supersymmetry transformation 
laws for the component fields of 
$\Phi$, either by using 
$\sqrt{2} \delta_\epsilon \Phi = -i
(\epsilon \hat Q + \epsilon^\dagger \hat 
Q^\dagger) \Phi$, or by plugging 
eqs.~(\ref{eq:specSPhione})-(\ref{eq:specSPhitwo}) into the 
results for a general superfield, 
eqs.~(\ref{eq:gensuperfieldtransa})-(\ref{eq:gensuperfieldtransd}). 
The results are
\beq
\delta_\epsilon \phi &=& \epsilon \psi,\\
\delta_\epsilon \psi_\alpha &=& 
-i (\sigma^\mu \epsilon^\dagger)_\alpha \partial_\mu \phi + 
\epsilon_\alpha F
,
\\ 
\delta_\epsilon F &=& -i \epsilon^\dagger \sigmabar^\mu \partial_\mu \psi,
\label{eq:runningupthathill}
\eeq
in agreement with eqs.~(\ref{phitrans}), (\ref{Ftrans}), (\ref{fermiontrans}).

One way to construct a chiral superfield (or an anti-chiral superfield) is 
\beq
\Phi \,=\, 
\Dcon\Dcon S
\,\equiv\, 
\Dcon_{\dot\alpha} \Dcon^{\dot \alpha} S,
\qquad\qquad
\Phi^* \,=\, DDS^* \,\equiv\, D^\alpha D_\alpha S^*,
\label{eq:chiralfromgeneral}
\eeq 
where $S$ is any 
general superfield. The fact that these are chiral and anti-chiral, respectively, 
follows immediately from eq.~(\ref{eq:DDDeq0}).
The converse is also true; 
for every chiral superfield $\Phi$, one can find a superfield $S$ such that 
eq.~(\ref{eq:chiralfromgeneral}) is true. 

Another way to build a chiral superfield is as 
a function $W(\Phi_i)$ of other chiral superfields $\Phi_i$ 
but not anti-chiral superfields; in other words, $W$ is holomorphic in 
chiral superfields treated as complex variables. 
This fact follows immediately from 
the linearity and product rule properties of the differential operator 
$\Dcon_{\dot \alpha}$ appearing in the constraint 
eq.~(\ref{eq:leftchiralsuperfieldconstraint}). It will be useful below 
for constructing superspace Lagrangians.

\subsection{Vector superfields\label{subsec:vectorsuperfields}}
\setcounter{equation}{0}
\setcounter{footnote}{2}

A vector (or real) superfield $V$ is obtained by
imposing the constraint $V = V^*$. 
This is equivalent to imposing the following constraints 
on the components of the
general superfield eq.~(\ref{eq:gensuperfield}):
\beq
a \> = \> a^*,
\qquad
\chi^\dagger \>=\> \xi^\dagger,
\qquad
c \> = \> b^*,
\qquad
v_\mu \>=\> v_\mu^*,
\qquad
\zeta^\dagger \>=\> \eta^\dagger,
\qquad
d = d^*.
\label{eq:constrainvectorfromgeneral}
\eeq
It is also convenient and traditional to define:
\beq
\eta_\alpha \>=\> \lambda_\alpha - \frac{i}{2} (\sigma^\mu \partial_\mu \xi^\dagger)_\alpha ,
\qquad
v_\mu = A_\mu,
\qquad
d = \frac{1}{2} D \BDminus \frac{1}{4} \partial_\mu \partial^\mu a.
\label{eq:redefinevectorfromgeneral}
\eeq
The component expansion of the vector superfield is then
\beq
V(x,\theta,\theta^\dagger) &=& 
a
+ \theta \xi 
+ \theta^\dagger\hspace{-1pt} \xi^\dagger 
+ \theta\theta b 
+ \thdthd b^* 
+ \thetasigmamuthetadagger  A_\mu 
+ \thdthd \theta 
  (\lambda - \frac{i}{2} \sigma^\mu \partial_\mu \xi^\dagger) 
\nonumber
\\
&&
+ \theta \theta  \theta^\dagger\hspace{-1pt} 
  (\lambda^\dagger - \frac{i}{2} \sigmabar^\mu \partial_\mu \xi)   
+ \theta\theta\thdthd 
  ( \frac{1}{2} D \BDminus \frac{1}{4} \partial_\mu \partial^\mu a).
\label{eq:vectorsuperfieldexpansion}
\eeq
The supersymmetry transformations of these components can be obtained either from
$\sqrt{2} \delta_\epsilon V = -i(\epsilon \hat Q + \epsilon^\dagger \hat Q^\dagger) V$,
or by plugging 
eqs.~(\ref{eq:constrainvectorfromgeneral})-(\ref{eq:redefinevectorfromgeneral}) 
into the results for
a general superfield,
eqs.~(\ref{eq:gensuperfieldtransa})-(\ref{eq:gensuperfieldtransd}).
The results are:
\beq
\sqrt{2}\,
\delta_\epsilon a &=& \epsilon \xi + \epsilon^\dagger\hspace{-1pt}\xi^\dagger
\\
\sqrt{2}\,
\delta_\epsilon \xi_{\alpha} &=& 2 \epsilon_\alpha b
\BDplus (\sigma^\mu \epsilon^\dagger)_\alpha (A_\mu \BDminus i \partial_\mu a)
,
\\
\sqrt{2}\,
\delta_\epsilon b &=& \epsilon^\dagger\hspace{-1pt} \lambda^\dagger 
- i \epsilon^\dagger \sigmabar^\mu \partial_\mu \xi
,
\\
\sqrt{2}\,
\delta_\epsilon A^\mu &=& 
\BDneg i \epsilon \partial^\mu \xi 
\BDplus i \epsilon^\dagger \partial^\mu \xi^\dagger
+ \epsilon \sigma^\mu \lambda^\dagger 
- \epsilon^\dagger \sigmabar^\mu \lambda ,
\\
\sqrt{2}\,
\delta_\epsilon \lambda_\alpha &=&
\epsilon_\alpha D \BDminus \frac{i}{2} (\sigma^\mu \sigmabar^\nu \epsilon)_\alpha
(\partial_\mu A_\nu - \partial_\nu A_\mu)
,
\\
\sqrt{2}\,
\delta_\epsilon D &=&
- i \epsilon \sigma^\mu \partial_\mu \lambda^\dagger
- i \epsilon^\dagger \sigmabar^\mu \partial_\mu \lambda
\label{eq:sqrttwodeltaD}
\eeq

A superfield cannot be both chiral and real at the same time, 
unless it is identically constant (i.e., independent of $x^\mu$, $\theta$, 
and $\theta^\dagger$). This follows from 
eqs.~(\ref{eq:specSPhione})-(\ref{eq:specSPhitwo}), and
(\ref{eq:constrainvectorfromgeneral}).
However, if $\Phi$ is a chiral superfield, then $\Phi+\Phi^*$ and
$i (\Phi - \Phi^*)$ and $\Phi \Phi^*$ are all real (vector) superfields.

As the notation chosen in 
eq.~(\ref{eq:vectorsuperfieldexpansion}) suggests,
a vector superfield that is used to represent a gauge supermultiplet contains 
gauge boson, gaugino, and gauge auxiliary fields $A^\mu$, $\lambda$, $D$
as components. (Such a vector superfield $V$ must be dimensionless in order for the component fields to have the canonical mass dimensions.) 
However, there are other component fields in $V$ that did 
not appear in sections \ref{subsec:susylagr.gauge} and 
\ref{subsec:susylagr.gaugeinter}.
They are: a real scalar $a$, a two-component fermion $\xi$, 
and a complex scalar $b$, with mass dimensions
respectively 0, $1/2$, and 1. These are additional auxiliary fields, 
which can be ``supergauged" away.
To see this, suppose $V$ is the vector superfield for a $U(1)$ 
gauge symmetry, and consider the ``supergauge transformation":
\beq
V &\rightarrow & V + i (\Omega^* - \Omega),
\label{eq:u1superfieldgt}
\eeq
where $\Omega$ is a chiral superfield gauge transformation parameter,
$\Omega = \phi + \sqrt{2}\theta \psi + \theta\theta F+\ldots$. 
In components, this transformation is
\beq
a &\rightarrow& a + i (\phi^* - \phi),
\\
\xi_\alpha &\rightarrow& \xi_\alpha - i\sqrt{2} \psi_\alpha ,
\\
b &\rightarrow& b - iF ,
\\
A_\mu &\rightarrow& A_\mu \BDminus \partial_\mu (\phi + \phi^*) ,
\label{eq:AgaugetransfromLambda}
\\
\lambda_\alpha &\rightarrow& \lambda_\alpha ,
\\
D &\rightarrow& D.
\label{eq:DgaugetransfromLambda}
\eeq
Equation (\ref{eq:AgaugetransfromLambda}) 
shows that eq.~(\ref{eq:u1superfieldgt}) provides the vector 
boson field with the usual gauge transformation, with parameter 2Re$(\phi)$.
By requiring the gauge transformation to take a supersymmetric 
form, it follows that 
appropriate independent choices of 
Im$(\phi)$, $\psi_\alpha$, and $F$ can also change $a$, $\xi_\alpha$,
and $b$ arbitrarily.
Thus the supergauge transformation 
eq.~(\ref{eq:u1superfieldgt}) has ordinary gauge transformations as a 
special case. 

In particular, supergauge transformations can eliminate
the auxiliary fields $a$, $\xi_\alpha$,
and $b$ completely.
A superspace Lagrangian for a vector superfield
must be invariant under the 
supergauge transformation
eq.~(\ref{eq:u1superfieldgt}) in the Abelian case, or a 
suitable generalization given below for the non-Abelian case. After making a 
supergauge transformation to eliminate $a,\xi$, and $b$, 
the vector superfield is said to be in 
Wess-Zumino gauge, and is simply given by
\beq
V_{{\rm WZ}\>{\rm gauge}} \>=\> 
\thetasigmamuthetadagger  A_\mu 
+ \thdthd \theta \lambda 
+ \theta \theta  \theta^\dagger\hspace{-1pt} \lambda^\dagger  
+ \frac{1}{2} \theta\theta\thdthd D.
\label{eq:WZgauge}
\eeq
The restriction of the vector superfield to Wess-Zumino gauge 
is not consistent with the linear superspace version of supersymmetry 
transformations. This is because 
$\sqrt{2} \delta_{\epsilon} (V_{{\rm WZ}\>{\rm gauge}})$ 
contains 
$\BDneg \theta^\dagger \sigmabar^\mu \epsilon A_\mu \BDplus \theta \sigma^\mu \epsilon^\dagger A_\mu
+ \theta\theta \epsilon^\dagger \lambda^\dagger +
\theta^\dagger\theta^\dagger \epsilon \lambda$, 
and so the supersymmetry transformation of the Wess-Zumino gauge
vector superfield is not in Wess-Zumino gauge.
However, a supergauge transformation can always restore
$\delta_{\epsilon} (V_{{\rm WZ}\>{\rm gauge}})$ to Wess-Zumino gauge.
Adopting Wess-Zumino gauge is equivalent to partially fixing 
the supergauge, while still maintaining the full freedom 
to do ordinary gauge transformations.

\subsection{How to make a Lagrangian in superspace\label{superspacelagr}}
\setcounter{equation}{0}
\setcounter{footnote}{2}

So far, we have been concerned with the structural features of fields in 
superspace. We now turn to the dynamical issue of how to construct 
manifestly supersymmetric actions. A key observation is that the integral 
of any superfield over all of superspace is automatically 
invariant:
\beq
\delta_\epsilon A = 0,\quad\mbox{for}\quad
A = \int d^4x \int d^2\theta d^2 \theta^\dagger \> S(x, \theta, \theta^\dagger). 
\label{eq:defsuperspaceaction}
\eeq
This follows immediately from the fact that $\hat Q$ and $\hat Q^\dagger$ 
as defined in eqs.~(\ref{eq:defQhat}), (\ref{eq:defQdaggerhat}) are sums 
of total derivatives with respect to the superspace coordinates 
$x^\mu,\theta,\theta^\dagger$, so that $(\epsilon \hat Q + 
\epsilon^\dagger \hat Q^\dagger)S$ vanishes upon integration. As a check, 
eq.~(\ref{eq:gensuperfieldtransd}) shows that the 
$\theta\theta\thdthd$ component of a 
superfield transforms into a total spacetime derivative.

Therefore, the action governing the dynamics of a theory can have 
contributions of the form of eq.~(\ref{eq:defsuperspaceaction}), with 
reality of the action demanding that $S$ is some real (vector) superfield 
$V$. From eq.~(\ref{eq:supertranslation}), we see that the principle of 
global supersymmetric invariance is embodied in the requirement that the 
action should be an integral over superspace which is unchanged under 
rigid translations of the superspace coordinates. To obtain the 
Lagrangian density ${\cal L}(x)$, one integrates over only the fermionic 
coordinates. This is often written in the notation:
\beq
[V]_D \,\equiv\, \int d^2\theta d^2\theta^\dagger\> V(x,\theta,\theta^\dagger) 
\,=\, 
V(x,\theta,\theta^\dagger) \Bigl |_{\theta\theta\thdthd}
\,=\,
\frac{1}{2} D \BDminus \frac{1}{4} \partial_\mu \partial^\mu a
\label{eq:DtermLag}
\eeq
using eq.~(\ref{eq:heckuvajobtimmy}) and the form of $V$ 
in eq.~(\ref{eq:vectorsuperfieldexpansion}) for the last equality. This is 
referred to as a 
$D$-term contribution to the Lagrangian
(note that the $\partial_\mu \partial^\mu a$ part will 
vanish upon integration $\int d^4 x$). 

Another type of contribution to the action can be inferred from the fact 
that the $F$-term of a chiral superfield also transforms into a total 
derivative under a supersymmetry transformation, see 
eq.~(\ref{eq:runningupthathill}). 
This implies that 
one can have a 
contribution to the Lagrangian density of the form
\beq
[\Phi]_F \,\equiv\, \Phi \Bigl |_{\theta\theta} \,=\, 
\int d^2\theta\, \Phi \Bigl |_{\theta^\dagger =0} \,=\, 
\int d^2\theta d^2\theta^\dagger\,
\delta^{(2)}(\theta^\dagger)
\,
\Phi \>=\>F,
\label{eq:PhiF}
\eeq
using the form of $\Phi$ in eq.~(\ref{eq:Phiinxrep}) for the last equality.
This satisfies $\delta_{\epsilon} (\int d^4 x [\Phi]_F) = 0$.
The $F$-term of a chiral superfield 
is complex in general, but 
the action must be real, which can be ensured if  
this type of contribution to the Lagrangian is accompanied
by its complex conjugate:
\beq
[\Phi]_F + {\rm c.c.} \>=\> 
\int d^2\theta d^2\theta^\dagger\,
\left [ 
\delta^{(2)}(\theta^\dagger) 
\,
\Phi 
+
\delta^{(2)}(\theta) 
\,
\Phi^* 
\right ].
\label{eq:FtermLag}
\eeq
Note that the identification of the $F$-term component of a chiral superfield is 
the same in the $(x^\mu,\theta,\theta^\dagger)$ and 
$(y^\mu,\theta,\theta^\dagger)$ coordinates, in the sense that in both cases, one 
simply isolates the $\theta\theta$ component. This follows because the 
difference between $x^\mu$ and $y^\mu$ is higher order in 
$\theta^\dagger$. It is a useful trick, because many 
calculations involving chiral superfields are 
simpler to carry out in terms of $y^\mu$.

Another possible try would be to take the $D$-term of a chiral superfield. However,
this is a waste of time, because
\beq
[\Phi]_D 
\,=\, \int d^2\theta d^2\theta^\dagger\> \Phi 
\,=\, 
\Phi \Bigl |_{\theta\theta\thdthd}
\,=\,
\frac{1}{4} \partial_\mu \partial^\mu \phi,
\label{eq:tryPhiD}
\eeq
where the last equality follows from eq.~(\ref{eq:Phiinxrep}), and
$\phi$ is the scalar component of $\Phi$. Equation (\ref{eq:tryPhiD})
is a total derivative, so adding it (and its complex conjugate) to the Lagrangian
density has no effect.

Therefore, the two ways of making a supersymmetric Lagrangian are to take the
$D$-term component of a real superfield, and to take the $F$-term component of
a chiral superfield, plus the complex conjugate.
When building a Lagrangian, the real superfield $V$ used in 
eq.~(\ref{eq:DtermLag}) and the chiral superfield $\Phi$ used in 
eq.~(\ref{eq:FtermLag}) are usually composites, built out of more 
fundamental superfields. However, contributions from fundamental 
fields $V$ and $\Phi$ are allowed, when $V$ is the vector superfield for 
an Abelian gauge symmetry and when $\Phi$ is a singlet under all 
symmetries.

It is always possible to rewrite a $D$ term contribution to a Lagrangian
as an $F$ term contribution, by the trick of noticing that
\beq
\Dcon\Dcon (\theta^\dagger \theta^\dagger) &=& 
DD(\theta\theta) \>\,=\,\> -4,
\label{eq:DDthetatheta}
\eeq
and using the fact that $\delta^{(2)}(\theta^\dagger) = 
\theta^\dagger\theta^\dagger$ 
from eq.~(\ref{eq:deltathetatheta}).
Thus, by integrating by parts twice with respect to $\theta^\dagger$:
\beq
[V]_D
&=&
-\frac{1}{4} \int d^2\theta d^2\theta^\dagger\, V
\,
\Dcon\Dcon (\thdthd)
\>=\>
-\frac{1}{4} \int d^2\theta d^2\theta^\dagger\,
\delta^{(2)}(\theta^\dagger) \, 
\Dcon\Dcon V 
+ \ldots\phantom{xxxx}
\label{eq:rewriteDasFz}
\\
&=&
-\frac{1}{4} \bigl [\Dcon\Dcon V \bigr ]_F 
+ 
\ldots.
\label{eq:rewriteDasF}
\eeq
The $\ldots$ indicates total derivatives with respect to $x^\mu$,
coming from the two integrations by parts. As noted in section \ref{chiralsuperfields},
$\Dcon\Dcon V$ is always a chiral superfield.
If $V$ is real, then the imaginary part of eq.~(\ref{eq:rewriteDasF}) 
is a total derivative, and the result can be rewritten as
$-\frac{1}{8} \bigl [\Dcon\Dcon V \bigr ]_F + {\rm c.c.}$

\subsection{Superspace Lagrangians for chiral supermultiplets\label{superspacelagrchiral}}
\setcounter{equation}{0}
\setcounter{footnote}{2}

In section \ref{chiralsuperfields}, we verified that the chiral 
superfield components have the same supersymmetry transformations as the 
Wess-Zumino model fields. We now have the tools to complete the 
demonstration of equivalence by reconstructing the Lagrangian in 
superspace language. Consider the composite superfield
\beq
\Phi^{*i} \Phi_j 
&=& 
\phi^{*i} \phi_j 
+ \sqrt{2} \theta \psi_j \phi^{*i}
+ \sqrt{2} \theta^\dagger\hspace{-1pt} \psi^{\dagger i} \phi_j
+ \theta\theta \phi^{*i} F_j
+ \thdthd \phi_j F^{*i}
\nonumber \\ &&
+ \thetasigmamuthetadagger  \left [
\BDneg i \phi^{*i} \partial_\mu \phi_j 
\BDplus i \phi_j \partial_\mu \phi^{*i}
- \psi^{\dagger i} \sigmabar_\mu \psi_j
\right ]
\nonumber \\ &&
+ \frac{i}{\sqrt{2}} \theta\theta\theta^\dagger \sigmabar^\mu (
\psi_j \partial_\mu \phi^{*i} - \partial_\mu \psi_j \phi^{*i}) 
+ \sqrt{2} \theta\theta \theta^\dagger \psi^{\dagger i} F_j
\nonumber \\ &&
+ \frac{i}{\sqrt{2}} \thdthd\theta \sigma^\mu (
\psi^{\dagger i} \partial_\mu \phi_j 
- \partial_\mu \psi^{\dagger i} \phi_j) 
+ \sqrt{2} \thdthd \theta \psi_j F^{*i}
\nonumber \\ &&
+ \theta\theta\thdthd 
\Bigl [
F^{*i} F_j 
\BDplus \frac{1}{2}\partial^\mu \phi^{*i} \partial_\mu \phi_j  
\BDminus \frac{1}{4}\phi^{*i} \partial^\mu \partial_\mu \phi_j
\BDminus \frac{1}{4}\phi_j \partial^\mu \partial_\mu \phi^{*i}
\nonumber \\ && \quad
+ \frac{i}{2} \psi^{\dagger i} \sigmabar^\mu \partial_\mu \psi_j  
+ \frac{i}{2} \psi_j \sigma^\mu \partial_\mu \psi^{\dagger i} 
\Bigr ] .
\label{eq:PhistariPhij}
\eeq
where 
all fields 
are evaluated as functions of $x^\mu$ (not $y^\mu$ or $y^{\mu *}$). 
For $i=j$, eq.~(\ref{eq:PhistariPhij}) is a real (vector) superfield, and
the massless free-field 
Lagrangian for each chiral superfield is just obtained by taking the $\theta\theta\thdthd$ component:
\beq
[\Phi^* \Phi]_D = \int d^2\theta d^2\theta^\dagger \,\Phi^* \Phi
\>=\> \BDpos \partial^\mu \phi^* \partial_\mu \phi 
+ i \psi^\dagger \sigmabar^\mu \partial_\mu \psi
+ F^* F + \ldots .
\label{eq:freelagrphisuperspace}
\eeq 
The $\ldots$ indicates a total derivative part, which may be 
dropped since this is destined to be integrated $\int d^4 x$. Equation 
(\ref{eq:freelagrphisuperspace}) is exactly the Lagrangian density 
obtained in section \ref{subsec:susylagr.freeWZ} for the massless free 
Wess-Zumino model.

To obtain the superpotential interaction and mass terms, recall that products of 
chiral superfields are also superfields. For example,
\beq
\Phi_i \Phi_j &=& \phi_i \phi_j 
+ \sqrt{2}\theta (\psi_i \phi_j + \psi_j \phi_i)
+ \theta\theta (\phi_i F_j + \phi_j F_i - \psi_i \psi_j)
,
\label{eq:superfieldPhiPhi}
\\
\Phi_i \Phi_j \Phi_k &=& \phi_i \phi_j \phi_k 
+ \sqrt{2}\theta (\psi_i \phi_j \phi_k 
+ \psi_j \phi_i \phi_k + \psi_k \phi_i \phi_j)
\nonumber 
\\ &&
+ \,\theta\theta (\phi_i \phi_j F_k + \phi_i \phi_k F_j + 
\phi_j \phi_k F_i 
- \psi_i \psi_j \phi_k - \psi_i \psi_k \phi_j
- \psi_j \psi_k \phi_i
),
\label{eq:superfieldPhiPhiPhi}
\eeq
where the presentation has been simplified by taking the component fields 
on the right sides to be functions of $y^\mu$ as given in 
eq.~(\ref{eq:defineycoord}). More generally, any holomorphic function of
a chiral superfields is a chiral superfield.
So, one may form a complete Lagrangian as
\beq
{\cal L}(x) &=& [\Phi^{*i} \Phi_i]_D + 
\left ( \left  [ W(\Phi_i) \right ]_F
+ {\rm c.c.} \right ),
\label{eq:oslo}
\eeq
where $W(\Phi_i)$ 
can be any holomorphic function 
of the chiral superfields (but not anti-chiral superfields) 
taken as complex variables, and
coincides with the superpotential $W(\phi_i)$ that was treated 
in subsection \ref{subsec:susylagr.chiral} as a function of the scalar components.
For $W = \frac{1}{2} M^{ij} \Phi_i \Phi_j
+ \frac{1}{6} y^{ijk} \Phi_i \Phi_j \Phi_k$, the result of 
eq.~(\ref{eq:oslo})
is exactly the same as eq.~(\ref{lagrchiral}), after writing in 
component form using eqs.~(\ref{eq:freelagrphisuperspace}), 
(\ref{eq:superfieldPhiPhi}), 
(\ref{eq:superfieldPhiPhiPhi}) and integrating out the auxiliary fields. 

It is instructive to obtain the superfield equations of motion
from the Lagrangian eq.~(\ref{eq:oslo}). The quickest way to do this
is to first use the remarks at the very end of section \ref{superspacelagr}
to rewrite the Lagrangian density as:
\beq
{\cal L}(x) &=& 
\int d^2\theta \left [ -\frac{1}{4} \Dcon\Dcon \Phi^{*i} \Phi_i  +
W(\Phi_i) \right ]
\,+ \,
\int d^2\theta^\dagger \left [W(\Phi_i)\right ]^*.
\eeq
Now varying with respect to $\Phi_i$ immediately gives the superfield equation of
motion:
\beq
0 &=& -\frac{1}{4} \Dcon\Dcon \Phi^{*i} + \frac{\delta W}{\delta \Phi_i} ,
\label{eq:superfieldeqmot}
\eeq
and its complex conjugate,
\beq
0&=& -\frac{1}{4} DD \Phi_i + \frac{\delta W^*}{\delta \Phi^{*i}} .
\label{eq:superfieldeqmotCON}
\eeq
These are equivalent to the component-level equations of motion as can be found from the
Lagrangian in section \ref{subsec:susylagr.chiral}. To verify this,
it is easiest to write eq.~(\ref{eq:superfieldeqmot})
in the coordinate system $(y^\mu, \theta, \theta^\dagger)$, 
in which the first term has the simple form
\beq
-\frac{1}{4} \Dcon\Dcon \Phi^{*i}
= F^*(y) - i \sqrt{2} \theta\sigma^\mu \partial_\mu \psi^{\dagger i}(y)
\BDminus \theta\theta \partial_\mu \partial^\mu \phi^{*i}(y) .
\eeq
Because this is a chiral (not anti-chiral) superfield, it is simpler to
write the components as functions of $y^\mu$ as shown, 
not $y^{\mu *}$, even though the left-hand 
side involves $\Phi^*$. 

For an alternate method, consider a Lagrangian density $V$ on 
the full superspace, so that the action is
\beq
A &=& \int d^4 x \hspace{-0.5pt} \int d^2\theta d^2\theta^\dagger\, V,
\eeq
with $V(S_i, \, D_\alpha S_i, \, \Dcon_{\dot\alpha} S_i)$ assumed to be
a function of 
general dynamical superfields $S_i$ and their chiral and anti-chiral 
first derivatives. Then the superfield equations of motion obtained by 
variation of the action are
\beq
0 &=& \frac{\partial V}{\partial S_i}
- D_\alpha \left ( \frac{\partial V}{\partial (D_\alpha S_i)} 
\right ) 
- \Dcon_{\dot\alpha} \left (\frac{\partial V}{\partial 
(\Dcon_{\dot\alpha} S_i)} \right )
.
\label{eq:superspaceeqmo}
\eeq
In the case of the Lagrangian for chiral superfields eq.~(\ref{eq:oslo}), 
Lagrange multipliers 
$\Lambda^{*i\dot \alpha}$ and $\Lambda^{\alpha}_i$ can be introduced
to enforce the chiral and anti-chiral superfield constraints 
on $\Phi_i$ and $\Phi^{*i}$ respectively. 
The Lagrangian density on superspace is then:
\beq
V &=& 
\Lambda^{*i\dot \alpha} \Dcon_{\dot \alpha} \Phi_i
+ \Lambda^{\alpha}_i D_{\alpha} \Phi^{*i}
+ \Phi^{*i} \Phi_i 
+ \delta^{(2)}(\theta^\dagger) W(\Phi_i)
+ \delta^{(2)}(\theta) [W(\Phi_i)]^* .
\eeq
Variation with respect to the Lagrange multipliers just gives the 
constraints
$\Dcon_{\dot \alpha} \Phi_i = 0$ and
$D_{\alpha} \Phi^{*i} = 0$. Applying eq.~(\ref{eq:superspaceeqmo}) to the 
superfields $\Phi_i$ and $\Phi^{*i}$ leads to equations of motion:
\beq
0 &=&  
\Phi^{*i} +
\delta^{(2)}(\theta^\dagger) \frac{\delta W}{\delta \Phi_i}
- \Dcon_{\dot\alpha} \Lambda^{*i\dot \alpha},
\\
0 &=& 
\Phi_{i} +
\delta^{(2)}(\theta) \frac{\delta W^*}{\delta \Phi^{*i}}
- D_{\alpha} \Lambda^{\alpha}_i 
.
\eeq
Now acting on these equations 
with $-\frac{1}{4} \Dcon\Dcon$ and
$-\frac{1}{4} DD$ respectively, and applying 
eqs.~(\ref{eq:deltathetatheta}) and (\ref{eq:DDthetatheta}),
one again obtains eqs.~(\ref{eq:superfieldeqmot}) and (\ref{eq:superfieldeqmotCON}).

\subsection{Superspace Lagrangians for Abelian 
gauge theory\label{subsec:superspacelagrabelian}}
\setcounter{equation}{0}
\setcounter{footnote}{2}

Now consider the superspace Lagrangian for a gauge theory, treating the 
$U(1)$ case first for simplicity. The non-Abelian case will be considered 
in the next subsection.

The vector superfield $V(x,\theta,\theta^\dagger)$ of 
eq.~(\ref{eq:vectorsuperfieldexpansion})
contains the gauge potential $A^\mu$. 
Define corresponding gauge-invariant Abelian
field strength superfields by 
\beq
{\cal W}_\alpha = -\frac{1}{4} \Dcon\Dcon D_\alpha V,
\qquad\qquad
{\cal W}^\dagger_{\dot\alpha} = -\frac{1}{4} DD \Dcon_{\dot\alpha} V.
\label{eq:defineWalpha}
\eeq
These
are respectively chiral and anti-chiral by construction 
[see eq.~(\ref{eq:chiralfromgeneral})], and are examples of 
superfields that carry spinor indices and are anti-commuting. They carry 
dimension [mass]$^{3/2}$.
To see that ${\cal W}_\alpha$ is gauge invariant, note that under 
a supergauge transformation of the form eq.~(\ref{eq:u1superfieldgt}),
\beq
{\cal W}_\alpha \,\rightarrow\, 
-\frac{1}{4} \Dcon\Dcon D_\alpha \bigl [V
+ i (\Omega^* - \Omega) \bigr ]
\label{eq:Walphagione}
&=& 
{\cal W}_\alpha + \frac{i}{4} \Dcon\Dcon D_\alpha \Omega
\\
&=&
{\cal W}_\alpha - \frac{i}{4} \Dcon^{\dot\beta}\bigl \lbrace
\Dcon_{\dot\beta}, D_\alpha \bigr \rbrace \Omega \phantom{xxxxxx}
\\
&=&
{\cal W}_\alpha + \frac{1}{2} \sigma^\mu_{\alpha\dot\beta} 
\partial_\mu \Dcon^{\dot\beta}\Omega
\\
&=&
{\cal W}_\alpha
\label{eq:Walphagitwo}
\eeq
The first equality follows from eq.~(\ref{eq:anti-chiralsuperfieldconstraint})
because $\Omega^*$ is anti-chiral, the second and fourth equalities from
eq.~(\ref{eq:leftchiralsuperfieldconstraint}) because $\Omega$ is 
chiral, and the third from eq.~(\ref{eq:diffopderivs}).

To see how the component fields fit into ${\cal W}_\alpha$, it is convenient
to temporarily specialize to Wess-Zumino gauge as in eq.~(\ref{eq:WZgauge}), and
then convert to the coordinates $(y^\mu, \theta, \theta^\dagger)$ 
as defined in eq.~(\ref{eq:defineycoord}), with the result 
\beq
V(y^\mu, \theta,\theta^\dagger) &=&
\thetasigmamuthetadagger  A_\mu(y) 
+ \thdthd \theta \lambda(y) 
+ \theta \theta  \theta^\dagger\hspace{-1pt} \lambda^\dagger(y)  
+ \frac{1}{2} \theta\theta\thdthd 
\left [D(y)
+ i \partial_\mu A^\mu(y) \right ].\phantom{xxx}
\eeq
Now application of eqs.~(\ref{eq:Dinyrep}), (\ref{eq:Ddaggerinyrep}) 
yields
\beq
{\cal W}_\alpha(y,\theta,\theta^\dagger)
&=& 
\lambda_\alpha + \theta_\alpha D 
\BDminus \frac{i}{2} (\sigma^\mu\sigmabar^\nu\theta)_\alpha F_{\mu\nu}
+ i \theta\theta (\sigma^\mu\partial_\mu \lambda^\dagger)_\alpha ,
\label{eq:gottawearshades}
\\
{\cal W}^{\dagger\dot\alpha}(y^*,\theta,\theta^\dagger)
&=& 
\lambda^{\dagger\dot \alpha} + \theta^{\dagger\dot\alpha} D 
\BDplus \frac{i}{2} (\sigmabar^\mu\sigma^\nu \theta^\dagger)^{\dot\alpha} 
F_{\mu\nu}
+ i \thdthd (\sigmabar^\mu\partial_\mu \lambda  )^{\dot\alpha},
\label{eq:futuresobright}
\eeq
where 
all fields on the right side are understood to be functions of 
$y^\mu$ and $y^{\mu *}$ respectively, and
\beq
F_{\mu\nu} \,=\, \partial_\mu A_\nu - \partial_\nu A_\mu
\eeq
is the ordinary component field strength.
Although it was convenient to derive eqs.~(\ref{eq:gottawearshades}) and 
(\ref{eq:futuresobright}) in Wess-Zumino gauge, they must 
be true in general, because ${\cal W}_\alpha$ and ${\cal W}^{\dagger\dot\alpha}$ 
are supergauge invariant.

Equation (\ref{eq:gottawearshades}) implies
\beq
[{\cal W}^\alpha {\cal W}_\alpha]_F = D^2 +
2 i \lambda \sigma^\mu \partial_\mu \lambda^\dagger
-\frac{1}{2} F^{\mu\nu} F_{\mu\nu} 
+ \frac{i}{4} \epsilon^{\mu\nu\rho\sigma} F_{\mu\nu} F_{\rho\sigma} 
,
\label{eq:Bushwickblues}
\eeq
where now all fields on the right side are functions of $x^\mu$.
Integrating, and eliminating total derivative parts, one obtains
the action
\beq
\int d^4 x\, {\cal L} 
&=& 
\int d^4 x\, \frac{1}{4} [{\cal W}^\alpha {\cal W}_\alpha]_F + {\rm c.c.}
\,=\,
\int d^4 x\, 
\left [ 
\frac{1}{2} D^2  + 
i \lambda^\dagger \sigmabar^\mu \partial_\mu \lambda 
- \frac{1}{4} F^{\mu\nu} F_{\mu\nu} 
\right ],
\phantom{xx}
\eeq
in agreement with eq.~(\ref{lagrgauge}). 
Additionally,
the integral of the $D$-term component of $V$ itself is invariant  under 
both supersymmetry [see eq.~(\ref{eq:sqrttwodeltaD})] and supergauge [see 
eq.~(\ref{eq:DgaugetransfromLambda})]
transformations. Therefore, one can include a Fayet-Iliopoulos term
\beq
{\cal L}_{\mbox{FI}} &=& -2 \kappa [V]_D \>=\>-\kappa D,
\eeq
again dropping a total derivative. This type of term can play a role in spontaneous
supersymmetry breaking, as we will discuss in section \ref{subsec:origins.Dterm}.

It is also possible to write the Lagrangian density eq.~(\ref{eq:Bushwickblues}) 
as a $D$-term rather 
than an $F$-term. Since ${\cal W}^\alpha$ is a chiral superfield, with 
$\Dcon_{\dot \beta} {\cal W}^\alpha = 0$, one can use 
eq.~(\ref{eq:defineWalpha}) to write
\beq
{\cal W}^\alpha {\cal W}_\alpha = 
-\frac{1}{4} \Dcon\Dcon ({\cal W}^\alpha D_\alpha V).
\eeq
Therefore, using eq.~(\ref{eq:rewriteDasF}), 
the Lagrangian for $A^\mu$, $\lambda$, and $D$ can be rewritten as:
\beq
{\cal L}(x) = \int d^2\theta d^2\theta^\dagger \, \left [
\frac{1}{4}
\left (
{\cal W}^\alpha D_\alpha V 
+ {\cal W}^\dagger_{\dot\alpha} \Dcon^{\dot\alpha} V \right ) - 2 \kappa V
\right ].
\eeq

Next consider the coupling of the Abelian gauge field to a set of chiral 
superfields $\Phi_i$ carrying $U(1)$ charges $q_i$. Supergauge transformations, as in 
eqs.~(\ref{eq:u1superfieldgt})-(\ref{eq:DgaugetransfromLambda}), are 
parameterized by a non-dynamical chiral 
superfield $\Omega$,
\beq
\Phi_i &\rightarrow& e^{2ig q_i \Omega} \Phi_i ,
\qquad\qquad
\Phi^{*i} \>\rightarrow\> e^{-2ig q_i \Omega^*} \Phi^{*i} ,
\eeq
where $g$ is the gauge coupling. 
In the special case that $\Omega$ is 
just a real function $\phi(x)$, independent of $\theta$ and $\theta^\dagger$, 
this reproduces the usual gauge transformations with $A^\mu \rightarrow 
A^\mu \BDminus 2 \partial^\mu \phi$. 
The kinetic term 
from eq.~(\ref{eq:freelagrphisuperspace}) 
involves the superfield 
$\Phi^{*i} \Phi_i$, which is not supergauge invariant:
\beq
\Phi^{*i} \Phi_i &\rightarrow& 
e^{2ig q_i(\Omega - \Omega^*)} \Phi^{*i} \Phi_i .
\label{eq:gtofPhistarPhi}
\eeq
To remedy this, we modify the chiral superfield kinetic term in the Lagrangian to
\beq
\left [ \Phi^{*i} e^{2 g q_i V} \Phi_i \right ]_D .
\eeq
The gauge transformation of the 
$e^{2 g q_i V}$ factor, found from 
eq.~(\ref{eq:u1superfieldgt}),
exactly cancels that 
of eq.~(\ref{eq:gtofPhistarPhi}).

The presence of an exponential of $V$ in the Lagrangian is possible 
because $V$ is dimensionless. It might appear to be dangerous, because 
normally such a non-polynomial term would be non-renormalizable. However, 
the gauge dependence of $V$ comes to the rescue: the higher order 
terms can be supergauged away. In particular, evaluating $e^{2 g q_i V}$ 
in the Wess-Zumino gauge, the power series expansion of the exponential 
is simple and terminates, because
\beq
V^2 &=& \BDpos\frac{1}{2} 
\theta\theta\thdthd A_\mu A^\mu,
\\
V^n &=& 0\qquad(n\geq 3),
\eeq
so that 
\beq
e^{2 g q_i V} = 1 + 2 g q_i (\thetasigmamuthetadagger  A_\mu + 
\thdthd \theta \lambda 
+ \theta\theta \theta^\dagger\hspace{-1pt} \lambda^\dagger +
\frac{1}{2} \theta\theta\thdthd D )
\BDplus g^2 q_i^2 \theta\theta\thdthd A_\mu A^\mu.
\eeq
Using this, one can work out that, in Wess-Zumino gauge and up to total 
derivative terms,
\beq
\left [ \Phi^{*i} e^{2 g q_i V} \Phi_i \right ]_D 
&=&
F^{*i} F_i 
\BDplus \nablasubmu \phi^{*i} \nabla^\mu \phi_i
+ i \psi^{\dagger i} \sigmabar^\mu \nablasubmu \psi_i
- \sqrt{2} g q_i 
( \phi^{*i} \psi_i \lambda + \lambda^\dagger \psi^{\dagger i} \phi_i)
\phantom{xx}
\nonumber \\ &&
+ g q_i  \phi^{*i} \phi_i D,
\label{eq:superfieldchiralkinetic}
\eeq 
where $\nablasubmu$ is the gauge-covariant spacetime derivative:
\beq
\nablasubmu \phi_i &=& \partial_\mu \phi_i \BDplus i g q_i A_\mu \phi_i,
\qquad\qquad
\nablasubmu \phi^{*i} \>=\> \partial_\mu \phi^{*i} \BDminus i g q_i A_\mu 
\phi^{*i},
\\
\nablasubmu \psi_i &=& \partial_\mu \psi_i \BDplus i g q_i A_\mu \psi_i .
\eeq
Equation (\ref{eq:superfieldchiralkinetic}) agrees with the 
specialization of eq.~(\ref{gensusylagr}) to the Abelian case.

In summary, the superspace Lagrangian
\beq
{\cal L} &=& 
 \left [ \Phi^{*i} e^{2 g q_i V} \Phi_i \right ]_D
+ \left (\left [W(\Phi_i)\right ]_F + {\rm c.c.} \right )
+
\frac{1}{4}\left (\left [{\cal W}^\alpha {\cal W}_\alpha \right ]_F 
+ {\rm c.c.} \right )
-2 \kappa [V]_D
\eeq
reproduces the component form Lagrangian found in 
subsection \ref{subsec:susylagr.gaugeinter}
in the special case of matter fields coupled to each other and to a 
$U(1)$ gauge symmetry, plus a Fayet-Iliopoulos parameter $\kappa$.

\subsection{Superspace Lagrangians for general 
gauge theories\label{subsec:superspacelagrnonabelian}}
\setcounter{equation}{0}
\setcounter{footnote}{2}

Now consider a general gauge symmetry realized on chiral superfields $\Phi_i$ in a representation $R$
with matrix generators $T^{aj}_i$:
\beq
\Phi_i \rightarrow \bigl (e^{2 i g_a \Omega^a T^a} \bigr )_i{}^j 
\Phi_j,
\qquad\qquad
\Phi^{*i} \rightarrow \Phi^{*j}
\bigl (e^{-2 i g_a \Omega^a T^a} \bigr )_j{}^i.
\eeq
The gauge couplings for the irreducible components of the Lie algebra are $g_a$. As in the Abelian case, the supergauge transformation parameters are chiral superfields $\Omega^a$.
For each Lie algebra generator, there is a
vector superfield $V^a$, which contains the vector gauge boson and gaugino. 
The Lagrangian then contains a supergauge-invariant term
\beq
{\cal L} &=& \Bigl [ \Phi^{*i} (e^{2 g_a T^a V^a})_i{}^j \Phi_j \Bigr ]_D.
\eeq
It is convenient to define matrix-valued vector and gauge parameter 
superfields in the representation $R$:
\beq
V_i{}^j &=& 2 g_a T_i^{aj} V^a ,
\qquad\qquad
\Omega_i{}^j \>\,=\,\> 2 g_a T_i^{aj} \Omega^a ,
\label{eq:yourenotthebossofmenow}
\eeq
so that one can write
\beq
\Phi_i \rightarrow \bigl (e^{i \Omega} \bigr )_i{}^j 
\Phi_j,
\qquad\qquad
\Phi^{*i} \rightarrow \Phi^{*j}
\bigl (e^{-i \Omega^\dagger} \bigr )_j{}^i ,
\eeq
and
\beq
{\cal L} &=& \Bigl [ \Phi^{*i} (e^{V})_i{}^j \Phi_j \Bigr ]_D.
\label{eq:bigbadbeluva}
\eeq
For this to be supergauge invariant, 
the non-Abelian gauge transformation rule for the vector superfields must be
\beq
e^V &\rightarrow& e^{i \Omega^\dagger} e^V e^{-i\Omega}
.
\label{eq:kentuckyavenue}
\eeq
[Here chiral supermultiplet representation indices $i,j,\ldots$ are suppressed;
$V$ and $\Omega$ with no indices stand for the matrices defined in 
eq.~(\ref{eq:yourenotthebossofmenow}).]
Equation (\ref{eq:kentuckyavenue}) can be expanded, keeping terms linear in
$\Omega$, $\Omega^\dagger$, using the Baker-Campbell-Hausdorff formula, to find
\beq
V &\rightarrow & V + i (\Omega^\dagger - \Omega) 
- \frac{i}{2} \bigl [V,\, \Omega + \Omega^\dagger ] 
+ i\sum_{k=1}^{\infty}
\frac{B_{2k}}{(2k)!} \left [ V, \left [ V,\ldots \left [V,\, \Omega^\dagger - \Omega \right ]\ldots \right]\right]
,
\phantom{xxx}
\label{eq:nonAbelianVtransexpansion}
\eeq
where the $k$th term in the sum involves $k$ matrix commutators of 
$V$, and $B_{2k}$ are the Bernoulli numbers defined by
\beq
\frac{x}{e^x - 1} = \sum_{n=0}^\infty \frac{B_n}{n!} x^n.
\eeq
Equation~(\ref{eq:nonAbelianVtransexpansion}) is equivalent to
\beq
V^a &\rightarrow& V^a + i (\Omega^{a*} - \Omega^a)
+ g_a f^{abc} V^b (\Omega^{c*} + \Omega^c)
- \frac{i}{3} g_a^2 f^{abc} f^{cde} V^b V^d (\Omega^{e*} - \Omega^e) + 
\ldots \phantom{xxx}
\label{eq:nonAbelianVtransexpansiona}
\eeq
where eq.~(\ref{eq:yourenotthebossofmenow}) and $[T^a, T^b] = i f^{abc} T^c$ have been used.
This supergauge transformation includes ordinary gauge transformations as 
the special case
$\Omega^{a*} = \Omega^a$. 

Because the second term on the right side of eq.~(\ref{eq:nonAbelianVtransexpansiona}) is  
independent of $V^a$, one can always do a supergauge transformation to Wess-Zumino gauge
by choosing $\Omega^{a*} - \Omega^a$ appropriately, 
just as in the
Abelian case, so that
\beq
\bigl (V^a \bigr )_{\mbox{WZ gauge}}
&=& 
\thetasigmamuthetadagger  A^a_\mu
+ \thdthd \theta \lambda^a 
+ \theta \theta  \theta^\dagger\hspace{-1pt} \lambda^{\dagger a} 
+ \frac{1}{2} \theta\theta\thdthd D^a .
\eeq
After fixing the supergauge to Wess-Zumino gauge, one still has the freedom to do ordinary
gauge transformations.
In the Wess-Zumino gauge, the Lagrangian contribution eq.~(\ref{eq:bigbadbeluva}) is polynomial,
in agreement with what was found in component language in section 
\ref{subsec:susylagr.gaugeinter}:
\beq
\left [ \Phi^{*i} \bigl (e^{V} \bigr )_i{}^j \Phi_j \right ]_D 
&=&
F^{*i} F_i 
\BDplus \nablasubmu \phi^{*i} \nabla^\mu \phi_i
+ i \psi^{\dagger i} \sigmabar^\mu \nablasubmu \psi_i
- \sqrt{2} g_a (\phi^{*} T^a \psi) \lambda^a 
- \sqrt{2} g_a \lambda^\dagger (\psi^{\dagger } T^a \phi)
\phantom{xx}
\nonumber \\ &&
+ g_a  (\phi^{*} T^a \phi) D^a,
\eeq 
where $\nablasubmu$ is the gauge-covariant derivative
defined in eqs.~(\ref{ordtocovphi})-(\ref{ordtocovpsi}).

To make kinetic terms and self-interactions for the vector 
supermultiplets in the non-Abelian case,
define a field-strength chiral superfield
\beq
{\cal W}_\alpha \>=\> -\frac{1}{4} \Dcon\Dcon 
\left (e^{-V} D_\alpha e^V \right),
\label{eq:definenonabelianfieldstrengthW}
\eeq
generalizing the Abelian case.
Using eq.~(\ref{eq:kentuckyavenue}), 
one can show that it transforms under supergauge transformations as
\beq
{\cal W}_\alpha &\rightarrow& e^{i\Omega} {\cal W}_\alpha e^{-i\Omega}
.
\eeq
(The proof makes use of the fact that $\Omega$ is chiral and $\Omega^\dagger$ is anti-chiral, so that $\Dcon_{\dot\alpha} \Omega = 0$
and $D_\alpha \Omega^\dagger = 0$.) This implies that Tr$[W^\alpha W_\alpha]$
is a supergauge-invariant chiral superfield.
The contents of the parentheses in 
eq.~(\ref{eq:definenonabelianfieldstrengthW}) can be expanded as 
\beq
e^{-V} D_\alpha e^V &=& D_\alpha V - \frac{1}{2} [V, D_\alpha V]
+ \frac{1}{6} \left [V, \left [V, D_\alpha V\right ] \right ] + \ldots ,
\eeq
where again the commutators apply in the matrix sense, and only the first two terms 
contribute in Wess-Zumino gauge.

The field strength chiral superfield ${\cal W}_\alpha$ defined in 
eq.~(\ref{eq:definenonabelianfieldstrengthW}) 
is matrix-valued in the representation $R$. One can recover an adjoint representation
field strength superfield ${\cal W}^a_\alpha$ from the matrix-valued
one by writing
\beq
{\cal W}_\alpha = 2 g_a T^a {\cal W}^a_\alpha ,
\eeq
leading to
\beq
{\cal W}^a_\alpha &=& 
-\frac{1}{4} \Dcon\Dcon \left (
D_\alpha V^a - i g_a f^{abc} V^b D_\alpha V^c + \ldots \right ) .
\eeq
The terms shown explicitly are enough to evaluate this in components in 
Wess-Zumino gauge, with the result
\beq
({\cal W}^a_\alpha)_{\mbox{WZ gauge}}
&=&
\lambda^a_\alpha + \theta_\alpha D^a 
\BDminus \frac{i}{2} (\sigma^\mu 
\sigmabar^\nu \theta)_\alpha F^a_{\mu\nu} 
+ i \theta\theta (\sigma^\mu \nablasubmu \lambda^{\dagger a})_\alpha
,
\label{eq:GeneAmmons}
\eeq
where $F^a_{\mu\nu}$ is the non-Abelian field strength of eq.~(\ref{eq:YMfs})
and $\nablasubmu$ is the usual gauge covariant derivative from eq.~(\ref{ordtocovlambda}).

The kinetic terms and self-interactions for the gauge supermultiplet fields are obtained
from 
\beq
\frac{1}{4 k_a g_a^2} {\rm Tr}[{\cal W}^{\alpha} {\cal W}_\alpha]_F &=&
[{\cal W}^{a\alpha} {\cal W}^a_\alpha]_F,
\label{eq:foxforce}
\eeq
which is invariant under both supersymmetry and supergauge transformations.
Here the normalization of generators is assumed to be
$
{\rm Tr}[T^a T^b] = k_a \delta_{ab},
$
with $k_a$ usually set to $1/2$ by convention for the 
defining representations of simple groups.
Equation (\ref{eq:foxforce}) is most easily
evaluated in Wess-Zumino gauge using eq.~(\ref{eq:GeneAmmons}), yielding
\beq
[{\cal W}^{a\alpha} {\cal W}^a_\alpha]_F &=& D^{a}D^a +
2 i \lambda^a \sigma^\mu \nablasubmu \lambda^{\dagger a}
-\frac{1}{2} F^{a\mu\nu} F^a_{\mu\nu} 
+ \frac{i}{4} \epsilon^{\mu\nu\rho\sigma} F^a_{\mu\nu} F^a_{\rho\sigma} .
\label{eq:CharlieParker}
\eeq 
Since eq.~(\ref{eq:CharlieParker}) is supergauge invariant, the same expression is valid 
even outside of Wess-Zumino gauge.

Now we can write the general renormalizable Lagrangian for a supersymmetric gauge theory (including superpotential interactions for the chiral supermultiplets when allowed by gauge invariance):
\beq
{\cal L} &=& \left ( \frac{1}{4}  - i \frac{g_a^2 \Theta_a}{32 \pi^2} 
\right ) 
\left [
{\cal W}^{a\alpha} {\cal W}^a_\alpha 
\right ]_F
+ {\rm c.c.}
+
\left [
\Phi^{*i} (e^{2g_a T^a V^a})_i{}^j \Phi_j \right]_D
+ 
\left ( [W(\Phi_i)]_F + {\rm c.c.} \right ) .
\phantom{xxxx}
\eeq
This introduces and defines 
$\Theta_a$, a CP-violating parameter, whose effect is to include a total derivative term
in the Lagrangian density:
\beq
{\cal L}_{\Theta_a} &=& 
\frac{g_a^2 \Theta_a}{64\pi^2} \epsilon^{\mu\nu\rho\sigma} F^a_{\mu\nu} F^a_{\rho\sigma}.
\eeq
In the non-Abelian case, this can have physical
effects due to topologically non-trivial field configurations (instantons). 
For a globally non-trivial
gauge configuration with integer winding number $n$, one has 
$\int d^4 x\, \epsilon^{\mu\nu\rho\sigma} F^a_{\mu\nu} F^a_{\rho\sigma} 
= 64 \pi^2 n/g_a^2$ for a simple gauge group,
so that the contribution to the path integral is 
$\exp({i\int d^4 x\,{\cal L}_{\scriptstyle \Theta_a}})
=e^{i n \Theta_a}$.
Note that for non-Abelian gauge groups, a Fayet-Iliopoulos term $-2\kappa [V^a]_D$ is not 
allowed, because it is not a gauge singlet.

When the superfields are restricted to the Wess-Zumino gauge, the supersymmetry transformations 
are not realized linearly in superspace, but the Lagrangian is polynomial. The 
non-polynomial form of the superspace Lagrangian is thus seen to be a supergauge artifact. 
Within Wess-Zumino gauge, supersymmetry transformations are still realized, but 
non-linearly, as we found in sections \ref{subsec:susylagr.gauge} and 
\ref{subsec:susylagr.gaugeinter}.

The gauge coupling $g_a$ and CP-violating angle $\Theta_a$ are often 
combined into a single holomorphic
coupling:
\beq
\tau_a &=& \frac{1}{g_a^2} -i \frac{\Theta_a}{8 \pi^2}
\eeq
(There are several different normalization conventions for $\tau_a$ in the literature.)
Then, with redefined
vector and field strength superfields that include $g_a$ as part of their normalization,
\beq
\widehat V^{a} &\equiv& g_a V^{a},
\label{eq:holonorm}
\\
\widehat {\cal W}^{a}_{\alpha} &\equiv& 
g_a {\cal W}^{a}_{\alpha}
\>\,=\>\,
-\frac{1}{4} \Dcon\Dcon \left (
D_\alpha \widehat V^a - i f^{abc} \widehat V^b D_\alpha \widehat V^c + \ldots \right ) 
,
\label{eq:holonormW}
\eeq
the gauge part of the Lagrangian is written as
\beq
{\cal L} &=& 
\frac{1}{4} 
\left [ \tau_a \widehat{\cal W}^{a \alpha}\widehat{\cal W}^{a}_{\alpha} \right ]_F 
+ {\rm c.c.}  + 
\left [
\Phi^{*i} (e^{2 T^a \widehat V^a})_i{}^j \Phi_j \right]_D .
\label{eq:tauaholo}
\eeq
An advantage of this normalization convention is that 
when written in terms of $\widehat V^a$, the only appearance of the gauge coupling
and $\Theta_a$ is in the $\tau_a$ in eq.~(\ref{eq:tauaholo}).
It is then sometimes useful to treat the complex holomorphic coupling $\tau_a$ 
as a 
chiral superfield with an expectation value for its scalar component. 
An expectation value for the 
$F$-term component of $\tau_a$ will give gaugino masses; 
this is sometimes
a useful way to implement the effects of explicit soft supersymmetry breaking.

\subsection{Non-renormalizable 
supersymmetric Lagrangians\label{superspacenonrenorm}}
\setcounter{equation}{0}
\setcounter{footnote}{2}

So far, we have discussed only renormalizable supersymmetric Lagrangians. 
However, integrating out the effects
of heavy states will generally lead to
non-renormalizable interactions in the low-energy 
effective description. Furthermore,  
when any realistic supersymmetric theory is extended to include gravity, 
the resulting supergravity theory is 
non-renormalizable as a quantum field theory. Fortunately, 
the non-renormalizable interactions can be neglected 
for most phenomenological purposes, because they 
involve couplings of negative mass dimension, proportional to 
powers of $1/\MPlanck$ (or perhaps $1/\Lambda_{\rm UV}$, where 
$\Lambda_{\rm UV}$ is some other cutoff scale associated with new 
physics). This means that their effects at energy scales $E$ ordinarily 
accessible to experiment are typically suppressed by powers of 
${E/\MPlanck}$ (or $E/\Lambda_{\rm UV}$). For energies 
$E\lsim 1$ TeV, the consequences of non-renormalizable interactions are 
therefore usually far too small to be interesting.

Still, there are several reasons why one may need to include 
non-renormalizable contributions to supersymmetric Lagrangians. First, 
some very rare processes (like proton decay) might only be described using 
an effective MSSM Lagrangian that includes non-renormalizable terms. 
Second, one may be interested in understanding physics at very high 
energy scales where the suppression associated with non-renormalizable 
terms is not enough to stop them from being important. For example, this 
could be the case in the study of the very early universe, or in 
understanding how additional gauge symmetries get broken. Third, the 
non-renormalizable interactions may play a crucial role in understanding 
how supersymmetry breaking is transmitted to the MSSM. Finally, it is 
sometimes useful to treat strongly coupled supersymmetric gauge theories 
using non-renormalizable effective Lagrangians, in the same way that 
chiral effective Lagrangians are used to study hadron physics in QCD. 
Unfortunately, we will not be able to treat these subjects in any sort of 
systematic way. Instead, we will merely sketch a few of the key elements 
that go into defining a non-renormalizable supersymmetric Lagrangian. 
More detailed treatments and pointers to the literature may be found for example in 
refs.~\cite{WessBaggerbook,Westbook,BailinLovebook,Buchbinder:1998qv,
Weinbergbook,Freedman:2012zz,Shifman:2012zz,Nillesreview,GGRS,VNreview,Bertolini:2013via}.

A non-renormalizable gauge-invariant theory involving chiral and vector 
superfields can be constructed as:
\beq
{\cal L} &=& \left [ K(\Phi_i, \, \tilde \Phi^{*j}) \right ]_D
+ \left ( 
\left [\frac{1}{4}  f_{ab}(\Phi_i) \widehat{\cal W}^{a\alpha} \widehat{\cal W}^b_\alpha 
\,+\, W(\Phi_i)
\right ]_F
+ {\rm c.c.}
\right ),
\label{eq:Lnonren}
\eeq
where, in order to preserve supergauge invariance, we define
\beq
\tilde \Phi^{*j} \equiv \bigl ( \Phi^{*} e^{V} \bigr )^j ,
\eeq
with $V = 2 g_a T^a V^a = 2 T^a \widehat V^a$ as above,
and the hatted normalization of the field-strength superfields indicated in 
(\ref{eq:holonormW}) has been used.
Equation (\ref{eq:Lnonren})
depends on couplings encoded in three functions of the 
superfields:
\begin{itemize}
\item[$\bullet$]
The superpotential $W$, which we have already encountered in the
special case of renormalizable supersymmetric Lagrangians. More 
generally, it can be an arbitrary
holomorphic function of the chiral superfields treated as complex 
variables, and 
must be invariant under the gauge symmetries of the theory, and has 
dimension [mass]$^3$. 
\item[$\bullet$]
The {\em K\"ahler potential} $K$. 
Unlike the
superpotential, the K\"ahler potential is a function of both chiral and 
anti-chiral superfields, and includes the vector superfields in such a way 
as to be supergauge invariant. It is real, and has dimension 
[mass]$^2$. In the special
case of renormalizable theories, we did not have to discuss the K\"ahler
potential explicitly, because at tree-level it is always just 
$K = \Phi_i \tilde \Phi^{i*}$.
Any additive part of $K$ that is a chiral (or anti-chiral) 
superfield does not contribute to the 
action, since the $D$-term of a chiral superfield is a total derivative 
on spacetime.
\item[$\bullet$]
The {\it gauge kinetic function} $f_{ab}(\Phi_i)$. Like the
superpotential, it is itself a chiral superfield, and is a holomorphic 
function of the chiral superfields 
treated as
complex variables. It is dimensionless and symmetric under interchange of
its two indices $a,b$, which run over the adjoint representations of the
simple and Abelian component gauge groups of the model. For the 
non-Abelian components of the gauge 
group, it is always just proportional to $\delta_{ab}$, but if there are
two or more Abelian components, the gauge invariance of the field-strength 
superfield [see eqs.~(\ref{eq:Walphagione})-(\ref{eq:Walphagitwo})] 
allows kinetic mixing so that
$f_{ab}$ is not proportional to $\delta_{ab}$ in general.
In the special case of renormalizable
supersymmetric Lagrangians at tree level, it is 
independent of the chiral superfields, and just 
equal to $f_{ab} = \delta_{ab}(1/g_a^2 - i \Theta_a/8 \pi^2)$, 
(for fewer than two Abelian components in the gauge group). More generally, it also
encodes the non-renormalizable couplings of the gauge supermultiplets
to the chiral supermultiplets.
\end{itemize}

It should be emphasized that eq.~(\ref{eq:Lnonren}) is still not the 
most general non-renormalizable supersymmetric Lagrangian, 
even if one restricts to chiral and 
gauge vector superfields. One can also include chiral, anti-chiral, and spacetime 
derivatives acting on the superfields, so that for example the K\"ahler potential
can be generalized to include dependence on $D_\alpha \Phi_i$, 
$\Dcon_{\dot\alpha} \Phi^{*i}$, $DD\Phi_i$, $\Dcon\Dcon\Phi^{*i}$, etc. 
Such terms typically have an extra suppression at low energies 
compared to terms without derivatives,
because of the positive mass dimension of the chiral covariant derivatives.
I will not discuss these possibilities below, but will only
make a remark on how supergauge invariance is maintained. 
The chiral covariant derivative of a chiral superfield, 
$D_\alpha \Phi_i$ is not gauge covariant unless $\Phi_i$ is a gauge singlet;
the ``covariant" in the name refers to supersymmetry transformations, not gauge
transformations. However, one can define
a ``gauge covariant chiral covariant" derivative $\nabla_\alpha$, 
whose action on a chiral superfield $\Phi$ is defined by:
\beq
\nabla_\alpha \Phi &\equiv&
e^{-V} D_\alpha (e^{V} \Phi) ,
\eeq
where the representation indices $i$ are suppressed.
From eq.~(\ref{eq:kentuckyavenue}), the supergauge transformation 
for $e^{-V}$ is
\beq
e^{-V} \rightarrow e^{i \Omega} e^{-V} e^{-i \Omega^\dagger},
\eeq
so that
\beq
e^{-V} D_\alpha (e^{V} \Phi) &\rightarrow&
e^{i \Omega} e^{-V} e^{-i \Omega^\dagger} D_\alpha (e^{i \Omega^\dagger} e^{V} \Phi)
\>=\> e^{i \Omega} e^{-V} D_\alpha (e^{V} \Phi),
\eeq
where the equality follows from the fact that $\Omega^\dagger$ is anti-chiral,
and thus ignored by $D_\alpha$. This is the correct covariant transformation law
under supergauge transformations.
So, using $\nabla_\alpha \Phi_i$ 
as a building block instead of $D_\alpha \Phi_i$, one can 
maintain supergauge covariance along with manifest supersymmetry.
Similarly, one can define building blocks:
\beq
\overline\nabla_{\dot\alpha} \Phi^* &\equiv& \Dcon_{\dot\alpha} (\Phi^*e^V ) e^{-V} 
,
\\
\nabla\nabla \Phi &\equiv& e^{-V} DD (e^V \Phi)
\\
\overline\nabla\overline\nabla \Phi^* &\equiv& \Dcon\Dcon (\Phi^*e^{V} ) e^{-V} 
\eeq
which each have covariant supergauge transformation rules.
 
Returning to the globally supersymmetric non-renormalizable theory defined 
by eq.~(\ref{eq:Lnonren}), with no extra derivatives,
the part of the Lagrangian coming from the superpotential is
\beq
\left [ W(\Phi_i) \right ]_F = W^i F_i  - \frac{1}{2} W^{ij} \psi_i \psi_j ,
\eeq
with
\beq
W^i = \frac{\delta W}{\delta\Phi_i} 
\biggl |_{\Phi_i \rightarrow \phi_i},
\qquad\qquad
W^{ij} = \frac{\delta^{2} W}{\delta\Phi_i\delta \Phi_j}
\biggl |_{\Phi_i \rightarrow \phi_i}
,
\eeq
where the superfields have been replaced by their scalar components after 
differentiation.
[Compare eqs.~(\ref{tryint}), (\ref{expresswij}), (\ref{wiwiwi}) and the 
surrounding discussion.] After integrating out the auxiliary 
fields $F_i$, the part of the scalar potential coming from the 
superpotential is
\beq
V = W^i W_j^* (K^{-1})_i^j,
\eeq
where $K^{-1}$ is the inverse matrix of the K\"ahler metric:
\beq
K^i_j = \frac{\delta^2 K}{\delta \Phi_i \delta \tilde \Phi^{*j}} \biggl |_{\Phi_i 
\rightarrow \phi_i,\> \tilde \Phi^{*i}
\rightarrow \phi^{*i}}.
\eeq
More generally, the whole component field Lagrangian after integrating 
out the auxiliary fields is determined in 
terms of the functions $W$, $K$ and $f_{ab}$ and their derivatives with 
respect to the chiral superfields, with the remaining chiral superfields
replaced by their scalar components. 
The complete form of this is straightforward to evaluate, but somewhat 
complicated. In supergravity, there are additional contributions, some of 
which are discussed in section \ref{subsec:origins.sugra} below.

\subsection{$R$ symmetries\label{Rsymmetry}}
\setcounter{equation}{0}
\setcounter{footnote}{2}

Some supersymmetric Lagrangians are also invariant under a global 
$U(1)_R$ symmetry. The defining feature of a continuous $R$ symmetry is 
that the anti-commuting coordinates $\theta$ and $\theta^\dagger$ 
transform under it with charges $+1$ and $-1$ respectively, so
\beq
\theta \> \rightarrow \> e^{i \alpha} \theta,
\qquad\qquad
\theta^\dagger \> \rightarrow \> e^{-i \alpha} \theta^\dagger
\label{eq:Rtranstheta}
\eeq
where $\alpha$ parameterizes the global $R$ transformation. It follows that 
\beq
\hat Q \> \rightarrow \> e^{-i \alpha} \hat Q,
\qquad\qquad
\hat Q^\dagger \> \rightarrow \> e^{i \alpha} \hat Q^\dagger,
\eeq
which in turn implies that the supersymmetry generators have 
$U(1)_R$ charges $-1$ and $+1$, and so 
they do not commute
with the $R$ symmetry generator:
\beq
[R, Q] = -Q, \qquad\qquad
[R, Q^\dagger] = Q^\dagger
\eeq
Thus the distinct components within a superfield always have different $R$ charges.

If the theory is invariant under an $R$ symmetry, then each superfield $S(x,\theta,\theta^\dagger)$ can be assigned an $R$ charge, denoted $r_S$, defined by its transformation rule
\beq
S(x,\,\theta,\,\theta^\dagger) &\rightarrow& 
e^{i r_S \alpha} S(x,\, e^{-i \alpha} \theta,\, e^{i\alpha}\theta^\dagger).
\eeq
The $R$ charge of a product of superfields is the sum of the individual $R$ charges.
For a chiral superfield $\Phi$ with $R$ charge $r_\Phi$, the $\phi$, $\psi$, and $F$ 
components transform with charges $r_\Phi$, $r_\Phi -1$, and $r_\Phi-2$, respectively:
\beq
\phi \rightarrow e^{ir_\Phi\alpha} \phi,\qquad\quad
\psi \rightarrow e^{i(r_\Phi - 1)\alpha} \psi,\qquad\quad
F \rightarrow e^{i(r_\Phi - 2)\alpha} F.
\eeq
The components of $\Phi^*$ carry the opposite charges.

Gauge vector superfields will always have vanishing $U(1)_R$ charge, since they are 
real. It follows that the components that are non-zero in Wess-Zumino 
gauge transform as:
\beq
A^\mu \rightarrow A^\mu,\qquad\quad
\lambda \rightarrow e^{i\alpha} \lambda,\qquad\quad
D \rightarrow D.\qquad\quad
\label{eq:Rchargesvector}
\eeq
and so have $U(1)_R$ charges $0$, $1$, and $0$ respectively.
Therefore, a Majorana gaugino mass term $\frac{1}{2} M_\lambda \lambda\lambda$, 
which will appear when supersymmetry is broken, also always breaks the 
continuous $U(1)_R$ symmetry.
The superspace integration measures $d^2\theta$ and $d^2\theta^\dagger$ 
and the chiral covariant derivatives $D_\alpha$ and 
$\Dcon_{\dot\alpha}$ carry $U(1)_R$ charges $-2$, $+2$, $-1$, and $+1$ 
respectively. It follows that the gauge field-strength superfield 
${\cal W}_\alpha$ carries $U(1)_R$ charge $+1$. (The $U(1)_R$ charges of various
objects are collected in Table \ref{table:Rcharges}.) 
It is then not hard to check that all 
supersymmetric Lagrangian terms found above that involve gauge superfields are 
automatically and necessarily $R$-symmetric, including the couplings to 
chiral superfields. This is also true of the canonical K\"ahler potential
contribution.%
\renewcommand{\arraystretch}{1.4}
\begin{table}[tb]
\begin{center}
\begin{tabular}{|c|c|c|c|c|c|c|c|c|c|c|c|c|c|}
\hline
& $\theta_\alpha$ 
& $\theta^\dagger_{\dot\alpha}$
& $d^2\theta$
& $\hat Q_\alpha$ 
& $D_\alpha$ 
& ${\cal W}_\alpha$
& $A^\mu$ 
& $\lambda_\alpha$ 
& $D$ 
& $W$
& $\phi$ 
& $\psi_\alpha$
& $F_\Phi$
\\ \hline 
$U(1)_R$ charge 
& $+1$ 
& $-1$ 
& $-2$ 
& $-1$ 
& $-1$ 
& $+1$ 
& $0$ 
& $+1$ 
& $0$ 
& $+2$ 
& $r_\Phi$ 
& $r_\Phi-1$ 
& $r_\Phi-2$
\\ \hline
\end{tabular}
\caption{$U(1)_R$ charges of various objects.\label{table:Rcharges}} 
\end{center}
\end{table}

However, the superpotential $W(\Phi_i)$ must carry 
$U(1)_R$ charge $+2$ in order to conserve the $R$ symmetry, 
and this is certainly not 
automatic, and often not true. As a simple toy example, 
with a single gauge-singlet superfield $\Phi$, the allowed renormalizable 
terms in the superpotential are $W(\Phi) = L \Phi + \frac{M}{2} \Phi^2 + 
\frac{y}{6} \Phi^3$. If one wants to impose a continuous $U(1)_R$ symmetry, 
then one can have at most one of these terms; $L$ is allowed only if 
$r_\Phi = 2$, $M$ is allowed only if $r_\Phi = 1$, and $y$ is allowed 
only if $r_\Phi = 2/3$. The MSSM superpotential 
does turn out to conserve a global $U(1)_R$ symmetry, but it is both anomalous
and broken by Majorana gaugino masses and other supersymmetry breaking effects.

Since continuous $R$ symmetries do not commute 
with supersymmetry, and are not conserved in the MSSM after anomalies 
and supersymmetry breaking effects are
included, one might wonder why they are considered at all. 
Perhaps the most important answer to this involves the role of $U(1)_R$ 
symmetries in models that break global supersymmetry spontaneously, as 
will be discussed in section \ref{subsec:origins.Fterm} below. 
It is also possible to extend the particle content of the MSSM in such a way as to 
preserve a continuous, non-anomalous $U(1)_R$ symmetry, 
but at the cost of introducing Dirac gaugino masses and extra Higgs 
fields \cite{Kribs:2007ac}.

Another possibility is that a superpotential could 
have a discrete $Z_n$ $R$ symmetry, which can be 
obtained by restricting the transformation parameter $\alpha$ in 
eqs.~(\ref{eq:Rtranstheta})-(\ref{eq:Rchargesvector}) to integer 
multiples of $2\pi/n$. The $Z_n$ $R$ charges of all fields are then 
integers modulo $n$. However, note that the case $n=2$ is always trivial, in the 
sense that any $Z_2$ $R$ symmetry is exactly equivalent to a 
corresponding ordinary (non-$R$) $Z_2$ symmetry under which all components of 
each supermultiplet transform the same way. This is because when $\alpha$ 
is an integer multiple of $\pi$, then both $\theta$ and $\theta^\dagger$ 
always just transform by changing sign, which means that fermionic fields just  
change sign relative to their bosonic partners. The number of fermionic fields 
in any Lagrangian term, in any theory, is always even, so the extra sign 
change for fermionic fields has no effect.

\section{Soft supersymmetry breaking interactions}\label{sec:soft}
\renewcommand{\theequation}{\arabic{section}.\arabic{equation}}
\setcounter{equation}{0}
\setcounter{figure}{0}
\setcounter{table}{0}
\setcounter{footnote}{1}

A realistic phenomenological model must contain supersymmetry breaking.
From a theoretical perspective, we expect that supersymmetry, if it exists
at all, should be an exact symmetry that is broken spontaneously. In other
words, the underlying model should have a Lagrangian density that is
invariant under supersymmetry, but a vacuum state that is not. In this
way, supersymmetry is hidden at low energies in a manner analogous
to the fate of the electroweak symmetry in the ordinary Standard Model. 

Many models of spontaneous symmetry breaking have indeed been proposed and
we will mention the basic ideas of some of them in section
\ref{sec:origins}. These always involve extending the MSSM to include new
particles and interactions at very high mass scales, and there is no
consensus on exactly how this should be done. However, from a practical
point of view, it is extremely useful to simply parameterize our ignorance
of these issues by just introducing extra terms that break supersymmetry
explicitly in the effective MSSM Lagrangian. As was argued in the
Introduction, the supersymmetry-breaking couplings should be soft (of
positive mass dimension) in order to be able to naturally maintain a
hierarchy between the electroweak scale and the Planck (or any other very
large) mass scale. This means in particular that dimensionless 
supersymmetry-breaking couplings should be absent.

The possible soft supersymmetry-breaking terms in the Lagrangian of a
general theory are
\beq
\lagr_{\rm soft}\! &=& \!
-\left (
\half M_a\, \lambda^a\lambda^a 
+ {1\over 6}a^{ijk} \phi_i\phi_j\phi_k 
+ \half b^{ij} \phi_i\phi_j 
+ t^i \phi_i \right )
+ \conj 
- (m^2)_j^i \phi^{j*} \phi_i , \phantom{xxxx}
\label{lagrsoft}
\\
\lagr_{{\rm maybe}\>\,{\rm soft}}\! &=& \!
-{1\over 2}c_i^{jk} \phi^{*i}\phi_j\phi_k + \conj
\label{lagrsoftprime}
\eeq
They consist of gaugino masses $M_a$ for each gauge group, scalar
squared-mass terms $(m^2)_i^j$ and $b^{ij}$, and (scalar)$^3$ couplings
$a^{ijk}$ and $c_i^{jk}$, and ``tadpole" couplings $t^i$. The last of
these requires $\phi_i$ to be a gauge singlet, and so $t^i$ does not occur 
in the MSSM. One might wonder why we have not included possible soft mass
terms for the chiral supermultiplet fermions, like ${\cal L} = -\half
m^{ij} \psi_i \psi_j + {\rm c.c.}$~~Including such terms would be
redundant; they can always be absorbed into a redefinition of the
superpotential and the terms $(m^2)_j^{i}$ and $c_i^{jk}$. 

It has been shown rigorously that a softly broken supersymmetric theory
with $\lagr_{\rm soft}$ as given by eq.~(\ref{lagrsoft}) is indeed free of
quadratic divergences in quantum corrections to scalar masses, to all
orders in perturbation theory \cite{softterms}. The situation is slightly
more subtle if one tries to include the non-holomorphic (scalar)$^3$
couplings in $\lagr_{{\rm maybe}\>\,{\rm soft}}$. If any of the chiral
supermultiplets in the theory are singlets under all gauge symmetries,
then non-zero $c_i^{jk}$ terms can lead to quadratic divergences, despite
the fact that they are formally soft. Now, this constraint need not apply
to the MSSM, which does not have any gauge-singlet chiral supermultiplets.
Nevertheless, the possibility of $c_i^{jk}$ terms is nearly always
neglected. The real reason for this is that it is difficult to
construct models of spontaneous supersymmetry breaking in which the
$c_i^{jk}$ are not negligibly small.  
In the special case of a theory that has 
chiral supermultiplets that are singlets or 
in the adjoint representation of a simple factor of the gauge group,
then there are also possible soft supersymmetry-breaking Dirac mass
terms between the corresponding fermions $\psi_a$ and the gauginos
\cite{Polchinski:1982an}-\cite{Fox:2002bu}:
\beq
{\cal L} \,=\,  -M_{\rm Dirac}^a \lambda^a \psi_a + {\rm c.c.}
\label{eq:Diracgauginos}
\eeq
This is not relevant for the MSSM with minimal field 
content, which does not have adjoint representation chiral 
supermultiplets. Therefore, equation (\ref{lagrsoft}) is usually taken to 
be the general form of the soft supersymmetry-breaking Lagrangian. For 
some interesting exceptions, see 
refs.~\cite{Polchinski:1982an}-\cite{Plehn:2008ae}.

The terms in $\lagr_{\rm soft}$ clearly do break supersymmetry, 
because they involve
only scalars and gauginos and not their respective superpartners. In fact,
the soft terms in $\lagr_{\rm soft}$ are capable of giving masses to all
of the scalars and gauginos in a theory, even if the gauge bosons and
fermions in chiral supermultiplets are massless (or relatively light). The
gaugino masses $M_a$ are always allowed by gauge symmetry. The $(m^2)_j^i$
terms are allowed for $i,j$ such that $\phi_i$, $\phi^{j*}$ transform in
complex conjugate representations of each other under all gauge
symmetries; in particular this is true of course when $i=j$, so every
scalar is eligible to get a mass in this way if supersymmetry is broken. 
The remaining soft terms may or may not be allowed by the symmetries. 
The $a^{ijk}$, $b^{ij}$, and $t^i$ 
terms have the same form as the $y^{ijk}$, $M^{ij}$, and $L^i$ terms in 
the superpotential [compare eq.~(\ref{lagrsoft}) to 
eq.~(\ref{superpotentialwithlinear})
or eq.~(\ref{superpot})], so they will each be allowed by gauge 
invariance if and only if a corresponding superpotential term is allowed. 

\begin{figure}
\begin{center}
\begin{picture}(72,56)(0,-4)
\SetScale{1.1}
\Photon(0,12)(36,12){2.25}{4}
\Photon(72,12)(36,12){2.25}{4}
\SetWidth{0.85}
\ArrowLine(0,12)(36,12)
\ArrowLine(72,12)(36,12)
\Line(33,9)(39,15)
\Line(39,9)(33,15)
\Text(39.6,-13.2)[c]{(a)}
\end{picture}
\hspace{1.2cm}
\begin{picture}(72,56)(0,-4)
\SetScale{1.1}
\SetWidth{0.85}
\DashLine(0,12)(36,12){4}
\DashLine(72,12)(36,12){4}
\ArrowLine(17.99,12)(18,12)
\ArrowLine(55,12)(55.01,12)
\Line(33,9)(39,15)
\Line(39,9)(33,15)
\Text(0.5,20.8)[c]{$i$}
\Text(77.5,21.2)[c]{$j$}
\Text(39.6,-13.2)[c]{(b)}
\end{picture}
\hspace{1.2cm}
\begin{picture}(72,56)(0,-4)
\SetScale{1.1}
\SetWidth{0.85}
\DashLine(0,12)(36,12){4}
\DashLine(72,12)(36,12){4}
\ArrowLine(17.99,12)(18,12)
\ArrowLine(55.01,12)(55,12)
\Line(33,9)(39,15)
\Line(39,9)(33,15)
\Text(0.5,20.8)[c]{$i$}
\Text(77.5,21.2)[c]{$j$}
\Text(39.6,-13.2)[c]{(c)}
\end{picture}
\hspace{1.2cm}
\begin{picture}(66,56)(0,-4)
\SetScale{1.1}
\SetWidth{0.85}
\DashLine(33,52.5)(33,12){4}
\DashLine(0,0)(33,12){4}
\DashLine(66,0)(33,12){4}
\ArrowLine(33,32.2501)(33,32.25)
\ArrowLine(16.5,6)(16.5165,6.006)
\ArrowLine(49.5,6)(49.4835,6.006)
\Text(1,10.7)[c]{$j$}
\Text(70,9.7)[c]{$k$}
\Text(30.5,53)[c]{$i$}
\Text(36.3,-13.2)[c]{(d)}
\end{picture}
\end{center}
\caption{Soft supersymmetry-breaking terms:
(a) Gaugino mass $M_a$;
(b) non-holomorphic scalar squared mass $(m^2)_j^i$;
(c) holomorphic scalar squared mass $b^{ij}$;
and
(d) scalar cubic coupling $a^{ijk}$.
\label{fig:soft}}
\end{figure}
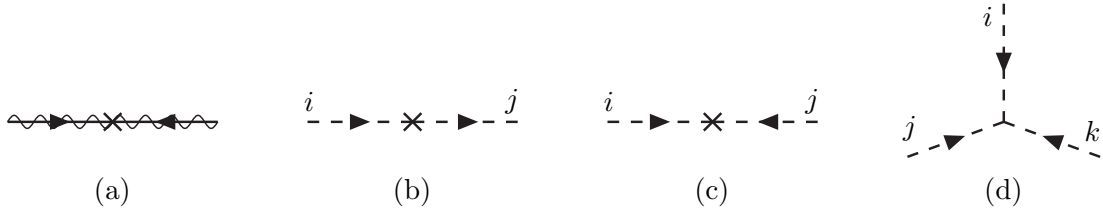

The Feynman diagram interactions corresponding to the allowed soft terms
in eq.~(\ref{lagrsoft}) are shown in Figure~\ref{fig:soft}.%
For each of
the interactions in Figures~\ref{fig:soft}a,c,d there is another with all
arrows reversed, corresponding to the complex conjugate term in the
Lagrangian. We will apply these general results to the specific case of
the MSSM in the next section. 

\section{The Minimal Supersymmetric Standard Model}\label{sec:mssm}
\renewcommand{\theequation}{\arabic{section}.\arabic{subsection}.\arabic{equation}}
\setcounter{equation}{0}
\setcounter{figure}{0}
\setcounter{table}{0}
\setcounter{footnote}{1}

In sections \ref{sec:susylagr} and \ref{sec:soft}, we have found a general
recipe for constructing Lagrangians for softly broken supersymmetric
theories. We are now ready to apply these general results to the MSSM. The
particle content for the MSSM was described in the Introduction. In this
section we will complete the model by specifying the superpotential and
the soft supersymmetry-breaking terms. 

\subsection{The superpotential and supersymmetric
interactions}\label{subsec:mssm.superpotential}
\setcounter{equation}{0}
\setcounter{footnote}{1}

The superpotential for the MSSM is 
\beq
W_{\rm MSSM} =
\sbar u {\bf y_u} Q H_u -
\sbar d {\bf y_d} Q H_d -
\sbar e {\bf y_e} L H_d +
\mu H_u H_d \> .
\label{MSSMsuperpot}
\eeq
The objects $H_u$, $H_d$, $Q$, $L$, $\sbar u$, $\sbar d$, $\sbar e$
appearing here are chiral superfields corresponding to the chiral
supermultiplets in Table \ref{tab:chiral}.  (Alternatively, they can be
just thought of as the corresponding scalar fields, as was done in section
\ref{sec:susylagr}, but we prefer not to put the tildes on $Q$, $L$,
$\sbar u$, $\sbar d$, $\sbar e$ in order to reduce clutter.) The
dimensionless Yukawa coupling parameters ${\bf y_u}, {\bf y_d}, {\bf y_e}$
are 3$\times 3$ matrices in family space. All of the gauge [$SU(3)_C$
color and $SU(2)_L$ weak isospin] and family indices in
eq.~(\ref{MSSMsuperpot}) are suppressed. The ``$\mu$ term", as it is
traditionally called, can be written out as $\mu (H_u)_\alpha (H_d)_\beta
\epsilon^{\alpha\beta}$, where $\epsilon^{\alpha\beta}$ is used to tie
together $SU(2)_L$ weak isospin indices $\alpha,\beta=1,2$ in a
gauge-invariant way. Likewise, the term $\sbar u {\bf y_u} Q H_u$ can be
written out as $\sbar u^{ia}\, {({\bf y_u})_i}^j\, Q_{j\alpha a}\,
(H_u)_\beta \epsilon^{\alpha \beta}$, where $i=1,2,3$ is a family index,
and $a=1,2,3$ is a color index which is lowered (raised) in the $\bf 3$
($\bf \overline 3$) representation of $SU(3)_C$. 

The $\mu$ term in eq.~(\ref{MSSMsuperpot}) is the supersymmetric version
of the Higgs boson mass in the Standard Model. It is unique, because terms
$H_u^* H_u$ or $H_d^* H_d$ are forbidden in the superpotential, which must
be holomorphic in the chiral superfields (or equivalently in the scalar
fields) treated as complex variables, as shown in section
\ref{subsec:susylagr.chiral}. We can also see from the form of
eq.~(\ref{MSSMsuperpot}) why both $H_u$ and $H_d$ are needed in order to
give Yukawa couplings, and thus masses, to all of the quarks and leptons.
Since the superpotential must be holomorphic, the $\sbar u Q H_u $ Yukawa
terms cannot be replaced by something like $\sbar u Q H_d^*$. Similarly,
the $\sbar d Q H_d$ and $\sbar e L H_d$ terms cannot be replaced by
something like $\sbar d Q H_u^*$ and $\sbar e L H_u^*$. The analogous
Yukawa couplings would be allowed in a general non-supersymmetric two
Higgs doublet model, but are forbidden by the structure of supersymmetry.
So we need both $H_u$ and $H_d$, even without invoking the argument based
on anomaly cancellation mentioned in the Introduction. 

The Yukawa matrices determine the current masses and CKM mixing angles of
the ordinary quarks and leptons, after the neutral scalar components of
$H_u$ and $H_d$ get VEVs. Since the top quark, bottom quark and tau lepton
are the heaviest fermions in the Standard Model, it is often useful to
make an approximation that only the $(3,3)$ family components of each of
${\bf y_u}$, ${\bf y_d}$ and ${\bf y_e}$ are important:
\beq
{\bf y_u} \approx \pmatrix{0&0&0\cr 0&0&0 \cr 0&0&y_t},\qquad
{\bf y_d} \approx \pmatrix{0&0&0\cr 0&0&0 \cr 0&0&y_b},\qquad
{\bf y_e} \approx \pmatrix{0&0&0\cr 0&0&0 \cr 0&0&y_\tau}.\>\>{}
\label{heavytopapprox}
\eeq
In this limit, only the third family and Higgs fields contribute to the
MSSM superpotential. It is instructive to write the superpotential in
terms of the separate $SU(2)_L$ weak isospin components [$Q_3 = (t\, b)$,
$L_3 = (\nu_\tau\, \tau)$, $H_u = (H_u^+\, H_u^0)$, $H_d = (H_d^0\,
H_d^-)$, $\sbar u_3 = \sbar t$, $\sbar d_3 = \sbar b$, $\sbar e_3 = \sbar
\tau$], so: 
\beq
W_{\rm MSSM}\! &\approx & \!
y_t (\sbar t t H_u^0 - \sbar t b H_u^+) -
y_b (\sbar b t H_d^- - \sbar b b H_d^0) -
y_\tau (\sbar \tau \nu_\tau H_d^- - \sbar \tau \tau H_d^0)
\> \nonumber \\
&& +
\mu (H_u^+ H_d^- - H_u^0 H_d^0).
\label{Wthird}
\eeq
The minus signs inside the parentheses appear because of the antisymmetry
of the $\epsilon^{\alpha\beta}$ symbol used to tie up the $SU(2)_L$
indices. The other minus signs in eq.~(\ref{MSSMsuperpot}) were chosen 
(as a convention) so
that the terms $y_t \sbar t t H_u^0$, $y_b \sbar b b H_d^0$, and $y_\tau
\sbar \tau \tau H_d^0$, which will become the top, bottom and tau masses
when $H_u^0$ and $H_d^0$ get VEVs, each have overall positive signs in
eq.~(\ref{Wthird}). 

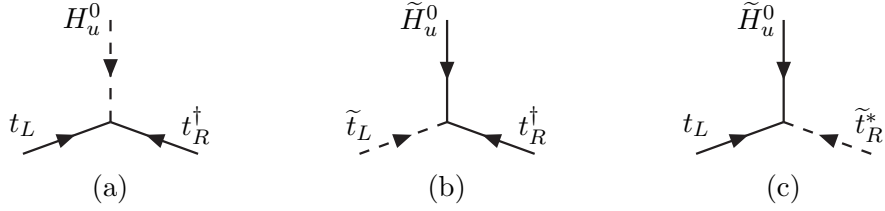
\begin{figure}
\begin{center}
\begin{picture}(66,60)(0,0)
\SetWidth{0.85}
\ArrowLine(0,0)(33,12)  
\ArrowLine(66,0)(33,12)
\DashLine(33,50.5)(33,12){4}
\ArrowLine(33,32.2501)(33,32.25)
\Text(0,11)[c]{$t_L$}
\Text(65.5,11)[c]{$t_R^\dagger$}
\Text(22.5,51)[c]{$H_u^0$}
\Text(33,-13)[c]{(a)}
\end{picture}
\hspace{1.9cm}
\begin{picture}(66,60)(0,0)
\SetWidth{0.85}
\DashLine(0,0)(33,12){4}  
\ArrowLine(66,0)(33,12)
\ArrowLine(33,50.5)(33,12)
\ArrowLine(16.5,6)(16.5165,6.006)
\Text(0,11)[c]{$\stilde t_L$}
\Text(65.5,11)[c]{$t_R^\dagger$}
\Text(22.7,51)[c]{$\widetilde H_u^0$}
\Text(33,-13)[c]{(b)}
\end{picture}
\hspace{1.9cm}
\begin{picture}(66,60)(0,0)
\SetWidth{0.85}
\ArrowLine(0,0)(33,12)  
\DashLine(66,0)(33,12){4}
\ArrowLine(33,50.5)(33,12)
\ArrowLine(49.5,6)(49.4835,6.006)
\Text(0,11)[c]{$t_L$}
\Text(65.5,11)[c]{$\stilde t_R^*$}
\Text(22.7,51)[c]{$\widetilde H_u^0$}
\Text(33,-13)[c]{(c)}
\end{picture}
\end{center}
\caption{The top-quark Yukawa coupling (a) and its ``supersymmetrizations"
(b), (c), all of strength~$y_t$.\label{fig:topYukawa}}
\end{figure}

Since the Yukawa interactions $y^{ijk}$ in a general supersymmetric theory
must be completely symmetric under interchange of $i,j,k$, we know that
${\bf y_u}$, ${\bf y_d}$ and ${\bf y_e}$ imply not only Higgs-quark-quark
and Higgs-lepton-lepton couplings as in the Standard Model, but also
squark-Higgsino-quark and slepton-Higgsino-lepton interactions. To
illustrate this, Figures~{\ref{fig:topYukawa}}a,b,c show some of the
interactions involving the top-quark Yukawa coupling $y_t$.%
Figure \ref{fig:topYukawa}a is the Standard Model-like coupling of the top
quark to the neutral complex scalar Higgs boson, which follows from the
first term in eq.~(\ref{Wthird}). For variety, we have used $t_L$ and
$t_R^\dagger$ in place of their synonyms $t$ and $\sbar t$ 
(see the discussion near the end of section
\ref{sec:notations}). In Figure~\ref{fig:topYukawa}b, we have the coupling
of the left-handed top squark $\stilde t_L$ to the neutral higgsino field
${\stilde H}_u^0$ and right-handed top quark, while in
Figure~\ref{fig:topYukawa}c the right-handed top anti-squark field (known
either as $\stilde {\sbar t}$ or $\stilde t_R^*$ depending on taste)
couples to ${\stilde H}^0_u$ and $t_L$. For each of the three
interactions, there is another with $H_u^0\rightarrow H_u^+$ and $t_L
\rightarrow -b_L$ (with tildes where appropriate), corresponding to the
second part of the first term in eq.~(\ref{Wthird}). All of these
interactions are required by supersymmetry to have the same strength
$y_t$. These couplings are dimensionless and can be modified by the
introduction of soft supersymmetry breaking only through finite (and
small) radiative corrections, so this equality of interaction strengths is
also a prediction of softly broken supersymmetry. A useful mnemonic is
that each of Figures~{\ref{fig:topYukawa}}a,b,c can be obtained from any
of the others by changing two of the particles into their superpartners. 

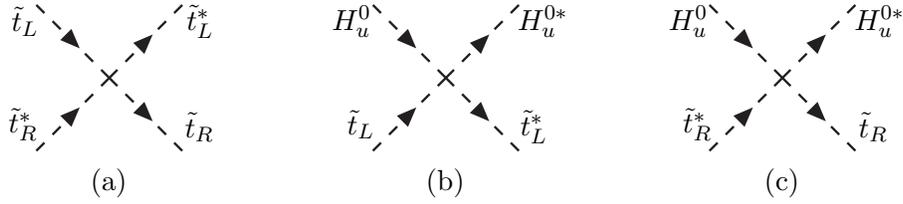
\begin{figure}
\begin{center}
\begin{picture}(56,55)(0,0)
\SetScale{1.1}
\SetWidth{0.8}
\DashLine(0,0)(25,25){4}
\DashLine(50,0)(25,25){4}
\DashLine(0,50)(25,25){4}
\DashLine(50,50)(25,25){4}  
\ArrowLine(12.5,12.5)(12.501,12.501)
\ArrowLine(37.499,12.501)(37.5,12.5)
\ArrowLine(12.499,37.501)(12.5,37.5)
\ArrowLine(37.5,37.5)(37.501,37.501)
\Text(-5,10)[c]{$\tilde t_R^*$}
\Text(61.75,9)[c]{$\tilde t_R$}
\Text(-4.25,49)[c]{$\tilde t_L$}
\Text(62,49)[c]{$\tilde t_L^*$}   
\Text(27.5,-12.1)[c]{(a)}
\end{picture}
\hspace{2.25cm}
\begin{picture}(56,55)(0,0)
\SetScale{1.1}
\SetWidth{0.8}
\DashLine(0,0)(25,25){4}
\DashLine(50,0)(25,25){4}
\DashLine(0,50)(25,25){4}
\DashLine(50,50)(25,25){4}  
\ArrowLine(12.5,12.5)(12.501,12.501)
\ArrowLine(37.499,12.501)(37.5,12.5)
\ArrowLine(12.499,37.501)(12.5,37.5)
\ArrowLine(37.5,37.5)(37.501,37.501)
\Text(-4.5,10)[c]{$\tilde t_L$}
\Text(61,9)[c]{$\tilde t_L^*$}
\Text(-8,49)[c]{$H_u^0$}
\Text(64.3,49)[c]{$H_u^{0*}$}   
\Text(27.5,-12.1)[c]{(b)}
\end{picture}
\hspace{2.25cm}
\begin{picture}(56,55)(0,0)
\SetScale{1.1}
\SetWidth{0.8}
\DashLine(0,0)(25,25){4}
\DashLine(50,0)(25,25){4}
\DashLine(0,50)(25,25){4}
\DashLine(50,50)(25,25){4}  
\ArrowLine(12.5,12.5)(12.501,12.501)
\ArrowLine(37.499,12.501)(37.5,12.5)
\ArrowLine(12.499,37.501)(12.5,37.5)
\ArrowLine(37.5,37.5)(37.501,37.501)
\Text(-4.5,10)[c]{$\tilde t_R^*$}
\Text(62,9)[c]{$\tilde t_R$}
\Text(-8,49)[c]{$H_u^0$}
\Text(64.3,49)[c]{$H_u^{0*}$}   
\Text(27.5,-12.1)[c]{(c)}
\end{picture}
\end{center}
\caption{Some of the (scalar)$^4$ interactions with strength
proportional to $y_t^2$.
\label{fig:stop}}
\end{figure}

There are also scalar quartic interactions with strength proportional to
$y_t^2$, as can be seen from Figure~\ref{fig:dim0}c or the last term
in eq.~(\ref{ordpot}). Three of them are shown in Figure~{\ref{fig:stop}}.%
Using eq.~(\ref{ordpot}) and eq.~(\ref{Wthird}), one can see that 
there are
five more, which can be obtained by replacing $\stilde t_L \rightarrow
\stilde b_L$ and/or $H_u^0 \rightarrow H_u^+$ in each vertex. This
illustrates the remarkable economy of supersymmetry; there are many
interactions determined by only a single parameter. In a similar way, the
existence of all the other quark and lepton Yukawa couplings in the
superpotential eq.~(\ref{MSSMsuperpot}) leads not only to
Higgs-quark-quark and Higgs-lepton-lepton Lagrangian terms as in the
ordinary Standard Model, but also to squark-higgsino-quark and
slepton-higgsino-lepton terms, and scalar quartic couplings [(squark)$^4$,
(slepton)$^4$, (squark)$^2$(slepton)$^2$, (squark)$^2$(Higgs)$^2$, and
(slepton)$^2$(Higgs)$^2$]. If needed, these can all be obtained in terms
of the Yukawa matrices $\bf y_u$, $\bf y_d$, and $\bf y_e$ as outlined
above. 

\begin{figure}
\begin{center}
\begin{picture}(66,60)(0,0)
\Photon(0,0)(33,12){1.8}{4}  
\SetWidth{0.85}
\ArrowLine(33,12)(0,0)  
\ArrowLine(33,12)(66,0)
\DashLine(33,50.5)(33,12){4}
\ArrowLine(33,32.2501)(33,32.25)
\Text(0,11)[c]{$\stilde g$}
\Text(65.5,10)[c]{$q$}
\Text(40,51)[c]{$\stilde q$}
\Text(33,-13)[c]{(a)}
\end{picture}
\hspace{1.75cm}
\begin{picture}(124,60)(0,0)
\Photon(0,0)(33,12){1.8}{4}  
\SetWidth{0.85}
\ArrowLine(33,12)(0,0)  
\ArrowLine(33,12)(66,0)
\DashLine(33,50.5)(33,12){4}
\ArrowLine(33,32.2501)(33,32.25)
\Text(0,11)[c]{$\stilde W$}
\Text(63.5,11)[l]{$q_L$, $\ell_L$, $\stilde H_u$, $\stilde H_d$}
\Text(38,51)[l]{$\stilde q_L$, $\stilde \ell_L$, $H_u$, $H_d$}
\Text(33,-13)[c]{(b)}
\end{picture}
\hspace{1.8cm}
\begin{picture}(124,60)(0,0)
\Photon(0,0)(33,12){1.8}{4}  
\SetWidth{0.85}
\ArrowLine(33,12)(0,0)  
\ArrowLine(33,12)(66,0)
\DashLine(33,50.5)(33,12){4}
\ArrowLine(33,32.2501)(33,32.25)
\Text(0,11)[c]{$\stilde B$}
\Text(63.5,11)[l]{$q$, $\ell$, $\stilde H_u$, $\stilde H_d$}
\Text(38,51)[l]{$\stilde q$, $\stilde \ell$, $H_u$, $H_d$}
\Text(33,-13)[c]{(c)}
\end{picture}
\end{center}
\caption{Couplings of the gluino, wino, and bino to MSSM (scalar,
fermion) pairs.
\label{fig:gaugino}}
\end{figure}
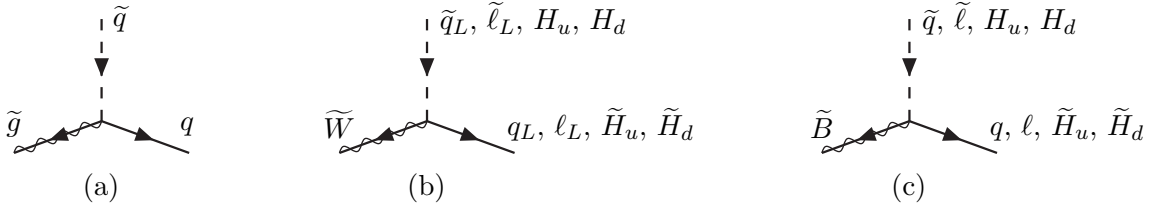

However, the dimensionless interactions determined by the superpotential
are usually not the most important ones of direct interest for
phenomenology. This is because the Yukawa couplings are already known to
be very small, except for those of the third family (top, bottom, tau).
Instead, production and decay processes for superpartners in the MSSM are
typically dominated by the supersymmetric interactions of gauge-coupling
strength, as we will explore in more detail in sections \ref{sec:decays}
and \ref{sec:signals}. The couplings of the Standard Model gauge bosons
(photon, $W^\pm$, $Z^0$ and gluons) to the MSSM particles are determined
completely by the gauge invariance of the kinetic terms in the Lagrangian.
The gauginos also couple to (squark, quark) and (slepton, lepton) and
(Higgs, higgsino) pairs as illustrated in the general case in
Figure~\ref{fig:gauge}g,h and the first two terms in the second line in
eq.~(\ref{gensusylagr}). For instance, each of the squark-quark-gluino
couplings is given by $\sqrt{2} g_3 (\stilde q \, T^{a} q \stilde g +
\conj)$ where $T^a = \lambda^a/2$ ($a=1\ldots 8$) are the matrix
generators for $SU(3)_C$. The Feynman diagram for this interaction is
shown in Figure~\ref{fig:gaugino}a.%
In Figures~\ref{fig:gaugino}b,c we
show in a similar way the couplings of (squark, quark), (lepton, slepton)
and (Higgs, higgsino) pairs to the winos and bino, with strengths
proportional to the electroweak gauge couplings $g$ and $g^\prime$
respectively. For each of these diagrams, there is another with all arrows
reversed. Note that the winos only couple to the left-handed squarks and
sleptons, and the (lepton, slepton) and (Higgs, higgsino) pairs of course
do not couple to the gluino. The bino coupling to each (scalar, fermion)
pair is also proportional to the weak hypercharge $Y$ as 
given in Table \ref{tab:chiral}.
The interactions shown in Figure~\ref{fig:gaugino} provide, for example,
for decays $\stilde q \rightarrow q\stilde g$ and $\stilde q \rightarrow
\stilde W q^\prime$ and $\stilde q \rightarrow \stilde B q$ when the final
states are kinematically allowed to be on-shell. However, a complication
is that the $\stilde W$ and $\stilde B$ states are not mass eigenstates,
because of splitting and mixing due to electroweak symmetry breaking, as
we will see in section \ref{subsec:MSSMspectrum.inos}. 

There are also various scalar quartic interactions in the MSSM that are
uniquely determined by gauge invariance and supersymmetry, according to
the last term in eq.~(\ref{fdpot}), as illustrated in
Figure~\ref{fig:gauge}i. Among them are (Higgs)$^4$ terms proportional to
$g^2$ and $g^{\prime 2}$ in the scalar potential. These are the direct
generalization of the last term in the Standard Model Higgs potential,
eq.~(\ref{higgspotential}), to the case of the MSSM. We will have occasion
to identify them explicitly when we discuss the minimization of the MSSM
Higgs potential in section \ref{subsec:MSSMspectrum.Higgs}. 

The dimensionful couplings in the supersymmetric part of the MSSM
Lagrangian are all dependent on $\mu$. Using the general result of
eq.~(\ref{lagrchiral}), $\mu$ provides for higgsino fermion
mass terms
\beq
-\lagr_{\mbox{higgsino mass}}= \mu (\stilde H_u^+ \stilde H_d^- - \stilde 
H_u^0 \stilde
H_d^0)+ \conj,
\label{poody}
\eeq
as well as Higgs squared-mass terms in the scalar potential
\beq
-\lagr_{\mbox{supersymmetric Higgs mass}} \,=\, |\mu|^2 
\bigl 
(
|H_u^0|^2 + |H_u^+|^2 + |H_d^0|^2 + |H_d^-|^2 \bigr ).
\label{movie}
\eeq
Since eq.~(\ref{movie}) is non-negative with a minimum at
$H_u^0 = H_d^0 = 0$, we cannot understand
electroweak symmetry breaking without including a negative
supersymmetry-breaking squared-mass soft term for the Higgs scalars. An
explicit treatment of the Higgs scalar potential will therefore have to
wait until we have introduced the soft terms for the MSSM. However, we can
already see a puzzle: we expect that $\mu$ should be roughly of order
$10^2$ or $10^3$ GeV, in order to allow a Higgs VEV of order 174 GeV
without too much miraculous cancellation between $|\mu|^2$ and the
negative soft squared-mass terms that we have not written down yet. But
why should $|\mu|^2$ be so small compared to, say, $\MPlanck^2$, and in
particular why should it be roughly of the same order as $m^2_{\rm soft}$?
The scalar potential of the MSSM seems to depend on two types of
dimensionful parameters that are conceptually quite distinct, namely the
supersymmetry-respecting mass $\mu$ and the supersymmetry-breaking soft
mass terms. Yet the observed value for the electroweak breaking scale
suggests that without miraculous cancellations, both of these apparently
unrelated mass scales should be within an order of magnitude or so of 100
GeV. This puzzle is called ``the $\mu$ problem". Several different
solutions to the $\mu$ problem have been proposed, involving extensions of
the MSSM of varying intricacy. They all work in roughly the same way; the
$\mu$ term is required or assumed to be absent at tree-level before
symmetry breaking, and then it arises from the VEV(s) of some new
field(s). These VEVs are in turn determined by minimizing a potential that
depends on soft supersymmetry-breaking terms. In this way, the value of
the effective parameter $\mu$ is no longer conceptually distinct from the
mechanism of supersymmetry breaking; if we can explain why $m_{\rm soft}
\ll \MPlanck$, we will also be able to understand why $\mu$ is of the same
order. In sections \ref{subsec:variations.NMSSM} and
\ref{subsec:variations.munonrenorm}
we will study three such
mechanisms: the Next-to-Minimal Supersymmetric Standard Model, the Kim-Nilles mechanism 
\cite{KimNilles}, and the Giudice-Masiero 
mechanism \cite{GiudiceMasiero}. 
Another solution based on loop effects was
proposed in ref.~\cite{muproblemGMSB}. 
From the point of view of the MSSM, however, 
we can just treat $\mu$ as an independent parameter, without
committing to a specific mechanism. 

\begin{figure}
\begin{center}
\begin{picture}(66,62)(0,0)
\SetWidth{0.85}
\DashLine(33,52.5)(33,12){4}
\DashLine(0,0)(33,12){4}
\DashLine(66,0)(33,12){4}
\ArrowLine(33,32.25)(33,32.2501)
\ArrowLine(16.5,6)(16.5165,6.006)
\ArrowLine(49.5,6)(49.4835,6.006)
\Text(0,11)[c]{$\stilde t_L$}
\Text(66.5,10)[c]{$\stilde t_R^*$}
\Text(23,56)[c]{$H_d^{0*}$}
\Text(33,-12)[c]{(a)}
\end{picture}
\hspace{1.7cm}
\begin{picture}(66,62)(0,0)
\SetWidth{0.85}
\DashLine(33,52.5)(33,12){4}
\DashLine(0,0)(33,12){4}
\DashLine(66,0)(33,12){4}
\ArrowLine(33,32.25)(33,32.2501)
\ArrowLine(16.5,6)(16.5165,6.006)
\ArrowLine(49.5,6)(49.4835,6.006)
\Text(0,11)[c]{$\stilde b_L$}
\Text(66.5,10)[c]{$\stilde b_R^*$}
\Text(23,56)[c]{$H_u^{0*}$}
\Text(33,-12)[c]{(b)}
\end{picture}
\hspace{1.7cm}
\begin{picture}(66,62)(0,0)
\SetWidth{0.85}
\DashLine(33,52.5)(33,12){4}
\DashLine(0,0)(33,12){4}
\DashLine(66,0)(33,12){4}
\ArrowLine(33,32.25)(33,32.2501)
\ArrowLine(16.5,6)(16.5165,6.006)
\ArrowLine(49.5,6)(49.4835,6.006)
\Text(0,11)[c]{$\stilde \tau_L$}
\Text(66.5,10)[c]{$\stilde \tau_R^*$}
\Text(23,56)[c]{$H_u^{0*}$}
\Text(33,-12)[c]{(c)}
\end{picture}
\end{center}
\caption{Some of the supersymmetric (scalar)$^3$ couplings proportional to
$\mu^* y_t$, $\mu^* y_b$, and $\mu^* y_\tau$. When $H_u^0$ and $H_d^0$ get
VEVs, these contribute to 
(a) $\stilde t_L,\stilde t_R$ mixing,
(b) $\stilde b_L,\stilde b_R$ mixing,
and (c) $\stilde \tau_L,\stilde \tau_R$ mixing.
\label{fig:stri}}
\end{figure}
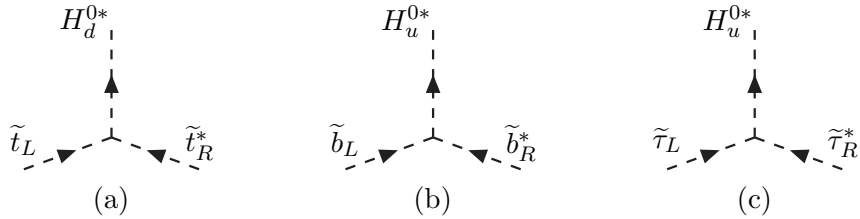

The $\mu$-term and the Yukawa couplings in the superpotential
eq.~(\ref{MSSMsuperpot}) combine to yield (scalar)$^3$ couplings [see the
second and third terms on the right-hand side of eq.~(\ref{ordpot})] of
the form
\beq
\lagr_{\mbox{supersymmetric (scalar)$^3$}} &= &
\mu^* (
 {\stilde{\sbar u}} {\bf y_u} \stilde u H_d^{0*}
+ {\stilde{\sbar d}} {\bf y_d} \stilde d H_u^{0*}
+ {\stilde{\sbar e}} {\bf y_e} \stilde e H_u^{0*}
\cr
&&
+{\stilde{\sbar u}} {\bf y_u} \stilde d H_d^{-*}
+{\stilde{\sbar d}} {\bf y_d} \stilde u H_u^{+*}
+{\stilde{\sbar e}} {\bf y_e} \stilde \nu H_u^{+*}
)
+ \conj
\label{striterms}
\eeq
Figure~\ref{fig:stri} shows some of these couplings, 
proportional to $\mu^* y_t$, $\mu^* y_b$, and $\mu^* y_\tau$ respectively.
These play an important role in determining the mixing of top squarks,
bottom squarks, and tau sleptons, as we will see in section
\ref{subsec:MSSMspectrum.sfermions}. 

\subsection{$R$-parity (also known as matter parity) and its
consequences}\label{subsec:mssm.rparity}
\setcounter{equation}{0}

The superpotential eq.~(\ref{MSSMsuperpot}) is minimal in the sense that
it is sufficient to produce a phenomenologically viable model. However,
there are other terms that one can write that are gauge-invariant
and holomorphic in the chiral superfields, but are not included in the MSSM
because they violate either baryon number ($\Baryon$) or total lepton
number ($\Lepton$). The most general gauge-invariant and renormalizable
superpotential would include not only eq.~(\ref{MSSMsuperpot}), but also
the terms
\beq
W_{\Delta {\rm L} =1} &=&
{1\over 2} \lambda^{ijk} L_iL_j{\sbar e_k}
+ \lambda^{\prime ijk} L_i Q_j {\sbar d_k}
+ \mu^{\prime i} L_i H_u
\label{WLviol} \\
W_{\Delta {\rm B}= 1} &=& {1\over 2} \lambda^{\prime\prime ijk}
{\sbar u_i}{\sbar d_j}{\sbar d_k}
\label{WBviol}
\eeq
where family indices $i=1,2,3$ have been restored. The chiral
supermultiplets carry baryon number assignments $\Baryon=+1/3$ for $Q_i$;
$\Baryon=-1/3$ for $\sbar u_i, \sbar d_i$; and $\Baryon=0$ for all others.
The total lepton number assignments are $\Lepton=+1$ for $L_i$,
$\Lepton=-1$ for $\sbar e_i$, and $\Lepton=0$ for all others. Therefore,
the terms in eq.~(\ref{WLviol}) violate total lepton number by 1 unit (as
well as the individual lepton flavors) and those in eq.~(\ref{WBviol})
violate baryon number by 1 unit. 

The possible existence of such terms might seem rather disturbing, since
corresponding $\Baryon$- and $\Lepton$-violating processes have not been
seen experimentally. The most obvious experimental constraint comes from
the non-observation of proton decay, which would violate both $\Baryon$
and $\Lepton$ by 1 unit. If both $\lambda^\prime$ and
$\lambda^{\prime\prime}$ couplings were present and unsuppressed, then
the lifetime of the proton would be extremely short.
For example, Feynman diagrams like the one in 
Figure~\ref{fig:protondecay}\footnote{In this diagram and 
others below, the arrows
on propagators are often
omitted for simplicity, and external fermion labels refer to physical
particle states rather than 2-component fermion fields.}
would lead to%
\begin{figure}
\begin{minipage}[]{0.41\linewidth}
\caption{Squarks would mediate disastrously rapid proton
decay if $R$-parity were violated by both $\Delta {\rm B} = 1$ and 
$\Delta {\rm L} = 1$ interactions. This example shows 
$p \rightarrow e^+ \pi^0$ mediated by a strange (or bottom) squark.
\label{fig:protondecay}}
\end{minipage}
\begin{minipage}[]{0.585\linewidth}
\begin{picture}(200,60)(-56,0)
\SetScale{1.5}
\Line(5,0)(125,0)
\Line(5,12)(42.5,26)
\Line(5,40)(42.5,26)
\Line(125,12)(87.5,26)
\Line(125,40)(87.5,26)
\DashLine(42.5,26)(87.5,26){4}
\Text(4,5)[c]{$u$}
\Text(4,26)[c]{$u$}
\Text(4,53)[c]{$d$}
\Text(98,50)[c]{$\tilde s_R^*$}
\Text(-18,30)[c]{$p^+$ 
$\displaystyle \left\{\vphantom{\displaystyle 
{{\Biggl \{ \Biggr \}}\atop{\Biggl \{ \Biggr \}}}}\right.$}
\Text(204,10)[c]{$\biggr \}{\pi^0}$}
\Text(185,5)[c]{$u$}
\Text(185,28)[c]{$u^*$}
\Text(198,58)[c]{$e^+$}
\Text(65,27)[c]{$ \lambda^{\prime\prime*}_{112}$}
\Text(140,27)[c]{$ \lambda'_{112}$}
\end{picture}
\end{minipage}
\end{figure}
$p^+ \rightarrow e^+ \pi^0$ (shown) or  
$\mu^+\pi^0$ or $\bar\nu \pi^+$ or $\bar\nu K^+$ etc.~depending on which
components of $\lambda^{\prime}$ and $\lambda^{\prime\prime}$ are
largest.\footnote{The coupling $\lambda^{\prime\prime}$ must be
antisymmetric in its last two flavor indices, since the color indices are
combined antisymmetrically. That is why the squark in
Figure~\ref{fig:protondecay} can be $\stilde{s}$ 
or $\stilde{b}$, but not $\stilde{d}$, for $u,d$ quarks in the proton.} 
Also, diagrams with $t$-channel squark exchange can lead to final states
$e^+ K^0$, $\mu^+ K^0$, $\nu\pi^+$, or $\nu K^+$, 
with the last two relying on left-right squark mixing.
As a rough estimate based on dimensional analysis, for example,
\beq
\Gamma_{p \rightarrow e^+ \pi^0} \>\sim\> m_{{\rm proton}}^5 \sum_{i=2,3}  
|\lambda^{\prime 11i}\lambda^{\prime\prime 11i}|^2/m_{\stilde d_i}^4,
\eeq
which would be a tiny fraction of a second if the couplings were of order 
unity and the squarks have masses of order 1 TeV.  In contrast, the decay 
time of the proton into lepton+meson final states is known experimentally 
to be in excess of $10^{32}$ years. Therefore, at least one of 
$\lambda^{\prime ijk}$ or $\lambda^{\prime\prime 11k}$ for each of 
$i=1,2$; $j=1,2$; $k=2,3$ must be extremely small. Many other processes 
also give strong constraints on the violation of lepton and baryon 
numbers \cite{rparityconstraints,RPVreviews}.

One could simply try to take $\Baryon$ and $\Lepton$ conservation as a
postulate in the MSSM. However, this is clearly a step backward from the
situation in the Standard Model, where the conservation of these quantum
numbers is {\it not} assumed, but is rather a pleasantly ``accidental"
consequence of the fact that there are no possible renormalizable
Lagrangian terms that violate $\Baryon$ or $\Lepton$. Furthermore, there
is a quite general obstacle to treating $\Baryon$ and $\Lepton$ as
fundamental symmetries of Nature, since they are known to be necessarily
violated by non-perturbative electroweak effects \cite{tHooft} 
(even though those
effects are calculably negligible for experiments at ordinary energies).
Therefore, in the MSSM one adds a new symmetry, which has the effect of
eliminating the possibility of $\Baryon$ and $\Lepton$ violating terms in
the renormalizable superpotential, while allowing the good terms in
eq.~(\ref{MSSMsuperpot}). This new symmetry is called ``$R$-parity"
\cite{Rparity} or equivalently ``matter parity" \cite{matterparity}.

Matter parity is a multiplicatively conserved quantum number defined as
\beq
P_M = (-1)^{3 (\Baryon-\Lepton)}
\label{defmatterparity}
\eeq
for each particle in the theory. It follows that the quark and
lepton supermultiplets all have $P_M=-1$, while the Higgs supermultiplets
$H_u$ and $H_d$ have $P_M=+1$. The gauge bosons and gauginos of course do
not carry baryon number or lepton number, so they are assigned matter
parity $P_M=+1$. The symmetry principle to be enforced is that a candidate
term in the Lagrangian (or in the superpotential) is allowed only if the
product of $P_M$ for all of the fields in it is $+1$. It is easy to see
that each of the terms in eqs.~(\ref{WLviol}) and (\ref{WBviol}) is thus
forbidden, while the good and necessary terms in eq.~(\ref{MSSMsuperpot})
are allowed. This discrete symmetry commutes with supersymmetry, as all
members of a given supermultiplet have the same matter parity. The
advantage of matter parity is that it can in principle be an {\it exact}
and fundamental symmetry, which B and L themselves cannot, since they are
known to be violated by non-perturbative electroweak effects. So even with
exact matter parity conservation in the MSSM, one expects that baryon
number and total lepton number violation can occur in tiny amounts, due to
non-renormalizable terms in the Lagrangian. However, the MSSM does not have
renormalizable interactions that violate B or L, with the standard
assumption of matter parity conservation.

It is often useful to recast matter parity in terms of $R$-parity,
defined for each particle as
\beq
P_R = (-1)^{3(\Baryon-\Lepton) + 2 s}
\label{defRparity}
\eeq
where $s$ is the spin of the particle. Now, matter parity conservation and
$R$-parity conservation are precisely equivalent, since the product of
$(-1)^{2s}$ for the particles involved in any interaction vertex in a
theory that conserves angular momentum is always equal to $+1$. However,
particles within the same supermultiplet do not have the same $R$-parity.
In general, symmetries with the property that fields within the same
supermultiplet have different transformations are called $R$ symmetries;  
they do not commute with supersymmetry.  Continuous $U(1)$ $R$ symmetries
were described in section \ref{Rsymmetry}, and
are often encountered in the model-building literature; they should not be
confused with $R$-parity, which is a discrete $Z_2$ symmetry. In fact, the
matter parity version of $R$-parity makes clear that there is really
nothing intrinsically ``$R$" about it; in other words it secretly does
commute with supersymmetry, so its name is somewhat suboptimal.
Nevertheless, the $R$-parity assignment is very useful for phenomenology
because all of the Standard Model particles and the Higgs bosons have even
$R$-parity ($P_R=+1$), while all of the squarks, sleptons, gauginos, and
higgsinos have odd $R$-parity ($P_R=-1$).

The $R$-parity odd particles are known as ``supersymmetric particles" or
``sparticles" for short, and they are distinguished by a tilde (see Tables
\ref{tab:chiral} and \ref{tab:gauge}). 
If $R$-parity is exactly conserved, then there can be no mixing
between the sparticles and the $P_R=+1$ particles. Furthermore, every
interaction vertex in the theory contains an even number of $P_R=-1$
sparticles. This has three extremely important phenomenological
consequences:
\begin{itemize}
\item[$\bullet$] The lightest sparticle with $P_R=-1$, called the
``lightest supersymmetric particle" or LSP, must be absolutely stable. If
the LSP is electrically neutral, it interacts only weakly with ordinary
matter, and so can make an attractive candidate \cite{neutralinodarkmatter} 
for the non-baryonic dark matter that seems to be required by cosmology.
\item[$\bullet$] Each sparticle other than the LSP must eventually decay
into a state that contains an odd number of LSPs (usually just one).
\item[$\bullet$] In collider experiments, sparticles can only be produced
in even numbers (usually two-at-a-time).
\end{itemize}

We {\it define} the MSSM to conserve $R$-parity or equivalently matter
parity. While this decision seems to be well-motivated phenomenologically
by proton decay constraints and the hope that the LSP will provide a good
dark matter candidate, it might appear somewhat artificial from a
theoretical point of view. After all, the MSSM would not suffer any
internal inconsistency if we did not impose matter parity conservation.
Furthermore, it is fair to ask why matter parity should be exactly
conserved, given that the discrete symmetries in the Standard Model
(ordinary parity $P$, charge conjugation $C$, time reversal $T$, etc.) are
all known to be inexact symmetries. Fortunately, it {\it is} sensible to
formulate matter parity as a discrete symmetry that is exactly conserved.
In general, exactly conserved, or ``gauged" discrete symmetries \cite{KW}
can exist provided that they satisfy certain anomaly cancellation
conditions \cite{discreteanomaly} (much like continuous gauged
symmetries). One particularly attractive way this could occur is if B$-$L
is a continuous gauge symmetry that is spontaneously broken at some
very high energy scale. A continuous $U(1)_{\Baryon-\Lepton}$
forbids the renormalizable terms that violate B and L 
\cite{Rparityoriginone,Rparityorigintwo},
but this gauge symmetry must be spontaneously broken, since there
is no corresponding massless vector boson.
However, if gauged $U(1)_{\Baryon - \Lepton}$ is only broken
by scalar VEVs (or other order parameters) that carry even integer
values of $3($B$-$L$)$, then $P_M$ will automatically survive as an
exactly conserved discrete remnant subgroup \cite{Rparityorigintwo}. 
A variety of extensions of the MSSM in which exact $R$-parity 
conservation is guaranteed in just this way have been proposed
(see for example \cite{Rparityorigintwo,Rparityoriginthree}). 

It may also be possible to have gauged discrete symmetries that do not owe
their exact conservation to an underlying continuous gauged symmetry, but
rather to some other structure such as can occur in string theory. It is
also possible that $R$-parity is broken, or is replaced by some
alternative discrete symmetry. We will briefly consider these as
variations on the MSSM in section \ref{subsec:variations.RPV}.

\subsection{Soft supersymmetry breaking in the
MSSM}\label{subsec:mssm.soft}
\setcounter{equation}{0}

To complete the description of the MSSM, we need to specify the soft
supersymmetry breaking terms. In section \ref{sec:soft}, we learned how to
write down the most general set of such terms in any supersymmetric
theory. Applying this recipe to the MSSM, we have: 
\beq
\lagr_{\rm soft}^{\rm MSSM} &=& -\half\left ( M_3 \stilde g\stilde g
+ M_2 \stilde W \stilde W + M_1 \stilde B\stilde B 
+\conj \right )
\nonumber
\\
&&
-\left ( \stilde {\sbar u} \,{\bf a_u}\, \stilde Q H_u
- \stilde {\sbar d} \,{\bf a_d}\, \stilde Q H_d
- \stilde {\sbar e} \,{\bf a_e}\, \stilde L H_d
+ \conj \right ) 
\nonumber
\\
&&
-\stilde Q^\dagger \, {\bf m^2_{Q}}\, \stilde Q
-\stilde L^\dagger \,{\bf m^2_{L}}\,\stilde L
-\stilde {\sbar u} \,{\bf m^2_{{\sbar u}}}\, {\stilde {\sbar u}}^\dagger
-\stilde {\sbar d} \,{\bf m^2_{{\sbar d}}} \, {\stilde {\sbar d}}^\dagger
-\stilde {\sbar e} \,{\bf m^2_{{\sbar e}}}\, {\stilde {\sbar e}}^\dagger
\nonumber \\
&&
- \, m_{H_u}^2 H_u^* H_u - m_{H_d}^2 H_d^* H_d
- \left ( b H_u H_d + \conj \right ) .
\label{MSSMsoft}
\eeq
In eq.~(\ref{MSSMsoft}), $M_3$, $M_2$, and $M_1$ are the gluino, wino, and
bino mass terms. Here, and from now on, we suppress the adjoint
representation gauge indices on the wino and gluino fields, and the gauge
indices on all of the chiral supermultiplet fields. The second line in
eq.~(\ref{MSSMsoft}) contains the (scalar)$^3$ couplings [of the type
$a^{ijk}$ in eq.~(\ref{lagrsoft})]. Each of ${\bf a_u}$, ${\bf a_d}$,
${\bf a_e}$ is a complex $3\times 3$ matrix in family space, with
dimensions of [mass]. They are in one-to-one correspondence with the
Yukawa couplings of the superpotential. The third line of
eq.~(\ref{MSSMsoft}) consists of squark and slepton mass terms of the
$(m^2)_i^j$ type in eq.~(\ref{lagrsoft}). Each of ${\bf m^2_{ Q}}$, ${\bf
m^2_{{\sbar u}}}$, ${\bf m^2_{{\sbar d}}}$, ${\bf m^2_{L}}$, ${\bf
m^2_{{\sbar e}}}$ is a $3\times 3$ matrix in family space that can have
complex entries, but they must be hermitian so that the Lagrangian is
real. (To avoid clutter, we do not put tildes on the $\bf Q$ in $\bf
m^2_Q$, etc.) Finally, in the last line of eq.~(\ref{MSSMsoft}) we have
supersymmetry-breaking contributions to the Higgs potential; $m_{H_u}^2$
and $m_{H_d}^2$ are squared-mass terms of the $(m^2)_i^j$ type, while $b$
is the only squared-mass term of the type $b^{ij}$ in eq.~(\ref{lagrsoft})
that can occur in the MSSM.\footnote{The parameter called $b$ here is often
seen elsewhere as $B\mu$ or $m_{12}^2$ or $m_3^2$.} As argued in
the Introduction, we expect
\beq
&&\!\!\!\!\! M_1,\, M_2,\, M_3,\, {\bf a_u},\, {\bf a_d},\, {\bf a_e}\,
\sim\, m_{\rm soft},\\
&&\!\!\!\!\! {\bf m^2_{ Q}},\,
{\bf m^2_{L}},\,
{\bf m^2_{{\sbar u}}},\,
{\bf m^2_{{\sbar d}}},\,
{\bf m^2_{{\sbar e}}},\, m_{H_u}^2,\, m_{H_d}^2,\, b\, \sim \,
m_{\rm soft}^2 ,
\eeq
with a characteristic mass scale $m_{\rm soft}$ that is not much larger
than $10^3$ GeV. The expression eq.~(\ref{MSSMsoft}) is the most general
soft supersymmetry-breaking Lagrangian of the form eq.~(\ref{lagrsoft})
that is compatible with gauge invariance and matter parity conservation in
the MSSM. 

Unlike the supersymmetry-preserving part of the Lagrangian, the above
$\lagr_{\rm soft}^{\rm MSSM}$ introduces many new parameters that were not
present in the ordinary Standard Model. A careful count \cite{dimsut}
reveals that there are 105 masses, phases and mixing angles in the MSSM
Lagrangian that cannot be rotated away by redefining the phases and flavor
basis for the quark and lepton supermultiplets, and that have no
counterpart in the ordinary Standard Model. Thus, in principle,
supersymmetry {\em breaking} (as opposed to supersymmetry itself) appears
to introduce a tremendous arbitrariness in the Lagrangian. 

\subsection{Hints of an Organizing Principle}\label{subsec:mssm.hints}
\setcounter{equation}{0}

Fortunately, there is already good experimental evidence that some
powerful organizing principle must govern the soft supersymmetry breaking
Lagrangian. This is because most of the new parameters in
eq.~(\ref{MSSMsoft}) imply flavor mixing or CP violating processes of the
types that are severely restricted by experiment
\cite{FCNCs}-\cite{Ciuchini:2002uv}. 

\begin{figure}
\begin{center}
\begin{picture}(138,58)(0,-5)
\SetScale{1.05}
\Photon(36,0)(93.6,0){1.75}{5}
\SetWidth{0.8}
\DashCArc(64.8,0)(28.8,0,180){5}
\Photon(85.1647,20.3647)(117.748,52.9481){2.4}{5}
\Line(0,0)(36,0)
\Line(93.6,0)(129.6,0)
\Line(93.6,0)(36,0)
\Line(62,26)(67.6,31.6)
\Line(62,31.6)(67.6,26)
\Text(68,-16.5)[c]{(a)}
\Text(132,48)[c]{${\gamma}$}
\Text(136,9)[c]{$e^-$}
\Text(6,9)[c]{$\mu^-$}
\Text(68,10)[c]{$\stilde B$}
\Text(46,34)[c]{$\stilde\mu_R$}
\Text(84,34)[c]{$\stilde e_R$}
\end{picture}
\hspace{0.75cm}
\begin{picture}(138,58)(0,-5)
\SetScale{1.05}
\PhotonArc(64.8,0)(28.8,0,180){2}{8}
\SetWidth{0.8}
\Photon(85.1647,20.3647)(117.748,52.9481){2.4}{5}
\CArc(64.8,0)(28.8,0,180)
\Line(0,0)(36,0)
\Line(93.6,0)(129.6,0)
\DashLine(93.6,0)(36,0){5}
\Line(62,-2.5)(67.6,2.5)
\Line(62,2.5)(67.6,-2.5)
\Text(68,-16.5)[c]{(b)}
\Text(132,48)[c]{${\gamma}$}
\Text(136,9)[c]{$e^-$}
\Text(6,9)[c]{$\mu^-$}
\Text(68.5,42.5)[c]{$\stilde W^-$}
\Text(56,8)[c]{$\stilde\nu_\mu$}
\Text(82,8)[c]{$\stilde\nu_e$}
\end{picture}
\hspace{0.75cm}
\begin{picture}(138,58)(0,-5)
\SetScale{1.05}
\Photon(36,0)(93.6,0){1.75}{5}
\SetWidth{0.8}
\DashCArc(64.8,0)(28.8,0,180){5}
\Photon(85.1647,20.3647)(117.748,52.9481){2.4}{5}
\Line(0,0)(36,0)
\Line(93.6,0)(129.6,0)
\Line(93.6,0)(36,0)
\Line(62,26)(67.6,31.6)
\Line(62,31.6)(67.6,26)
\Text(68,-16.5)[c]{(c)}
\Text(132,48)[c]{${\gamma}$}
\Text(136,9)[c]{$e^-$}
\Text(6,9)[c]{$\mu^-$}
\Text(68,10)[c]{$\stilde B$}
\Text(46,34)[c]{$\stilde\mu_L$}
\Text(84,34)[c]{$\stilde e_R$}
\end{picture}
\end{center}
\caption{Some of the diagrams that contribute to the 
process $\mu^- \rightarrow e^- \gamma$ in 
models with lepton flavor-violating 
soft supersymmetry breaking parameters (indicated by $\times$).
Diagrams (a), (b), and (c) contribute to constraints on the 
off-diagonal elements of ${\bf m^2_{\sbar e}}$,
${\bf m^2_{L}}$, and ${\bf a_e}$, respectively.
\label{fig:flavormuegamma}}
\end{figure}
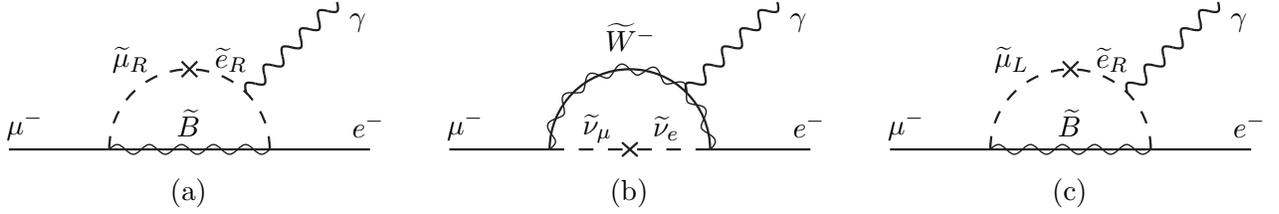

For example, suppose that $\bf m_{\sbar e}^2$ is not diagonal in the basis
$(\stilde e_R, \stilde \mu_R, \stilde \tau_R)$ of sleptons whose
superpartners are the right-handed parts of the Standard Model mass
eigenstates $e,\mu,\tau$. In that case, slepton mixing occurs, so the
individual lepton numbers will not be conserved, even for
processes that only involve the sleptons as virtual particles. A
particularly strong limit on this possibility comes from the experimental
bound on the process 
$\mu\rightarrow e \gamma$, which could arise from the one-loop
diagram shown in Figure~\ref{fig:flavormuegamma}a.
The symbol ``$\times$" on the slepton line
represents an insertion coming from 
$-({\bf m^2_{\sbar e}})_{21}\stilde \mu^*_R\stilde e_R$ in 
$\lagr_{\rm soft}^{\rm MSSM}$, and
the slepton-bino vertices are determined by the weak hypercharge gauge
coupling [see Figures~\ref{fig:gauge}g,h and eq.~(\ref{gensusylagr})].%
The result of calculating this diagram gives \cite{muegamma,muegammatwo},
approximately,
\beq 
\Branching (\mu \rightarrow e \gamma) 
&=& 
\left ( \frac{ |m^2_{\tilde \mu^*_R \tilde e_R}| }{m^2_{\tilde \ell_R}} 
\right )^2
\left (\frac{100\>{\rm GeV}}{m_{\tilde \ell_R}} \right )^4 
10^{-6} \times
\left \{ \begin{array}{ll}
15 & {\rm for}\>\, m_{\tilde B} \ll m_{\tilde \ell_R},
\\[-5pt]
5.6 & {\rm for}\>\, m_{\tilde B} = 0.5 m_{\tilde \ell_R},
\\[-5pt]
1.4 & {\rm for}\>\, m_{\tilde B} = m_{\tilde \ell_R},
\\[-5pt]
0.13 & {\rm for}\>\, m_{\tilde B} = 2 m_{\tilde \ell_R},
\end{array}
\right. 
\phantom{xxx}
\label{eq:muegamma}
\eeq
where it is assumed for simplicity that both $\tilde e_R$ and $\tilde
\mu_R$ are nearly mass eigenstates with almost degenerate squared masses
$m^2_{\tilde \ell_R}$, that $m^2_{\tilde \mu_R^* \tilde e_R} \equiv ({\bf
m^2_{\sbar e}})_{21} = [({\bf m^2_{\sbar e}})_{12}]^*$ can be treated as a
perturbation, and that the bino $\stilde B$ is nearly a mass eigenstate. This
result is to be compared to the present experimental upper limit 
$\Branching (\mu \rightarrow e \gamma)_{\rm exp} < 5.7\times 10^{-13}$ from
\cite{muegammaexperiment}. So, if the right-handed slepton squared-mass
matrix ${\bf m^2_{\sbar e}}$ were ``random", with all entries of
comparable size, then the prediction for $\Branching (\mu\rightarrow e\gamma)$
would be too large even if the sleptons and bino masses were at 1 TeV. 
For lighter superpartners, the constraint on $\tilde \mu_R, \tilde e_R$
squared-mass mixing becomes correspondingly more severe. There are also
contributions to $\mu \rightarrow e \gamma$ that depend on the
off-diagonal elements of the left-handed slepton squared-mass matrix $\bf
m^2_L$, coming from the 
diagram shown in fig.~\ref{fig:flavormuegamma}b 
involving the charged wino and the sneutrinos, as well as
diagrams just like fig.~\ref{fig:flavormuegamma}a but with left-handed
sleptons and either $\stilde B$ or $\stilde W^0$ exchanged.
Therefore, the slepton
squared-mass matrices must not have significant mixings for $\stilde
e_L,\stilde\mu_L$ either. 

Furthermore, after the Higgs scalars get VEVs, the $\bf a_e$ matrix could
imply squared-mass terms that mix left-handed and right-handed sleptons 
with different lepton flavors. For example, $\lagr_{\rm soft}^{\rm MSSM}$
contains ${\stilde{\sbar e}} {\bf a_e} {\stilde L} H_d + \conj$ which
implies terms $ -\langle H_d^0 \rangle ({\bf a_e})_{12} \stilde e_R^*
\stilde \mu_L -\langle H_d^0 \rangle ({\bf a_e})_{21} \stilde \mu_R^*
\stilde e_L + \conj$~~These also contribute to $\mu \rightarrow e \gamma$,
as illustrated in fig.~\ref{fig:flavormuegamma}c.
So the magnitudes of $({\bf a_e})_{12}$ and $({\bf a_e})_{21}$ are also
constrained by experiment to be small, but in a way that is more strongly
dependent on other model parameters \cite{muegammatwo}.  Similarly,
$({\bf a_e})_{13}, ({\bf a_e})_{31}$ and $({\bf a_e})_{23}, ({\bf
a_e})_{32}$ are constrained, although more weakly \cite{flavorreview}, by
the experimental limits on $\Branching (\tau \rightarrow e \gamma)$ and 
$\Branching (\tau \rightarrow \mu \gamma)$. 

\begin{figure}
\begin{center}
\begin{picture}(144,58)(0,-2)
\SetScale{0.8}
\Photon(40,0)(40,55){1.8}{5}
\Photon(104,0)(104,55){1.8}{5}
\SetWidth{0.96}
\DashLine(40,0)(104,0){5}
\DashLine(40,55)(104,55){5}
\Line(40,0)(0,0)
\Line(144,0)(104,0)
\Line(0,55)(40,55)
\Line(104,55)(144,55)
\Line(104,55)(104,0)
\Line(40,0)(40,55)
\Line(69.2,-2.8)(74.8,2.8)
\Line(69.2,2.8)(74.8,-2.8)
\Line(69.2,52.2)(74.8,57.8)
\Line(69.2,57.8)(74.8,52.2)
\Text(25.5,23.7)[c]{${\stilde g}$}
\Text(89,24.7)[c]{${\stilde g}$}
\Text(46,8.5)[c]{${\tilde d_R}$}
\Text(72.5,8)[c]{${\tilde s_R}$}
\Text(46,52)[c]{${\tilde s_R}^*$}
\Text(72.5,52.5)[c]{${\tilde d_R}^*$}
\Text(0,7)[c]{${d}$}
\Text(115,6)[c]{${s}$}
\Text(0,50)[c]{$\bar{s}$}
\Text(115,52)[c]{$\bar{d}$}
\Text(57.6,-16.5)[c]{(a)}
\end{picture}
\begin{picture}(144,58)(0,-2)
\SetScale{0.8}
\Photon(40,0)(40,55){1.8}{5}
\Photon(104,0)(104,55){1.8}{5}
\SetWidth{0.96}
\DashLine(40,0)(104,0){5}
\DashLine(40,55)(104,55){5}
\Line(40,0)(0,0)
\Line(144,0)(104,0)
\Line(0,55)(40,55)
\Line(104,55)(144,55)
\Line(104,55)(104,0)
\Line(40,0)(40,55)
\Line(69.2,-2.8)(74.8,2.8)
\Line(69.2,2.8)(74.8,-2.8)
\Line(69.2,52.2)(74.8,57.8)
\Line(69.2,57.8)(74.8,52.2)
\Text(25.5,23.7)[c]{${\stilde g}$}
\Text(89,24.7)[c]{${\stilde g}$}
\Text(46,8.5)[c]{${\tilde d_L}$}
\Text(72.5,8)[c]{${\tilde s_L}$}
\Text(46,52)[c]{${\tilde s_R}^*$}
\Text(72.5,52.5)[c]{${\tilde d_R}^*$}
\Text(0,7)[c]{${d}$}
\Text(115,6)[c]{${s}$}
\Text(0,50)[c]{$\bar{s}$}
\Text(115,52)[c]{$\bar{d}$}
\Text(57.6,-16.5)[c]{(b)}
\end{picture}
\begin{picture}(144,58)(0,-2)
\SetScale{0.8}
\Photon(40,0)(40,55){1.8}{5}
\Photon(104,0)(104,55){1.8}{5}
\SetWidth{0.96}
\DashLine(40,0)(104,0){5}
\DashLine(40,55)(104,55){5}
\Line(40,0)(0,0)
\Line(144,0)(104,0)
\Line(0,55)(40,55)
\Line(104,55)(144,55)
\Line(104,55)(104,0)
\Line(40,0)(40,55)
\Line(69.2,-2.8)(74.8,2.8)
\Line(69.2,2.8)(74.8,-2.8)
\Line(69.2,52.2)(74.8,57.8)
\Line(69.2,57.8)(74.8,52.2)
\Text(25.5,23.7)[c]{${\stilde g}$}
\Text(89,24.7)[c]{${\stilde g}$}
\Text(46,8.5)[c]{${\tilde d_L}$}
\Text(72.5,8)[c]{${\tilde s_R}$}
\Text(46,52)[c]{${\tilde s_R}^*$}
\Text(72.5,52.5)[c]{${\tilde d_L}^*$}
\Text(0,7)[c]{${d}$}
\Text(115,6)[c]{${s}$}
\Text(0,50)[c]{$\bar{s}$}
\Text(115,52)[c]{$\bar{d}$}
\Text(57.6,-16.5)[c]{(c)}
\end{picture}
\end{center}
\caption{Some of the diagrams that contribute to 
$K^0\leftrightarrow \overline K^0$ mixing in 
models with strangeness-violating 
soft supersymmetry breaking parameters (indicated by $\times$).
These diagrams contribute to constraints on the 
off-diagonal elements of (a) ${\bf m^2_{\sbar d}}$, (b) 
the combination of ${\bf m^2_{\sbar d}}$ and ${\bf m^2_{Q}}$, and 
(c) ${\bf a_d}$.
\label{fig:flavor}}
\end{figure}
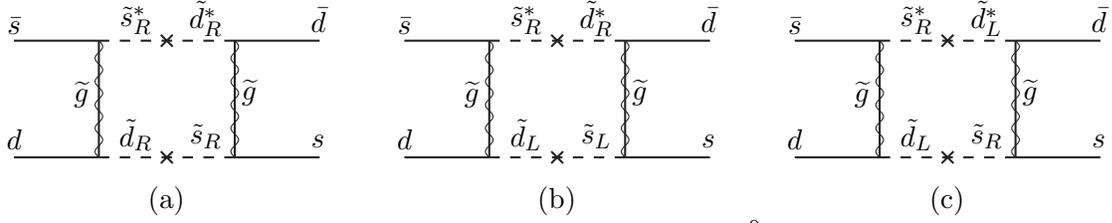

There are also important experimental constraints on the squark
squared-mass matrices. The strongest of these come from the neutral kaon
system. The effective Hamiltonian for $K^0\leftrightarrow \overline K^0$
mixing gets contributions from the diagrams in Figure~\ref{fig:flavor},
among others, if $\lagr_{\rm soft}^{\rm MSSM}$ contains terms
that mix down squarks and strange squarks.
The gluino-squark-quark
vertices in Figure~\ref{fig:flavor} are all fixed by supersymmetry to be
of QCD interaction strength.  (There are similar diagrams in which the
bino and winos are exchanged, which can be important depending on the
relative sizes of the gaugino masses.) For example, suppose that there is
a non-zero right-handed down-squark squared-mass mixing 
$({\bf m^2_{\sbar d}})_{21}$ 
in the basis corresponding to the quark mass eigenstates.%
Assuming that the supersymmetric correction to $\Delta m_K \equiv m_{K_L}
- m_{K_S}$ following from fig.~\ref{fig:flavor}a and others 
does not exceed, in absolute value, the experimental value 
$3.5 \times 10^{-12}$ MeV, ref.~\cite{Ciuchini:1998ix} obtains: 
\beq
\frac{|{\rm Re}[(m^2_{\tilde s_R^* \tilde d_R})^2]|^{1/2}}{
m^2_{\tilde q}}
&<&
\left ( \frac{m_{\tilde q}}{1000 \>{\rm GeV}} \right )
\times 
\left \{ \begin{array}{ll}
0.04 & {\rm for}\>\, m_{\tilde g} = 0.5 m_{\tilde q},
\\[-5pt]
0.10 & {\rm for}\>\, m_{\tilde g} = m_{\tilde q},
\\[-5pt]
0.22 & {\rm for}\>\, m_{\tilde g} = 2 m_{\tilde q}.
\end{array}
\right. 
\eeq
Here nearly degenerate squarks with mass $m_{\tilde q}$ are assumed for
simplicity, with $m^2_{\tilde s_R^* \tilde d_R} = ({\bf m^2_{\sbar
d}})_{21}$ treated as a perturbation. The same limit applies when
$m^2_{\tilde s_R^* \tilde d_R}$ is replaced by $m^2_{\tilde s_L^* \tilde
d_L} = ({\bf m^2_{Q}})_{21}$, in a basis corresponding to the
down-type quark mass eigenstates.  An even more striking limit applies to
the combination of both types of flavor mixing when they are comparable in
size, from diagrams including fig.~\ref{fig:flavor}b. 
The numerical constraint is \cite{Ciuchini:1998ix}: 
\beq
\frac{|{\rm Re}[m^2_{\tilde s_R^* \tilde d_R}
m^2_{\tilde s_L^* \tilde d_L}]|^{1/2}}{
m^2_{\tilde q}}
&<&
\left ( \frac{m_{\tilde q}}{1000 \>{\rm GeV}} \right )
\times
\left \{ 
\begin{array}{ll}
0.0016 & {\rm for}\>\, m_{\tilde g} = 0.5 m_{\tilde q},
\\[-5pt]
0.0020 & {\rm for}\>\, m_{\tilde g} = m_{\tilde q},
\\[-5pt]
0.0026 & {\rm for}\>\, m_{\tilde g} = 2 m_{\tilde q}.
\end{array}
\right.
\label{eq:striking}
\eeq
An off-diagonal contribution from ${\bf a_d}$ would cause flavor mixing 
between left-handed and right-handed squarks, just as discussed above for 
sleptons, resulting in a strong constraint from diagrams like 
fig.~\ref{fig:flavor}c. More generally, limits on $\Delta m_K$ and 
$\epsilon$ and $\epsilon'/\epsilon$ appearing in the neutral kaon 
effective Hamiltonian severely restrict the amounts of $\stilde 
d_{L,R},\,\stilde s_{L,R}$ squark mixings (separately and in various 
combinations), and associated CP-violating complex phases, that one can 
tolerate in the soft squared masses.

Weaker, but still interesting, constraints come from the $D^0, \overline 
D^0$ system, which limits the amounts of $\stilde u,\stilde c$ mixings 
from ${\bf m_{\sbar u}^2}$, ${\bf m_Q^2}$ and ${\bf a_u}$. The $B_d^0, 
\overline B_d^0$ and $B_s^0, \overline B_s^0$ systems similarly limit the 
amounts of $\stilde d,\stilde b$ and $\stilde s,\stilde b$ squark mixings 
from soft supersymmetry-breaking sources.  More constraints follow from 
rare $\Delta F=1$ meson decays, notably those involving the parton-level 
processes $b\rightarrow s\gamma$ and $b \rightarrow s \ell^+ \ell^-$ and 
$c \rightarrow u \ell^+ \ell^-$ and $s \rightarrow d e^+ e^-$ and $s 
\rightarrow d \nu \bar \nu$, all of which can be mediated by flavor mixing 
in soft supersymmetry breaking. There are also strict constraints on 
CP-violating phases in the gaugino masses and (scalar)$^3$ soft couplings 
following from limits on the electric dipole moments of the neutron and 
electron \cite{demon}. Detailed limits can be found in the literature
\cite{FCNCs}-\cite{Ciuchini:2002uv}, 
but the essential lesson from experiment is that the soft 
supersymmetry-breaking Lagrangian cannot be arbitrary or random.

\vspace{0.6mm}

All of these potentially dangerous flavor-changing and CP-violating
effects in the MSSM can be evaded if one assumes (or can explain!) that
supersymmetry breaking is suitably ``universal". Consider an idealized
limit in which the squark and slepton squared-mass matrices are
flavor-blind, each proportional to the $3\times 3$ identity matrix in
family space: 
\beq
{\bf m^2_{Q}} = m^2_{Q} {\bf 1},
\qquad\!\!\!\!
{\bf m^2_{\sbar u}} = m^2_{\sbar u} {\bf 1},
\qquad\!\!\!\!
{\bf m^2_{\sbar d}} = m^2_{\sbar d} {\bf 1},
\qquad\!\!\!\!
{\bf m^2_{L}} = m^2_{L} {\bf 1},
\qquad\!\!\!\!
{\bf m^2_{\sbar e}} = m^2_{\sbar e} {\bf 1}
.\>\>\>\>{}
\label{scalarmassunification}
\eeq
Then all squark and slepton mixing angles are rendered trivial, because
squarks and sleptons with the same electroweak quantum numbers will be
degenerate in mass and can be rotated into each other at will.
Supersymmetric contributions to flavor-changing neutral current processes
will therefore be very small in such an idealized limit, up to mixing
induced by $\bf a_u$, $\bf a_d$, $\bf a_e$. Making the further assumption
that the (scalar)$^3$ couplings are each proportional to the corresponding
Yukawa coupling matrix,
\beq
{\bf a_u} = A_{u0} \,{\bf y_u}, \>\>\>\qquad
{\bf a_d} = A_{d0} \,{\bf y_d}, \>\>\>\qquad
{\bf a_e} = A_{e0} \,{\bf y_e},
\label{aunification}
\eeq
will ensure that only the squarks and sleptons of the third family can
have large (scalar)$^3$ couplings. Finally, one can avoid disastrously
large CP-violating effects by assuming that the soft parameters do not
introduce new complex phases. This is automatic for $m_{H_u}^2$ and
$m_{H_d}^2$, and for $m_Q^2$, $m_{\sbar u}^2$, etc.~if
eq.~(\ref{scalarmassunification}) is assumed; if they were not real
numbers, the Lagrangian would not be real. One can also fix $\mu$ in the
superpotential and $b$ in eq.~(\ref{MSSMsoft}) to be real, by appropriate
phase rotations of fermion and scalar components of the $H_u$ and $H_d$
supermultiplets. If one then assumes that
\beq
{\rm Im} (M_1),\, {\rm Im} (M_2),\, {\rm Im} (M_3),\,
{\rm Im} (A_{u0}),\, {\rm Im} (A_{d0}),\, {\rm Im} (A_{e0}) 
\>\> = \>\>
0,
\qquad{}
\label{commonphase}
\eeq
then the only CP-violating phase in the theory will be the usual CKM phase
found in the ordinary Yukawa couplings. Together, the conditions
eqs.~(\ref{scalarmassunification})-(\ref{commonphase}) make up a rather
weak version of what is often called the hypothesis of {\it soft
supersymmetry-breaking universality}. The MSSM with these flavor- and
CP-preserving relations imposed has far fewer parameters than the most
general case. Besides the usual Standard Model gauge and Yukawa coupling
parameters, there are 3 independent real gaugino masses, only 5 real
squark and slepton squared mass parameters, 3 real scalar cubic coupling
parameters, and 4 Higgs mass parameters (one of which can be traded for
the known electroweak breaking scale). 

There are at least three other 
possible types of explanations for the suppression of flavor violation in 
the MSSM that could replace the universality hypothesis of 
eqs.~(\ref{scalarmassunification})-(\ref{commonphase}). They can be 
referred to as the ``irrelevancy", ``alignment", and ``$R$-symmetry" 
hypotheses for the soft masses. The ``irrelevancy" idea is that the 
sparticles masses are {\it extremely} heavy, so that their contributions 
to flavor-changing and CP-violating diagrams like 
Figures~\ref{fig:flavor}a,b are suppressed, as can be seen for example in 
eqs.~(\ref{eq:muegamma})-(\ref{eq:striking}).  In practice, however, 
if there is no flavor-blind structure, the 
degree of suppression needed typically requires $m_{\rm soft}$ much 
larger than 1 TeV for at least some of the scalar masses. This seems to 
go directly against the motivation for supersymmetry as a cure for the 
hierarchy problem as discussed in the Introduction. Nevertheless, it has 
been argued that this is a sensible possibility 
\cite{Moreminimal,splitsusy}. The fact that the LHC searches conducted so 
far have eliminated many models with lighter squarks anyway tends to make these 
models seem more attractive. Perhaps a combination of approximate flavor 
blindness and heavy superpartner masses is the true explanation 
for the suppression of flavor-violating effects.  

The ``alignment" idea is that the squark 
squared-mass matrices do not have the flavor-blindness indicated in 
eq.~(\ref{scalarmassunification}), but are arranged in flavor space to be 
aligned with the relevant Yukawa matrices in just such a way as to avoid 
large flavor-changing effects \cite{cterms,alignmentmodels}. The 
alignment models typically require rather special flavor symmetries. 

The third possibility is that the theory is (approximately) invariant under a 
continuous $U(1)_R$ symmetry \cite{Kribs:2007ac}. This requires that the 
MSSM is supplemented, as in \cite{Fox:2002bu}, by additional chiral 
supermultiplets in the adjoint representations of $SU(3)_c$, $SU(2)_L$, 
and $U(1)_Y$, as well as an additional pair of Higgs chiral 
supermultiplets. The gaugino masses in this theory are purely Dirac, of 
the type in eq.~(\ref{eq:Diracgauginos}), and the couplings $\bf a_u$, 
$\bf a_d$, and $\bf a_e$ are absent. This implies a very efficient 
suppression of flavor-changing effects 
\cite{Kribs:2007ac,Blechman:2008gu}, even if the squark and slepton mass 
eigenstates are light, non-degenerate, and have large mixings in the 
basis determined by the Standard Model quark and lepton mass eigenstates. 
This can lead to unique and intriguing collider signatures 
\cite{Kribs:2007ac,Plehn:2008ae}. However, we will not consider these 
possibilities further here.

The soft-breaking universality relations
eqs.~(\ref{scalarmassunification})-(\ref{commonphase}), or stronger (more
special) versions of them, can be presumed to be the result of some
specific model for the origin of supersymmetry breaking, although there is
no consensus among theorists as to what the specific model
should actually be. In any case, they are indicative of an assumed
underlying simplicity or symmetry of the Lagrangian at some very high
energy scale $Q_0$. If we used this Lagrangian to compute masses and
cross-sections and decay rates for experiments at ordinary energies near
the electroweak scale, the results would involve large logarithms of order
ln$(Q_0/m_Z)$ coming from loop diagrams. As is usual in quantum field
theory, the large logarithms can be conveniently resummed using
renormalization group (RG) equations, by treating the couplings and masses
appearing in the Lagrangian as running parameters. Therefore,
eqs.~(\ref{scalarmassunification})-(\ref{commonphase}) should be
interpreted as boundary conditions on the running soft parameters at the
scale $Q_0$, which is likely very far removed from direct experimental
probes. We must then RG-evolve all of the soft parameters, the
superpotential parameters, and the gauge couplings down to the electroweak
scale or comparable scales where humans perform experiments. 

At the electroweak scale, eqs.~(\ref{scalarmassunification}) and
(\ref{aunification}) will no longer hold, even if they were exactly true
at the input scale $Q_0$.  However, to a good approximation, key flavor-
and CP-conserving properties remain. This is because, as we will see in
section \ref{subsec:RGEs} below, RG corrections due to gauge interactions
will respect the form of
eqs.~(\ref{scalarmassunification}) and (\ref{aunification}),
while RG corrections due to Yukawa interactions are quite small except for
couplings involving the top, bottom, and tau flavors. Therefore, the
(scalar)$^3$ couplings and scalar squared-mass mixings should be quite
negligible for the squarks and sleptons of the first two families.
Furthermore, RG evolution does not introduce new CP-violating phases.
Therefore, if universality can be arranged to hold at the input scale,
supersymmetric contributions to flavor-changing and CP-violating
observables can be acceptably small in comparison to present limits
(although quite possibly measurable in future experiments). 

\begin{figure}
\begin{minipage}[]{0.355\linewidth} 
\caption{Two-loop renormalization group evolution of the inverse gauge 
couplings 
$\alpha_a^{-1}(Q)$ in
the Standard Model (dashed lines) and the MSSM (solid lines). In the MSSM
case, the sparticle masses are treated as a common threshold varied 
between 750 GeV and 2.5 TeV, and $\alpha_3(m_Z)$ is varied between 
$0.117$ and $0.120$. 
\label{fig:gaugeunification}}
\end{minipage}
\begin{minipage}[]{0.64\linewidth} 
\hspace{.06\linewidth}
{\psfig{figure=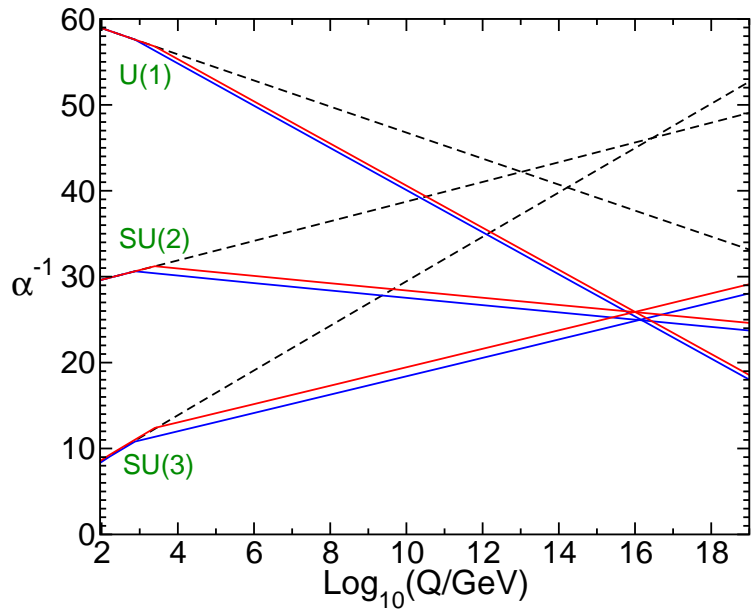,width=0.9\linewidth}} 
\end{minipage}
\end{figure}

One good reason to be optimistic that such a program can succeed is the
celebrated apparent unification of gauge couplings in the MSSM
\cite{gaugeunification}. The 1-loop RG equations for the Standard Model
gauge couplings $g_1, g_2, g_3$ are 
\beq
\beta_{g_a} \equiv {d\over dt} g_a =  {1\over 16\pi^2} b_a g_a^3, 
\qquad\quad 
(b_1, b_2, b_3) = 
\left \{ \begin{array}{ll}
(41/10,\> -19/6,\> -7) & \mbox{Standard Model}\\
(33/5,\> 1,\> -3) & \mbox{MSSM}
\end{array}
\right.
\label{mssmg}
\eeq
where $t= {\rm ln} (Q/Q_0)$, with $Q$ the RG scale. The MSSM 
coefficients are larger because of the extra MSSM
particles in loops. The normalization for $g_1$ here is chosen to agree
with the canonical covariant derivative for grand unification of the gauge
group $SU(3)_C \times SU(2)_L\times U(1)_Y$ into $SU(5)$ or $SO(10)$. Thus
in terms of the conventional electroweak gauge couplings $g$ and
$g^\prime$ with $e = g\sin\theta_W = g^\prime \cos\theta_W$, one has
$g_2=g$ and $g_1 = \sqrt{5/3} g^\prime$. The quantities $\alpha_a =
g_a^2/4\pi$ have the nice property that their reciprocals run linearly
with RG scale at one-loop order:
\beq
{d\over dt} \alpha_a^{-1} = -{b_a\over 2\pi} \qquad\qquad (a=1,2,3)
\qquad
\label{mssmgrecip}
\eeq  
Figure \ref{fig:gaugeunification} compares the RG evolution of the 
$\alpha_a^{-1}$, including two-loop effects, in the Standard Model (dashed 
lines) and the MSSM (solid lines). 
Unlike the Standard Model, the MSSM 
includes just the right particle content to ensure that the gauge 
couplings can unify, at a scale $M_U \sim 1.5\times 10^{16}$ GeV. 
This unification is of course not perfect; $\alpha_3$ tends to be 
slightly smaller than the common value of $\alpha_1(M_U) = \alpha_2(M_U)$
at the point where they meet, which is often taken to be the definition 
of $M_U$. However, this small difference can easily be ascribed to 
threshold corrections due to whatever new particles exist near $M_U$.
Note that $M_U$ decreases slightly as the superpartner masses are 
raised. While 
the apparent approximate unification of gauge couplings at $M_U$ might be 
just an 
accident, it may also be taken as a strong hint in favor of a grand 
unified theory (GUT) or superstring models, both of which can naturally 
accommodate gauge coupling unification below $\MPlanck$. Furthermore, if 
this hint is taken seriously, then we can reasonably expect to be able to 
apply a similar RG analysis to the other MSSM couplings and soft masses as 
well. The next section discusses the form of the necessary RG equations.

\subsection{Renormalization Group equations for the MSSM\label{subsec:RGEs}}
\setcounter{footnote}{1}
\setcounter{equation}{0}

In order to translate a set of predictions at an input scale into 
physically meaningful quantities that describe physics near the 
electroweak scale, it is necessary to evolve the gauge couplings, 
superpotential parameters, and soft terms using their renormalization 
group (RG) equations. This ensures that the loop expansions for 
calculations of observables will not suffer from very large logarithms.

As a technical aside, some care is required in choosing
regularization and renormalization procedures in supersymmetry.
The most popular regularization method for computations of radiative
corrections within the Standard Model is dimensional regularization
(DREG), in which the number of spacetime dimensions is continued to
$d=4-2\epsilon$. Unfortunately, DREG introduces a spurious violation of
supersymmetry, because it has a mismatch between the numbers of gauge
boson degrees of freedom and the gaugino degrees of freedom off-shell.
This mismatch is only $2\epsilon$, but can be multiplied by factors up to
$1/\epsilon^n$ in an $n$-loop calculation. In DREG, supersymmetric
relations between dimensionless coupling constants (``supersymmetric Ward
identities") are therefore not explicitly respected by radiative
corrections involving the finite parts of one-loop graphs and by the
divergent parts of two-loop graphs.  Instead, one may use the slightly
different scheme known as regularization by dimensional reduction, or
DRED, which does respect supersymmetry \cite{DRED}. In the DRED method,
all momentum integrals are still performed in $d=4-2\epsilon$ dimensions,
but the vector index $\mu$ on the gauge boson fields $A^a_\mu$ now runs
over all 4 dimensions to maintain the match with the gaugino degrees of
freedom. Running couplings are then renormalized using DRED with modified
minimal subtraction ($\drbar$) rather than the usual DREG with modified
minimal subtraction ($\msbar$). In particular, the boundary conditions at
the input scale should presumably be applied in a supersymmetry-preserving
scheme like $\drbar$.  One loop $\beta$-functions are always the same in
these two schemes, but it is important to realize that the $\msbar$ scheme
does violate supersymmetry, so that $\drbar$ is preferred\footnote{Even
the DRED scheme may not provide a supersymmetric regulator, because of
either ambiguities or inconsistencies (depending on the precise method)
appearing at five-loop order at the latest \cite{DREDdies}. Fortunately,
this does not seem to cause practical difficulties
\cite{JJperspective,Stockinger}. See also ref.~\cite{Woodard} for an 
interesting proposal that avoids doing violence to the number of spacetime
dimensions.} from that point of view. (The NSVZ scheme \cite{Shifman} also
respects supersymmetry and has some very useful properties, but with a
less obvious connection to calculations of physical observables. It is
also possible, but not always very practical, to work consistently within
the $\overline{\rm MS}$ scheme, as long as one translates all
$\overline{\rm DR}$ couplings and masses into their $\overline{\rm MS}$
counterparts 
\cite{mstodrone}-\cite{mstodrmore}.)

A general and powerful result known as the {\it supersymmetric
non-renormalization theorem} \cite{nonrentheo} governs the form of the
renormalization group equations for supersymmetric theories. This theorem
implies that the logarithmically divergent contributions to a particular
process can always be written in terms of wave-function renormalizations,
without any coupling vertex renormalization.\footnote{Actually, there {\it
is} vertex renormalization working in a supersymmetric gauge theory in
which auxiliary fields have been integrated out, but the sum of divergent
contributions for a process always has the form of wave-function
renormalization. This is related to the fact that the anomalous dimensions
of the superfields differ, by gauge-fixing dependent terms, from the
anomalous dimensions of the fermion and boson component fields
\cite{Jonesreview}.} It can be proved most easily using superfield
techniques. For the parameters appearing in the superpotential
eq.~(\ref{superpotentialwithlinear}), the implication is that
\beq
\beta_{y^{ijk}} \equiv \frac{d}{dt}y^{ijk} \!&=&\! 
\gamma^i_n y^{njk} + \gamma^j_n y^{ink} + \gamma^k_n y^{ijn},
\label{eq:genyrge}
\\
\beta_{M^{ij}} \equiv \frac{d}{dt}M^{ij} \!&=&\! 
\gamma^i_n M^{nj} + \gamma^j_n M^{in},
\label{eq:genMrge}
\\
\beta_{L^{i}} \equiv \frac{d}{dt}L^{i} \!&=&\! 
\gamma^i_n L^{n},
\eeq
where the $\gamma^i_j$ are anomalous dimension matrices associated with
the superfields, which generally have to be calculated in a perturbative
loop expansion. [Recall $t = \ln(Q/Q_0)$, where $Q$ is the renormalization
scale, and $Q_0$ is a reference scale.] The anomalous dimensions and RG
equations for softly broken supersymmetry are now known up to 3-loop
order, with some partial 4-loop results; they have been given in
refs.~\cite{rges2gauge}-\cite{fourloops}. 
There are also relations, good to all orders in perturbation
theory, that give the RG equations for soft supersymmetry couplings in
terms of those for the supersymmetric couplings \cite{Shifman,allorders}.
Here, for simplicity, only the 1-loop approximation will be shown
explicitly. 

In general, at 1-loop order, 
\beq
\gamma^i_j \,=\, \frac{1}{16 \pi^2} \left [
\half y^{imn} y^*_{jmn} - 2 g_a^2 C_a(i) 
\delta_j^i
\right ],
\label{eq:gengamma}
\eeq
where $C_a(i)$ are the quadratic Casimir group theory invariants for the
superfield $\Phi_i$, defined in terms of the Lie algebra generators $T^a$
by 
\beq
(T^aT^a)_i{}^{j}= C_a(i) \delta_i^j 
\label{eq:defCasimir}
\eeq 
with gauge couplings $g_a$. 
Explicitly, for the MSSM supermultiplets: 
\beq
&&
C_3(i) =
\Biggl \{ \begin{array}{ll}
4/3 & {\rm for}\>\,\Phi_i = Q, \sbar u, \sbar d,
\\
0 & {\rm for}\>\,\Phi_i = L, \sbar e, H_u, H_d,
\end{array}
\label{defC3}
\\
&&
C_2(i) =
\Biggl \{ \begin{array}{ll}
3/4 & {\rm for}\>\,\Phi_i = Q, L, H_u, H_d,\\
0 & {\rm for}\>\,\Phi_i = \sbar u, \sbar d, \sbar e
,\end{array}
\\
&&
C_1(i) = \>
3 Y_i^2/5 \>\>\>{\rm for~each}\>\,\Phi_i\>\,{\rm
with~weak~hypercharge}\>\, Y_i.
\label{defC1}
\eeq
For the one-loop renormalization of gauge couplings, one has in
general
\beq
\beta_{g_a} = 
{d\over dt} g_a 
\!&=&\! 
{1\over 16\pi^2} g_a^3 \Bigl [\sum_i I_a(i) - 3 C_a(G) \Bigr ],
\eeq
where $C_a(G)$ is the quadratic Casimir invariant of the
group [0 for $U(1)$, and $N$ for $SU(N)$], and
$I_a(i)$ is the Dynkin index of the chiral supermultiplet $\phi_i$
[normalized to $1/2$ for each fundamental representation of $SU(N)$ and
to $3 Y_i^2/5$ for $U(1)_Y$]. Equation (\ref{mssmg})
is a special case of this.

The 1-loop renormalization group equations for the
general soft supersymmetry breaking Lagrangian parameters appearing in
eq.~(\ref{lagrsoft}) are:
\beq
\beta_{M_a} \equiv 
{d\over dt} M_a 
\!&=&\! 
{1\over 16\pi^2} g_a^2 \Bigl [2 \sum_n I_a(n) - 6 C_a(G) \Bigr ] M_a
,
\\
\beta_{a^{ijk}} \equiv \frac{d}{dt} a^{ijk}
\!&=&\!
\frac{1}{16 \pi^2} \left [
\frac{1}{2} a^{ijp} y^*_{pmn} y^{kmn} 
+ y^{ijp} y^*_{pmn} a^{mnk}
+ g_a^2 C_a(i) (4 M_a y^{ijk}  - 2 a^{ijk}) 
\right ]\phantom{xxxx}
\nonumber \\ &&
+ (i \leftrightarrow k)
+ (j \leftrightarrow k)
,\\
\beta_{b^{ij}} \equiv \frac{d}{dt} b^{ij}
\!&=&\!
\frac{1}{16 \pi^2} 
\biggl [
\frac{1}{2} b^{ip} y^*_{pmn} y^{jmn} 
+ \frac{1}{2} y^{ijp} y^*_{pmn} b^{mn} 
+ M^{ip} y^*_{pmn} a^{mnj}
\nonumber \\ &&
+  g_a^2 C_a(i) (4 M_a M^{ij}  - 2 b^{ij})
\biggr ]
+ (i \leftrightarrow j) 
,
\\
\beta_{t^{i}} \equiv \frac{d}{dt} t^{i}
\!&=&\!
\frac{1}{16 \pi^2} \left [
\frac{1}{2} y^{imn} y^*_{mnp} t^p 
+ a^{imn} y^*_{mnp} L^p
+ M^{ip} y^*_{pmn} b^{mn} 
\right ],
\\
\beta_{(m^2)_{i}^j} \equiv \frac{d}{dt} (m^2)_{i}^j
\!&=&\!
\frac{1}{16 \pi^2} \biggl [
\frac{1}{2} y_{ipq}^* y^{pqn} (m^2)_n^j
+ \frac{1}{2} y^{jpq} y_{pqn}^* (m^2)_i^n
+ 2 y^*_{ipq} y^{jpr} (m^2)_r^q
\nonumber \\ &&
+ a^*_{ipq} a^{jpq}
- 8 g_a^2 C_a(i) |M_a|^2 \delta_i^j
+ 2 g_a^2 (T^a)_i{}^j
{\rm Tr}(T^a m^2)
\biggr ]
.
\eeq

Applying the above results to the special case of the MSSM, we will use the 
approximation that only the third-family Yukawa
couplings are significant, as in eq.~(\ref{heavytopapprox}). Then the
Higgs and third-family superfield anomalous dimensions are diagonal
matrices, and from eq.~(\ref{eq:gengamma}) they are, at 1-loop
order: 
\beq
\gamma_{H_u} \!&=&\! 
\frac{1}{16 \pi^2} \left [
3 y_t^* y_t - \frac{3}{2} g_2^2 - \frac{3}{10} g_1^2
\right ],
\label{eq:gammaHu}
\\
\gamma_{H_d} \!&=&\! \frac{1}{16 \pi^2} \left [
3 y_b^* y_b + y_\tau^* y_\tau - \frac{3}{2} g_2^2 
- \frac{3}{10} g_1^2
\right ],
\\
\gamma_{Q_3} \!&=&\! \frac{1}{16 \pi^2} \left [
y_t^* y_t + y_b^* y_b - \frac{8}{3} g_3^2
- \frac{3}{2} g_2^2 - \frac{1}{30} g_1^2
\right ],
\\
\gamma_{\sbar{u}_3} \!&=&\! \frac{1}{16 \pi^2} \left [
2 y_t^* y_t  -\frac{8}{3} g_3^2 -\frac{8}{15} 
g_1^2
\right ],
\\
\gamma_{\sbar{d}_3} \!&=&\! \frac{1}{16 \pi^2} \left [
2 y_b^* y_b  -\frac{8}{3} g_3^2 -\frac{2}{15} 
g_1^2
\right ],
\\
\gamma_{L_3} \!&=&\! \frac{1}{16 \pi^2} \left [
y_\tau^* y_\tau - \frac{3}{2} g_2^2 - \frac{3}{10} 
g_1^2
\right ],
\\
\gamma_{\sbar{e}_3} \!&=&\! \frac{1}{16 \pi^2} \left [
2 y_\tau^* y_\tau  -\frac{6}{5} g_1^2
\right ].
\label{eq:gammae}
\eeq
[The first and second family anomalous dimensions in the approximation of
eq.~(\ref{heavytopapprox}) follow by setting $y_t$, $y_b$, and $y_\tau$ to
$0$ in the above.] Putting these into eqs.~(\ref{eq:genyrge}),
(\ref{eq:genMrge}) gives the running of the superpotential parameters with
renormalization scale:
\beq
\beta_{y_t} \equiv
{d\over dt} y_t \!&=&\! {y_t \over 16 \pi^2} \Bigl [ 6 y_t^* y_t + y_b^* y_b
- {16\over 3} g_3^2 - 3 g_2^2 - {13\over 15} g_1^2 \Bigr ],
\label{eq:betayt}
\\
\beta_{y_b} \equiv
{d\over dt} y_b \!&=&\! {y_b \over 16 \pi^2} 
\Bigl [ 6 y_b^* y_b + y_t^* y_t + 
y_\tau^* y_\tau - {16\over 3} g_3^2 - 3 g_2^2 - {7\over 15} g_1^2 \Bigr ],
\\
\beta_{y_\tau} \equiv
{d\over dt} y_\tau \!&=&\! {y_\tau \over 16 \pi^2} 
\Bigl [ 4 y_\tau^* y_\tau 
+ 3 y_b^* y_b - 3 g_2^2 - {9\over 5} g_1^2 \Bigr ],
\\
\beta_{\mu} \equiv
{d\over dt} \mu \!&=&\! {\mu \over 16 \pi^2} 
\Bigl [ 3 y_t^* y_t + 3 y_b^* y_b
+ y_\tau^* y_\tau - 3 g_2^2 - {3\over 5} g_1^2 \Bigr ].
\label{eq:betamu}
\eeq
The one-loop RG equations for the gauge couplings $g_1$, $g_2$, and $g_3$
were already listed in eq.~(\ref{mssmg}).  The presence of soft
supersymmetry breaking does not affect eqs.~(\ref{mssmg}) and
(\ref{eq:betayt})-(\ref{eq:betamu}). As a result of the
supersymmetric non-renormalization theorem, the $\beta$-functions  
for each supersymmetric
parameter are proportional to the parameter itself.  One consequence of
this is that once we have a theory that can explain why $\mu$ is of order
$10^2$ or $10^3$ GeV at tree-level, we do not have to worry about $\mu$
being made very large by radiative corrections involving the masses of
some very heavy unknown particles; all such RG corrections to $\mu$ will
be directly proportional to $\mu$ itself and to some combinations of
dimensionless couplings. 

The one-loop RG equations for the three gaugino mass parameters in the
MSSM are determined by the same quantities $b_a^{\rm MSSM}$ that appear in
the gauge coupling RG eqs.~(\ref{mssmg}): 
\beq
\beta_{M_a} \equiv
{d\over dt} M_a \,=\, {1\over 8\pi^2} b_a g_a^2 M_a\qquad\>\>\>
(b_a = 33/5, \>1,\>-3)
\label{gauginomassrge}
\eeq
for $a=1,2,3$. It follows that the three ratios $M_a/g_a^2$ are each
constant (RG scale independent) up to small two-loop corrections.  Since
the gauge couplings are observed to unify at $Q = M_U = 1.5 \times 10^{16}$
GeV, it is a popular assumption that the gaugino masses also
unify\footnote{In GUT models, it is automatic that the gauge couplings
and gaugino masses are unified at all scales $Q\geq M_U$, because in the
unified theory the gauginos all live in the same representation of the
unified gauge group. In many superstring models, this can also be a good
approximation.} near that scale, with a value called $m_{1/2}$. 
If so, then it follows that
\beq
{M_1 \over g_1^2} =
{M_2 \over g_2^2} =
{M_3 \over g_3^2} = {m_{1/2} \over g_U^2}
\label{gauginomassunification}
\eeq
at any RG scale, up to small (and known) two-loop effects and possibly
much larger (and unknown) threshold effects near $M_U$. Here $g_U$ is
the unified gauge coupling at $Q = M_U$. The hypothesis of
eq.~(\ref{gauginomassunification}) is particularly powerful because the
gaugino mass parameters feed strongly into the RG equations for all of the
other soft terms, as we are about to see. 

Next we consider the 1-loop RG equations for the holomorphic soft parameters
${\bf a_u}$, ${\bf a_d}$, ${\bf a_e}$. In models obeying
eq.~(\ref{aunification}), these matrices start off proportional to the
corresponding Yukawa couplings at the input scale. The RG evolution
respects this property. With the approximation of
eq.~(\ref{heavytopapprox}), one can therefore also write, at any RG scale,
\beq
{\bf a_u} \approx \pmatrix{0&0&0\cr 0&0&0 \cr 0&0&a_t},\qquad\!\!
{\bf a_d} \approx \pmatrix{0&0&0\cr 0&0&0 \cr 0&0&a_b},\qquad\!\!
{\bf a_e} \approx \pmatrix{0&0&0\cr 0&0&0 \cr 0&0&a_\tau},\>\>{}
\label{heavyatopapprox}
\eeq
which defines\footnote{Rescaled soft parameters $A_t = a_t/y_t$,
$A_b=a_b/y_b$, and $A_\tau=a_\tau/y_\tau$ are often used in the
literature. We do not follow this notation, because it cannot be
generalized beyond the approximation of eqs. (\ref{heavytopapprox}),
(\ref{heavyatopapprox}) without introducing horrible complications such as
non-polynomial RG equations, and because $a_t$, $a_b$ and $a_\tau$ are the
couplings that actually appear in the Lagrangian anyway.} running
parameters $a_t$, $a_b$, and $a_\tau$. In this approximation, 
the RG equations for these
parameters and $b$ are
\beq
16\pi^2 {d\over dt} a_t \!&=&\! a_t \Bigl [ 18 y_t^* y_t + y_b^* y_b
- {16\over 3} g_3^2 - 3 g_2^2 - {13\over 15} g_1^2 \Bigr ]
+ 2 a_b y_b^* y_t
\nonumber\\ && 
\!+ y_t \Bigl [ {32\over 3} g_3^2 M_3 + 6 g_2^2 M_2 + {26\over 15} g_1^2 M_1
\Bigr ],
\label{atrge}
\\
16\pi^2{d\over dt} a_b \!&=&\! a_b \Bigl [ 18 y_b^* y_b + y_t^* y_t +
y_\tau^* y_\tau
- {16\over 3} g_3^2 - 3 g_2^2 - {7\over 15} g_1^2 \Bigr ]
+ 2 a_t y_t^* y_b + 2 a_\tau y_\tau^* y_b
\phantom{xxxx}
\nonumber \\&&
\!+ y_b \Bigl [ {32\over 3} g_3^2 M_3 + 6 g_2^2 M_2 + {14 \over 15} g_1^2 M_1 
\Bigr ],\qquad{}
\\
16\pi^2{d\over dt} a_\tau \!&=&\! a_\tau \Bigl [ 12 y_\tau^* y_\tau 
+ 3 y_b^* y_b - 3 g_2^2 - {9\over 5} g_1^2 \Bigr ]
+ 6 a_b y_b^* y_\tau
+ y_\tau \Bigl [ 6 g_2^2 M_2 + {18\over 5} g_1^2 M_1 \Bigr ],
\\
16\pi^2{d\over dt} b \!&=&\! b \Bigl [ 3 y_t^* y_t + 3 y_b^* y_b
+ y_\tau^* y_\tau - 3 g_2^2 - {3\over 5} g_1^2 \Bigr ]
\nonumber \\ && 
\!+ \mu \Bigl [ 6 a_t y_t^* + 6 a_b y_b^* + 2 a_\tau y_\tau^* +
6 g_2^2 M_2 + {6\over 5} g_1^2 M_1 \Bigr ] .
\label{brge}
\eeq
The $\beta$-function for each of these soft
parameters is {\it not} proportional to the parameter itself, because
couplings that violate supersymmetry are not protected by the
supersymmetric non-renormalization theorem. So, even if $a_t$, $a_b$,
$a_\tau$ and $b$ vanish at the input scale, the RG corrections
proportional to gaugino masses appearing in
eqs.~(\ref{atrge})-(\ref{brge}) ensure that they will not vanish at
the electroweak scale. 

Next let us consider the RG equations for the scalar squared masses in the
MSSM. In the approximation of eqs.~(\ref{heavytopapprox}) and
(\ref{heavyatopapprox}), the squarks and sleptons of the first two
families have only gauge interactions. This means that if the scalar
squared masses satisfy a boundary condition like
eq.~(\ref{scalarmassunification}) at an input RG scale, then when
renormalized to any other RG scale, they will still be almost diagonal,
with the approximate form
\beq
{\bf m_Q^2} \approx \pmatrix{
m_{Q_1}^2 & 0 & 0\cr
0 & m_{Q_1}^2 & 0 \cr
0 & 0 & m_{Q_3}^2 \cr},\qquad\>\>\>
{\bf m_{\sbar u}^2} \approx \pmatrix{
m_{\sbar u_1}^2 & 0 & 0\cr
0 & m_{\sbar u_1}^2 & 0 \cr
0 & 0 & m_{\sbar u_3}^2 \cr},
\eeq
etc. The first and second family squarks and sleptons with given gauge
quantum numbers remain very nearly degenerate, but the third-family
squarks and sleptons feel the effects of the larger Yukawa couplings and
so their squared masses get renormalized differently.  The one-loop RG
equations for the first and second family squark and slepton squared
masses are
\beq
16 \pi^2 {d\over dt} m_{\phi_i}^2 \>\,=\>\, 
-\!\sum_{a=1,2,3} 8 C_a (i) g_a^2 |M_a|^2
+ \frac{6}{5} Y_i g^{2}_1 S
\label{easyscalarrge}
\eeq
for each scalar $\phi_i$, where the $\sum_a$ is over the three gauge
groups $U(1)_Y$, $SU(2)_L$ and $SU(3)_C$, with Casimir invariants $C_a(i)$
as in eqs.~(\ref{defC3})-(\ref{defC1}), and $M_a$ are the corresponding
running gaugino mass parameters. Also,
\beq
S \equiv {\rm Tr}[Y_j m^2_{\phi_j}] =
m_{H_u}^2 - m_{H_d}^2 + {\rm Tr}[
{\bf m^2_Q} - {\bf m^2_L} - 2 {\bf m^2_{\overline u}}
+ {\bf m^2_{\overline d}} + {\bf m^2_{\overline e}}] .
\label{eq:defS}
\eeq
An important feature of eq.~(\ref{easyscalarrge}) is that the terms on the
right-hand sides proportional to gaugino squared masses are negative,
so\footnote{The contributions 
proportional to $S$ are
relatively small in most known realistic models.} the scalar squared-mass
parameters grow as they are RG-evolved from the input scale down to the
electroweak scale. Even if the scalars have zero or very small masses at
the input scale, they can obtain large positive squared masses at the
electroweak scale, thanks to the effects of the gaugino masses. 

\setcounter{footnote}{1}

The RG equations for the squared-mass parameters of the Higgs scalars and
third-family squarks and sleptons get the same gauge contributions as in
eq.~(\ref{easyscalarrge}), but they also have contributions due to the
large Yukawa ($y_{t,b,\tau}$) and soft ($a_{t,b,\tau}$) couplings. At
one-loop order, these only appear in three combinations: 
\beq
X_t \!&=&\!  2 |y_t|^2 (m_{H_u}^2 + m_{Q_3}^2 + m_{\sbar u_3}^2) +2 |a_t|^2,
\\
X_b \!&=& \! 2 |y_b|^2 (m_{H_d}^2 + m_{Q_3}^2 + m_{\sbar d_3}^2) +2 |a_b|^2,
\\
X_\tau\! &=&\!  2 |y_\tau|^2 (m_{H_d}^2 + m_{L_3}^2 + m_{\sbar e_3}^2)
+ 2 |a_\tau|^2.
\eeq
In terms of these quantities, the RG equations for the soft Higgs
squared-mass parameters $m_{H_u}^2$ and $m_{H_d}^2$ are
\beq
16 \pi^2 {d\over dt} m_{H_u}^2 \!&=&\!
3 X_t - 6 g_2^2 |M_2|^2 - {6\over 5} g_1^2 |M_1|^2 + \frac{3}{5} g^{2}_1 S
,
\label{mhurge}
\\
16\pi^2{d\over dt} m_{H_d}^2 \!&=&\!
3 X_b + X_\tau - 6 g_2^2 |M_2|^2 - {6\over 5} g_1^2 |M_1|^2 - \frac{3}{5} 
g^{2}_1 S
.
\label{mhdrge}
\eeq
Note that $X_t$, $X_b$, and $X_\tau$ are generally positive, so their
effect is to decrease the Higgs squared masses as one evolves the RG equations
down from the input scale to the electroweak scale. If $y_t$ is the
largest of the Yukawa couplings, as suggested by the experimental fact
that the top quark is heavy, then $X_t$ will typically be much larger than
$X_b$ and $X_\tau$. This can cause the RG-evolved $m_{H_u}^2$ to run
negative near the electroweak scale, helping to destabilize the point $H_u
= H_d = 0$ and so provoking a Higgs VEV (for a linear combination of $H_u$
and $H_d$, as we will see in section \ref{subsec:MSSMspectrum.Higgs}),
which is just what we want.\footnote{One should think of ``$m_{H_u}^2$" as
a parameter unto itself, and not as the square of some mythical real
number $m^{\phantom{2}}_{H_u}$. So there is nothing strange about having
$m_{H_u}^2 < 0$.
} Thus a large top Yukawa coupling favors
the breakdown of the electroweak symmetry breaking because it induces
negative radiative corrections to the Higgs squared mass. 

The third-family squark and slepton squared-mass parameters also get
contributions that depend on $X_t$, $X_b$ and $X_\tau$. Their RG equations
are given by
\beq 
16\pi^2{d\over dt} m_{Q_3}^2 \!&=&\!
X_t +X_b-{32\over 3} g_3^2 |M_3|^2 -6 g_2^2 |M_2|^2 -{2\over 15} g_1^2 |M_1|^2
+ \frac{1}{5} g^{2}_1 S ,
\label{mq3rge} 
\\
16\pi^2 {d\over dt} m_{\sbar u_3}^2 \!&=&\!
2 X_t - {32\over 3} g_3^2 |M_3|^2 - {32\over 15} g_1^2|M_1|^2
- \frac{4}{5} g^{2}_1 S ,
\label{mtbarrge}
\\
16\pi^2 {d\over dt} m_{\sbar d_3}^2 \!&=&\!
2 X_b - {32\over 3} g_3^2 |M_3|^2 - {8\over 15} g_1^2|M_1|^2 
+ \frac{2}{5} g^{2}_1 S ,
\label{md3rge}
\\
16\pi^2 {d\over dt} m_{L_3}^2 \!&=&\!
X_\tau  - 6 g_2^2 |M_2|^2 - {6\over 5} g_1^2 |M_1|^2 - \frac{3}{5} g^{2}_1 S,
\\
16\pi^2 {d\over dt} m_{\sbar e_3}^2 \!&=&\!
2 X_\tau - {24\over 5} g_1^2 |M_1|^2  + \frac{6}{5} g^{2}_1 S .
\label{mstaubarrge}
\eeq
In eqs.~(\ref{mhurge})-(\ref{mstaubarrge}), the terms proportional to
$|M_3|^2$, $|M_2|^2$, $|M_1|^2$, and $S$ are just the same ones as in
eq.~(\ref{easyscalarrge}). Note that the terms proportional to $X_t$ and
$X_b$ appear with smaller numerical coefficients in the $m^2_{Q_3}$,
$m^2_{\sbar u_3}$, $m^2_{\sbar d_3}$ RG equations than they did for the
Higgs scalars, and they do not appear at all in the $m^2_{L_3}$ and
$m^2_{\sbar e_3}$ RG equations. Furthermore, the third-family squark
squared masses get a large positive contribution proportional to $|M_3|^2$
from the RG evolution, which the Higgs scalars do not get. These facts
make it plausible that the Higgs scalars in the MSSM get VEVs, while the
squarks and sleptons, having large positive squared mass, do not. 

An examination of the RG equations (\ref{atrge})-(\ref{brge}),
(\ref{easyscalarrge}), and (\ref{mhurge})-(\ref{mstaubarrge}) reveals that
if the gaugino mass parameters $M_1$, $M_2$, and $M_3$ are non-zero at the
input scale, then all of the other soft terms will be generated too.  This
implies that models in which gaugino masses dominate over all other
effects in the soft supersymmetry breaking Lagrangian at the input scale
can be viable. On the other hand, if the gaugino masses were to vanish at
tree-level, then they would not get any contributions to their masses at
one-loop order; in that case the gauginos would be extremely light and the
model would not be phenomenologically acceptable. 

Viable models for the origin of supersymmetry breaking typically make
predictions for the MSSM soft terms that are refinements of
eqs.~(\ref{scalarmassunification})-(\ref{commonphase}). These predictions
can then be used as boundary conditions for the RG equations listed above. 
In the next section we will study the ideas that go into making such
predictions, before turning to their implications for the MSSM spectrum in
section \ref{sec:MSSMspectrum}. 

\section{Origins of supersymmetry breaking}\label{sec:origins}
\subsection{General considerations for spontaneous
supersymmetry breaking}\label{subsec:origins.general}
\setcounter{equation}{0}
\setcounter{figure}{0}
\setcounter{table}{0}
\setcounter{footnote}{1}

In the MSSM, supersymmetry breaking is simply introduced explicitly.
However, we have seen that the soft parameters cannot be arbitrary. In
order to understand how patterns like eqs.~(\ref{scalarmassunification}),
(\ref{aunification}) and (\ref{commonphase}) can emerge, it is necessary
to consider models in which supersymmetry is spontaneously broken. By
definition, this means that the vacuum state $\vac$ is not invariant under
supersymmetry transformations, so $Q_\alpha \vac \not= 0$ and
$Q^\dagger_{\dot{\alpha}}\vac \not=0$. Now, in global supersymmetry, the
Hamiltonian operator $H$ is related to the supersymmetry generators
through the algebra eq.~(\ref{nonschsusyalg1}):
\beq
H=P^0 =
{1\over 4}( Q_1 Q_{{1}}^\dagger + Q_{{1}}^\dagger Q_1 
+ Q_2 Q_{{2}}^\dagger + Q_{{2}}^\dagger Q_2 ) .
\eeq
If supersymmetry is unbroken in the vacuum state, it follows that $H\vac =
0$ and the vacuum has zero energy. Conversely, if supersymmetry is
spontaneously broken in the vacuum state, then the vacuum must have
positive energy, since
\beq
\antivac H \vac = {1\over 4} \Bigl (\| Q_1^\dagger \vac \|^2 +
\| Q_{{1}} \vac \|^2
+ \| Q^\dagger_{2} \vac \|^2
+ \| Q_{{2}} \vac \|^2
\Bigr ) > 0
\eeq
if the Hilbert space is to have positive norm. If spacetime-dependent
effects and fermion condensates can be neglected, then $\antivac H\vac =
\antivac V \vac $, where $V$ is the scalar potential in eq.~(\ref{fdpot}).
Therefore, supersymmetry will be spontaneously broken if the 
expectation value of $F_i$ and/or
$D^a$ does not vanish in the vacuum state.

\begin{figure}
\begin{picture}(130,130)(0,0)
\SetScale{0.85}
\LongArrow(75,30)(150,30)
\LongArrow(75,30)(0,30)
\LongArrow(75,30)(75,150)
\LongArrow(75,30)(75,10)
\Text(48,128)[c]{$V(\phi)$}
\Text(130,35)[c]{$\phi$}
\SetWidth{1.1}
\Curve{(0,150)(75,30)(150,150)}
\rText(66,-4)[][]{(a)}
\end{picture}
\hspace{1.1cm}
\begin{picture}(130,130)(0,0)
\SetScale{0.85}
\LongArrow(75,30)(150,30)
\LongArrow(75,30)(0,30)
\LongArrow(75,30)(75,150)
\LongArrow(75,30)(75,10)
\Text(48,128)[c]{$V(\phi)$}
\Text(130,35)[c]{$\phi$}
\SetWidth{1.1}
\Curve{(5,150)(75,56)(145,150)}
\rText(66,-4)[][]{(b)}
\end{picture}
\hspace{1.1cm}
\begin{picture}(130,130)(0,0)
\SetScale{0.85}
\LongArrow(75,30)(150,30)
\LongArrow(75,30)(0,30)
\LongArrow(75,30)(75,150)
\LongArrow(75,30)(75,10)
\Text(48,128)[c]{$V(\phi)$}
\Text(130,35)[c]{$\phi$}
\SetWidth{1.1}
\Curve{(5,150)(40,87)(86,60)(100,80)(120,30.5)(150,150)}
\rText(66,-4)[][]{(c)}
\end{picture}
\vspace{0.15cm}
\caption{Scalar potentials for (a) unbroken supersymmetry, 
(b) spontaneously broken supersymmetry, and (c) metastable supersymmetry 
breaking, as functions of an order parameter $\phi$. 
\label{fig:susybreakingpotentials}}
\end{figure}
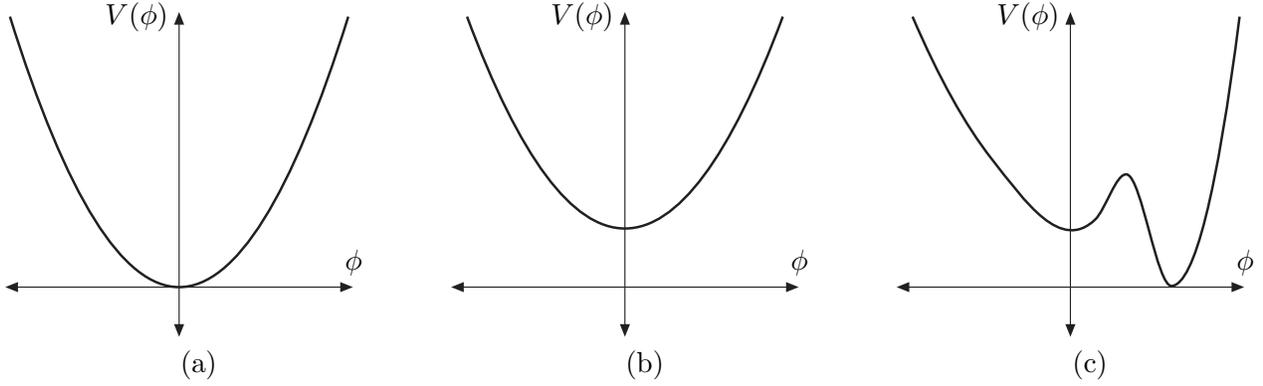

If any state exists in which all $F_i$ and $D^a$ vanish, then it will 
have zero energy, implying that supersymmetry is not spontaneously broken 
in the true ground state. Conversely, one way to guarantee spontaneous 
supersymmetry breaking is to look for models in which the equations 
$F_i=0$ and $D^a=0$ cannot all be simultaneously satisfied for {\it any} 
values of the fields. Then the true ground state necessarily has broken 
supersymmetry, as does the vacuum state we live in (if it is different).
However, another possibility is that the vacuum state in which we live is 
not the true ground state (which may preserve supersymmetry), but is 
instead a higher energy metastable supersymmetry-breaking state with 
lifetime at least of order the present age of the universe 
\cite{Ellis:1982vi}-\cite{Intriligator:2006dd}. Finite temperature 
effects can indeed cause the early universe to prefer the metastable 
supersymmetry-breaking local minimum of the potential over the 
supersymmetry-breaking global minimum \cite{metastableearlyuniverse}.
Scalar potentials for the
three possibilities are illustrated qualitatively in 
Figure \ref{fig:susybreakingpotentials}.

Regardless of whether the vacuum state is stable or metastable,
the spontaneous breaking of a global symmetry always implies a
massless Nambu-Goldstone mode with the same quantum numbers as the broken
symmetry generator. In the case of global supersymmetry, the broken
generator is the fermionic charge $Q_\alpha$, so the Nambu-Goldstone
particle ought to be a massless neutral Weyl fermion, called the {\it
goldstino}.  To prove it, consider a general supersymmetric model with
both gauge and chiral supermultiplets as in section \ref{sec:susylagr}.  
The fermionic degrees of freedom consist of gauginos ($\lambda^a$) and
chiral fermions ($\psi_i$). After some of the scalar fields in the theory
obtain VEVs, the fermion mass matrix has the form:
\beq
{\bf m}_{\rm F} =
\pmatrix{
0 & \sqrt{2} g_b  (\langle \phi^{*}\rangle T^b)^i    
\cr
\sqrt{2} g_a  (\langle \phi^{*}\rangle T^a)^j  & \langle W^{ji} \rangle
}
\label{eq:MFwithSSB}
\eeq
in the $(\lambda^a,\,\psi_i)$ basis. [The off-diagonal entries in this
matrix come from the first term in the second line of
eq.~(\ref{gensusylagr}), and the lower right entry can be seen in
eq.~(\ref{noFlagr}).] Now observe that ${\bf m}_{\rm F}$ annihilates the
vector
\beq
{\stilde G} \,=\, \pmatrix{
{\langle D^a \rangle /\sqrt{2}} \cr \langle F_i\rangle }.
\label{explicitgoldstino}
\eeq
The first row of ${\bf m}_{\rm F}$ annihilates $\stilde G$ by virtue of
the requirement eq.~(\ref{wgaugeinvar}) that the superpotential is gauge
invariant, and the second row does so because of the condition $ \langle
{\partial V/ \partial \phi_i} \rangle = 0 , $ which must be satisfied at any
local minimum of the scalar potential. Equation (\ref{explicitgoldstino}) 
is therefore proportional to the goldstino wavefunction; it is 
non-trivial if
and only if at least one of the auxiliary fields has a VEV, breaking
supersymmetry. So we have proved that if global supersymmetry is
spontaneously broken, then there must be a massless goldstino, and that
its components among the various fermions in the theory are just
proportional to the corresponding auxiliary field VEVs.

There is also a useful sum rule that governs the tree-level squared masses
of particles in theories with spontaneously broken supersymmetry. For a
general theory of the type discussed in section \ref{sec:susylagr}, the
squared masses of the real scalar degrees of freedom are the eigenvalues
of the matrix
\beq
{\bf m}^2_{\rm S} =
\pmatrix{
W_{jk}^* W^{ik} + g^2_a (T^a \phi)_j (\phi^* T^a)^i -g_a T^{ai}_j D^a
&
W_{ijk}^* W^k + g^2_a (T^a \phi)_i (T^a \phi)_j
\vspace{0.1cm}
\cr
W^{ijk} W_k^* + g^2_a (\phi^* T^a)^i (\phi^* T^a)^j
&
W_{ik}^* W^{jk} + g^2_a (T^a \phi)_i (\phi^* T^a)^j - g_a T^{aj}_i D^a
},\phantom{x}
\label{eq:rescalarmasssq}
\eeq
which can be obtained from writing the quadratic part of the tree-level potential as 
\beq
V \,=\, \frac{1}{2} \pmatrix{\phi^{*j} \!\!& \phi_j} {\bf m}^2_{\rm S}
\pmatrix{\phi_{i} \cr \phi^{*i}} .
\eeq
In eq.~(\ref{eq:rescalarmasssq}), 
$W^{ijk} = \delta^3 W/\delta \phi_i\delta \phi_j\delta \phi_k$, and
the scalar fields 
are understood to be replaced by their VEVs. It follows that the sum of
the real scalar squared-mass eigenvalues is
\beq
{\rm Tr}({\bf m}^2_{\rm S}) \,=\, 
2 W_{ik}^* W^{ik} + 2 g^2_a C_a (i) \phi^{*i} \phi_i 
-2 g_a {\rm Tr}(T^{a}) D^a ,
\eeq
with the Casimir invariants $C_a(i)$ defined by eq.~(\ref{eq:defCasimir}).
Meanwhile, the squared masses of the two-component fermions are given by
the eigenvalues of
\beq
{\bf m}_{\rm F}^\dagger {\bf m}_{\rm F}\,=\,
\pmatrix{
2 g_a g_b (\phi^* T^a T^b \phi)
&
\sqrt{2} g_b (T^b \phi)_k W^{ik}
\vspace{0.1cm}
\cr
\sqrt{2} g_a (\phi^* T^a)^k W_{jk}^*
&
\phantom{x} W_{jk}^* W^{ik}  + 2 g^2_c (T^c \phi)_j (\phi^* T^c)^i},
\label{eq:fermionmasssquared}
\eeq
so the sum of the two-component fermion squared masses is
\beq
{\rm Tr}({\bf m}_{\rm F}^\dagger {\bf m}_{\rm F}) \,=\,
W_{ik}^* W^{ik} + 4 g^2_a C_a (i) \phi^{*i} \phi_i .
\eeq
Finally, the vector squared masses are:
\beq
{\bf m}^2_{\rm V} \,=\, g_a^2 (\phi^* \{T^a, T^b\} \phi),
\eeq
so 
\beq
{\rm Tr}({\bf m}^2_{\rm V}) \,=\, 2 g_a^2 C_a(i) \phi^{*i}\phi_i .
\eeq
It follows that the {\em supertrace} of the tree-level squared-mass
eigenvalues, defined in general by a weighted sum over all particles with
spin $j$:
\beq
{\rm STr}(m^2) 
\,\equiv\, 
\sum_j (-1)^{2j} (2j +1) {\rm Tr}(m^2_j), 
\label{eq:supertracedef}
\eeq
satisfies the sum rule
\beq
{\rm STr}(m^2) 
\,=\, 
{\rm Tr}({\bf m^2_{\rm S}})
- 2 {\rm Tr}({\bf m}_{\rm F}^\dagger {\bf m}_{\rm F})
+ 3 {\rm Tr}({\bf m^2_{\rm V}})
\,=\, -2g_a {\rm Tr}(T^{a}) D^a\> =\> 0.
\label{eq:supertracesumrule}
\eeq
The last equality assumes that the traces of the $U(1)$ charges over the
chiral superfields are 0.  This holds for $U(1)_Y$ in the MSSM, and more
generally for any non-anomalous gauge symmetry. The sum rule
eq.~(\ref{eq:supertracesumrule})  is often a useful check on models of
spontaneous supersymmetry breaking.

\subsection{Fayet-Iliopoulos ($D$-term) supersymmetry breaking
\label{subsec:origins.Dterm}}
\setcounter{equation}{0}
\setcounter{footnote}{1}

Supersymmetry breaking with a non-zero $D$-term VEV can occur through the
Fayet-Iliopoulos mechanism \cite{FayetIliopoulos}. If the gauge symmetry
includes a $U(1)$ factor, then, as noted in section \ref{subsec:superspacelagrabelian}, 
one can introduce a term linear in the
auxiliary field of the corresponding gauge supermultiplet,
\beq
\lagr_{\rm FI} \,=\, -\kappa D,
\label{FI}
\eeq
where $\kappa$ is a constant with dimensions of [mass]$^2$. This
term is gauge-invariant and supersymmetric by itself. [Note that for a
$U(1)$ gauge symmetry, the supersymmetry transformation $\delta D$ in
eq.~(\ref{Dtransf}) is a total derivative.] If we include it in the
Lagrangian, then $D$ may be forced to get a non-zero VEV. To see this,
consider the relevant part of the scalar potential from
eqs.~(\ref{lagrgauge}) and (\ref{gensusylagr}):
\beq
V = \kappa D -{1\over 2} D^2 - g D \sum_i q_i |\phi_i|^2 .
\eeq
Here the $q_i$ are the charges of the scalar fields $\phi_i$ under the
$U(1)$ gauge group in question. The presence of the Fayet-Iliopoulos term
modifies the equation of motion eq.~(\ref{solveforD}) to
\beq
D = \kappa - g \sum_i q_i |\phi_i|^2.
\label{booya}
\eeq
Now suppose that the scalar fields $\phi_i$ that are charged under the
$U(1)$ all have non-zero superpotential masses $m_i$. (Gauge invariance
then requires that they come in pairs with opposite charges.) Then the
potential will have the form
\beq
V  = \sum_i |m_i|^2 |\phi_i|^2 +
     {1\over 2} (\kappa -g \sum_i q_i |\phi_i|^2)^2 .
\eeq
Since this cannot vanish, supersymmetry must be broken; one can check that
the minimum always occurs for non-zero $D$. For the simplest case in which
$|m_i|^2 > g q_i \kappa$ for each $i$, the minimum is realized for all
$\phi_i=0$ and $D = \kappa$, with the $U(1)$ gauge symmetry unbroken.  As
further evidence that supersymmetry has indeed been spontaneously broken,
note that the scalars then have squared masses $|m_i|^2 - g q_i \kappa$,
while their fermion partners have squared masses $|m_i|^2$. The gaugino
remains massless, as can be understood from the fact that
it is the goldstino, as argued on general
grounds in section \ref{subsec:origins.general}.

For non-Abelian gauge groups, the analog of eq.~(\ref{FI}) would not be
gauge-invariant and is therefore not allowed, so only $U(1)$ $D$-terms can
drive spontaneous symmetry breaking. In the MSSM, one might imagine that
the $D$ term for $U(1)_Y$ has a Fayet-Iliopoulos term as the principal
source of supersymmetry breaking. Unfortunately, this cannot work, because
the squarks and sleptons do not have superpotential mass terms. So, at
least some of them would just get non-zero VEVs in order to make
eq.~(\ref{booya}) vanish. That would break color and/or electromagnetism,
but not supersymmetry. Therefore, a Fayet-Iliopoulos term for $U(1)_Y$
must be subdominant compared to other sources of supersymmetry breaking in
the MSSM, if not absent altogether. One could instead attempt to trigger
supersymmetry breaking with a Fayet-Iliopoulos term for some other $U(1)$
gauge symmetry, which is as yet unknown because it is spontaneously broken
at a very high mass scale or because it does not couple to the Standard
Model particles. However, if this is the dominant source for supersymmetry
breaking, it proves difficult to give appropriate masses to all of the
MSSM particles, especially the gauginos. In any case, we will not discuss
$D$-term breaking as the ultimate origin of supersymmetry violation any
further (although it may not be ruled out \cite{dtermbreakingmaywork}).

\subsection{O'Raifeartaigh ($F$-term) supersymmetry breaking
\label{subsec:origins.Fterm}}
\setcounter{equation}{0}
\setcounter{footnote}{1}

Models where spontaneous supersymmetry breaking is ultimately due to a
non-zero $F$-term VEV, called O'Rai\-f\-ear\-taigh models
\cite{ORaifeartaigh}, have brighter phenomenological prospects. The idea
is to pick a set of chiral supermultiplets $\Phi_i\supset (\phi_i, \psi_i,
F_i)$ and a superpotential $W$ in such a way that the equations $F_i =
-\delta W^*/\delta \phi^{*i} = 0$ have no simultaneous solution within some
compact domain. Then
$V=\sum_i |F_i|^2$ will have to be positive at its minimum, ensuring that
supersymmetry is broken. The supersymmetry breaking minimum may be 
a global minimum of the potential as in Figure \ref{fig:susybreakingpotentials}(b), or only
a local minimum as in Figure \ref{fig:susybreakingpotentials}(c).

The simplest example with a supersymmetry breaking 
global minimum has three chiral supermultiplets $\Phi_{1,2,3}$, with superpotential
\beq
W = -k \Phi_1 + m \Phi_2 \Phi_3 + {y\over 2} \Phi_1 \Phi_3^2 .
\label{oraif}
\eeq
Note that $W$ contains a linear term, with $k$ having dimensions of
[mass]$^2$.  Such a term is allowed if the corresponding chiral
supermultiplet is a gauge singlet.  In fact, a linear term is necessary to
achieve $F$-term breaking at tree-level in renormalizable
superpotentials,\footnote{Non-polynomial superpotential terms, which
arise from non-perturbative effects in strongly coupled gauge theories, 
avoid this requirement.} since
otherwise setting all $\phi_i=0$ will always give a supersymmetric global
minimum with all $F_i=0$. Without loss of generality, we can choose $k$,
$m$, and $y$ to be real and positive (by phase rotations of the fields).
The scalar potential following from eq.~(\ref{oraif}) is
\beq
&& V_{\rm tree-level} = |F_1|^2 + |F_2|^2 + |F_3|^2, \\
&& F_1 =
k - {y\over 2} \phi_3^{*2} ,\qquad
F_2 = -m \phi^*_3 ,\qquad
F_3 = -m \phi^*_2 - y \phi^*_1 \phi^*_3 .
\eeq
Clearly, $F_1=0$ and $F_2=0$ are not compatible, so supersymmetry must
indeed be broken. If $m^2 > yk$ (which we assume from now on), then the
absolute minimum of the classical potential is at $\phi_2=\phi_3=0$ with $\phi_1$
undetermined, so $F_1 = k$ and $V_{\rm tree-level}=k^2$ at the minimum. The fact that
$\phi_1$ is undetermined at tree level is an example of a ``flat direction" in the
scalar potential; this is a common feature of supersymmetric
models.\footnote{More generally, flat directions, also known as moduli, are non-compact lines
and surfaces in the space of scalar fields along which the scalar
potential vanishes. The classical renormalizable scalar potential of the MSSM would have
many flat directions if supersymmetry were not broken 
\cite{flatdirections}.}

The flat direction parameterized by $\phi_1$ is an accidental feature of
the classical scalar potential, and in this case it is removed (``lifted")
by quantum corrections. This can be seen by computing the Coleman-Weinberg
one-loop effective potential \cite{ColemanWeinberg}. In a loop expansion,
the effective potential can be written as
\beq
V_{\rm eff} &=& V_{\rm tree-level} + V_{\rm 1-loop} + \ldots
\label{eq:Veffexp}
\eeq
where the one-loop contribution is a supertrace over the
scalar-field-dependent squared-mass eigenstates labeled $n$, with spin $s_n$:
\beq
V_{\rm 1-loop} &=& \sum_n (-1)^{2 s_n} (2 s_n + 1) h(m_n^2),
\\
h(z) &\equiv& \frac{1}{64 \pi^2} z^2 \left [\ln(z/Q^2) + a \right].
\label{eq:ColemanWeinberg}
\eeq
Here $Q$ is the renormalization scale and 
$a$ is a renormalization scheme-dependent constant.\footnote{Actually, $a$ can be 
different for the different spin contributions, if one chooses a renormalization 
scheme that does not respect 
supersymmetry. For example, in 
the $\msbar$ scheme, $a = -3/2$ for the spin-0 and spin-$1/2$ contributions, 
but $a = -5/6$ for 
the spin-1 contributions. See ref.~\cite{twoloopEP} 
for a discussion, and the extension to two-loop order.}  
In the $\drbar$ scheme based on dimensional reduction, $a=-3/2$.
Using eqs.~(\ref{eq:rescalarmasssq}) and
(\ref{eq:fermionmasssquared}), the squared mass eigenvalues for the 6 real scalar and 
3 two-component fermion 
states are found to be, as a function of varying $x = |\phi_1|^2$, with $\phi_2=\phi_3=0$:
\beq
\mbox{scalars:}
\quad&&
0
,\>\>
0
,\>\>\> 
m^2 + \frac{y}{2} \Bigl (y x - k + \sqrt{4 m^2 x + (y x - k)^2} \Bigr ),
\nonumber
\\
&& m^2 + \frac{y}{2} \Bigl (y x + k - \sqrt{4 m^2 x + (y x + k)^2} \Bigr ),
\nonumber
\\
&& m^2 + \frac{y}{2} \Bigl (y x - k - \sqrt{4 m^2 x + (y x - k)^2} \Bigr ),
\nonumber
\\ 
&&
m^2 + \frac{y}{2} \Bigl (y x + k + \sqrt{4 m^2 x + (y x + k)^2} \Bigr ),
\\
\mbox{fermions}:\quad&&
0
,\>\>\>
m^2 + \frac{y}{2} \Bigl (y x + \sqrt{4 m^2 x + y^2 x^2} \Bigr )
,\>\>\>
m^2 + \frac{y}{2} \Bigl (y x - \sqrt{4 m^2 x + y^2 x^2} \Bigr ).\phantom{xxxx}
\eeq
[Note that the sum rule
eq.~(\ref{eq:supertracesumrule}) is indeed satisfied by these squared
masses.] 
Now, plugging these into eq.~(\ref{eq:ColemanWeinberg}), one finds that the global minimum of the 
one-loop 
effective potential is at $x=0$, so $\phi_1 = \phi_2 = \phi_3 = 0$. 
The tree-level mass
spectrum of the theory at this point in field space simplifies to
\beq
0,\>\> 0,\>\> m^2,\>\> m^2,\>\> m^2 - yk,\>\> m^2 + yk ,
\label{ORscalars}
\eeq
for the scalars, and
\beq
0,\>\> m^2,\>\> m^2
\label{ORfermions}
\eeq
for the fermions.
The non-degeneracy of scalars and fermions is a clear check that
supersymmetry has been spontaneously broken. 

The 0 eigenvalues in eqs.~(\ref{ORscalars}) and
(\ref{ORfermions}) correspond to the complex scalar $\phi_1$ and its
fermionic partner $\psi_1$. However, $\phi_1$ and $\psi_1$ have different
reasons for being massless. The masslessness of $\phi_1$ corresponds to
the existence of the classical flat direction, since any value of $\phi_1$ gives the
same energy at tree-level. 
The one-loop potential 
lifts this flat direction, so that $\phi_1$ gains a 
mass once 
quantum corrections are included. Expanding $V_{\rm 1-loop}$ to first order in 
$x$, one finds that
the complex scalar $\phi_1$ receives a
positive-definite squared mass equal to
\beq
m_{\phi_1}^2 = {y^2 m^2\over 16 \pi^2} \left [
\ln (1 - r^2) - 1 + \frac{1}{2} \left ( r + 1/r \right )
\ln \left ( \frac{1 + r}{1-r} \right )
\right ],
\label{eq:ORmsq}
\eeq
where $r = y k/m^2$. [This reduces to $m_{\phi_1}^2 =
y^4 k^2/48 \pi^2 m^2$ in the limit $yk\ll m^2$.] 
In contrast, the Weyl
fermion $\psi_1$ remains exactly massless, to all orders in perturbation theory,
because it is the goldstino, as
predicted in section \ref{subsec:origins.general}.

The O'Rai\-f\-ear\-taigh superpotential eq.~(\ref{oraif}) yields a Lagrangian that 
is invariant under a
$U(1)_R$ symmetry (see section \ref{Rsymmetry}) with charge assignments
\beq
r_{\Phi_1} \,=\, r_{\Phi_2} \,=\, 2,\qquad
r_{\Phi_3} \,=\, 0.
\eeq
This illustrates a general result, the Nelson-Seiberg theorem \cite{Nelson:1993nf}, 
which says that
if a theory has a scalar potential with a global minimum that breaks supersymmetry by a non-zero 
$F$-term, and the superpotential is generic 
(contains all terms not forbidden by symmetries), 
then the theory must have an exact $U(1)_R$
symmetry. If the $U(1)_R$ symmetry remains unbroken when supersymmetry breaks, 
as is the case in the O'Rai\-f\-ear\-taigh model discussed above, then 
there is a 
problem of explaining how gauginos get masses, because non-zero gaugino mass
terms have $R$-charge 
$2$. On the other hand, if the $U(1)_R$ symmetry is spontaneously broken,
then there results a pseudo-Nambu-Goldstone boson (the $R$-axion) which is problematic
experimentally, although gravitational effects may give it a large enough mass to avoid being 
ruled out \cite{Raxion}. 

If the supersymmetry breaking vacuum is only metastable, then one does not need an exact $U(1)_R$ 
symmetry. This can be illustrated by adding to the O'Rai\-f\-ear\-taigh superpotential 
eq.~(\ref{oraif}) a term $\Delta W$ that explicitly breaks the continuous $R$ symmetry. For example, consider 
\cite{Intriligator:2007py}:
\beq
\Delta W &=& \frac{1}{2} \epsilon m \Phi_2^2,
\eeq
where $\epsilon$ is a small dimensionless parameter, so that the tree-level scalar potential is
\beq
&&
V_{\rm tree-level} \>=\> |F_1|^2 + |F_2|^2 + |F_3|^2,
\\
&& F_1 = k - {y\over 2} \phi_3^{*2} ,\qquad
F_2 = -\epsilon m \phi_2^* -m \phi^*_3 ,\qquad
F_3 = -m \phi^*_2 - y \phi^*_1 \phi^*_3 .
\eeq  
In accord with the Nelson-Seiberg theorem,
there are now (two) supersymmetric minima, with
\beq
\phi_1 \,=\, m/\epsilon y,
\qquad
\phi_2 \,=\, \pm \frac{1}{\epsilon} \sqrt{2k/y},
\qquad
\phi_3 \,=\, \mp \sqrt{2k/y}.
\eeq
However, for small enough $\epsilon$, the local supersymmetry-breaking minimum at 
$\phi_1=\phi_2=\phi_3=0$
is also still present and stabilized by 
the one-loop effective potential, with potential barriers 
between it and the supersymmetric minima, so the situation is qualitatively like
Figure \ref{fig:susybreakingpotentials}(c).
As $\epsilon \rightarrow 0$, the supersymmetric global minima 
move off to infinity in field space,
and there is negligible effect on the supersymmetry-breaking local minimum.
One can show \cite{Intriligator:2007py} that the lifetime of the metastable vacuum state 
due to quantum tunneling can be made 
arbitrarily large. The same effect can be realized by a variety of other 
perturbations to the O'Rai\-f\-ear\-taigh model; by eliminating the continuous
$R$ symmetry using small additional contributions to the Lagrangian, 
the stable supersymmetry breaking vacuum is converted to a metastable one. 
(In some cases, the Lagrangian remains invariant under a discrete $R$ symmetry.)

The O'Rai\-f\-ear\-taigh superpotential determines the mass scale of
supersymmetry breaking $\sqrt{F_1}$ in terms of a dimensionful parameter
$k$ put in by hand. This appears somewhat artificial, since $k$ will have
to be tiny compared to $\MPlanck^2$ in order to give the right order of
magnitude for the MSSM soft terms. It may be more plausible to have a mechanism that
can instead generate such scales naturally. This can be done in models of
dynamical supersymmetry breaking, in which the small 
mass scales associated with supersymmetry breaking arise by
dimensional transmutation. In other words, they generally feature a new
asymptotically free non-Abelian gauge symmetry with a gauge coupling $g$
that is perturbative at $\MPlanck$ and gets strong in the infrared at some
smaller scale $\Lambda \sim e^{-8\pi^2/|b| g_0^2} \MPlanck$, where $g_0$
is the running gauge coupling at $\MPlanck$ with negative beta function $-
|b| g^3/16 \pi^2$. Just as in QCD, it is perfectly natural for $\Lambda$
to be many orders of magnitude below the Planck scale. Supersymmetry
breaking may then be best described in terms of the effective dynamics of
the strongly coupled theory. 
Supersymmetry is still broken by the VEV of an $F$ field, but it may be
the auxiliary field of a composite chiral supermultiplet built out
of fields that are charged under the new strongly coupled gauge group.

The construction of such models that break supersymmetry through strong-coupling dynamics is 
non-trivial if one wants a stable supersymmetry-breaking ground state. In addition to the 
argument from the Nelson-Seiberg theorem that a $U(1)_R$ symmetry should be present, one can 
prove using the Witten index \cite{Wittenindex,AffleckDineSeiberg} that any strongly coupled 
gauge theory with only vectorlike, massive matter cannot spontaneously break supersymmetry in 
its true ground state. However, things are easier if one only requires a local (metastable) 
minimum of the potential. Intriligator, Seiberg, and Shih showed \cite{Intriligator:2006dd} that 
supersymmetric Yang-Mills theories with vectorlike matter can have metastable vacuum states with 
non-vanishing $F$-terms that break supersymmetry, and lifetimes that can be arbitrarily long. The 
simplest model that does this is remarkably economical; it is just supersymmetric $SU(N_c)$ gauge 
theory, with $N_f$ massive flavors of quark and antiquark supermultiplets, with $N_c + 1 \leq N_f 
< 3 N_c/2$. The recognition of the advantages of a metastable vacuum state 
opens up many new model building possibilities and ideas 
\cite{Intriligator:2006dd,Intriligator:2007py,metastablemodels}.

The topic of known ways of breaking supersymmetry spontaneously through
strongly coupled gauge theories is a big subject that is in danger of becoming vast, and is beyond 
the scope of this primer. Fortunately,
there are several excellent reviews, including 
\cite{susybreakingrevs} for the more recent developments and \cite{dynamicalsusybreaking}
for older models with stable vacua.
Finding the ultimate cause of supersymmetry breaking is one of the
most important goals for the future. However, for many purposes,
one can simply assume that an $F$-term has obtained a VEV, without
worrying about the specific dynamics that caused it. For
understanding collider phenomenology, the most immediate concern is usually
the nature of the couplings of the $F$-term VEV to the MSSM fields. 
This is the subject we turn to next.

\subsection{The need for a separate supersymmetry-breaking sector
\label{subsec:origins.sector}}
\setcounter{equation}{0}
\setcounter{footnote}{1}

It is now clear that spontaneous supersymmetry breaking (dynamical or not)
requires us to extend the MSSM. The ultimate supersymmetry-breaking order
parameter cannot belong to any of the MSSM supermultiplets; a $D$-term VEV
for $U(1)_Y$ does not lead to an acceptable spectrum, and there is no
candidate gauge singlet whose $F$-term could develop a VEV. Therefore one
must ask what effects {\it are} responsible for spontaneous supersymmetry
breaking, and how supersymmetry breakdown is ``communicated" to the MSSM
particles.  It is very difficult to achieve the latter in a
phenomenologically viable way working only with renormalizable
interactions at tree-level, even if the model is extended to involve new
supermultiplets including gauge singlets. First, on general grounds it 
would be
problematic to give masses to the MSSM gauginos, because the results of
section \ref{sec:susylagr} inform us that renormalizable supersymmetry
never has any (scalar)-(gaugino)-(gaugino) couplings that could turn into
gaugino mass terms when the scalar gets a VEV.  Second, at least some of
the MSSM squarks and sleptons would have to be unacceptably light, and
should have been discovered already. This can be understood from the
existence of sum rules that can be obtained in the same way as
eq.~(\ref{eq:supertracesumrule}) when the restrictions imposed by flavor
symmetries are taken into account. For example, in the limit in which
lepton flavors are conserved, the selectron mass eigenstates $\tilde e_1$
and $\tilde e_2$ could in general be mixtures of $\tilde e_L$ and $\tilde
e_R$. But if they do not mix with other scalars, then part of the sum rule
decouples from the rest, and one obtains:
\beq
m_{\tilde e_1}^2 + m_{\tilde e_2}^2 = 2 m_e^2,
\label{eq:sumrulee}
\eeq
which is of course ruled out by experiment. Similar sum rules follow for
each of the fermions of the Standard Model, at tree-level and in the
limits in which the corresponding flavors are conserved. In principle, the
sum rules can be evaded by introducing flavor-violating mixings, but it is
very difficult to see how to make a viable model in this way.  Even
ignoring these problems, there is no obvious reason why the resulting MSSM
soft supersymmetry-breaking terms in this type of model should satisfy
flavor-blindness conditions like eqs.~(\ref{scalarmassunification}) or
(\ref{aunification}).

\begin{figure}
\centerline{\psfig{figure=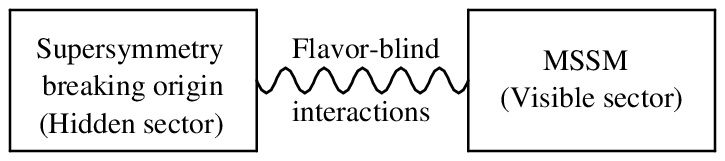,height=.8in}}
\caption{The presumed schematic structure for supersymmetry breaking.
\label{fig:structure}}
\end{figure}

For these reasons, we expect that the MSSM soft terms arise indirectly or
radiatively, rather than from tree-level renormalizable couplings to the
supersymmetry-breaking order parameters. Supersymmetry breaking evidently
occurs in a ``hidden sector" of particles that have no (or only very
small) direct couplings to the ``visible sector" chiral supermultiplets of
the MSSM. However, the two sectors do share some interactions that are
responsible for mediating supersymmetry breaking from the hidden sector to
the visible sector, resulting in the MSSM soft terms.
(See Figure~\ref{fig:structure}.)%
In this scenario, the tree-level squared 
mass sum rules need not hold, even approximately, for the physical masses 
of the visible sector fields, so that a phenomenologically viable 
superpartner mass spectrum is, in principle, achievable. As a bonus, if 
the mediating interactions are flavor-blind, then the soft terms appearing 
in the MSSM will automatically obey conditions like 
eqs.~(\ref{scalarmassunification}), (\ref{aunification}) and 
(\ref{commonphase}).

There have been two main competing proposals for what the mediating
interactions might be. The first (and historically the more popular) is
that they are gravitational. More precisely, they are associated with the
new physics, including gravity, that enters near the Planck scale. In this
``gravity-mediated", or {\it Planck-scale-mediated supersymmetry breaking}
(PMSB) scenario, if supersymmetry is broken in the hidden sector by a VEV
$\langle F\rangle$, then the soft terms in the visible sector should be
roughly
\beq
m_{\rm soft} \>\sim\> {\langle F \rangle / \MPlanck},
\label{mgravusual}
\eeq
by dimensional analysis. This is because we know that $m_{\rm soft}$ must
vanish in the limit $\langle F \rangle \rightarrow 0$ where supersymmetry
is unbroken, and also in the limit $\MPlanck \rightarrow \infty$
(corresponding to $G_{\rm Newton} \rightarrow 0$) in which gravity becomes
irrelevant. For $m_{\rm soft}$ of order a few hundred GeV, one would
therefore expect that the scale associated with the origin of
supersymmetry breaking in the hidden sector should be roughly
${\sqrt{\langle F\rangle}} \sim 10^{10}$ or $10^{11}$ GeV. 

A second possibility is that the flavor-blind mediating interactions for
supersymmetry breaking are the ordinary electroweak and QCD gauge
interactions. In this {\it gauge-mediated supersymmetry breaking} (GMSB)
scenario, the MSSM soft terms come from loop diagrams involving some {\it
messenger} particles.  The messengers are new chiral supermultiplets that
couple to a supersymmetry-breaking VEV $\langle F\rangle$, and also have
$SU(3)_C \times SU(2)_L \times U(1)_Y$ interactions, which provide the
necessary connection to the MSSM. Then, using dimensional analysis, one 
estimates for the MSSM soft terms
\beq
m_{\rm soft} \sim {\alpha_a\over 4\pi} {\langle F \rangle
\over M_{\rm mess}}
\label{mgravgmsb}
\eeq
where the $\alpha_a/4\pi$ is a loop factor for Feynman diagrams involving
gauge interactions, and $M_{\rm mess}$ is a characteristic scale of the
masses of the messenger fields. So if $M_{\rm mess}$ and $\sqrt{\langle
F\rangle}$ are roughly comparable, then the scale of supersymmetry
breaking can be as low as about ${\sqrt{\langle F\rangle}} \sim 10^4$ GeV
(much lower than in the gravity-mediated case!) to give $m_{\rm soft}$ of
the right order of magnitude.

\subsection{The goldstino and the gravitino}\label{subsec:origins.gravitino}
\setcounter{equation}{0}
\setcounter{footnote}{1}

As shown in section \ref{subsec:origins.general}, the spontaneous breaking
of global supersymmetry implies the existence of a massless Weyl fermion,
the goldstino.  The goldstino is the fermionic component of the
supermultiplet whose auxiliary field obtains a VEV.

\begin{figure}
\begin{center}
\begin{picture}(66,60)(0,0)
\SetWidth{0.85}
\ArrowLine(0,0)(33,12)
\ArrowLine(66,0)(33,12)
\DashLine(33,52.5)(33,12){4}
\ArrowLine(33,32.2501)(33,32.25)
\Text(0,10)[c]{$\psi$}
\Text(66,10)[c]{$\stilde G$}
\Text(26,52)[c]{$\phi$}
\Text(33,-12)[c]{(a)}
\end{picture}
\hspace{2.5cm}
\begin{picture}(66,60)(0,0)
\Photon(0,0)(33,12){1.75}{4}
\SetWidth{0.85}
\Photon(33,52.5)(33,12){2}{4.5}
\ArrowLine(0,0)(33,12)
\ArrowLine(66,0)(33,12)
\ArrowLine(33,32.2501)(33,32.25)
\Text(0,11)[c]{$\lambda$}
\Text(66,10)[c]{$\stilde G$}
\Text(25,52)[c]{$A$}
\Text(33,-12)[c]{(b)}
\end{picture}
\end{center}
\caption{Goldstino/gravitino $\tilde G$ interactions with superpartner 
pairs $(\phi,\psi)$ and $(\lambda,A)$.
\label{fig:goldstino}}
\end{figure}
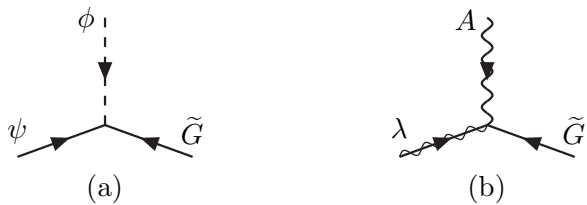

We can derive an important property of the goldstino by considering the
form of the conserved supercurrent eq.~(\ref{supercurrent}). Suppose for
simplicity\footnote{More generally, if supersymmetry is spontaneously
broken by VEVs for several auxiliary fields $F_i$ and $D^a$, then one
should make the replacement $\langle F \rangle \rightarrow ( \sum_i
|\langle F_i \rangle|^2 + {1\over 2} \sum_a \langle D^a \rangle^2 )^{1/2}$
everywhere in the following.} that the only non-vanishing auxiliary field
VEV is $\langle F \rangle$ with goldstino superpartner $\stilde G$. Then
the supercurrent conservation equation tells us that
\beq
0 = \partial_\mu J^\mu_\alpha =
-i \langle F \rangle (\sigma^\mu \partial_\mu \stilde G^\dagger)_\alpha +
\partial_\mu j^\mu_\alpha + \ldots
\label{beezlebub}
\eeq
where $j^\mu_\alpha$ is the part of the supercurrent that involves all of
the other supermultiplets, and the ellipses represent other contributions
of the goldstino supermultiplet to $\partial_\mu J^\mu_\alpha$, which we
can ignore. [The first term in eq.~(\ref{beezlebub}) comes from the second
term in eq.~(\ref{supercurrent}), using the equation of motion $F_i =
-W^{*}_i$ for the goldstino's auxiliary field.] This equation of motion
for the goldstino field allows us to write an effective Lagrangian
\beq
\lagr_{\rm goldstino}
=  i \stilde G^\dagger \sigmabar^\mu \partial_\mu \stilde G 
- {1\over \langle F \rangle}(\stilde G \partial_\mu j^\mu
+ \conj) ,
\label{goldstinointeraction}
\eeq
which describes the interactions of the goldstino with all of the other
fermion-boson pairs \cite{Fayetsupercurrent}. In particular, since
$j^\mu_\alpha = 
(\sigma^\nu\sigmabar^\mu \psi_i)_\alpha \partial_\nu\phi^{*i} 
\BDplus \sigma^\nu \sigmabar^\rho \sigma^\mu \lambda^{\dagger a}
F_{\nu\rho}^a/2\sqrt{2} + \ldots$, there are goldstino-scalar-chiral
fermion and goldstino-gaugino-gauge boson vertices as shown in
Figure~\ref{fig:goldstino}. Since this derivation depends only on
supercurrent conservation, eq.~(\ref{goldstinointeraction}) holds
independently of the details of how supersymmetry breaking is communicated
from $\langle F \rangle$ to the MSSM sector fields $(\phi_i,\psi_i)$ and
$(\lambda^a, A^a)$. It may appear strange at first that the interaction
couplings in eq.~(\ref{goldstinointeraction}) get larger in the limit
$\langle F \rangle$ goes to zero. However, the interaction term $\stilde G
\partial_\mu j^\mu$ contains two derivatives, which turn out to always
give a kinematic factor proportional to the squared-mass difference of the
superpartners when they are on-shell, i.e.~$m_{\phi}^2 - m_{\psi}^2$
and $m^2_{\lambda} - m_{A}^2$ for Figures~\ref{fig:goldstino}a and
\ref{fig:goldstino}b respectively. These can be non-zero only by virtue of
supersymmetry breaking, so they must also vanish as $\langle F\rangle
\rightarrow 0$, and the interaction is well-defined in that limit.
Nevertheless, for fixed values of $m_{\phi}^2 - m_{\psi}^2$ and
$m^2_{\lambda} - m_{A}^2$, the interaction term in
eq.~(\ref{goldstinointeraction}) can be phenomenologically important if
$\langle F \rangle $ is not too large
\cite{Fayetsupercurrent}-\cite{AKKMM2}.

The above remarks apply to the breaking of global supersymmetry. However, 
taking into account gravity, supersymmetry must be promoted to a local 
symmetry. This means that the spinor parameter $\epsilon^\alpha$, which 
first appeared in section \ref{subsec:susylagr.freeWZ}, is no longer a 
constant, but can vary from point to point in spacetime. The resulting 
locally supersymmetric theory is called {\it supergravity} 
\cite{supergravity,superconformalsupergravity}. 
It necessarily unifies the spacetime 
symmetries of ordinary general relativity with local supersymmetry 
transformations. In supergravity, the spin-2 graviton has a spin-3/2 
fermion superpartner called the gravitino, which we will denote $\stilde 
\Psi_\mu^\alpha$. The gravitino has odd $R$-parity ($P_R=-1$), as can be 
seen from the definition eq.~(\ref{defRparity}). It carries both a vector 
index ($\mu$) and a spinor index ($\alpha$), and transforms 
inhomogeneously under local supersymmetry transformations:
\beq
\delta \stilde\Psi_\mu^\alpha \>=\> 
\partial_\mu\epsilon^\alpha +\ldots
\label{gravitinoisgauge}
\eeq
Thus the gravitino should be thought of as the ``gauge" field of local
supersymmetry transformations [compare eq.~(\ref{Agaugetr})]. As long as
supersymmetry is unbroken, the graviton and the gravitino are both
massless, each with two spin helicity states. Once supersymmetry is
spontaneously broken, the gravitino acquires a mass by absorbing
(``eating") the goldstino, which becomes its longitudinal (helicity $\pm
1/2$) components. This is called the {\it super-Higgs} mechanism, and it
is analogous to the ordinary Higgs mechanism for gauge theories, by which the
$W^\pm$ and $Z^0$ gauge bosons in the Standard Model gain mass by
absorbing the Nambu-Goldstone bosons associated with the spontaneously
broken electroweak gauge invariance. The massive spin-3/2 gravitino now
has four helicity states, of which two were originally assigned to the
would-be goldstino. The gravitino mass is traditionally called $m_{3/2}$,
and in the case of $F$-term breaking it can be estimated as
\cite{gravitinomassref}
\beq
m_{3/2} \>\sim\> {\langle F \rangle / \MPlanck},
\label{gravitinomass}
\eeq
This follows simply from dimensional analysis, since $m_{3/2}$ must vanish 
in the limits that supersymmetry is restored ($\langle F \rangle 
\rightarrow 0$) and that gravity is turned off ($\MPlanck \rightarrow 
\infty$). 
Equation (\ref{gravitinomass}) implies very different expectations for the 
mass of the gravitino in gravity-mediated and in gauge-mediated models, 
because they usually make very different predictions for $\langle F 
\rangle$.

In the Planck-scale-mediated supersymmetry breaking case, the gravitino 
mass is comparable to the masses of the MSSM sparticles [compare 
eqs.~(\ref{mgravusual}) and (\ref{gravitinomass})]. Therefore $m_{3/2}$ is 
expected to be at least of order 100 GeV or so. Its interactions will be 
of gravitational strength, so the gravitino will not play any role in 
collider physics, but it can be important in cosmology 
\cite{cosmogravitino}. If it is the LSP, then it is stable and its 
primordial density could easily exceed the critical density, causing the 
universe to become matter-dominated too early. Even if it is not the LSP, 
the gravitino can cause problems unless its density is diluted by 
inflation at late times, or it decays sufficiently rapidly.

In contrast, gauge-mediated supersymmetry breaking models predict that the 
gravitino is much lighter than the MSSM sparticles as long as $M_{\rm 
mess} \ll \MPlanck$. This can be seen by comparing eqs.~(\ref{mgravgmsb}) 
and (\ref{gravitinomass}). The gravitino is almost certainly the LSP in 
this case, and all of the MSSM sparticles will eventually decay into final 
states that include it. Naively, one might expect that these decays are 
extremely slow. However, this is not necessarily true, because the 
gravitino inherits the non-gravitational interactions of the goldstino it 
has absorbed. This means that the gravitino, or more precisely its 
longitudinal (goldstino) components, can play an important role in 
collider physics experiments. The mass of the gravitino can generally be 
ignored for kinematic purposes, as can its transverse (helicity $\pm 3/2$) 
components, which really do have only gravitational interactions. 
Therefore in collider phenomenology discussions one may interchangeably 
use the same symbol $\stilde G$ for the goldstino and for the gravitino of 
which it is the longitudinal (helicity $\pm 1/2$) part. By using the 
effective Lagrangian eq.~(\ref{goldstinointeraction}), one can compute 
that the decay rate of any sparticle $\stilde X$ into its Standard Model 
partner $X$ plus a goldstino/gravitino $\stilde G$ is 
\beq
\Gamma(\stilde X \rightarrow X\stilde G) \,=\,
{m^5_{\stilde X} \over 16 \pi \langle F \rangle^2}
\left ( 1 - {m_X^2/ m_{\stilde X}^2} \right )^4 .
\qquad\>\>{}
\label{generalgravdecay}
\eeq
This corresponds to either Figure~\ref{fig:goldstino}a or 
\ref{fig:goldstino}b, with $(\stilde X,X) = (\phi,\psi)$ or $(\lambda,A)$ 
respectively. One factor $(1 - m_X^2/m_{\stilde X}^2 )^2$ came from the 
derivatives in the interaction term in eq.~(\ref{goldstinointeraction}) 
evaluated for on-shell final states, and another such factor comes from 
the kinematic phase space integral with $m_{3/2} \ll m_{\stilde X}, m_X$.

If the supermultiplet containing the goldstino and $\langle F \rangle$ has
canonically normalized kinetic terms, and the tree-level
vacuum energy is required to
vanish, then the estimate eq.~(\ref{gravitinomass}) is
sharpened to
\beq
m_{3/2} \>=\> {\langle F \rangle / \sqrt{3} \MPlanck} .
\label{gravitinomassbogus}
\eeq
In that case, one can rewrite eq.~(\ref{generalgravdecay}) as
\beq
\Gamma(\stilde X \rightarrow X\stilde G) \,=\,
{m_{\stilde X}^5 \over 48 \pi \MPlanck^2 m_{3/2}^2}
\left ( 1 - {m_X^2/ m_{\stilde X}^2} \right )^4 ,
\qquad\>\>{}
\label{specificgravdecay}
\eeq
and this is how the formula is sometimes presented, although it is less 
general since it assumes eq.~(\ref{gravitinomassbogus}). The 
decay width is larger for smaller $\langle F \rangle$, or equivalently for 
smaller $m_{3/2}$, if the other masses are fixed.  If $\stilde X$ is a 
mixture of superpartners of different Standard Model particles $X$, then 
each partial width in eq.~(\ref{generalgravdecay}) should be multiplied by a 
suppression factor equal to the square of the cosine of the appropriate 
mixing angle. If $m_{\stilde X}$ is of order 100 GeV or more, and $\sqrt{ 
\langle F\rangle } \lsim$ few $\times 10^6$ GeV [corresponding to 
$m_{3/2}$ less than roughly 1 keV according to 
eq.~(\ref{gravitinomassbogus})], then the decay $\stilde X \rightarrow X 
\stilde G$ can occur quickly enough to be observed in a modern collider 
detector. This implies some interesting phenomenological signatures, 
which we will discuss further in sections \ref{subsec:decays.gravitino} 
and \ref{sec:signals}.

We now turn to a more systematic analysis of the way in which the MSSM 
soft terms arise.

\subsection{Planck-scale-mediated supersymmetry breaking
models}\label{subsec:origins.sugra}
\setcounter{equation}{0}
\setcounter{footnote}{1}

Consider models in which the spontaneous supersymmetry breaking sector 
connects with our 
MSSM sector mostly through gravitational-strength interactions, including the 
effects of supergravity \cite{MSUGRA,rewsbtwo}. 
Let $X$ be the chiral superfield whose $F$ term 
auxiliary field breaks supersymmetry, and consider first a globally supersymmetric 
effective Lagrangian, with the Planck scale suppressed effects 
that communicate between the two sectors included as 
non-renormalizable 
terms of 
the types discussed in section \ref{superspacenonrenorm}. The superpotential, the K\"ahler 
potential, and the gauge kinetic function, expanded for large $\MPlanck$, are:
\beq
W &=& W_{\rm MSSM} - \frac{1}{\MPlanck} \left (
\frac{1}{6} y^{Xijk} X \Phi_i \Phi_j \Phi_k +
\frac{1}{2} \mu^{Xij} X \Phi_i \Phi_j \right ) + \ldots
,
\label{eq:WnonrenormX}
\\
K &=& \Phi^{*i} \Phi_i 
+ \frac{1}{\MPlanck} \bigl (
n_i^j X  + \overline n_i^j X^* \bigr ) \Phi^{*i} \Phi_j
- \frac{1}{\MPlanck^2} k_i^j X X^* \Phi^{*i} 
\Phi_j + 
\ldots ,
\label{eq:KnonrenormX}
\\
f_{ab} &=& \frac{\delta_{ab}}{g_a^2} 
\Bigl (1 - \frac{2}{\MPlanck} f_a X + \ldots 
\Bigr ) .
\label{eq:fnonrenormX}
\eeq
Here $\Phi_i$ represent the chiral superfields of the MSSM or an 
extension of it, 
and $y^{Xijk}$, $k_i^j$, $n_i^j$, $\overline n_i^j$ and $f_a$ are dimensionless 
couplings while $\mu^{Xij}$ has the dimension of mass. 
The leading term in the K\"ahler potential is chosen to give canonically normalized 
kinetic terms.
The matrix $k_i^j$ must be Hermitian, and $\overline n_i^j = (n^i_j)^*$, 
in order for the Lagrangian to be real. 
To find the resulting soft supersymmetry breaking terms in the low-energy 
effective theory, one can apply the superspace formalism of section 
\ref{sec:superfields}, treating $X$ as a ``spurion" by making the 
replacements:
\beq
X \rightarrow \theta\theta F ,\qquad\qquad
X^* \rightarrow \theta^\dagger\theta^\dagger F^*,
\eeq
where $F$ denotes $\langle F_X 
\rangle$. The resulting supersymmetry-breaking Lagrangian, after integrating out the auxiliary fields in $\Phi_i$, is:
\beq 
\lagr_{\rm soft} &=& 
-{F\over 2\MPlanck} f_a  \lambda^a \lambda^a
- {F \over  6\MPlanck} y^{Xijk} \phi_i \phi_j \phi_k
- {F\over 2 \MPlanck} \mu^{Xij}\phi_i \phi_j
- {F \over \MPlanck} n_i^j \phi_j W^i_{\rm MSSM} 
+ \conj
\phantom{xxx}
\nonumber
\\
&&
- \frac{|F|^2}{\MPlanck^2} ( k^i_j  + n^i_p \overline n^p_j) \phi^{*j} \phi_i ,
\label{hiddengrav}
\eeq
where $\phi_i$ and $\lambda^a$ are the scalar and gaugino 
fields in the MSSM sector. 
Now if one 
assumes that $\sqrt{F} \sim 10^{10}$ or $10^{11}$ GeV, 
then eq.~(\ref{hiddengrav}) 
has the same form as eq.~(\ref{lagrsoft}), with MSSM-sector soft terms of 
order $m_{\rm soft} \sim F/\MPlanck$, perhaps of order a few hundred 
GeV. 
In particular, if we write the visible sector superpotential as
\beq
W_{\rm MSSM} &=& \frac{1}{6} y^{ijk} \Phi_i \Phi_j \Phi_k
+ \frac{1}{2} \mu^{ij} \Phi_i \Phi_j,
\eeq
then the soft terms in that sector, in the 
notation of eq.~(\ref{lagrsoft}), are:
\beq
M_a &=& {F\over \MPlanck} f_a,
\\
a^{ijk} &=& {F\over \MPlanck}( y^{Xijk}
+ n^i_p y^{pjk}+ n^j_p y^{pik}+ n^k_p y^{pij} )
,
\\
b^{ij} &=& {F\over \MPlanck} (\mu^{Xij} + n^i_p \mu^{pj}+ n^j_p \mu^{pi}),
\\
(m^2)^i_j &=& \frac{|F|^2}{\MPlanck^2} (k^i_j + n^i_p \overline n^p_j)
.
\label{eq:hiddengravcoup}
\eeq

Note that couplings of the form $\lagr_{\rm maybe~soft}$ in
eq.~(\ref{lagrsoftprime}) do not arise from eq.~(\ref{hiddengrav}). 
Although they
actually are expected to occur, the largest possible sources for them 
are non-renormalizable K\"ahler potential terms, which lead to:
\beq
{\cal L} &=& -{|F|^2 \over \MPlanck^3} x^{jk}_i 
\phi^{*i} \phi_j \phi_k 
 + {\rm c.c.},
\eeq
where $x^{jk}_i$ is dimensionless.
This explains why, at least within this model framework, the couplings $c_i^{jk}$ 
in eq.~(\ref{lagrsoftprime}) are of order 
$|F|^2/\MPlanck^3 
\sim m^2_{\rm soft}/\MPlanck$, and therefore 
negligible.

In principle, the parameters $f_a$, $k^i_j$, $n_i^j$, $y^{Xijk}$ and $\mu^{Xij}$ 
ought to be 
determined by the fundamental underlying theory. The familiar flavor 
blindness of 
gravity expressed in Einstein's equivalence principle does not, by 
itself, tell us anything about their form.
Therefore, the requirement of approximate flavor blindness 
in ${\cal L}_{\rm soft}$ is a new 
assumption in this framework, and is not guaranteed without further 
structure. Nevertheless, it has 
historically been popular to make a dramatic simplification by assuming a 
``minimal" form for the normalization of kinetic terms and gauge 
interactions in the non-renormalizable Lagrangian. 
Specifically, it is often assumed that there is a common $f_a=f$ for the 
three gauginos, that $k_i^j = k \delta_i^j$ and $n_i^j = n \delta_i^j$ 
are the same for all scalars, with $k$ and $n$ real,
and that the other couplings are proportional to the corresponding 
superpotential parameters, so that $y^{Xijk} = \alpha y^{ijk}$ and 
$\mu^{Xij} = \beta \mu^{ij}$ with universal real dimensionless constants 
$\alpha$ and $\beta$. Then the soft terms in $\lagr_{\rm soft}^{\rm MSSM}$ 
are all determined by just four parameters:
\beq
m_{1/2} = f{\langle \FX \rangle\over \MPlanck},\qquad\!\!\!
m^2_{0} = (k + n^2) {|\langle \FX \rangle|^2\over \MPlanck^2},\qquad\!\!\!
A_0 = (\alpha + 3n){\langle \FX \rangle\over \MPlanck},\qquad\!\!\!
B_0 = (\beta + 2 n) {\langle \FX \rangle\over \MPlanck}.\>\>\phantom{xx}
\label{sillyassumptions}
\eeq
In terms of these, the parameters appearing in
eq.~(\ref{MSSMsoft}) are:
\beq
&&\!\!\!\! M_3 = M_2 = M_1 = m_{1/2},
\label{gauginounificationsugra}
\\
&&\!\!\!\! {\bf m^2_{Q}} =
{\bf m^2_{{\sbar u}}} =
{\bf m^2_{{\sbar d}}} =
{\bf m^2_{ L}} =
{\bf m^2_{{\sbar e}}} =
m_0^2\, {\bf 1},
\>\>\>\>\>\>\>\> \> m_{H_u}^2 = m^2_{H_d} = m_0^2, \>\>\>\qquad\qquad{}
\label{scalarunificationsugra}
\\
&&\!\!\!\! {\bf a_u} = A_0 {\bf y_u},\qquad
{\bf a_d} = A_0 {\bf y_d},\qquad
{\bf a_e} = A_0 {\bf y_e},
\label{aunificationsugra}
\\
&&\!\!\!\! b = B_0 \mu ,
\label{bsilly}
\eeq
at a renormalization scale $Q \approx \MPlanck$. It is a matter of some 
controversy whether the assumptions going into this parameterization are 
well-motivated on purely theoretical grounds, but from 
a phenomenological perspective they are clearly very nice. This framework 
successfully evades the most dangerous types of flavor changing and CP 
violation as discussed in section \ref{subsec:mssm.hints}. In particular, 
eqs.~(\ref{scalarunificationsugra}) and (\ref{aunificationsugra}) are just 
stronger versions of eqs.~(\ref{scalarmassunification}) and 
(\ref{aunification}), respectively. If $m_{1/2}$, $A_0$ and $B_0$ all have 
the same complex phase, then eq.~(\ref{commonphase}) will also be 
satisfied.

Equations (\ref{gauginounificationsugra})-(\ref{bsilly}) also have the 
virtue of being extraordinarily predictive, at least in principle. 
[Of course, 
eq.~(\ref{bsilly}) is 
content-free unless one can relate $B_0$ to the other parameters in some 
non-trivial way.] As discussed in sections \ref{subsec:mssm.hints} and 
\ref{subsec:RGEs}, they should be applied as RG boundary conditions at the 
scale $\MPlanck$. The RG evolution of the soft parameters down to the 
electroweak scale will then allow us to predict the entire MSSM spectrum 
in terms of just five parameters $m_{1/2}$, $m_0^2$, $A_0$, $B_0$, and 
$\mu$ (plus the already-measured gauge and Yukawa couplings of the MSSM). 
A popular approximation is to start this RG running from the unification 
scale $M_U\approx 1.5\times 10^{16}$ GeV instead of $\MPlanck$. The reason 
for this is more practical than principled; the apparent unification of 
gauge couplings gives us a strong hint that we know something about how 
the RG equations behave up to $M_U$, but unfortunately gives us little 
guidance about what to expect at scales between $M_U$ and $\MPlanck$. The 
errors made in neglecting these effects are proportional to a loop 
suppression factor times ln$(\MPlanck/M_U)$. These corrections hopefully 
can be partly absorbed into a redefinition of $m_0^2$, $m_{1/2}$, $A_0$ 
and $B_0$ at $M_U$, but in many cases will lead to other important effects 
\cite{PP} that are difficult to anticipate. 

The framework described in the previous two paragraphs has been 
the subject of the bulk of phenomenological and experimental studies of supersymmetry,
and has become a benchmark scenario for experimental collider 
search limits. It 
is sometimes referred to as the {\it minimal supergravity} (MSUGRA) 
or {\it Constrained Minimal Supersymmetric Standard Model} (CMSSM) 
scenario for the soft terms. 

Particular models of gravity-mediated supersymmetry breaking can be even 
more predictive, relating some of the parameters $m_{1/2}$, $m_0^2$, $A_0$ 
and $B_0$ to each other and to the mass of the gravitino $m_{3/2}$. For 
example, three popular kinds of models for the soft terms are:

\vspace{.08in}

$\bullet$
Dilaton-dominated: \cite{dilatondominated}~~~$m^2_0 =
m^2_{3/2}$,~~~~$m_{1/2} = -A_0 = {\sqrt 3} m_{3/2}$.

\vspace{.02in}

$\bullet$
Polonyi: \cite{polonyi}
{}~~~$m^2_0 = m^2_{3/2}$,
{}~~~~$A_0 = (3 -{\sqrt 3}) m_{3/2}$,
{}~~~~$m_{1/2} = {\cal O}(m_{3/2})$.

\vspace{.08in}

$\bullet$ ``No-scale": \cite{noscale}~~~$m_{1/2} \gg
m_0, A_0, m_{3/2}$.

\vspace{.1in}

\noindent Dilaton domination arises in a particular limit of 
superstring theory. While it appears to be highly predictive, it can 
easily be generalized in other limits \cite{stringsoft}. The Polonyi model 
has the advantage of being the simplest possible model for supersymmetry 
breaking in the hidden sector, but it is rather {\it ad hoc} and does not 
seem to have a special place in grander schemes like superstrings. The 
``no-scale" limit may appear in a low-energy limit of superstrings in which 
the gravitino mass scale is undetermined at tree-level (hence the name). 
It implies that the gaugino masses dominate over other sources of 
supersymmetry breaking near $\MPlanck$. As we saw in section 
\ref{subsec:RGEs}, RG evolution feeds the gaugino masses into the squark, slepton, 
and Higgs squared-mass parameters with sufficient magnitude to give 
acceptable phenomenology at the electroweak scale. More recent versions of 
the no-scale scenario, however, also can give significant $A_0$ and 
$m_0^2$ at the input scale. In many cases $B_0$ can also be predicted in terms 
of the other parameters, but this is quite sensitive to model assumptions. 
For phenomenological studies, $m_{1/2}$, $m_0^2$, $A_0$ and $B_0$ are 
usually just taken to be convenient but imperfect (and perhaps 
downright misleading) parameterizations 
of our ignorance of the supersymmetry breaking mechanism.
In a more perfect world, experimental searches might be conducted and reported
using something like the larger 15-dimensional flavor-blind parameter 
space of eqs.~(\ref{scalarmassunification})-(\ref{commonphase}), 
but such a higher dimensional parameter space is difficult 
to simulate comprehensively,
for practical reasons.

Let us now review in a little more detail how the soft supersymmetry 
breaking terms can arise in supergravity models. The part of the scalar 
potential that does not depend on the gauge kinetic function can be found 
as follows. First, one may define the real, dimensionless {\em K\"ahler 
function} in terms of the K\"ahler potential and superpotential with the 
chiral superfields replaced by their scalar components:
\beq
G \>=\> 
{K/\MPlanck^2} + {\rm ln}({W/\MPlanck^3}) +{\rm ln}{(W^*/\MPlanck^3)}.
\label{eq:defKahlerfun}
\eeq
Many references use units with $\MPlanck=1$,
which simplifies the expressions but can slightly obscure the
correspondence with the global supersymmetry limit of large $\MPlanck$.
From $G$, one can construct its derivatives with respect to the scalar
fields and their complex conjugates: $G^i = {\delta G/\delta \phi_i}$; 
$G_i = {\delta G/\delta \phi^{* i}}$; and $G_i^j = {\delta^2
G/\delta\phi^{* i}\delta\phi_j}$. As in section 
\ref{subsec:susylagr.chiral}, raised (lowered) indices $i$
correspond to derivatives with respect to $\phi_i$ ($\phi^{*i}$). 
Note that $G_i^j =
K_i^j/\MPlanck^2$, which is often called the K\"ahler metric, does not
depend on the superpotential.  The inverse of this matrix is denoted
$(G^{-1})_i^j$, or equivalently $\MPlanck^2 (K^{-1})_i^j$, so that
$(G^{-1})^k_i G_k^j = (G^{-1})^j_k G_i^k = \delta_i^j$. In terms of these
objects, the generalization of the $F$-term contribution to the
scalar potential in ordinary renormalizable global supersymmetry turns out
\cite{supergravity,superconformalsupergravity} to be: 
\beq
V_F \>=\> \MPlanck^4 \, e^G \Bigl [ G^i (G^{-1})_i^j G_j -3 \Bigr ]
\label{vsugra}
\eeq
in supergravity. It can be rewritten as
\beq
V_F \>=\> K_i^j F_j F^{*i} - 3 e^{K/\MPlanck^2} WW^*/\MPlanck^2 ,
\label{compactvsugra}
\eeq
where
\beq
F_i \>=\> -\MPlanck^2\, e^{G/2} \, (G^{-1})_i^j G_j 
    \>=\> -e^{K/2 \MPlanck^2}\, (K^{-1})_i^j
\Bigl ( W^*_j + W^* K_j/\MPlanck^2 \Bigr ),
\label{fisugra}
\eeq
with $K^i = \delta K/\delta \phi_i$ and $K_j = \delta K/\delta \phi^{*j}$.
The $F_i$ are order parameters for supersymmetry breaking in supergravity
(generalizing the auxiliary fields in the renormalizable global
supersymmetry case). In other words, local supersymmetry will be broken if
one or more of the $F_i$ obtain a VEV. The gravitino then absorbs the
would-be goldstino and obtains a squared mass
\beq
m^2_{3/2} \,=\, \langle K_j^i F_i F^{*j}\rangle/3\MPlanck^2.
\label{sugragravitinomass}
\eeq

Taking a minimal K\"ahler potential $ K = \phi^{*i} \phi_i $, 
one has $K_i^j=(K^{-1})_i^j = \delta_i^j$, so that expanding 
eqs.~(\ref{compactvsugra}) and (\ref{fisugra}) to lowest order in 
$1/\MPlanck$ just reproduces the results $F_i = -W^*_i$ and $V = F_i 
F^{*i} = W^i W_i^*$, which were found in section 
\ref{subsec:susylagr.chiral} for renormalizable global supersymmetric 
theories [see eqs.~(\ref{replaceF})-(\ref{ordpot})]. 
Equation~(\ref{sugragravitinomass}) also reproduces the expression for the 
gravitino mass that was quoted in eq.~(\ref{gravitinomass}).

The scalar potential eq.~(\ref{vsugra}) does not include the $D$-term 
contributions from gauge interactions, which are given by
\beq
V_D \>=\> {1\over 2}{\rm Re}[f_{ab}\, {\widehat D}^a {\widehat 
D}^b],
\label{eq:defVD}
\eeq
with $\widehat D^a = f_{ab}^{-1} \widetilde D^b$, where 
\beq
{\widetilde D}^a \equiv 
-G^i (T^a)_i{}^j \phi_j = -\phi^{*j} (T^a)_j{}^i G_i =
-K^i (T^a)_i{}^j \phi_j = -\phi^{*j} (T^a)_j{}^i K_i, 
\eeq
are real order parameters of supersymmetry breaking, with the last three 
equalities following from the gauge invariance of $W$ and $K$. 
Note that in the tree-level global supersymmetry case $f_{ab} = 
\delta_{ab}/g_a^2$ and $K^i = \phi^{*i}$, eq.~(\ref{eq:defVD}) reproduces 
the result of section \ref{subsec:susylagr.gaugeinter} for the 
renormalizable global supersymmetry $D$-term scalar potential, with 
$\widehat{D}^a = g_a D^a$ (no sum on $a$).
The full scalar 
potential is
\beq
V = V_F + V_D,
\eeq
and it depends on $W$ and $K$ only through the combination $G$ in 
eq.~(\ref{eq:defKahlerfun}). There are many other contributions to the 
supergravity Lagrangian involving fermions and vectors, which 
can be found in 
ref.~\cite{supergravity,superconformalsupergravity}, and
also turn 
out to depend 
only on $f_{ab}$ and 
$G$. This allows one 
to consistently 
redefine $W$ and $K$ so that there are no purely holomorphic or purely 
anti-holomorphic terms appearing in the latter.

Unlike in the case of global supersymmetry, the scalar potential in 
supergravity is {\it not} necessarily non-negative, because of the $-3$ 
term in eq.~(\ref{vsugra}). Therefore, in principle, one can have 
supersymmetry breaking with a positive, negative, or zero vacuum energy. 
Results in experimental cosmology \cite{cosmokramer} imply 
a positive vacuum energy associated with the acceleration of the
scale factor of the observable universe,
\beq
\rho_{\rm vac}^{\rm observed} = \frac{\Lambda}{8\pi G_{\rm Newton}} \approx 
(2.3 \times 10^{-12}\>{\rm GeV})^4,
\eeq
but this is also certainly tiny compared to the scales associated with 
supersymmetry breaking. Therefore, it is tempting to simply assume that 
the vacuum energy is 0 within the approximations pertinent for working out 
the supergravity effects on particle physics at collider energies. However, it 
is notoriously unclear {\em why} the terms in the scalar potential in a 
supersymmetry-breaking vacuum should conspire to give $\langle V \rangle 
\approx 0$ at the minimum.  A naive estimate, without miraculous 
cancellations, would give instead $ \langle V \rangle$ of order $|\langle F 
\rangle|^2$, so at least roughly ($10^{10}$ GeV)$^4$ for Planck-scale 
mediated supersymmetry breaking, or ($10^4$ GeV)$^4$ for gauge-mediated 
supersymmetry breaking. Furthermore, while $\rho_{\rm vac} = \langle V 
\rangle$ classically, the former is a very large-distance scale measured 
quantity, while the latter is associated with effective field theories at 
length scales comparable to and shorter than those familiar to high energy 
physics. So, in the absence of a compelling explanation for the tiny value 
of $\rho_{\rm vac}$, it is not at all clear that $\langle V \rangle 
\approx 0$ is really the right condition to impose \cite{cosmock}. 
Nevertheless, with $\langle V \rangle = 0$ imposed as a 
constraint, 
eqs.~(\ref{compactvsugra})-(\ref{sugragravitinomass}) tell us that $ 
\langle K_j^i F_i F^{*j} \rangle = 3 \MPlanck^4 e^{\langle G \rangle} = 3 
e^{\langle K \rangle/\MPlanck^2} |\langle W \rangle|^2/\MPlanck^2$, and an 
equivalent formula for the gravitino mass is therefore $m_{3/2} = 
e^{\langle G\rangle/2} \MPlanck$.

An interesting special case arises if we assume a minimal K\"ahler
potential and divide the fields $\phi_i$ into a visible sector including
the MSSM fields $\varphi_i$, and a hidden sector containing a field $X$
that breaks supersymmetry for us (and other fields that we need not treat
explicitly). In other words, suppose that the superpotential and the
K\"ahler potential 
have the forms
\beq
W &=& W_{\rm vis}(\varphi_i) + W_{\rm hid}(X),
\label{minw}\\
K &=& \varphi^{*i} \varphi_i + X^* X .
\label{mink}
\eeq
Now let us further assume that the dynamics of the hidden sector fields
provides non-zero VEVs
\beq
\langle X \rangle = x \MPlanck,\qquad
\langle W_{\rm hid}\rangle = w \MPlanck^2,\qquad
\langle \delta W_{\rm hid}/\delta X \rangle = w^\prime \MPlanck ,
\eeq
which define a dimensionless quantity $x$, and $w$, $w^\prime$ with 
dimensions of [mass]. Requiring\footnote{We do this only 
to follow popular example; as just 
noted we cannot endorse this imposition.} $\langle V \rangle = 0$ yields $|w^\prime 
+ x^* w|^2 = 3 |w|^2$, and
\beq
m_{3/2} \>=\> {|\langle F_X \rangle |/\sqrt{3} \MPlanck} \>=\> e^{|x|^2/2}|w|.
\eeq
Now we suppose that it is valid to expand the scalar potential in powers
of the dimensionless quantities $w/\MPlanck$, $w^\prime/\MPlanck$,
$\varphi_i/\MPlanck$, etc., keeping only terms that depend on the visible
sector fields $\varphi_i$. In
leading order the result is: 
\beq
V &=& (W^*_{\rm vis})_i (W_{\rm vis})^i + m_{3/2}^2
\varphi^{*i}\varphi_{i}
\nonumber \\ && \!\!
+ e^{|x|^2/2} \left [w^* \varphi_i (W_{\rm vis})^i\, +\,
(x^* w^{\prime *} + |x|^2 w^* - 3 w^*) W_{\rm vis} + \conj \right
].\qquad{}
\label{yapot}
\eeq
A tricky point here is that we have rescaled the visible sector
superpotential $W_{\rm vis} \rightarrow e^{-|x|^2/2} W_{\rm vis}$
everywhere, in order that the first term in eq.~(\ref{yapot}) is the
usual, properly normalized, $F$-term contribution in global supersymmetry.
The next term is a universal soft scalar squared mass of the form
eq.~(\ref{scalarunificationsugra}) with
\beq
m_0^2 \>=\> {|\langle F_X \rangle|^2/ 3 \MPlanck^2}
\>=\> m_{3/2}^2 .
\eeq
The second line of eq.~(\ref{yapot}) just gives soft (scalar)$^3$ and 
(scalar)$^2$ holomorphic couplings of the form
eqs.~(\ref{aunificationsugra}) and (\ref{bsilly}), with
\beq
A_0 \,=\, - x^* {\langle F_X \rangle / \MPlanck},
\qquad\>\>
B_0 \,=\, \Bigl (
{1\over x + w^{\prime *}/w^*} -x^*\Bigr ){\langle F_X \rangle / \MPlanck}
\qquad{}
\label{a0b0x}
\eeq
since $\varphi_i (W_{\rm vis})^i$ is equal to $3 W_{\rm vis}$ for the
cubic part of $W_{\rm vis}$, and to $2 W_{\rm vis}$ for the quadratic
part. [If the complex phases of $x$, $w$, $w^\prime$ can be rotated away,
then eq.~(\ref{a0b0x}) implies $B_0 = A_0 - m_{3/2}$, but there are many
effects that can ruin this prediction.] The Polonyi model mentioned in
section \ref{subsec:origins.sugra} is just the special case of this
exercise in which $W_{\rm hid}$ is assumed to be linear in $X$. 

However, there is no  reason why $W$ and $K$ must have the 
simple form eq.~(\ref{minw}) and eq.~(\ref{mink}). In general, the 
superpotential and K\"ahler potential will have terms coupling $X$ to the MSSM 
fields as in eqs.~(\ref{eq:WnonrenormX}) and (\ref{eq:KnonrenormX}).
If one now plugs such terms into eq.~(\ref{vsugra}), one obtains a
general form like eq.~(\ref{hiddengrav}) for the soft terms. It is only
when special assumptions are made [like eqs.~(\ref{minw}), (\ref{mink})]
that one gets the phenomenologically desirable results in
eqs.~(\ref{sillyassumptions})-(\ref{bsilly}). Thus
supergravity by itself does not guarantee universality or 
even flavor-blindness of the soft
terms.

\subsection{Gauge-mediated supersymmetry breaking
models}\label{subsec:origins.gmsb}
\setcounter{equation}{0}
\setcounter{footnote}{1}

In gauge-mediated supersymmetry breaking (GMSB) models
\cite{oldgmsb,newgmsb}, the ordinary gauge interactions, rather than
gravity, are responsible for the appearance of soft supersymmetry breaking
in the MSSM.  The basic idea is to introduce some new chiral
supermultiplets, called messengers, that couple to the ultimate source of
supersymmetry breaking, and also couple indirectly to the (s)quarks and
(s)leptons and higgs(inos) of the MSSM through the ordinary $SU(3)_C\times
SU(2)_L\times U(1)_Y$ gauge boson and gaugino interactions.  There is
still gravitational communication between the MSSM and the source of
supersymmetry breaking, of course, but that effect is now relatively
unimportant compared to the gauge interaction effects. 

In contrast to Planck-scale mediation, GMSB can be understood entirely in 
terms of loop effects in a renormalizable framework. In the simplest 
such model, the messenger fields are a set of left-handed chiral 
supermultiplets $q$, $\overline q$, $\ell$, $\overline \ell$
transforming under $SU(3)_C\times SU(2)_L\times U(1)_Y$ as
\beq
q\sim({\bf 3},{\bf 1}, -{1\over 3}),\qquad\!\!\!
\overline q\sim({\bf \overline 3},{\bf 1}, {1\over 3}),\qquad\!\!\!
\ell \sim({\bf 1},{\bf 2}, {1\over 2}),\qquad\!\!\!
\overline \ell\sim({\bf 1},{\bf 2}, -{1\over 2}).\qquad\!\!\!{}
\label{minimalmess}
\eeq
These supermultiplets contain messenger quarks $\psi_q, \psi_{\overline
q}$ and scalar quarks $q, \overline q$ and messenger leptons $\psi_\ell,
\psi_{\overline \ell}$ and scalar leptons $\ell, \overline \ell$. All of
these particles must get very large masses so as not to have been
discovered already. Assume they do so by coupling to a gauge-singlet
chiral supermultiplet $S$ through a superpotential: 
\beq
W_{\rm mess} \, = \, y_2 S \ell \overline \ell + y_3 S q \overline q .
\eeq
The scalar component of $S$ and its auxiliary ($F$-term) component are 
each supposed to acquire VEVs, denoted $\langle S \rangle $ and $\langle 
F_S \rangle $ respectively. This can be accomplished either by putting $S$ 
into an O'Rai\-f\-ear\-taigh-type model \cite{oldgmsb}, or by a dynamical 
mechanism \cite{newgmsb}. Exactly how this happens is an interesting and 
important question, with many possible answers but no clear 
favorite at present.  Here, we will 
simply parameterize our ignorance of the precise mechanism of 
supersymmetry breaking by asserting that $S$ participates in another part 
of the superpotential, call it $W_{\rm breaking}$, which provides for the 
necessary spontaneous breaking of supersymmetry.

Let us now consider the mass spectrum of the messenger fermions and
bosons. The fermionic messenger fields pair up to get mass terms: 
\beq
\lagr &=& 
- y_2 \langle S \rangle \psi_\ell \psi_{\overline \ell}
- y_3 \langle S \rangle \psi_q \psi_{\overline q} + \conj 
\label{messfermass}
\eeq
as in eq.~(\ref{lagrchiral}). Meanwhile, their scalar messenger partners
$\ell,\overline\ell$ and $q,\overline q$ have a scalar potential given by
(neglecting $D$-term contributions, which do not affect the following
discussion): 
\beq
V &=& 
\left | {\delta \Wmess \over \delta \ell} \right |^2 +
\left | {\delta \Wmess \over \delta \overline\ell} \right |^2 +
\left | {\delta \Wmess \over \delta q} \right |^2 +
\left | {\delta \Wmess \over \delta \overline q} \right |^2 +
\left | {\delta \over \delta S}  (\Wmess + W_{\rm breaking}) \right |^2
\>\phantom{xxx}
\eeq
as in eq.~(\ref{ordpot}). Now, suppose that, at the minimum of the 
potential,
\beq
\langle S \rangle &\not=& 0,\\
\langle \delta W_{\rm breaking}/\delta S \rangle  
&=& -\langle F_S^* \rangle \>\not=\> 0,\\
\langle \delta \Wmess /\delta S  \rangle  &=& 0.
\eeq
Replacing $S$ and $F_S$ by their VEVs, one finds quadratic mass terms in
the potential for the messenger scalar leptons: 
\beq
V &=& 
|y_2 \langle S \rangle|^2 \bigl ( |\ell|^2 + 
|\overline \ell |^2 \bigr ) +
|y_3 \langle S \rangle|^2 \bigl ( |q|^2 + 
|\overline q|^2 \bigr )
\nonumber \\
&&-\left (y_2 \langle F_S \rangle \ell\overline \ell 
+ y_3 \langle F_S \rangle q\overline q + \conj \right )
\nonumber
\\&& +\> {\rm quartic}\> {\rm terms}.
\label{nosteenkinlabel}
\eeq
The first line in eq.~(\ref{nosteenkinlabel}) represents supersymmetric
mass terms that go along with eq.~(\ref{messfermass}), while the second
line consists of soft supersymmetry-breaking masses. The complex scalar
messengers $\ell,\overline\ell$ thus obtain a squared-mass matrix equal
to: 
\beq
\pmatrix{ |y_2 \langle S \rangle |^2 
& -y^*_2 \langle F^*_S \rangle \cr
-y_2 \langle F_S \rangle & \phantom{x}|y_2 \langle S \rangle |^2 }
\eeq
with squared mass eigenvalues $|y_2 \langle S\rangle |^2 \pm |y_2 \langle
F_S \rangle |$. In just the same way, the scalars $q,\overline q$ get
squared masses $|y_3 \langle S\rangle |^2 \pm |y_3 \langle F_S \rangle |$. 

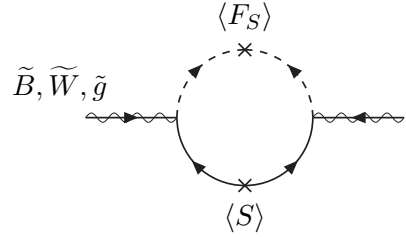
\begin{figure}
\begin{minipage}[]{0.55\linewidth}
\caption{Contributions to the MSSM gaugino masses in gauge-mediated
supersymmetry breaking models come from one-loop graphs involving
virtual messenger particles.\label{fig:1loop}}
\end{minipage}
\hspace{0.01\linewidth}
\begin{minipage}[]{0.4\linewidth} 
\begin{picture}(170,74)(-100,-44)
\SetScale{0.8}
\SetWidth{0.475}
\Photon(-75,0)(-32,0){2.2}{4}
\Photon(75,0)(32,0){2.2}{4}
\SetWidth{0.8}
\Line(-3,35)(3,29)
\Line(3,35)(-3,29)
\Line(-3,-35)(3,-29)
\Line(3,-35)(-3,-29)
\ArrowLine(-75,0)(-32,0)
\ArrowLine(75,0)(32,0)
\ArrowArc(0,0)(32,270,360)
\ArrowArcn(0,0)(32,270,180)
\DashArrowArc(0,0)(32,0,90){4}
\DashArrowArcn(0,0)(32,180,90){4}
\Text(-70,13)[c]{$\stilde B, \stilde W, \tilde g$}
\Text(0,38)[c]{$\langle F_S \rangle$}
\Text(0,-38)[c]{$\langle S \rangle$}
\end{picture}
\end{minipage}
\end{figure}

So far, we have found that the effect of supersymmetry breaking is to
split each messenger supermultiplet pair apart: 
\beq
\ell,\overline\ell : \qquad & m_{\rm fermions}^2 = |y_2 \langle S\rangle
|^2\, ,
\qquad & m_{\rm scalars}^2 = |y_2 \langle S\rangle |^2
\pm |y_2 \langle F_S \rangle | \, , \\
q,\overline q : \qquad & m_{\rm fermions}^2 = |y_3 \langle S\rangle
|^2\, ,
\qquad & m_{\rm scalars}^2 = |y_3 \langle S\rangle |^2
\pm |y_3 \langle F_S \rangle | \> .
\eeq 
The supersymmetry violation apparent in this messenger spectrum for 
$\langle F_S \rangle \not= 0$ is communicated to the MSSM sparticles 
through radiative corrections. The MSSM gauginos obtain masses 
from the 1-loop Feynman diagram shown in Figure~\ref{fig:1loop}.%
The scalar and fermion lines in the loop are messenger fields.  Recall that 
the interaction vertices in Figure~\ref{fig:1loop} are of gauge coupling 
strength even though they do not involve gauge bosons; compare 
Figure~\ref{fig:gauge}g. In this way, gauge-mediation provides that 
$q,\overline q$ messenger loops give masses to the gluino and the bino, 
and $\ell,\overline \ell$ messenger loops give masses to the wino and bino 
fields. Computing the 1-loop diagrams, one finds \cite{newgmsb} that the 
resulting MSSM gaugino masses are given by
\beq
M_a \,=\, {\alpha_a\over 4\pi} \Lambda , \qquad\>\>\>(a=1,2,3) ,
\label{gauginogmsb}
\eeq
in the normalization for $\alpha_a$ discussed in section
\ref{subsec:mssm.hints}, where we have introduced a mass parameter
\beq
\Lambda \,\equiv\, \langle F_S\rangle/\langle S \rangle \> .
\label{defLambda}
\eeq
(Note that if $\langle F_S\rangle$ were 0, then $\Lambda=0$ and the 
messenger scalars would be degenerate with their fermionic superpartners 
and there would be no contribution to the MSSM gaugino masses.) In 
contrast, the corresponding MSSM gauge bosons cannot get a corresponding 
mass shift, since they are protected by gauge invariance.  So 
supersymmetry breaking has been successfully communicated to the MSSM 
(``visible sector"). To a good approximation, eq.~(\ref{gauginogmsb}) 
holds for the running gaugino masses at an RG scale $Q_0$ corresponding to 
the average characteristic mass of the heavy messenger particles, roughly 
of order $M_{\rm mess} \sim y_I \langle S \rangle$ for $I = 2,3$. The 
running mass parameters can then be RG-evolved down to the electroweak 
scale to predict the physical masses to be measured by future experiments.

\begin{figure} 
\vspace{0.2cm}
\centerline{\psfig{figure=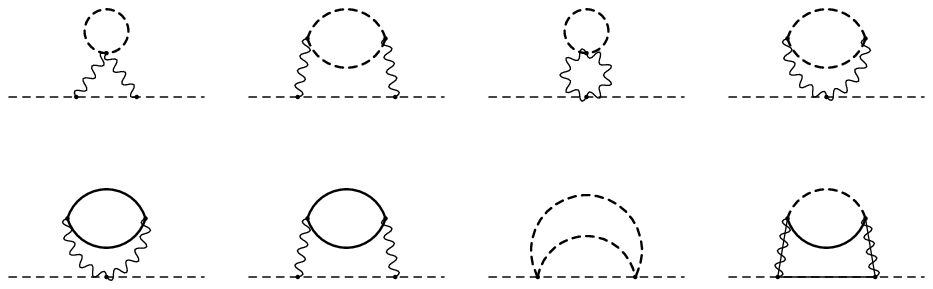,height=1.55in}}
\caption{MSSM scalar squared masses in gauge-mediated
supersymmetry breaking models arise in leading order from these two-loop
Feynman graphs. The heavy dashed lines are messenger scalars, the solid
lines are messenger fermions, the wavy lines are ordinary Standard Model
gauge bosons, and the solid lines with wavy lines superimposed are the
MSSM gauginos. \label{fig:2loops}} 
\end{figure}

The scalars of the MSSM do not get any radiative corrections to their
masses at one-loop order. The leading contribution to their masses comes
from the two-loop graphs shown in Figure~\ref{fig:2loops}, with the
messenger fermions (heavy solid lines) and messenger scalars (heavy dashed
lines) and ordinary gauge bosons and gauginos running around the loops.%
By computing these graphs,
one finds that each MSSM scalar $\phi_i$ gets a squared mass given by: 
\beq  
m^2_{\phi_i} \,=\,
2 {\Lambda^2}
\left [ \left ({\alpha_3\over 4\pi}\right )^2 C_3(i) +
\left ({\alpha_2\over
4 \pi}\right )^2 C_2(i) +
\left ({\alpha_1\over 4 \pi}\right )^2 C_1(i) \right ] ,
\label{scalargmsb}
\eeq
with the quadratic Casimir invariants $C_a(i)$ as in 
eqs.~(\ref{eq:defCasimir})-(\ref{defC1}). The squared masses in 
eq.~(\ref{scalargmsb}) are positive (fortunately!).

The terms $\bf a_u$, $\bf a_d$, $\bf a_e$ arise first at two-loop order, 
and are suppressed by an extra factor of $\alpha_a/4 \pi$ compared to the 
gaugino masses. So, to a very good approximation one has, at the messenger 
scale,
\beq
{\bf a_u} = {\bf a_d} = {\bf a_e} = 0,
\label{aaagmsb}
\eeq
a significantly stronger condition than eq.~(\ref{aunification}). Again, 
eqs.~(\ref{scalargmsb}) and (\ref{aaagmsb}) should be applied at an RG 
scale equal to the average mass of the messenger fields running in the 
loops. However, evolving the RG equations down to the electroweak scale 
generates non-zero $\bf a_u$, $\bf a_d$, and $\bf a_e$ proportional to the 
corresponding Yukawa matrices and the non-zero gaugino masses, as 
indicated in section \ref{subsec:RGEs}. These will only be large for the 
third-family squarks and sleptons, in the approximation of 
eq.~(\ref{heavytopapprox}). The parameter $b$ may also be taken to vanish 
near the messenger scale, but this is quite model-dependent, and in any 
case $b$ will be non-zero when it is RG-evolved to the electroweak scale. 
In practice, $b$ can be fixed in terms of the other parameters
by the requirement of correct electroweak 
symmetry breaking, as discussed below in section 
\ref{subsec:MSSMspectrum.Higgs}.

Because the gaugino masses arise at {\it one}-loop order and the scalar 
squared-mass contributions appear at {\it two}-loop order, both 
eq.~(\ref{gauginogmsb}) and (\ref{scalargmsb}) correspond to the estimate 
eq.~(\ref{mgravgmsb}) for $m_{\rm soft}$, with $M_{\rm mess} \sim y_I 
\langle S \rangle$. Equations (\ref{gauginogmsb}) and (\ref{scalargmsb}) 
hold in the limit of small $\langle F_S \rangle /y_I\langle S \rangle^2$, 
corresponding to mass splittings within each messenger supermultiplet that 
are small compared to the overall messenger mass scale. The sub-leading 
corrections in an expansion in $\langle F_S \rangle /y_I\langle S 
\rangle^2$ turn out \cite{gmsbcorrA}-\cite{gmsbcorrC} 
to be quite small unless there 
are very large messenger mass splittings.

The model we have described so far is often called the minimal model of 
gauge-mediated supersymmetry breaking. Let us now generalize it to a more 
complicated messenger sector. Suppose that $q, \overline q$ and $\ell, 
\overline \ell $ are replaced by a collection of messengers 
$\Phi_I,\overline \Phi_I$ with a superpotential
\beq
W_{\rm mess} \,=\, \sum_I y_I S \Phi_I \overline \Phi_I . 
\eeq
The bar is used to indicate that the left-handed chiral superfields 
$\overline \Phi_I$ transform as the complex conjugate representations of 
the left-handed chiral superfields $\Phi_I$. Together they are said to 
form a ``vectorlike" (self-conjugate) representation of the Standard Model gauge 
group. As before, the fermionic components of each pair $\Phi_I$ and 
$\overline\Phi_I$ pair up to get squared masses $|y_I \langle S 
\rangle|^2$ and their scalar partners mix to get squared masses $|y_I 
\langle S \rangle|^2 \pm |y_I \langle F_S \rangle | $. The MSSM gaugino 
mass parameters induced are now
\beq
M_a \,=\, {\alpha_a\over 4\pi} \Lambda \sum_I n_a(I) \qquad\>\>\>(a=1,2,3)
\label{gauginogmsbgen}
\eeq
where $n_a(I)$ is the Dynkin index for each $\Phi_I+\overline \Phi_I$, in 
a normalization where $n_3 = 1$ for a ${\bf 3} + {\bf \overline 3}$ of 
$SU(3)_C$ and $n_2 = 1$ for a pair of doublets of $SU(2)_L$. For $U(1)_Y$, 
one has $n_1 = 6Y^2/5$ for each messenger pair with weak hypercharges $\pm 
Y$.  In computing $n_1$ one must remember to add up the contributions for 
each component of an $SU(3)_C$ or $SU(2)_L$ multiplet. So, for example, 
$(n_1, n_2, n_3) = (2/5, 0, 1)$ for $q+\overline q$ and $(n_1, n_2, n_3) = 
(3/5, 1, 0)$ for $\ell+\overline \ell$. Thus the total is $\sum_I (n_1, 
n_2, n_3) = (1, 1, 1)$ for the minimal model, so that 
eq.~(\ref{gauginogmsbgen}) is in agreement with eq.~(\ref{gauginogmsb}). 
On general group-theoretic grounds, $n_2$ and $n_3$ must be integers, and 
$n_1$ is always an integer multiple of $1/5$ if fractional electric 
charges are confined.

The MSSM scalar masses in this generalized gauge mediation framework are
now:
\beq  
m^2_{\phi_i} \,=\,
2 \Lambda^2
\left [ \left ({\alpha_3\over 4\pi}\right )^2 C_3(i) \sum_I n_3(I) +
\left ({\alpha_2\over 4 \pi}\right )^2 C_2(i) \sum_I n_2(I)+
\left ({\alpha_1\over 4 \pi}\right )^2 C_1(i) \sum_I n_1(I)
\right ] .\phantom{xx}
\label{scalargmsbgen}
\eeq
In writing eqs.~(\ref{gauginogmsbgen}) and (\ref{scalargmsbgen}) as simple
sums, we have implicitly assumed that the messengers are all approximately
equal in mass, with
\beq
M_{\rm mess} \,\approx\, y_I \langle S \rangle .
\eeq
Equation (\ref{scalargmsbgen}) is still not a bad approximation if the 
$y_I$ are not very different from each other, because the dependence of 
the MSSM mass spectrum on the $y_I$ is only logarithmic (due to RG 
running) for fixed $\Lambda$. However, if large hierarchies in the 
messenger masses are present, then the additive contributions to the 
gaugino masses and scalar squared masses from each individual messenger 
multiplet $I$ should really instead be incorporated at the mass scale of 
that messenger multiplet. Then RG evolution is used to run these various 
contributions down to the electroweak or TeV scale; the individual 
messenger contributions to scalar and gaugino masses as indicated above 
can be thought of as threshold corrections to this RG running.

Messengers with masses far below the GUT scale will affect the running of 
gauge couplings and might therefore be expected to ruin the apparent 
unification shown in Figure~\ref{fig:gaugeunification}. However, if the 
messengers come in complete multiplets of the $SU(5)$ global 
symmetry\footnote{This $SU(5)$ may or may not be promoted to a local gauge 
symmetry at the GUT scale. For our present purposes, it is used only as 
a classification scheme, since the global $SU(5)$ symmetry is only 
approximate in the effective theory at the (much lower) messenger mass 
scale where gauge mediation takes place.} that contains the Standard Model 
gauge group, and are not very different in mass, then approximate 
unification of gauge couplings will still occur when they are extrapolated 
up to the same scale $M_U$ (but with a larger unified value for the gauge 
couplings at that scale). For this reason, a popular class of models is 
obtained by taking the messengers to consist of $\nmess$ copies of the 
${\bf 5}+{\bf \overline 5}$ of $SU(5)$, resulting in
\beq
\sum_I n_1(I) = \sum_I n_2(I) =\sum_I n_3(I) = \nmess\> .
\eeq  
Equations~(\ref{gauginogmsbgen}) and
(\ref{scalargmsbgen}) then reduce to 
\beq
&&M_a \,=\, {\alpha_a \over 4 \pi} \Lambda \nmess ,
\label{gmsbgauginonmess}\\
&&m^2_{\phi_i} \,=\, 2 \Lambda^2 \nmess
\sum_{a=1}^3 C_a(i) \left ({\alpha_a\over 4\pi}\right )^2 ,
\label{gmsbnmess}
\eeq
since now there are $\nmess$ copies of the minimal messenger sector 
particles running around the loops. For example, the minimal model in 
eq.~(\ref{minimalmess}) corresponds to $\nmess = 1$. A single copy of 
${\bf 10} + {\bf \overline{ 10}}$ of $SU(5)$ has Dynkin indices $\sum_I 
n_a(I) = 3$, and so can be substituted for 3 copies of ${\bf 5}+{\bf 
\overline 5}$. (Other combinations of messenger multiplets can also 
preserve the apparent unification of gauge couplings.) Note that the 
gaugino masses scale like $\nmess$, while the scalar masses scale like 
$\sqrt{\nmess}$. This means that sleptons and squarks will tend to be 
lighter relative to the gauginos for larger values of $\nmess$ in 
non-minimal models. However, if $\nmess$ is too large, then the running 
gauge couplings will diverge before they can unify at $M_U$. For messenger 
masses of order $10^6$ GeV or less, for example, one needs $\nmess\leq 4$.

There are many other possible generalizations of the basic gauge-mediation 
scenario as described above; see for example 
refs.~\cite{gmsbcorrB}-\cite{GGMb}. The common feature that makes all such models 
attractive is that the masses of the squarks and sleptons depend only on 
their gauge quantum numbers, leading automatically to the degeneracy of 
squark and slepton masses needed for suppression of flavor-changing 
effects. But the most distinctive phenomenological prediction of 
gauge-mediated models may be the fact that the gravitino is the LSP. This 
can have crucial consequences for both cosmology and collider physics, as 
we will discuss further in sections \ref{subsec:decays.gravitino} and 
\ref{sec:signals}.

\subsection{Extra-dimensional and anomaly-mediated
supersymmetry breaking}\label{subsec:origins.amsb}
\setcounter{equation}{0}
\setcounter{footnote}{1}

It is also possible to take the partitioning of the MSSM and supersymmetry 
breaking sectors shown in fig.~\ref{fig:structure} seriously as geography. 
This can be accomplished by assuming that there are extra spatial 
dimensions of 
the Kaluza-Klein or warped type \cite{warped}, so that a physical distance 
separates the visible and hidden\footnote{The name ``sequestered" is often 
used instead of ``hidden" in this context.} sectors.  This general idea 
opens up numerous possibilities, which are hard to classify in a detailed 
way. For example, string theory suggests six such extra dimensions, with 
a staggeringly huge number of possible solutions.

Many of the popular models used to explore this 
extra-dimensional mediated supersymmetry breaking (the acronym XMSB is 
tempting) use just one single hidden extra dimension with the MSSM chiral 
supermultiplets confined to one 4-dimensional spacetime brane and the 
supersymmetry-breaking sector confined to a parallel brane a distance 
$R_5$ away, separated by a 5-dimensional bulk, as in 
fig.~\ref{fig:branes}.
\begin{figure}
\begin{minipage}[]{0.51\linewidth}
\caption{The separation of the supersymmetry-breaking sector from the MSSM
sector could take place along a hidden spatial dimension, as in the simple
example shown here. The branes are 4-dimensional parallel spacetime
hypersurfaces in
a 5-dimensional spacetime.\label{fig:branes}}
\end{minipage}
\hspace{0.025\linewidth}
\begin{minipage}[]{0.435\linewidth}
\begin{picture}(135,115)(-50,0)
\SetScale{1.25}
\rText(43,66)[][]{``the bulk"}
\rText(49,116)[][]{$R_5 $}
\Line(5,20)(5,75)
\Line(15,40)(15,95)
\Line(5,20)(15,40)
\Line(5,75)(15,95)
\Line(65,20)(65,75)
\Line(75,40)(75,95)
\Line(65,20)(75,40)
\Line(65,75)(75,95)
\LongArrow(40,99)(74,99)
\LongArrow(40,99)(16.6,99)
\rText(-1,13)[][]{MSSM brane}
\rText(-1,-1)[][]{(we live here)}
\rText(89,13)[][]{Hidden brane}
\rText(84,-1)[][]{$\langle F \rangle \not= 0$}
\end{picture}
\end{minipage}
\end{figure}%
Using this as an illustration, the dangerous flavor-violating terms 
proportional to $y^{Xijk}$ and 
$k^i_j$ in eq.~(\ref{hiddengrav}) are suppressed by 
factors like $e^{-R_5 M_5}$, where $R_5$ is the size of the 5th dimension 
and $M_5$ is the 5-dimensional fundamental (Planck) scale, and it is 
assumed that the MSSM chiral supermultiplets are confined to their brane. 
Therefore, it should be enough to require that $R_5 M_5 \gg 1$, in other 
words that the size of the 5th dimension (or, more generally, the volume 
of the compactified space) is relatively large in units of the fundamental 
length scale. Thus the suppression of flavor-violating effects does not
require any fine-tuning or extreme hierarchies, because it is exponential.

One possibility is that the gauge supermultiplets of the MSSM propagate in 
the bulk, and so mediate supersymmetry breaking 
\cite{MirabelliPeskin}-\cite{deconstructedgauginomediation}. 
This mediation is direct for gauginos, with
\beq
M_a \sim \frac{\langle F \rangle}{M_5(R_5 M_5)} ,
\eeq
but is loop-suppressed for the soft terms involving
scalars. This implies that in the simplest 
version of the idea, often called ``gaugino mediation", soft supersymmetry 
breaking is dominated by the gaugino masses. The phenomenology is 
therefore quite similar to that of the ``no-scale" boundary conditions 
mentioned in section \ref{subsec:origins.sugra} in the context of PMSB 
models. Scalar squared masses and the scalar cubic couplings come from 
renormalization group running down to the electroweak scale.  It is useful 
to keep in mind that gaugino mass dominance is really the essential 
feature that defeats flavor violation, so it may well 
turn out to be more robust 
than any particular model that provides it.

It is also possible that the gauge supermultiplet fields are also confined 
to the MSSM brane, so that the transmission of supersymmetry breaking is 
due entirely to supergravity effects. This leads to
anomaly-mediated supersymmetry breaking (AMSB) \cite{AMSB}, so-named 
because the resulting MSSM soft terms can be understood in terms of the 
anomalous violation of a local superconformal invariance, an extension of 
scale invariance. In one formulation of supergravity 
\cite{superconformalsupergravity}, Newton's constant (or equivalently, the 
Planck mass scale) is set by the VEV of a scalar field $\phi$ that is part 
of a non-dynamical chiral supermultiplet (called the ``conformal 
compensator"). As a gauge fixing, this field obtains a VEV of $\langle 
\phi \rangle = 1$, spontaneously breaking the local superconformal 
invariance. Now, in the presence of spontaneous supersymmetry breaking 
$\langle F \rangle \not= 0$, for example on the hidden brane, the 
auxiliary field component also obtains a non-zero VEV, with
\beq
\langle F_\phi \rangle 
\,\sim\, 
\frac{\langle F \rangle}{\MPlanck} 
\,\sim\,
m_{3/2} .
\label{eq:AMSBgravitino}
\eeq
The non-dynamical conformal compensator field $\phi$ is taken to 
be dimensionless, so that $F_\phi$ has dimensions of [mass].
 
In the classical limit, there is still no supersymmetry breaking in the 
MSSM sector, due to the exponential suppression provided by the extra 
dimensions.\footnote{AMSB can also be realized without invoking extra 
dimensions. The suppression of flavor-violating MSSM soft terms can 
instead be achieved using a strongly-coupled conformal field theory near 
an infrared-stable fixed point \cite{AMSBinfourd}.} However, there is an 
anomalous violation of superconformal (scale) invariance manifested in the 
running of the couplings. This causes supersymmetry breaking to show up in 
the MSSM by virtue of the non-zero beta functions and anomalous dimensions 
of the MSSM brane couplings and
fields. The resulting soft terms are \cite{AMSB} (using 
$\mAMSB$ to denote its VEV from now on):
\beq
M_a &=& \mAMSB \beta_{g_a}/g_a ,
\label{eq:AMSBgauginos}
\\
(m^2)_j^i &=& \frac{1}{2} |\mAMSB|^2 \frac{d}{dt} \gamma_j^i
\>=\> \frac{1}{2} |\mAMSB|^2 \left [
\beta_{g_a} \frac{\partial}{\partial g_a} 
+ \beta_{y^{kmn}} \frac{\partial}{\partial y^{kmn}}
+ \beta_{y^*_{kmn}} \frac{\partial}{\partial y^*_{kmn}}  \right]  \gamma_j^i,
\label{eq:AMSBscalars}
\\
a^{ijk} &=& -\mAMSB \beta_{y^{ijk}},
\label{eq:AMSBscalar}
\eeq
where the anomalous dimensions $\gamma^i_j$ are normalized as in 
eqs.~(\ref{eq:gengamma}) and (\ref{eq:gammaHu})-(\ref{eq:gammae}). As in 
the GMSB scenario of the previous subsection, gaugino masses arise at 
one-loop order, but scalar squared masses arise at two-loop order. Also, 
these results are approximately flavor-blind for the first two families, 
because the non-trivial flavor structure derives only from the MSSM Yukawa 
couplings.

There are several unique features of the AMSB scenario. First, there is no 
need to specify at which renormalization scale 
eqs.~(\ref{eq:AMSBgauginos})-(\ref{eq:AMSBscalar}) should be applied as 
boundary conditions. This is because they hold at every renormalization 
scale, exactly, to all orders in perturbation theory. In other words, 
eqs.~(\ref{eq:AMSBgauginos})-(\ref{eq:AMSBscalar}) are not just boundary 
conditions for the renormalization group equations of the soft parameters, 
but solutions as well. (These AMSB renormalization group trajectories can 
also be found from this renormalization group invariance property alone 
\cite{AMSBtrajectories}, without reference to the supergravity 
derivation.) In fact, even if there are heavy supermultiplets in the 
theory that have to be decoupled, the boundary conditions hold both above 
and below the arbitrary decoupling scale. This remarkable insensitivity to 
ultraviolet physics in AMSB ensures the absence of flavor 
violation in the low-energy MSSM soft terms. Another interesting 
prediction is that the gravitino mass $m_{3/2}$ in these models is 
actually much larger than the scale $m_{\rm soft}$ of the MSSM soft terms, 
since the latter are loop-suppressed compared to 
eq.~(\ref{eq:AMSBgravitino}).

There is only one unknown parameter, $\mAMSB$, among the MSSM soft terms 
in AMSB. Unfortunately, this exemplary falsifiability is 
marred by the fact that it is already falsified. The dominant 
contributions to the first-family squark and slepton squared masses are:
\beq
m^2_{\tilde q} &= & \frac{|\mAMSB|^2}{(16 \pi^2)^2} 
\left (8 g_3^4  + \ldots \right ),
\\
m^2_{\tilde e_L} &= & -\frac{|\mAMSB|^2}{(16 \pi^2)^2} 
\left (\frac{3}{2} g_2^4 + \frac{99}{50} g_1^4 \right )
\\
m^2_{\tilde e_R} &= & 
-\frac{|\mAMSB|^2}{(16 \pi^2)^2} \frac{198}{25} g_1^4
\eeq
The squarks have large positive squared masses, but the sleptons have 
negative squared masses, so the AMSB model in its simplest form is not 
viable. These signs come directly from those of the beta functions of the 
strong and electroweak gauge interactions, as can be seen from the 
right side of eq.~(\ref{eq:AMSBscalars}).

The characteristic ultraviolet insensitivity to physics at high mass 
scales also makes it somewhat non-trivial to modify the theory to escape 
this tachyonic slepton problem by deviating from the AMSB trajectory. There 
can be large deviations from AMSB provided by supergravity 
\cite{isAMSBrobust}, but then in general the flavor-blindness is also 
forfeit. One way to modify AMSB is to introduce additional supermultiplets 
that contain supersymmetry-breaking mass splittings that are large 
compared to their average mass \cite{AMSBlightstates}. Another way is to 
combine AMSB with gaugino mediation \cite{AMSBhybrid}. Some other 
proposals can be found in \cite{otherAMSBattempts}. Finally, there is a 
perhaps less motivated approach in which a common 
parameter $m_0^2$ is added to all of the scalar squared masses at some 
scale, and chosen large enough to allow the sleptons to have positive 
squared masses above bounds from the CERN LEP $e^+e^-$ collider. 
This allows the phenomenology to be 
studied in a framework conveniently parameterized by just:
\beq
\mAMSB,\, m^2_0,\, \tan\beta,\, {\rm arg}(\mu),
\eeq 
with $|\mu|$ and $b$ determined by requiring correct electroweak symmetry 
breaking as described in the next section. (Some sources use $m_{3/2}$ or 
$M_{\rm aux}$ to denote $ \mAMSB$.) The MSSM gaugino masses at the leading 
non-trivial order are unaffected by the {\it ad hoc} addition of $m_0^2$:
\beq
M_1 &=& \frac{\mAMSB}{16 \pi^2} \frac{33}{5} g_1^2
\\
M_2 &=& \frac{\mAMSB}{16 \pi^2} g_2^2
\\
M_3 &=& -\frac{\mAMSB}{16 \pi^2} 3 g_3^2
\eeq
This implies that $|M_2| \ll |M_1| \ll |M_3|$, so the lightest neutralino 
is actually mostly wino, with a lightest chargino that is only of order 
200 MeV heavier, depending on the values of $\mu$ and $\tan\beta$.
The decay $\stilde C_1^\pm \rightarrow \stilde N_1 \pi^\pm$ 
produces a very soft pion, implying
unique and difficult signatures in colliders 
\cite{Chen:1996ap}-\cite{AMSBphenothree}. 

Another large general class of models breaks supersymmetry using the 
geometric or topological 
properties of the extra dimensions. In the Scherk-Schwarz 
mechanism \cite{ScherkSchwarz}, the symmetry is broken by 
assuming different boundary 
conditions for the fermion and boson fields on the compactified space. In 
supersymmetric models where the size of the extra dimension is 
parameterized by a modulus (a massless or nearly massless excitation) 
called a radion, the $F$-term component of the radion chiral 
supermultiplet can obtain a VEV, which becomes a source for supersymmetry 
breaking in the MSSM. These two ideas turn out to be often related. 
Some of the variety of models proposed along these lines can be found in 
\cite{SSandradion}. These mechanisms 
can also be combined with gaugino-mediation and AMSB. 
It seems likely that the possibilities are not yet fully explored.

\section{The mass spectrum of the MSSM}\label{sec:MSSMspectrum}
\renewcommand{\theequation}{\arabic{section}.\arabic{subsection}.\arabic{equation}}
\setcounter{equation}{0}
\setcounter{figure}{0}
\setcounter{table}{0}
\setcounter{footnote}{1}

\subsection{Electroweak symmetry breaking and the Higgs
bosons}\label{subsec:MSSMspectrum.Higgs}
\setcounter{equation}{0}
\setcounter{footnote}{1}

In the MSSM, the description of electroweak symmetry breaking is slightly 
complicated by the fact that there are two complex Higgs doublets $H_u = 
(H_u^+,\> H_u^0)$ and $H_d = (H_d^0,\> H_d^-)$ rather than just one in the 
ordinary Standard Model. The classical scalar potential for the Higgs 
scalar fields in the MSSM is given by
\beq
V\! &=&\!
(|\mu|^2 + m^2_{H_u}) (|H_u^0|^2 + |H_u^+|^2)
+ (|\mu|^2 + m^2_{H_d}) (|H_d^0|^2 + |H_d^-|^2)
\nonumber \\ &&
+\, [b\, (H_u^+ H_d^- - H_u^0 H_d^0) + \conj]
\nonumber \\ &&
+ {1\over 8} (g^2 + g^{\prime 2})
(|H_u^0|^2 + |H_u^+|^2 - |H_d^0|^2 - |H_d^-|^2 )^2
+ \half g^2 |H_u^+ H_d^{0*} + H_u^0 H_d^{-*}|^2 .
\phantom{xxx}
\label{bighiggsv}
\eeq
The terms proportional to $|\mu |^2$ come from $F$-terms [see 
eq.~(\ref{movie})].  The terms proportional to $g^2$ and $g^{\prime 2}$ 
are the $D$-term contributions, obtained from the general 
formula eq.~(\ref{fdpot}) after some rearranging. Finally, the terms 
proportional to $m_{H_u}^2$, $m_{H_d}^2$, and $b$ are just a 
rewriting of the last three terms of eq.~(\ref{MSSMsoft}). The full scalar 
potential of the theory also includes many terms involving the squark and 
slepton fields that we can ignore here, since they do not get VEVs because 
they have large positive squared masses.

We now have to demand that the minimum of this potential should break 
electroweak symmetry down to electromagnetism $SU(2)_L\times U(1)_Y 
\rightarrow U(1)_{\rm EM}$, in accord with observation. We can use the 
freedom to make gauge transformations to simplify this analysis. First, 
the freedom to make $SU(2)_L$ gauge transformations allows us to rotate 
away a possible VEV for one of the weak isospin components of one of the 
scalar fields, so without loss of generality we can take $H_u^+=0$ at the 
minimum of the potential. Then one can check that a minimum of the potential 
satisfying $\partial V/\partial H_u^+=0$ must also have $H_d^- = 0$. This 
is good, because it means that at the minimum of the potential 
electromagnetism is necessarily unbroken, since the charged components of 
the Higgs scalars cannot get VEVs. After setting $H_u^+=H_d^-=0$, we are 
left to consider the scalar potential
\beq
V \!&=&\!
(|\mu|^2 + m^2_{H_u}) |H_u^0|^2 + (|\mu|^2 + m^2_{H_d}) |H_d^0|^2
- (b\, H_u^0 H_d^0 + \conj)
\nonumber \\ && 
+ {1\over 8} (g^2 + g^{\prime 2}) ( |H_u^0|^2 - |H_d^0|^2 )^2 .
\label{littlehiggsv}
\eeq
The only term in this potential that depends on the phases of the fields 
is the $b$-term. Therefore, a redefinition of the phase of $H_u$ or $H_d$ 
can absorb any phase in $b$, so we can take $b$ to be real and positive. 
Then it is clear that a minimum of the potential $V$ requires that $H_u^0 
H_d^0$ is also real and positive, so $\langle H_u^0\rangle$ and $\langle 
H_d^0\rangle$ must have opposite phases. We can therefore use a $U(1)_Y$ 
gauge transformation to make them both be real and positive without loss 
of generality, since $H_u$ and $H_d$ have opposite weak hypercharges ($\pm 
1/2$). It follows that CP cannot be spontaneously broken by the Higgs 
scalar potential, since the VEVs and $b$ can be simultaneously chosen 
real, as a convention. This implies that the Higgs scalar mass eigenstates 
can be assigned well-defined eigenvalues of CP, at least at tree-level. 
(CP-violating phases in other couplings can induce loop-suppressed CP 
violation in the Higgs sector, but do not change the fact that $b$, 
$\langle H_u^0 \rangle$, 
and $\langle H_d \rangle$ can always be chosen real and positive.)

In order for the MSSM scalar potential to be viable, we must first make 
sure that the potential is bounded from below for arbitrarily large values 
of the scalar fields, so that $V$ will really have a minimum. (Recall from 
the discussion in sections \ref{subsec:susylagr.chiral} and 
\ref{subsec:susylagr.gaugeinter} that scalar potentials in purely 
supersymmetric theories are automatically non-negative and so clearly 
bounded from below. But, now that we have introduced supersymmetry 
breaking, we must be careful.) The scalar quartic interactions in $V$ will 
stabilize the potential for almost all arbitrarily large values of $H_u^0$ 
and $H_d^0$. However, for the special directions in field space $|H_u^0| = 
|H_d^0|$, the quartic contributions to $V$ [the second line in 
eq.~(\ref{littlehiggsv})] are identically zero. Such directions in field 
space are called $D$-flat directions, because along them the part of the 
scalar potential coming from $D$-terms vanishes. In order for the 
potential to be bounded from below, we need the quadratic part of the 
scalar potential to be positive along the $D$-flat directions. This 
requirement amounts to
\beq
2 b < 2 |\mu |^2 + m^2_{H_u} + m^2_{H_d}.
\label{eq:boundedfrombelow}
\eeq

Note that the $b$-term always favors electroweak symmetry breaking. 
Requiring that one linear combination of $H_u^0$ and $H_d^0$ has a 
negative squared mass near $H_u^0=H_d^0=0$ gives
\beq
b^2 > (|\mu|^2 + m^2_{H_u} )(|\mu|^2 + m^2_{H_d}).
\label{eq:destabilizeorigin}
\eeq
If this inequality is not satisfied, then $H_u^0 = H_d^0 = 0$ will be a 
stable minimum of the potential (or there will be no stable minimum at 
all), and electroweak symmetry breaking will not occur.

Interestingly, if $m_{H_u}^2 = m_{H_d}^2$ then the constraints 
eqs.~(\ref{eq:boundedfrombelow}) and (\ref{eq:destabilizeorigin}) cannot 
both be satisfied. In models derived from the MSUGRA or 
GMSB boundary conditions, $m_{H_u}^2 = m_{H_d}^2$ is supposed to 
hold at tree level at the input scale, but the $X_t$ contribution to the 
RG equation for $m_{H_u}^2$ [eq.~(\ref{mhurge})] naturally pushes it to 
negative or small values $m_{H_u}^2 < m_{H_d}^2$ at the electroweak scale. 
Unless this effect is significant, the parameter space in which the 
electroweak symmetry is broken would be quite small. So, in these models 
electroweak symmetry breaking is actually driven by quantum corrections; 
this mechanism is therefore known as {\it radiative electroweak symmetry 
breaking}. Note that although a negative value for $|\mu|^2 + m_{H_u}^2$ 
will help eq.~(\ref{eq:destabilizeorigin}) to be satisfied, it is not 
strictly necessary. Furthermore, even if $m_{H_u}^2<0$, there may be no 
electroweak symmetry breaking if $|\mu|$ is too large or if $b$ is too 
small. Still, the large negative contributions to $m_{H_u}^2$ from the RG 
equation are an important factor in ensuring that electroweak symmetry 
breaking can occur in models with simple boundary conditions for the soft 
terms. The realization that this works most naturally with a large 
top-quark Yukawa coupling provides additional motivation for these models 
\cite{rewsbone,rewsbtwo}.

Having established the conditions necessary for $H_u^0$ and $H_d^0$ to get 
non-zero VEVs, we can now require that they are compatible with the 
observed phenomenology of electroweak symmetry breaking, $SU(2)_L \times 
U(1)_Y \rightarrow U(1)_{\rm EM}$. Let us write
\beq
v_u = \langle H_u^0\rangle,
\qquad\qquad
v_d = \langle H_d^0\rangle.
\label{defvuvd}
\eeq
These VEVs are related to the known mass of the $Z^0$ boson and the 
electroweak gauge couplings:
\beq
v_u^2 + v_d^2 = v^2 = 2 m_Z^2/(g^2 + g^{\prime 2}) \approx (174\>{\rm
GeV})^2.
\label{vuvdcon}
\eeq
The ratio of the VEVs is traditionally written as
\beq
\tan\beta \equiv v_u/v_d.
\label{deftanbeta}
\eeq
The value of $\tan\beta$ is not fixed by present experiments, but it 
depends on the Lagrangian parameters of the MSSM in a calculable way. 
Since $v_u = v \sin\beta$ and $v_d = v \cos\beta$ were taken to be real 
and positive by convention, 
we have $0 < \beta < \pi/2$, a requirement that will be 
sharpened below. Now one can write down the conditions $\partial 
V/\partial H_u^0= \partial V/\partial H_d^0 = 0$ under which the potential 
eq.~(\ref{littlehiggsv}) will have a minimum satisfying 
eqs.~(\ref{vuvdcon}) and (\ref{deftanbeta}):
\beq
&&m_{H_u}^2 + |\mu |^2 -b \cot\beta - (m_Z^2/2) \cos (2\beta) 
\>=\> 0 ,
\label{mubsub2}
\\
&&m_{H_d}^2 + |\mu |^2 -b \tan\beta + (m_Z^2/2) \cos (2\beta) \>=\> 0.
\label{mubsub1}
\eeq
It is easy to check that these equations indeed satisfy the necessary 
conditions eqs.~(\ref{eq:boundedfrombelow}) and 
(\ref{eq:destabilizeorigin}). They allow us to eliminate two of the 
Lagrangian parameters $b$ and $|\mu|$ in favor of $\tan\beta$, but do not 
determine the phase of $\mu$. Taking $|\mu|^2$, $b$, $m_{H_u}^2$ and 
$m_{H_d}^2$ as input parameters, and $m_Z^2$ and $\tan\beta$ as output 
parameters obtained by solving these two equations, one obtains:
\beq
\sin (2\beta) &=& \frac{2 b}{m^2_{H_u} + m^2_{H_d} + 2|\mu|^2},
\label{eq:solvesintwobeta}
\\
m_Z^2 &=& \frac{|m^2_{H_d} - m^2_{H_u}|}{\sqrt{1 - \sin^2(2\beta)}}
- m^2_{H_u} - m^2_{H_d} -2|\mu|^2
.
\label{eq:solvemzsq}
\eeq
Note that $\sin (2\beta)$ is always positive. If $m^2_{H_u} < m^2_{H_d}$, 
as is usually assumed, then $\cos(2\beta)$ is negative; otherwise it is 
positive.

As an aside, eqs.~(\ref{eq:solvesintwobeta}) and 
(\ref{eq:solvemzsq}) highlight the ``$\mu$ problem" already mentioned in 
section \ref{subsec:mssm.superpotential}.  Without miraculous 
cancellations, all of the input parameters ought to be within an order of 
magnitude or two of $m^2_Z$. However, in the MSSM, $\mu$ is a 
supersymmetry-respecting parameter appearing in the superpotential, while 
$b$, $m_{H_u}^2$, $m_{H_d}^2$ are supersymmetry-breaking parameters. This 
has lead to a widespread belief that the MSSM must be extended at very 
high energies to include a mechanism that relates the effective value of 
$\mu$ to the supersymmetry-breaking mechanism in some way; see sections 
\ref{subsec:variations.NMSSM} and 
\ref{subsec:variations.munonrenorm} and
ref.~\cite{muproblemGMSB} for examples.

Even if the value of $\mu$ is set by soft supersymmetry breaking, the 
cancellation needed by eq.~(\ref{eq:solvemzsq}) is often remarkable when 
evaluated in specific model frameworks, after constraints from direct 
searches for the superpartners are taken into account. 
For example, expanding for large $\tan\beta$, eq.~(\ref{eq:solvemzsq}) 
becomes
\beq
m_Z^2 = -2 (m^2_{H_u} + |\mu|^2) 
+ \frac{2}{\tan^2\beta} (m^2_{H_d} - m^2_{H_u})
+ {\cal O}(1/\tan^4\beta).
\eeq
Typical viable solutions for the MSSM have $-m^2_{H_u}$ and $|\mu|^2$ each 
much larger than $m_Z^2$, so that significant cancellation is needed. In 
particular, large top squark squared masses, needed to avoid having the 
Higgs boson mass turn out too small [see eq.~(\ref{hradcorr}) below] 
compared to the observed value of 125 GeV, will feed into $m^2_{H_u}$. 
The cancellation needed in the minimal model may therefore be at the 
several per cent level, or worse. It is impossible to objectively characterize 
whether this should be considered worrisome, but it certainly causes
subjective worry as the LHC bounds on superpartners increase.

Equations~(\ref{mubsub2})-(\ref{eq:solvemzsq}) 
are based on the tree-level potential, and involve 
running renormalized Lagrangian parameters, which depend on the choice of 
renormalization scale. In practice, one must include radiative corrections 
at one-loop order, at least, in order to get numerically stable results. 
To do this, one can compute the loop corrections $\Delta V$ to the 
effective potential $V_{\rm eff}(v_u, v_d) = V + \Delta V$ as a function 
of the VEVs. The impact of this is that the equations governing the VEVs 
of the full effective potential are obtained by simply replacing
\beq
m^2_{H_u} \rightarrow m^2_{H_u} + \frac{1}{2 v_u} 
\frac{\partial (\Delta V)}{\partial v_u},\qquad\qquad
m^2_{H_d} \rightarrow m^2_{H_d} + \frac{1}{2 v_d} 
\frac{\partial (\Delta V)}{\partial v_d}
\label{eq:Vradcor}
\eeq
in eqs.~(\ref{mubsub2})-(\ref{eq:solvemzsq}), treating $v_u$ and $v_d$ 
as real variables in the differentiation. 
The result for $\Delta V$ has now been obtained through two-loop 
order in the MSSM \cite{twoloopEP,twoloopEPMSSM}. The most important 
corrections come from 
the one-loop diagrams involving the top squarks and top quark, and 
experience shows that the validity of the tree-level approximation and 
the convergence of perturbation theory are therefore improved by choosing 
a renormalization scale roughly of order the average of the top squark masses.

The Higgs scalar fields in the MSSM consist of two complex 
$SU(2)_L$-doublet, or eight real, scalar degrees of freedom. When the 
electroweak symmetry is broken, three of them are the would-be 
Nambu-Goldstone bosons $G^0$, $G^\pm$, which become the longitudinal modes 
of the $Z^0$ and $W^\pm$ massive vector bosons. The remaining five Higgs 
scalar mass eigenstates consist of two CP-even neutral scalars $h^0$ and 
$H^0$, one CP-odd neutral scalar $A^0$, and a charge $+1$ scalar $H^+$ and 
its conjugate charge $-1$ scalar $H^-$. (Here we define $G^- = G^{+*}$ and 
$H^- = H^{+*}$. Also, by convention, $h^0$ is lighter than $H^0$.) The 
gauge-eigenstate fields can be expressed in terms of the mass eigenstate 
fields as:
\renewcommand{\arraystretch}{1.4}
\beq
\pmatrix{H_u^0 \cr H_d^0} &=&
\pmatrix{v_u \cr v_d} 
+ {1\over \sqrt{2}} R_\alpha \pmatrix{h^0 \cr H^0}
+ {i\over \sqrt{2}} R_{\beta_0} \pmatrix{G^0 \cr A^0}
\eeq
\beq
\pmatrix{H_u^+ \cr H_d^{-*}} &=& R_{\beta_\pm}  \pmatrix{G^+ \cr H^+}
\eeq
where the orthogonal rotation matrices
\beq
&&R_{\alpha} = \pmatrix{\cos\alpha & \sin\alpha \cr
                      -\sin\alpha & \cos\alpha},
\\
&&R_{\beta_0} = \pmatrix{\sin\beta_0 & \cos\beta_0 \cr
                      -\cos\beta_0 & \sin\beta_0},
\qquad\quad
R_{\beta_\pm} = \pmatrix{\sin\beta_\pm & \cos\beta_\pm \cr
                      -\cos\beta_\pm & \sin\beta_\pm}
,
\phantom{xxxxx}
\eeq
are chosen so that the quadratic part of the potential has diagonal
squared-masses:
\beq
V &=& 
\half m_{h^0}^2 (h^{0})^2 + \half m_{H^0}^2 (H^{0})^2 
+ \half m_{G^0}^2 (G^{0})^2 + \half m_{A^0}^2 (A^{0})^2 
\nonumber \\ &&
+  m_{G^\pm}^2 |G^+|^2 +  m_{H^\pm}^2 |H^+|^2  + \ldots ,
\phantom{xxx}
\eeq
Then, provided that $v_u,v_d$ minimize the tree-level 
potential,\footnote{It is often more
useful to expand around VEVs $v_u, v_d$ 
that do not minimize the tree-level potential, for example to minimize the 
loop-corrected effective potential instead. In that case, $\beta$, 
$\beta_0$, and $\beta_\pm$ are all slightly different.} one finds that 
$\beta_0 = \beta_\pm = \beta$, and $m^2_{G^0} = m^2_{G^\pm}=0$, and
\beq
m_{A^0}^2 \!\!&=&\!\! 2 b/\sin (2\beta)
\,=\, 2|\mu|^2 + m^2_{H_u} + m^2_{H_d}
\\
m^2_{h^0, H^0} \!\!\!&=\!\!& \half
\Bigl (
m^2_{A^0} + m_Z^2 \mp 
\sqrt{(m_{A^0}^2 - m_Z^2)^2 + 4 m_Z^2 m_{A^0}^2 \sin^2 (2\beta)} 
\Bigr ),\>\>\>\>\>{}
\label{eq:m2hH}
\\
m^2_{H^\pm} \!\!&=&\!\! m^2_{A^0} + m_W^2 .
\label{eq:m2Hpm}
\eeq
The mixing angle $\alpha$ is determined, at tree-level, by
\beq
{\sin 2\alpha\over \sin 2 \beta} \,=\, 
-\left ( {m_{H^0}^2 + m_{h^0}^2 \over 
                        m_{H^0}^2 - m_{h^0}^2} \right ) ,
\qquad\quad
{\tan 2\alpha\over \tan 2 \beta} \,=\, 
\left ( {m_{A^0}^2 + m_{Z}^2 \over 
                        m_{A^0}^2 - m_{Z}^2} \right ) ,
\eeq
and is traditionally chosen to be negative; it follows that $-\pi/2
<\alpha < 0$ (provided $m_{A^0} > m_Z$). The Feynman rules for couplings
of the mass eigenstate Higgs scalars to the Standard Model quarks and
leptons and the electroweak vector bosons, as well as to the various
sparticles, have been worked out in detail in ref.~\cite{GunionHaber,HHG,Haber:1997dt}.

\begin{figure}
\centerline{\psfig{figure=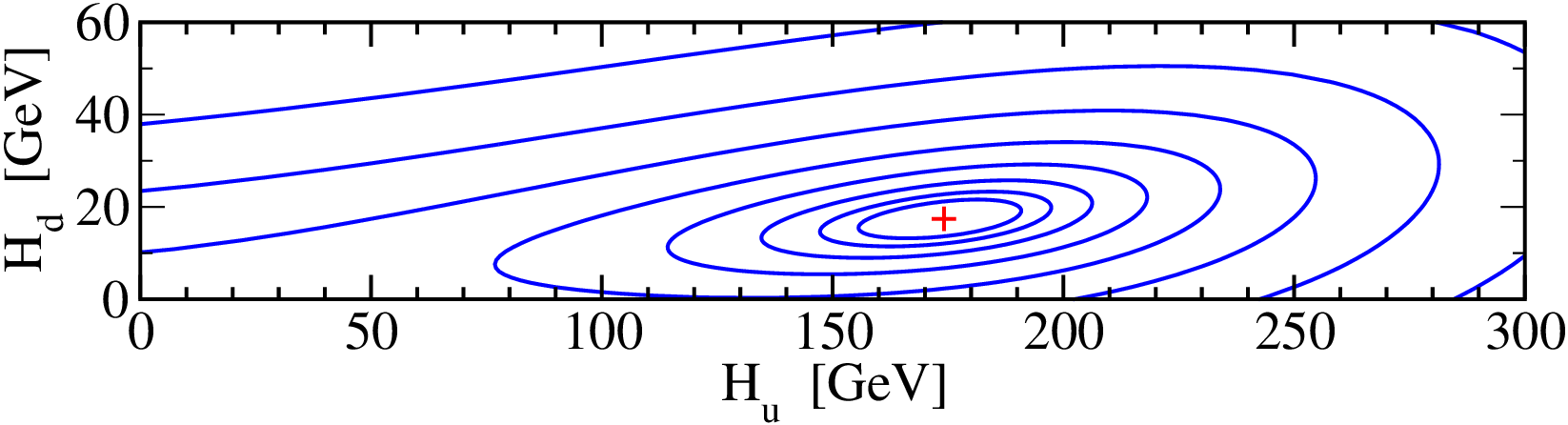,width=0.9\linewidth}}
\vspace{-0.32cm}
\caption{A contour map of the Higgs potential, for a typical case with
$\tan\beta \approx -\cot\alpha \approx 10$. The minimum of the potential
is marked by $+$, and the contours are equally spaced equipotentials.
Oscillations along the shallow direction, with $H^0_u/H_d^0 \approx 10$,
correspond to the mass eigenstate $h^0$, while the orthogonal steeper
direction corresponds to the mass eigenstate $H^0$.\label{fig:contourmap}}
\end{figure}

The masses of $A^0$, $H^0$ and $H^\pm$ can be arbitrarily
large, in principle, since they all grow with $b/\sin (2\beta)$. In contrast, the mass of
$h^0$ is bounded above. From eq.~(\ref{eq:m2hH}), one finds
at tree-level \cite{treelevelhiggsbound}:
\beq
m_{h^0} \><\> m_Z  |\cos (2\beta) | 
\label{eq:higgsineq}
\eeq
This corresponds to a shallow direction in the 
potential, along the 
direction $(H_u^0-v_u, H_d^0-v_d) \propto (\cos\alpha,-\sin\alpha)$. The 
existence of this shallow direction can be traced to the supersymmetric fact that the 
quartic Higgs couplings are small, being given by the squares of the electroweak gauge 
couplings, via the $D$-term. A contour map of the potential, for a typical 
case with $\tan\beta \approx -\cot\alpha \approx 10$, is shown in figure 
\ref{fig:contourmap}.%
If the tree-level inequality (\ref{eq:higgsineq}) were robust, the 
lightest Higgs boson of the MSSM would have been discovered in the previous century
at the CERN LEP2 $e^+e^-$ collider, and its mass obviously could not approach the 
observed value of 125 GeV.  
However, the tree-level formula for the squared mass of $h^0$ is subject 
to quantum corrections that are relatively drastic. The largest such 
contributions typically come from top and stop loops, as 
shown\footnote{In general, one-loop 1-particle-reducible 
tadpole diagrams should also be included.
However, they exactly cancel against the tree-level
tadpoles, and so both can be omitted,
if the VEVs $v_u$ and $v_d$ are taken at the minimum of the 
loop-corrected effective potential (see previous footnote).} 
in fig.~\ref{fig:MSSMhcorrections}.%
\begin{figure}[!t]
\begin{center}
\begin{picture}(338,38)(-38,-1)
\Text(-38,0)[c]{$\Delta(m_{h^0}^2)\> =\, $}
\SetWidth{0.7}
\DashLine(0,0)(25,0){5}
\DashLine(65,0)(90,0){5}
\SetWidth{1.3}
\CArc(45,0)(20,0,360)
\Text(3,8)[c]{$h^0$}
\Text(45,28)[c]{$t$}
\Text(105,0)[c]{$+$}
\SetWidth{0.7}
\DashLine(120,0)(145,0){4.5}
\DashLine(185,0)(210,0){4.5}
\SetWidth{1.3}
\DashCArc(165,0)(20,180,360){4}
\DashCArc(165,0)(20,0,180){4}
\Text(124,8)[c]{$h^0$}
\Text(165,28)[c]{$\tilde t$}
\SetWidth{0.7}
\Text(225,0)[c]{$+$}
\DashLine(240,-5)(275,-5){4.5}
\DashLine(310,-5)(275,-5){4.5}
\SetWidth{1.3}
\DashCArc(275,11)(16,-90,270){4}
\Text(243,3)[c]{$h^0$}
\Text(275,19)[c]{$\tilde t$}
\end{picture}
\end{center}
\caption{Contributions to the MSSM lightest Higgs squared mass from top-quark and 
top-squark one-loop diagrams. Incomplete cancellation, due to soft 
supersymmetry breaking, leads to a large positive correction to 
$m_{h^0}^2$ in the limit of heavy top squarks.\label{fig:MSSMhcorrections}}
\end{figure}

\begin{figure}
\begin{center}
\begin{picture}(435,30)(140,-6)
\SetWidth{0.7}
\DashLine(350,15)(380,15){5}
\DashLine(350,-15)(380,-15){5}
\DashLine(440,15)(410,15){5}
\DashLine(440,-15)(410,-15){5}
\SetWidth{1.3}
\DashLine(380,15)(380,-15){4}
\DashLine(410,15)(410,-15){4}
\DashLine(380,15)(410,15){4}
\DashLine(380,-15)(410,-15){4}
\Text(395,23)[c]{$\tilde t$}
\SetWidth{0.7}
\DashLine(235,15)(260,0){4}
\DashLine(235,-15)(260,0){4}
\DashLine(315,15)(285,15){5}
\DashLine(315,-15)(285,-15){5}
\SetWidth{1.3}
\DashLine(285,15)(285,-15){4}
\DashLine(260,0)(285,15){4}
\DashLine(260,0)(285,-15){4}
\Text(270,15.5)[c]{$\tilde t$}
\SetWidth{0.7}
\DashLine(122,15)(147,0){4}
\DashLine(122,-15)(147,0){4}
\DashLine(199,15)(179,0){4}
\DashLine(199,-15)(179,0){4}
\SetWidth{1.3}
\DashCArc(163,0)(16,0,180){4}
\DashCArc(163,0)(16,180,360){4}
\Text(165,24)[c]{$\tilde t$}
\SetWidth{0.7}
\DashLine(500,15)(530,15){5}
\DashLine(500,-15)(530,-15){5}
\DashLine(590,15)(560,15){5}
\DashLine(590,-15)(560,-15){5}
\SetWidth{1.3}
\Line(530,15)(530,-15)
\Line(560,15)(560,-15)
\Line(530,15)(560,15)
\Line(530,-15)(560,-15)
\Text(545,23)[c]{$t$}
%
%
\end{picture}
\end{center}
\caption{Contributions to the low-energy 
Standard Model effective Higgs quartic interaction. 
Integrating out the 
top squarks 
yields threshold contributions to the quartic Higgs coupling  
in the low-energy effective theory
from the first three one-loop diagrams. The last diagram, involving
the top quark, provides
renormalization group running of the low-energy effective
Higgs quartic coupling proportional to $y_t^4$.
\label{fig:MSSMVcorrections}}
\end{figure}
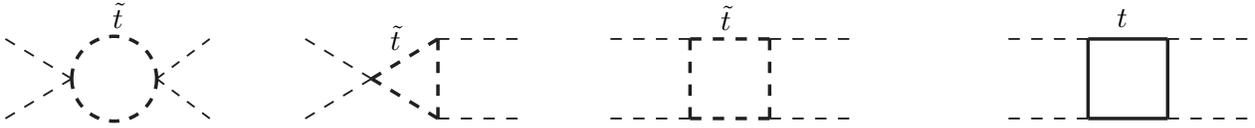

In the limit of top-squark
masses $m_{\stilde t_1}$, $m_{\stilde t_2}$ much greater than the top 
quark mass $m_t$, the largest radiative correction 
to $m_{h^0}^2$ in eq.~(\ref{eq:m2hH}) is:
\beq
\Delta (m^2_{h^0}) &=&
{3\over 4 \pi^2} \cos^2\!\alpha\>\, y_t^2 m_t^2
\left [
{\rm ln}(m_{\tilde t_1} m_{\tilde t_2} / m_t^2  )
+ \Delta_{\rm threshold} \right ], \phantom{xx}
\label{hradcorr}
\eeq
where 
\beq
\Delta_{\rm threshold} &=& 
c_{\tilde t}^2 s_{\tilde t}^2 [(m_{\tilde t_2}^2 - m_{\tilde t_1}^2)/m_t^2]
\, {\rm ln}(m_{\tilde t_2}^2/m_{\tilde t_1}^2) 
\nonumber
\\ 
&&
+ c_{\tilde t}^4 s_{\tilde t}^4 \left [
(m_{\tilde t_2}^2 - m_{\tilde t_1}^2)^2 - \half
(m_{\tilde t_2}^4 - m_{\tilde t_1}^4)
\, {\rm ln}(m_{\tilde t_2}^2/m_{\tilde t_1}^2) 
\right ]/m_t^4,
\label{eq:detDeltathreshold}
\eeq
with $c_{\tilde t}$ and $s_{\tilde t}$ equal to the cosine and sine of a top-squark mixing angle $\theta_{\tilde t}$, defined below following eq.~(\ref{pixies}).
One way to understand eq.~(\ref{hradcorr}) is by thinking in terms of the 
low energy effective Standard Model theory obtained by integrating out the top
squarks at a renormalization scale equal to the geometric mean of their masses.
Then 
$\Delta_{\rm threshold}$
comes from the finite threshold correction to the supersymmetric 
Higgs quartic coupling, via the first three
diagrams shown in fig.~\ref{fig:MSSMVcorrections}. 
The term with
${\rm ln}(m_{\tilde t_1} m_{\tilde t_2} / m_t^2)$
in eq.~(\ref{hradcorr}) then from comes the renormalization group running of
the Higgs quartic coupling (due to the last diagram in fig.~\ref{fig:MSSMVcorrections})
down to the top-quark mass scale, which turns out to be a good renormalization scale at
which to evaluate $m_{h^0}^2$ within the Standard Model effective theory. 
For small or moderate top-squark mixing, the logarithmic running term is largest, but 
$\Delta_{\rm threshold}$ can also be quite important. These corrections to the 
Higgs effective quartic coupling increase the steepness of the Higgs potential, thus 
raising $m_{h^0}$ compared to the naive tree-level prediction.

The term proportional to $c_{\tilde t}^2 s_{\tilde t}^2$ 
in eq.~(\ref{eq:detDeltathreshold}) is positive definite, 
while the term proportional to $c_{\tilde t}^4 s_{\tilde t}^4$ is negative definite. 
For fixed top-squark masses, the maximum possible $h^0$ mass therefore
occurs for rather large top-squark mixing, 
$c_{\tilde t}^2 s_{\tilde t}^2  = m_t^2/[m_{\tilde t_2}^2 + m_{\tilde t_1}^2
- 2 (m_{\tilde t_2}^2 - m_{\tilde t_1}^2)/{\rm ln}(m_{\tilde t_2}^2/m_{\tilde t_1}^2)
]$ or 1/4, whichever is less. This is often referred to as the ``maximal mixing" 
scenario for the MSSM Higgs sector. (What is being maximized 
is not the mixing, but $m_{h^0}$ with respect to the top-squark mixing.) 
It follows that the quantity in 
square brackets in eq.~(\ref{hradcorr}) is always less than 
$\ln(m_{\tilde t_2}^2/m_t^2) + 3$.

Equation (\ref{hradcorr}) already shows that $m_{h^0}$ 
can easily exceed the $Z$ boson mass, and the 
observed value of $m_{h^0} = 125$ GeV can in principle be accommodated. However, 
the above is a highly simplified account; to get reasonably accurate predictions for the Higgs scalar masses and mixings
for a given set of model parameters, one must also include
the remaining one-loop corrections and even the dominant two-loop and three-loop
effects
\cite{hcorrections}-\cite{Vega:2015fna}.
The theoretical 
uncertainties associated with the prediction of $m_{h^0}$, given all of the 
soft supersymmetry breaking parameters, are still quite large, especially
when the top-squarks are heavy, and are of order several GeV.
For a recent review, see \cite{DraperRzehak}.

Including such corrections, it had been estimated long
before the discovery of the 125 GeV Higgs boson that 
\beq
m_{h^0} \lsim 135\>{\rm GeV}
\label{mssmhiggsbound}
\eeq
in the MSSM. This prediction assumed that all of the sparticles that can contribute 
to $m_{h^0}^2$ in loops have masses that do not exceed 1 TeV, 
and the bound increases logarithmically with the top-squark masses. 
However, in many specific model
frameworks with small or moderate top-squark mixing,
the bound eq.~(\ref{mssmhiggsbound}) is very far from saturated, and 
it turns out to be a severe challenge to accommodate values even as large as the observed 
$m_{h^0} = 125$ GeV, unless the top squarks are extremely heavy, or else highly mixed. 
Unfortunately, it is difficult to make this statement very 
precise, due both to the high dimensionality of the 
supersymmetric parameter space and the theoretical errors in the $m_{h^0}$ prediction.

In the MSSM, the masses and CKM mixing angles of the quarks and leptons 
are determined not only by the Yukawa couplings of the superpotential but 
also the parameter $\tan\beta$. This is because the top, charm and up 
quark mass matrix is proportional to $v_u = v \sin\beta$ and the bottom, 
strange, and down quarks and the charge leptons get masses proportional to 
$v_d = v \cos\beta$. At tree-level,
\beq
m_t \,=\, y_t v \sin\beta
,
\qquad\quad
m_b \,=\, y_b v \cos\beta
,
\qquad\quad
m_\tau = y_\tau v \cos\beta .\phantom{xxx}
\label{eq:ytbtaumtbtau}
\eeq
These relations hold for the running masses rather than the physical pole 
masses, which are significantly larger for $t,b$ \cite{polecat}. Including 
those corrections, one can relate the Yukawa couplings to $\tan\beta$ and 
the known fermion masses and CKM mixing angles. It is now clear why we 
have not neglected $y_b$ and $y_\tau$, even though $m_b,m_\tau\ll m_t$. To 
a first approximation, $y_b/y_t = (m_b/m_t)\tan\beta$ and $y_\tau/y_t = 
(m_\tau/m_t)\tan\beta$, so that $y_b$ and $y_\tau$ cannot be neglected if 
$\tan\beta$ is much larger than 1. In fact, there are good theoretical 
motivations for considering models with large $\tan\beta$. For example, 
models based on the GUT gauge group $SO(10)$ 
can unify the running top, bottom and tau Yukawa couplings at the 
unification scale; this requires $\tan\beta$ to be very roughly of order 
$m_t/m_b$ \cite{so10,copw}.

Note that if one tries to make $\sin\beta$ too small, then $y_t$ will be 
nonperturbatively large. Requiring that $y_t$ does not blow up above the 
electroweak scale, one finds that $\tan\beta \gsim 1.2$ or so, depending 
on the mass of the top quark, the QCD coupling, and other details. In 
principle, there is also a constraint on $\cos\beta$ if one requires 
that $y_b$ and $y_\tau$ do not become nonperturbatively large. This 
gives a rough upper bound of $\tan\beta \lsim$ 65. However, this is 
complicated somewhat by the fact that the bottom quark mass gets 
significant one-loop non-QCD corrections in the large $\tan\beta$ limit 
\cite{copw}. One can obtain a stronger upper bound on $\tan\beta$ in some 
models where $m_{H_u}^2 = m_{H_d}^2$ at the input scale, by requiring that 
$y_b$ does not significantly exceed $y_t$. [Otherwise, $X_b$ would be 
larger than $X_t$ in eqs.~(\ref{mhurge}) and (\ref{mhdrge}), so one would 
expect $m_{H_d}^2 < m_{H_u}^2$ at the electroweak scale, and the minimum 
of the potential would have $\langle H_d^0 \rangle > \langle H_u^0 
\rangle$. This would be a contradiction with the supposition that 
$\tan\beta$ is large.] The parameter $\tan\beta$ also 
directly impacts the masses and mixings of the MSSM sparticles,
as we will see below.

It is interesting to write the dependences on the angles $\beta$ and $\alpha$ of
the tree-level couplings of the neutral MSSM Higgs bosons. The 
bosonic couplings are proportional to:
\beq
h^0 W^+W^-,\quad h^0 ZZ,\quad Z H^0 A^0,\quad W^\pm H^0 H^\mp &\propto& \sin(\beta-\alpha)
,
\\
H^0 W^+W^-,\quad H^0 ZZ,\quad
Z h^0 A^0,\quad W^\pm h^0 H^\mp\> &\propto& \cos(\beta-\alpha),
\eeq
and the couplings to fermions are proportional to
\beq
h^0b\bar b,\quad h^0\tau^+\tau^-\> &\propto&-\frac{\sin\alpha}{\cos\beta}\>=\>
\sin(\beta-\alpha) - \tan\beta \cos(\beta-\alpha),\phantom{xxxx}
\\
h^0t\bar t\> &\propto& \frac{\cos\alpha}{\sin\beta}\>=\>
\sin(\beta-\alpha) + \cot\beta \cos(\beta-\alpha),
\\
H^0b\bar b,\quad H^0\tau^+\tau^-\> &\propto&\frac{\cos\alpha}{\cos\beta}\>=\>
\cos(\beta-\alpha) + \tan\beta \sin(\beta-\alpha),
\\
H^0t\bar t\> &\propto& \frac{\sin\alpha}{\sin\beta}\>=\>
\cos(\beta-\alpha) - \cot\beta \sin(\beta-\alpha),
\\
A^0b\bar b,\quad A^0\tau^+\tau^-\> &\propto& \tan\beta,
\\
A^0t\bar t\> &\propto& \cot\beta.
\eeq
An important case, often referred to as the ``decoupling limit", occurs 
when $m_{A^0} \gg m_Z$. Then the tree-level prediction for 
$m_{h^0}$ saturates its upper bound 
mentioned above, with $m^2_{h^0} \approx m_Z^2 \cos^2 (2\beta) + $ loop 
corrections. The particles $A^0$, $H^0$, and $H^\pm$ will be much heavier 
and nearly degenerate, forming an isospin doublet that decouples from 
sufficiently low-energy processes. The angle $\alpha$ is very nearly 
$\beta-\pi/2$, with
\beq
\cos(\beta-\alpha) &=& 
\sin(2\beta) \cos(2\beta)\, m_Z^2/m_{A^0}^2+
{\cal O}(m_Z^4/m_{A^0}^4),
\\
\sin(\beta-\alpha) &=& 1 - {\cal O}(m_Z^4/m_{A^0}^4),
\eeq
so that 
$h^0$ has nearly the same couplings to quarks and leptons and 
electroweak gauge bosons as would the Higgs boson of the ordinary 
Standard Model without supersymmetry. Radiative corrections modify these tree-level
predictions, but model-building experiences 
have shown that it is not uncommon for $h^0$ to behave in a way nearly 
indistinguishable from a Standard Model-like Higgs boson, even if 
$m_{A^0}$ is not too huge. The measurements of the 125 GeV Higgs boson observed  
at the LHC are indeed consistent, so far, with the Standard Model predictions,
and it is sensible to identify this particle with $h^0$.
However, it should be kept in mind that the 
couplings of $h^0$ might still turn out to deviate in measurable ways from those 
of a Standard Model Higgs boson. After including the effects of radiative corrections,
the most significant effect for moderately 
large $m_{A^0}$ is a possible enhancement of the $h^0 b \overline b$ 
coupling compared to the value it would have in the Standard Model.

\subsection{Neutralinos and charginos}\label{subsec:MSSMspectrum.inos}
\setcounter{equation}{0}
\setcounter{footnote}{1}

The higgsinos and electroweak gauginos mix with each other because of the 
effects of electroweak symmetry breaking. The neutral higgsinos ($\stilde 
H_u^0$ and $\stilde H_d^0$) and the neutral gauginos ($\stilde B$, 
$\stilde W^0$) combine to form four mass eigenstates called {\it 
neutralinos}. The charged higgsinos ($\stilde H_u^+$ and $\stilde H_d^-$) 
and winos ($\stilde W^+$ and $\stilde W^-$) mix to form two mass 
eigenstates with charge $\pm 1$ called {\it charginos}. We will 
denote\footnote{Other common notations use $\stilde \chi_i^0$ or $\stilde 
Z_i$ for neutralinos, and $\stilde \chi^\pm_i$ or $\stilde W^\pm_i$ for 
charginos.} the neutralino and chargino mass eigenstates by $\stilde N_i$ 
($i=1,2,3,4$) and $\stilde C^\pm_i$ ($i=1,2$). By convention, these are 
labeled in ascending order, so that $m_{\stilde N_1} < m_{\stilde N_2} 
<m_{\stilde N_3} <m_{\stilde N_4}$ and $m_{\stilde C_1} < m_{\stilde 
C_2}$. The lightest neutralino, $\stilde N_1$, is usually assumed to be 
the LSP, unless there is a lighter gravitino or unless $R$-parity is not 
conserved, because it is the only MSSM particle that can make a good dark 
matter candidate. In this subsection, we will describe the mass spectrum 
and mixing of the neutralinos and charginos in the MSSM.

In the gauge-eigenstate basis $\psi^0 = (\stilde B, \stilde W^0, \stilde 
H_d^0, \stilde H_u^0)$, the neutralino mass part of the Lagrangian is
\beq
\lagr_{\mbox{neutralino mass}} = -\half (\psi^{0})^T {\bf M}_{\stilde N} 
\psi^0 + \conj ,
\eeq
where
\beq
{\bf M}_{\stilde N} \,=\, \pmatrix{
  M_1 & 0 & -g' v_d/\sqrt{2} & g' v_u/\sqrt{2} \cr
  0 & M_2 & g v_d/\sqrt{2} & -g v_u/\sqrt{2} \cr
  -g' v_d/\sqrt{2} & g v_d/\sqrt{2} & 0 & -\mu \cr
  g' v_u/\sqrt{2} & -g v_u/\sqrt{2}& -\mu & 0 \cr }.
\label{preneutralinomassmatrix}
\eeq
The entries $M_1$ and $M_2$ in this matrix come directly from the MSSM 
soft Lagrangian [see eq.~(\ref{MSSMsoft})], while the entries $-\mu$ are 
the supersymmetric higgsino mass terms [see eq.~(\ref{poody})]. The terms 
proportional to $g, g'$ are the result of Higgs-higgsino-gaugino couplings 
[see eq.~(\ref{gensusylagr}) and Figure~\ref{fig:gauge}g,h], with the 
Higgs scalars replaced by their VEVs [eqs.~(\ref{vuvdcon}), 
(\ref{deftanbeta})]. This can also be written as
\beq
{\bf M}_{\stilde N} \,=\, \pmatrix{
  M_1 & 0 & - \cbeta\, \sW\, m_Z & \sbeta\, \sW \, m_Z\cr
  0 & M_2 & \cbeta\, \cW\, m_Z & - \sbeta\, \cW\, m_Z \cr
  -\cbeta \,\sW\, m_Z & \cbeta\, \cW\, m_Z & 0 & -\mu \cr
  \sbeta\, \sW\, m_Z & - \sbeta\, \cW \, m_Z& -\mu & 0 \cr 
}.
\label{neutralinomassmatrix}
\eeq
Here we have introduced abbreviations $\sbeta = \sin\beta$, $\cbeta =
\cos\beta$, $\sW = \sin\theta_W$, and $\cW = \cos\theta_W$. The mass
matrix ${\bf M}_{\stilde N}$ can be diagonalized by a unitary matrix 
${\bf N}$ to obtain mass eigenstates:
\beq
\stilde N_i = {\bf N}_{ij} \psi^0_j, \phantom{xx}
\eeq
so that
\beq
{\bf N}^* {\bf M}_{\stilde N} {\bf N}^{-1}
\, =\,  
\pmatrix{
m_{\stilde N_1} & 0 & 0 & 0\cr
0 & m_{\stilde N_2} & 0 & 0 \cr
0 & 0 & m_{\stilde N_3} & 0\cr
0 & 0 & 0 & m_{\stilde N_4}
}\phantom{xxx}
\label{diagmN}
\eeq
has real positive entries on the diagonal. These are the magnitudes of the 
eigenvalues of ${\bf M}_{\stilde N}$, or equivalently the square roots of 
the eigenvalues of ${\bf M}^\dagger_{\stilde N}{\bf M}_{\stilde N}$. The 
indices $(i,j)$ on ${\bf N}_{ij}$ are (mass, gauge) eigenstate labels. The 
mass eigenvalues and the mixing matrix ${\bf N}_{ij}$ can be given in 
closed form in terms of the parameters $M_1$, $M_2$, $\mu$ and 
$\tan\beta$, by solving quartic equations, but the results are very 
complicated and not illuminating.

In general, the parameters $M_1$, $M_2$, and $\mu$ in the equations above 
can have arbitrary complex phases. A redefinition of the phases of 
$\stilde B$ and $\stilde W$ always allows us to choose a convention in 
which $M_1$ and $M_2$ are both real and positive. The phase of $\mu$ 
within that convention is then really a physical parameter and cannot be 
rotated away. [We have already used up the freedom to redefine the phases 
of the Higgs fields, since we have picked $b$ and $\langle H_u^0\rangle$ 
and $\langle H_d^0 \rangle$ to be real and positive, to guarantee that the 
off-diagonal entries in eq.~(\ref{neutralinomassmatrix}) proportional to 
$m_Z$ are real.] However, if $\mu$ is not real, then there can be 
potentially disastrous CP-violating effects in low-energy physics, 
including electric dipole moments for both the electron and the neutron. 
Therefore, it is usual [although not strictly mandatory, because of the 
possibility of nontrivial cancellations involving the phases of the 
(scalar)$^3$ couplings and the gluino mass] to assume that $\mu$ is real 
in the same set of phase conventions that make $M_1$, $M_2$, $b$, $\langle 
H_u^0\rangle$ and $\langle H_d^0 \rangle$ real and positive. The sign of 
$\mu$ is still undetermined by this constraint.

In models that satisfy eq.~(\ref{gauginomassunification}), one has the
nice prediction
\beq
M_1 \approx {5\over 3}\tan^2\theta_W \, M_2 \approx 0.5 M_2
\label{usualm1m2}
\eeq
at the electroweak scale. If so, then the neutralino masses and mixing 
angles depend on only three unknown parameters. This assumption is 
sufficiently theoretically compelling that it has been made in most 
phenomenological studies; nevertheless it should be recognized as an 
assumption, to be tested someday by experiment.

There is a not-unlikely limit in which electroweak symmetry breaking 
effects can be viewed as a small perturbation on the neutralino mass 
matrix. If
\beq
m_Z \,\ll\, |\mu \pm M_{1}|,\, |\mu \pm M_{2}| ,
\label{gauginolike}
\eeq
then the neutralino mass eigenstates are very nearly a ``bino-like" 
$\stilde N_1 \approx \stilde B$; a ``wino-like" $\stilde N_2 \approx 
\stilde W^0$; and ``higgsino-like" $\stilde N_3, \stilde N_4 \approx 
(\stilde H_u^0 \pm \stilde H_d^0)/\sqrt{2}$, with mass eigenvalues:
\beq
m_{{\stilde N}_1}\!\! &=&\!\! M_1 -
{ m_Z^2 s^2_W (M_1 + \mu \sin 2 \beta ) \over \mu^2 - M_1^2 }
+\ldots
\\
m_{{\stilde N}_2}\!\! &=&\!\! M_2 -
{ m_W^2 (M_2 + \mu \sin 2 \beta ) \over \mu^2 - M_2^2 }
+\ldots \qquad {}\\
m_{{\stilde N}_3}, m_{{\stilde N}_4}\!\! &=&\!\! |\mu|  +
{m_Z^2  (I - \sin 2 \beta) (\mu + M_1 c^2_W +M_2 s^2_W)
\over 2 (\mu + M_1) (\mu + M_2) }
+\ldots, \\
&&\!\!\! |\mu|  +
{m_Z^2  (I + \sin 2 \beta) (\mu - M_1 c^2_W - M_2 s^2_W)
\over 2 (\mu - M_1) (\mu - M_2) }
+\ldots \qquad {}
\eeq
where we have taken $M_1$ and $M_2$ real and positive by convention, and 
assumed $\mu$ is real with sign $I = \pm 1$. The subscript labels 
of the mass eigenstates may need to be rearranged depending on the 
numerical values of the parameters; in particular the above labeling of 
$\stilde N_1$ and $\stilde N_2$ assumes $M_1< M_2 \ll |\mu|$. This limit, 
leading to a bino-like neutralino LSP, often emerges from MSUGRA
boundary conditions on the soft 
parameters, which tend to 
require it in order to get correct electroweak symmetry breaking.

The chargino spectrum can be analyzed in a similar way. In the 
gauge-eigenstate basis $\psi^\pm = (\stilde W^+,\, \stilde H_u^+,\, 
\stilde W^- ,\, \stilde H_d^- )$, the chargino mass terms in the 
Lagrangian are
\beq
\lagr_{\mbox{chargino mass}} = -\half (\psi^\pm)^T {\bf M}_{\stilde C} 
\psi^\pm
+\conj
\eeq
where, in $2\times 2$ block form,
\beq
{\bf M}_{\stilde C}
&=& \pmatrix{{\bf 0}&{\bf X}^T\cr
             {\bf X} &{\bf 0}} ,
\phantom{xx}
\eeq
with
\beq
{\bf X} &=& \pmatrix{M_2 & g v_u\cr
                     g v_d & \mu \cr }
\>=\> \pmatrix{M_2 & \sqrt{2} \sbeta\, m_W\cr
               \sqrt{2} \cbeta\, m_W & \mu \cr }.
\label{charginomassmatrix}
\eeq
The mass eigenstates are related to the gauge eigenstates by two unitary 
2$\times$2 matrices $\bf U$ and $\bf V$ according to
\beq
\pmatrix{\stilde C^+_1\cr
         \stilde C^+_2} = {\bf V}
\pmatrix{\stilde W^+\cr
         \stilde H_u^+},\quad\qquad\>\>\>\>\>\>
\pmatrix{\stilde C^-_1\cr
         \stilde C^-_2} = {\bf U}
\pmatrix{\stilde W^-\cr
         \stilde H_d^-}.\phantom{xxx}
\eeq
Note that the mixing matrix for the positively charged left-handed 
fermions is different from that for the negatively charged left-handed 
fermions. They are chosen so that
\beq
{\bf U}^* {\bf X} {\bf V}^{-1} =
\pmatrix{m_{\stilde C_1} & 0\cr
              0   & m_{\stilde C_2}},
\eeq
with positive real entries $m_{\stilde C_i}$. Because these are only 
2$\times$2 matrices, it is not hard to solve for the masses analytically:
\beq
m^2_{{\stilde C}_{1}}, m^2_{{\stilde C}_{2}}
& = & {1\over 2} 
\Bigl [ |M_2|^2 + |\mu|^2 + 2m_W^2
\nonumber
\\
&&\mp
\sqrt{(|M_2|^2 + |\mu |^2 + 2 m_W^2 )^2 - 4 | \mu M_2 - m_W^2 \sin 2
\beta |^2 }
\Bigr ] .
\eeq
These are the (doubly degenerate) eigenvalues of the $4\times 4$ matrix 
${\bf M}_{\stilde C}^\dagger {\bf M}_{\stilde C}$, or equivalently the 
eigenvalues of ${\bf X}^\dagger {\bf X}$, since
\beq
{\bf V} {\bf X}^\dagger {\bf X} {\bf V}^{-1} =
{\bf U}^* {\bf X} {\bf X}^\dagger {\bf U}^{T} =
\pmatrix{m^2_{{\stilde C}_1} & 0 \cr 0 & m^2_{{\stilde C}_2}}.
\eeq
(But, they are {\it not} the squares of the eigenvalues of $\bf X$.) In 
the limit of eq.~(\ref{gauginolike}) with real $M_2$ and $\mu$, 
the chargino mass eigenstates consist of a wino-like $\stilde 
C_1^\pm$ and and a higgsino-like $\stilde C_2^\pm$, with masses
\beq
m_{{\stilde C}_1} &=& M_2 -
{ m_W^2 (M_2 + \mu \sin 2 \beta ) \over \mu^2 - M_2^2 } +\ldots
\\
m_{{\stilde C}_2}
&=& |\mu | + {I m_W^2 (\mu + M_2 \sin 2 \beta) \over \mu^2 - M^2_2 }
+\ldots .
\eeq
Here again the labeling assumes $M_2<|\mu|$, and $I$ is the sign of $\mu$. 
Amusingly, $\stilde C_1$ is degenerate with the neutralino $\stilde 
N_2$ in the approximation shown, but that is not an exact result. Their 
higgsino-like colleagues $\stilde N_3$, $\stilde N_4$ and $\stilde C_2$ 
have masses of order $|\mu|$. The case of $M_1 \approx 0.5 M_2 \ll |\mu|$ 
is not uncommonly found in viable models following from the boundary 
conditions in section \ref{sec:origins}, and it has been elevated to the 
status of a benchmark framework in many phenomenological studies. However 
it cannot be overemphasized that such expectations are not mandatory.

The Feynman rules involving neutralinos and charginos may be inferred in 
terms of $\bf N$, $\bf U$ and $\bf V$ from the MSSM Lagrangian as 
discussed above; they are collected in 
refs.~\cite{HaberKanereview}, \cite{GunionHaber}. 
Feynman rules 
based on two-component spinor notation have also  been given in 
\cite{DHM}.
In practice, the masses and mixing angles for the 
neutralinos and charginos are best computed numerically. Note that the 
discussion above yields the tree-level masses. Loop corrections to these 
masses can be significant, and have been found systematically at one-loop 
order in ref.~\cite{PBMZ}, with partial two-loop results in
\cite{GNCpoletwo,NCpole}.

\subsection{The gluino\label{subsec:MSSMspectrum.gluino}}
\setcounter{equation}{0}
\setcounter{footnote}{1}

The gluino is a color octet fermion, so it cannot mix with any other 
particle in the MSSM, even if $R$-parity is violated. In this regard, it 
is unique among all of the MSSM sparticles. In models with MSUGRA 
or GMSB boundary conditions, 
the gluino mass 
parameter $M_3$ is related to the bino and wino mass parameters $M_1$ and 
$M_2$ by eq.~(\ref{gauginomassunification}), so
\beq
M_3 \,=\, {\alpha_s\over \alpha} \sin^2\theta_W\, 
M_2 \,=\, {3\over 5} {\alpha_s \over \alpha} \cos^2\theta_W\, M_1
\label{eq:TiaEla}
\eeq
at any RG scale, up to small two-loop corrections. This implies a rough 
prediction
\beq
M_3 : M_2 : M_1 \approx 6:2:1
\eeq
near the TeV scale. It is therefore
reasonable to suspect that the gluino 
is considerably heavier than the lighter neutralinos and charginos
(even in many models where the gaugino mass unification condition is not
imposed).

For more precise estimates, one must take into account the fact that $M_3$ 
is really a running mass parameter with an implicit dependence on the RG 
scale $Q$. Because the gluino is a strongly interacting particle, $M_3$ 
runs rather quickly with $Q$ [see eq.~(\ref{gauginomassrge})]. A more 
useful quantity physically is the RG scale-independent mass $m_{\tilde g}$ 
at which the renormalized gluino propagator has a pole. Including one-loop 
corrections to the gluino propagator due to gluon exchange and 
quark-squark loops, one finds that the pole mass is given in terms of the 
running mass in the $\drbar$ scheme by \cite{gluinopolemass}
\beq
m_{\tilde g} = M_3(Q) \Bigl ( 1 + {\alpha_s\over 4 \pi}
\bigl [ 15 + 6\> {\rm ln}(Q/ M_3) + \sum A_{\tilde q}\bigr ] \Bigr )
\label{gluinopole}
\eeq
where
\beq
A_{\tilde q} \>=\> \int_0^1 \, dx \, x \, {\rm ln}
\bigl [
x m_{\tilde q}^2/M_3^2 + (1-x) m_{q}^2/M_3^2 - x(1-x) - i \epsilon
\bigr ].
\eeq
The sum in eq.~(\ref{gluinopole}) is over all 12 squark-quark 
supermultiplets, and we have neglected small effects due to squark mixing. 
[As a check, requiring $m_{\tilde g}$ to be independent of $Q$ in 
eq.~(\ref{gluinopole}) reproduces the one-loop RG equation for $M_3(Q)$ in 
eq.~(\ref{gauginomassrge}).] The correction terms proportional to 
$\alpha_s$ in eq.~(\ref{gluinopole}) can be quite significant,
because the gluino is strongly interacting, with a large group 
theory factor [the 15 in eq.~(\ref{gluinopole})] due to its color octet 
nature, and because it couples to all of the squark-quark pairs. The leading 
two-loop corrections to the gluino pole mass have also been found 
\cite{gluinopoletwo,GNCpoletwo,Martin:2006ub}, and are implemented in the latest
version of the {\tt SOFTSUSY} program \cite{SOFTSUSY}. 
They typically increase the prediction by another 1 or 2\%.

\subsection{The squarks and sleptons\label{subsec:MSSMspectrum.sfermions}}
\setcounter{equation}{0}
\setcounter{footnote}{1}

In principle, any scalars with the same electric charge, $R$-parity, and 
color quantum numbers can mix with each other. This means that with 
completely arbitrary soft terms, the mass eigenstates of the squarks and 
sleptons of the MSSM should be obtained by diagonalizing three $6\times 6$ 
squared-mass matrices for up-type squarks ($\stilde u_L$, $\stilde c_L$, 
$\stilde t_L$, $\stilde u_R$, $\stilde c_R$, $\stilde t_R$), down-type 
squarks ($\stilde d_L$, $\stilde s_L$, $\stilde b_L$, $\stilde d_R$, 
$\stilde s_R$, $\stilde b_R$), and charged sleptons ($\stilde e_L$, 
$\stilde \mu_L$, $\stilde \tau_L$, $\stilde e_R$, $\stilde \mu_R$, 
$\stilde \tau_R$), and one $3\times 3$ matrix for sneutrinos ($\stilde 
\nu_e$, $\stilde \nu_\mu$, $\stilde \nu_\tau$). Fortunately, the general 
hypothesis of flavor-blind soft parameters 
eqs.~(\ref{scalarmassunification}) and (\ref{aunification}) predicts that 
most of these mixing angles are very small. The third-family squarks and 
sleptons can have very different masses compared to their first- and 
second-family counterparts, because of the effects of large Yukawa ($y_t$, 
$y_b$, $y_\tau$) and soft ($a_t$, $a_b$, $a_\tau$) couplings in the RG 
equations (\ref{mq3rge})-(\ref{mstaubarrge}). Furthermore, they can have 
substantial mixing in pairs ($\stilde t_L$, $\stilde t_R$), ($\stilde 
b_L$, $\stilde b_R$) and ($\stilde \tau_L$, $\stilde \tau_R$). In 
contrast, the first- and second-family squarks and sleptons have 
negligible Yukawa couplings, so they end up in 7 very nearly degenerate, 
unmixed pairs $(\stilde e_R, \stilde \mu_R)$, $(\stilde \nu_e, \stilde 
\nu_\mu)$, $(\stilde e_L, \stilde \mu_L)$, $(\stilde u_R, \stilde c_R)$, 
$(\stilde d_R, \stilde s_R)$, $(\stilde u_L, \stilde c_L)$, $(\stilde d_L, 
\stilde s_L)$. As we have already discussed in section 
\ref{subsec:mssm.hints}, this avoids the problem of disastrously large 
virtual sparticle contributions to flavor-changing processes.

Let us first consider the spectrum of first- and second-family squarks and 
sleptons. In many models, including both MSUGRA 
[eq.~(\ref{scalarunificationsugra})] and GMSB 
[eq.~(\ref{scalargmsb})] boundary conditions, their running squared masses 
can be conveniently parameterized, to a good approximation, as:
\beq
m_{Q_1}^2 = m_{Q_2}^2  \!\!\!&=&\!\!\! m_0^2 + K_3 + K_2 + {1\over 36}K_1,
\label{mq1form} \\
m_{\sbar u_1}^2 = m_{\sbar u_2}^2 \!\!\! &=&\!\!\! m_0^2 + K_3
\qquad\>\>\>
+ {4\over 9} K_1,
\\
m_{\sbar d_1}^2 = m_{\sbar d_2}^2  \!\!\!&=&\!\!\! m_0^2 + K_3
\qquad\>\>\>
+ {1\over 9} K_1,
\\
m_{L_1}^2 = m_{L_2}^2 \!\!\!&=&\!\!\! m_0^2 \qquad\>\>\> + K_2 + {1\over
4} K_1,
\\
m_{\sbar e_1}^2 = m_{\sbar e_2}^2 \!\!\!&=&\!\!\! m_0^2
\qquad\qquad\>\>\>\>\>\> \, +\, K_1.
\label{me1form}
\eeq
A key point is that the same $K_3$, $K_2$ and $K_1$ appear everywhere in 
eqs.~(\ref{mq1form})-(\ref{me1form}), since all of the chiral 
supermultiplets couple to the same gauginos with the same gauge couplings. 
The different coefficients in front of $K_1$ just correspond to the 
various values of weak hypercharge squared for each scalar.

In MSUGRA models, $m_0^2$ is the same common scalar squared 
mass appearing in eq.~(\ref{scalarunificationsugra}). It can be very 
small, as in the ``no-scale" limit, but it could also be the dominant 
source of the scalar masses. The contributions $K_3$, $K_2$ and $K_1$ are 
due to the RG running\footnote{The quantity $S$ defined in 
eq.~(\ref{eq:defS}) vanishes at the input scale
for both MSUGRA and 
GMSB boundary conditions, and remains small under RG evolution.} 
proportional to the gaugino masses. Explicitly, they are found at one loop 
order by solving eq.~(\ref{easyscalarrge}):
\beq
K_a(Q) = \left\lbrace \matrix{{3/5}\cr {3/4} \cr {4/3}}
\right \rbrace \times
{1\over 2 \pi^2} \int^{{\rm ln} Q_{0}}_{{\rm ln}Q}dt\>\,
g^2_a(t) \,|M_a(t)|^2\qquad\>\>\> (a=1,2,3).
\label{kintegral}
\eeq
Here $Q_{0}$ is the input RG scale at which the MSUGRA 
boundary condition eq.~(\ref{scalarunificationsugra}) is applied, and $Q$ 
should be taken to be evaluated near the squark and slepton mass under 
consideration, presumably less than about 1 TeV. The running parameters 
$g_a(Q)$ and $M_a(Q)$ obey eqs.~(\ref{mssmg}) and 
(\ref{gauginomassunification}). If the input scale is approximated by the 
apparent scale of gauge coupling unification $Q_0 = M_U \approx 1.5 \times 
10^{16}$ GeV, one finds that numerically
\beq
K_1 \approx 0.15 m_{1/2}^2,\qquad
K_2 \approx 0.5 m_{1/2}^2,\qquad
K_3 \approx (4.5\>{\rm to}\> 6.5) m_{1/2}^2.
\label{k123insugra}
\eeq
for $Q$ near the electroweak scale. Here $m_{1/2}$ is the common gaugino 
mass parameter at the unification scale. Note that $K_3 \gg K_2 \gg K_1$; 
this is a direct consequence of the relative sizes of the gauge couplings 
$g_3$, $g_2$, and $g_1$. The large uncertainty in $K_3$ is due in part to 
the experimental uncertainty in the QCD coupling constant, and in part to 
the uncertainty in where to choose $Q$, since $K_3$ runs rather quickly 
below 1 TeV. If the gauge couplings and gaugino masses are unified between 
$M_U$ and $\MPlanck$, as would occur in a GUT model, then the effect of 
RG 
running for $M_U < Q < \MPlanck$ can be absorbed into a redefinition of 
$m_0^2$. Otherwise, it adds a further uncertainty roughly proportional to 
ln$(\MPlanck/M_U)$, compared to the larger contributions in 
eq.~(\ref{kintegral}), which go roughly like ln$(M_U/1$~TeV).

In gauge-mediated models, the same parameterization 
eqs.~(\ref{mq1form})-(\ref{me1form}) holds, but $m_0^2$ is always 0. At 
the input scale $Q_0$, each MSSM scalar gets contributions to its squared 
mass that depend only on its gauge interactions, as in 
eq.~(\ref{scalargmsb}). It is not hard to see that in general these 
contribute in exactly the same pattern as $K_1$, $K_2$, and $K_3$ in 
eq.~(\ref{mq1form})-(\ref{me1form}). The subsequent evolution of the 
scalar squared masses down to the electroweak scale again just yields more 
contributions to the $K_1$, $K_2$, and $K_3$ parameters. It is somewhat 
more difficult to give meaningful numerical estimates for these parameters 
in GMSB models than in the MSUGRA models without 
knowing the messenger mass scale(s) and the multiplicities of the 
messenger fields. However, in the gauge-mediated case one quite generally 
expects that the numerical values of the ratios $K_3/K_2$, $K_3/K_1$ and 
$K_2/K_1$ should be even larger than in eq.~(\ref{k123insugra}). There are 
two reasons for this. First, the running squark squared masses start off 
larger than slepton squared masses already at the input scale in 
gauge-mediated models, rather than having a common value $m_0^2$. 
Furthermore, in the gauge-mediated case, the input scale $Q_0$ is 
typically much lower than $\MPlanck$ or $M_U$, so that the RG evolution 
gives 
relatively more weight to RG scales closer to the electroweak scale, where 
the hierarchies $g_3>g_2>g_1$ and $M_3>M_2>M_1$ are already in effect.

In general, one therefore expects that the squarks should be considerably 
heavier than the sleptons, with the effect being more pronounced in 
gauge-mediated supersymmetry breaking models than in MSUGRA 
models. For any specific choice of model, this effect can be easily 
quantified with a numerical RG computation. 
The hierarchy $m_{\rm squark} > m_{\rm 
slepton}$ tends to hold even in models that do not fit neatly into any of 
the categories outlined in section \ref{sec:origins}, because the RG 
contributions to squark masses from the gluino are always present and 
usually quite large, since QCD has a larger gauge coupling than the 
electroweak interactions.

Regardless of the type of model,
there is also a ``hyperfine" splitting in the squark and slepton mass 
spectrum, produced by electroweak symmetry breaking. Each squark and 
slepton $\phi$ will get a contribution $\Delta_\phi$ to its squared mass, 
coming from the $SU(2)_L$ and $U(1)_Y$ $D$-term quartic interactions [see 
the last term in eq.~(\ref{fdpot})] of the form (squark)$^2$(Higgs)$^2$ 
and (slepton)$^2$(Higgs)$^2$, when the neutral Higgs scalars $H_u^0$ and 
$H_d^0$ get VEVs. They are model-independent for a given value of 
$\tan\beta$:
\beq
\Delta_\phi \>=\> \frac{1}{2}
(T_{3\phi} g^2 - Y_\phi g^{\prime 2}) (v_d^2 - v_u^2) \>=\> 
(T_{3\phi} - Q_\phi\sin^2\theta_W)
\cos (2\beta)\, m_Z^2 ,
\label{defDeltaphi}
\eeq
where $T_{3\phi}$, $Y_\phi$, and $Q_\phi$ are the third component of weak 
isospin, the weak hypercharge, and the electric charge of the left-handed 
chiral supermultiplet to which $\phi$ belongs. For example, 
$\Delta_{\tilde u_L} = ({1\over 2} - {2\over 3} \sin^2\theta_W)\cos 
(2\beta)\, m_Z^2$ and $\Delta_{\tilde d_L} = (-{1\over 2} + {1\over 3} 
\sin^2\theta_W)\cos (2\beta)\, m_Z^2$ and $\Delta_{\tilde u_R} = ({2\over 
3} \sin^2\theta_W)\cos (2\beta) \, m_Z^2$. These $D$-term contributions 
are typically smaller than the $m_0^2$ and $K_1$, $K_2$, $K_3$ 
contributions, but should not be neglected. They split apart the 
components of the $SU(2)_L$-doublet sleptons and squarks. Including them, 
the first-family squark and slepton masses are now given by:
\beq
m_{\tilde d_L}^2 \!\!&=&\!\! m_0^2 + K_3 + K_2 + {1\over 36} K_1 +
\Delta_{\tilde d_L},
\label{msdlform}
\\
m_{\tilde u_L}^2 \!\!&=&\!\! m_0^2 + K_3 + K_2 + {1\over 36} K_1 +
\Delta_{\tilde u_L},
\\
m_{\tilde u_R}^2\!\! &=&\!\! m_0^2 + K_3 \qquad\>\>\>  + {4\over 9} K_1 +
\Delta_{\tilde u_R},
\\
m_{\tilde d_R}^2 \!\!&=&\!\! m_0^2 + K_3 \qquad\>\>\>  + {1\over 9} K_1 +
\Delta_{\tilde d_R},
\label{msdrform}
\\
m_{\tilde e_L}^2 \!\!&=&\!\! m_0^2 \qquad\>\>\> + K_2 + {1\over 4} K_1 +
\Delta_{\tilde e_L},
\label{mselform}
\\
m_{\tilde \nu}^2 \!&=& \!\! m_0^2 \qquad\>\>\> + K_2 + {1\over 4} K_1 +
\Delta_{\tilde \nu},
\\
m_{\tilde e_R}^2 \!\!&=&\!\! m_0^2 \qquad\qquad\>\>\>\>\>\> \, +\,
K_1
\, + \Delta_{\tilde e_R},
\label{mserform}
\eeq
with identical formulas for the second-family squarks and sleptons. The 
mass splittings for the left-handed squarks and sleptons are governed by 
model-independent sum rules
\beq
m_{\tilde e_L}^2 -m_{\tilde \nu_e}^2 \,=\, 
m_{\tilde d_L}^2 -m_{\tilde u_L}^2 \,=\, 
g^2 (v_u^2 - v_d^2)/2 \,=\,  -\cos (2\beta)\, m_W^2  .
\eeq
In the allowed range $\tan\beta>1$, it follows that $m_{\tilde e_L} > 
m_{\tilde \nu_e}$ and $m_{\tilde d_L} > m_{\tilde u_L}$, with the 
magnitude of the splittings constrained by electroweak symmetry breaking.

Let us next consider the masses of the top squarks, for which there are 
several non-negligible contributions. First, there are squared-mass terms 
for $\stilde t^*_L \stilde t_L$ and $\stilde t_R^* \stilde t_R$ that are 
just equal to $m^2_{Q_3} + \Delta_{\tilde u_L}$ and $m^2_{\sbar u_3} + 
\Delta_{\tilde u_R}$, respectively, just as for the first- and 
second-family squarks. Second, there are contributions equal to $m_t^2$ 
for each of $\stilde t^*_L \stilde t_L$ and $\stilde t_R^* \stilde t_R$. 
These come from $F$-terms in the scalar potential of the form $y_t^2 
H_u^{0*} H_u^0 \stilde t_L^* \stilde t_L$ and $y_t^2 H_u^{0*} H_u^0 
\stilde t_R^* \stilde t_R$ (see Figures~\ref{fig:stop}b and 
\ref{fig:stop}c), with the Higgs fields replaced by their VEVs. (Of 
course, similar contributions are present for all of the squarks and 
sleptons, but they are too small to worry about except in the case of the 
top squarks.) Third, there are contributions to the scalar potential from 
$F$-terms of the form $-\mu^* y_t \stilde{\sbar t} \stilde t H_d^{0*} 
+\conj$; see eqs.~(\ref{striterms}) and Figure~\ref{fig:stri}a. These 
become $-\mu^* v y_t \cos\beta\, \stilde t^*_R \stilde t_L + \conj$ when 
$H_d^0$ is replaced by its VEV. Finally, there are contributions to the 
scalar potential from the soft (scalar)$^3$ couplings $a_t \stilde{\sbar 
t} \stilde Q_3 H_u^0 + \conj$ [see the first term of the second line of 
eq.~(\ref{MSSMsoft}), and eq.~(\ref{heavyatopapprox})], which become $ a_t 
v \sin\beta\, \stilde t_L \stilde t_R^* + \conj$ when $H_u^0$ is replaced 
by its VEV. Putting these all together, we have a squared-mass matrix for 
the top squarks, which in the gauge-eigenstate basis ($\stilde t_L$, 
$\stilde t_R$) is given by
\beq
\lagr_{\mbox{stop masses}} =  -\pmatrix{\stilde t_L^* & \stilde t_R^*}
{\bf m_{\stilde t}^2} \pmatrix{\stilde t_L \cr \stilde t_R}
\eeq
where
\beq
{\bf m_{\stilde t}^2} =
\pmatrix{
m^2_{Q_3} + m_t^2 + \Delta_{\tilde u_L} & 
v(a_t^* \sin\beta - \mu y_t\cos\beta )\cr
v (a_t \sin\beta - \mu^* y_t\cos\beta ) & 
m^2_{\sbar u_3} + m_t^2 + \Delta_{\tilde u_R}
} .
\label{mstopmatrix}
\eeq
This hermitian matrix can be diagonalized by a unitary matrix to give mass 
eigenstates:
\beq
\pmatrix{\stilde t_1\cr\stilde t_2} =
\pmatrix{ c_{\tilde t} & -s_{\tilde t}^* \cr
          s_{\tilde t} & c_{\tilde t}^*}
\pmatrix{\stilde t_L \cr \stilde t_R} .
\label{pixies}
\eeq
Here $m^2_{\tilde t_1}< m^2_{\tilde t_2}$ are the eigenvalues of 
eq.~(\ref{mstopmatrix}), and $|c_{\tilde t}|^2 + |s_{\tilde t}|^2 = 1$. If 
the off-diagonal elements of eq.~(\ref{mstopmatrix}) are real, then 
$c_{\tilde t}$ and $s_{\tilde t}$ are the cosine and sine of a stop mixing 
angle $\theta_{\tilde t}$, which can be chosen in the range $0\leq 
\theta_{\tilde t} < \pi$. Because of the large RG effects proportional to 
$X_t$ in eq.~(\ref{mq3rge}) and eq.~(\ref{mtbarrge}), in MSUGRA and GMSB and similar models one finds that $m_{\sbar u_3}^2 < m_{Q_3}^2$ at the electroweak 
scale, and both of these 
quantities are usually significantly smaller than the squark squared 
masses for the first two families. The diagonal terms $m_t^2$ in 
eq.~(\ref{mstopmatrix}) can mitigate this effect slightly, but only slightly,
and the off-diagonal entries will typically induce a significant mixing, which 
always reduces the lighter top-squark squared-mass eigenvalue. Therefore, 
models often predict that $\stilde t_1$ is the lightest squark of all, and 
that it is predominantly $\tilde t_R$.

A very similar analysis can be performed for the bottom squarks and 
charged tau sleptons, which in their respective gauge-eigenstate bases 
($\stilde b_L$, $\stilde b_R$) and ($\stilde \tau_L$, $\stilde \tau_R$) 
have squared-mass matrices:
\beq
{\bf m_{\stilde b}^2} =
\pmatrix{
m^2_{Q_3} + \Delta_{\tilde d_L} & 
v (a_b^* \cos\beta - \mu y_b\sin\beta )\cr
v (a_b \cos\beta - \mu^* y_b\sin\beta ) & 
m^2_{\sbar d_3} + \Delta_{\tilde d_R} },
\label{msbottommatrix}
\eeq
\beq
{\bf m_{\stilde \tau}^2} =
\pmatrix{
m^2_{L_3} + \Delta_{\tilde e_L} & 
v (a_\tau^* \cos\beta -\mu y_\tau\sin\beta )\cr
v (a_\tau \cos\beta - \mu^* y_\tau\sin\beta ) & 
m^2_{\sbar e_3} + \Delta_{\tilde e_R}}.
\label{mstaumatrix}
\eeq
These can be diagonalized to give mass eigenstates $\stilde b_1, \stilde 
b_2$ and $\stilde \tau_1, \stilde \tau_2$ in exact analogy with 
eq.~(\ref{pixies}).

The magnitude and importance of mixing in the sbottom and stau sectors 
depends on how big $\tan\beta$ is. If $\tan\beta$ is not too large (in 
practice, this usually means less than about $10$ or so, depending on the 
situation under study), the sbottoms and staus do not get a very large 
effect from the mixing terms and the RG effects due to $X_b$ and $X_\tau$, 
because $y_b,y_\tau \ll y_t$ from eq.~(\ref{eq:ytbtaumtbtau}). In that 
case the mass eigenstates are very nearly the same as the gauge 
eigenstates $\stilde b_L$, $\stilde b_R$, $\stilde \tau_L$ and $\stilde 
\tau_R$. The latter three, and $\stilde \nu_\tau$, will be nearly 
degenerate with their first- and second-family counterparts with the same 
$SU(3)_C \times SU(2)_L \times U(1)_Y$ quantum numbers. However, even in 
the case of small $\tan\beta$, $\stilde b_L$ will feel the effects of the 
large top Yukawa coupling because it is part of the doublet containing 
$\stilde t_L$. In particular, from eq.~(\ref{mq3rge}) we see that $X_t$ 
acts to decrease $m_{Q_3}^2$ as it is RG-evolved down from the input scale 
to the electroweak scale. Therefore the mass of ${\stilde b_L}$ can be 
significantly less than the masses of $\stilde d_L$ and $\stilde s_L$.

For larger values of $\tan\beta$, the mixing in 
eqs.~(\ref{msbottommatrix}) and (\ref{mstaumatrix}) can be quite 
significant, because $y_b$, $y_\tau$ and $a_b$, $a_\tau$ are 
non-negligible. Just as in the case of the top squarks, the lighter 
sbottom and stau mass eigenstates (denoted $\stilde b_1$ and $\stilde 
\tau_1$) can be significantly lighter than their first- and second-family 
counterparts. Furthermore, ${\stilde \nu_\tau}$ can be significantly 
lighter than the nearly degenerate ${\stilde \nu_e}$, $\stilde \nu_\mu$.

The requirement that the third-family squarks and sleptons should all have 
positive squared masses implies limits on the magnitudes of 
$a_t^*\sin\beta -\mu y_t \cos\beta$ and $a_b^*\cos\beta - \mu y_b 
\sin\beta$ and and $a_\tau^* \cos\beta - \mu y_\tau \sin\beta$. If they 
are too large, then the smaller eigenvalue of eq.~(\ref{mstopmatrix}), 
(\ref{msbottommatrix}) or (\ref{mstaumatrix}) will be driven negative, 
implying that a squark or charged slepton gets a VEV, breaking $SU(3)_C$ 
or electromagnetism. Since this is clearly unacceptable, one can put 
bounds on the (scalar)$^3$ couplings, or equivalently on the parameter 
$A_0$ in MSUGRA models. Even if all of the squared-mass 
eigenvalues are positive, the presence of large (scalar)$^3$ couplings can 
yield global minima of the scalar potential, with non-zero squark and/or 
charged slepton VEVs, which are disconnected from the vacuum that 
conserves $SU(3)_C$ and electromagnetism \cite{badvacua}. However, it is 
not always immediately clear whether the mere existence of such 
disconnected global minima should really disqualify a set of model
parameters, 
because the tunneling rate from our ``good" vacuum to the ``bad"  vacua 
can easily be longer than the age of the universe \cite{kusenko}.

Radiative corrections to the squark and slepton masses are potentially important, and are given at one-loop order in ref.~\cite{PBMZ}.
For squarks, the leading two-loop corrections have been found in 
refs.~\cite{Martin:2005eg,Martin:2006ub}, and are implemented in the latest
version of the {\tt SOFTSUSY} code \cite{SOFTSUSY}.

\subsection{Summary: the MSSM sparticle
spectrum}\label{subsec:MSSMspectrum.summary}
\setcounter{equation}{0}
\setcounter{footnote}{1}

In the MSSM, there are 32 distinct masses corresponding to undiscovered 
particles, not including the gravitino. Above, we have explained 
how the masses and mixing angles for these particles can be computed, 
given an underlying model for the soft terms at some input scale. 
The mass eigenstates of the MSSM are listed in Table
\ref{tab:undiscovered}, assuming 
only that the mixing of first- and second-family squarks and sleptons is 
negligible.%
\renewcommand{\arraystretch}{1.4}
\begin{table}[tb]
\begin{center}
\begin{tabular}{|c|c|c|c|c|}
\hline
Names & Spin & $P_R$ & Gauge Eigenstates & Mass Eigenstates \\
\hline\hline
Higgs bosons & 0 & $+1$ & 
$H_u^0\>\> H_d^0\>\> H_u^+ \>\> H_d^-$ 
& 
$h^0\>\> H^0\>\> A^0 \>\> H^\pm$
\\ \hline
& & &${\stilde u}_L\>\> {\stilde u}_R\>\> \stilde d_L\>\> \stilde d_R$&(same)
\\
squarks& 0&$-1$& ${\stilde s}_L\>\> {\stilde s}_R\>\> \stilde c_L\>\>
\stilde c_R$& (same) \\
& & &
$\stilde t_L \>\>\stilde t_R \>\>\stilde b_L\>\> \stilde b_R$ 
&
${\stilde t}_1\>\> {\stilde t}_2\>\> \stilde b_1\>\> \stilde b_2$
\\ \hline
& & &${\stilde e}_L\>\> {\stilde e}_R \>\>\stilde \nu_e$&(same) 
\\
sleptons& 0&$-1$&${\stilde \mu}_L\>\>{\stilde \mu}_R\>\>\stilde\nu_\mu$&(same)
\\
& & &
$\stilde \tau_L\>\> \stilde \tau_R \>\>\stilde \nu_\tau$ 
&
${\stilde \tau}_1 \>\>{\stilde \tau}_2 \>\>\stilde \nu_\tau$
\\
\hline
neutralinos & $1/2$&$-1$ & 
$\stilde B^0 \>\>\>\stilde W^0\>\>\> \stilde H_u^0\>\>\> \stilde H_d^0$   
&
$\stilde N_1\>\> \stilde N_2 \>\>\stilde N_3\>\> \stilde N_4$ 
\\
\hline
charginos & $1/2$&$-1$ & 
$\stilde W^\pm\>\>\> \stilde H_u^+ \>\>\>\stilde H_d^-$ 
&
$\stilde C_1^\pm\>\>\>\stilde C_2^\pm $ 
\\
\hline
gluino & $1/2$&$-1$ &$\stilde g$  &(same) \\
\hline
${\rm goldstino}\atop{\rm (gravitino)}$ & ${1/2}\atop{(3/2)}$&$-1$&$\stilde 
G$  &(same) \\
\hline
\end{tabular}
\caption{The undiscovered particles in the Minimal Supersymmetric Standard 
Model (with sfermion mixing for the first two families assumed to be 
negligible). 
\label{tab:undiscovered}}
\vspace{-0.4cm}
\end{center}
\end{table}%
A complete set of Feynman rules for the 
interactions of these particles with each other and with the Standard 
Model quarks, leptons, and gauge bosons can be found in 
refs.~\cite{HaberKanereview,GunionHaber}. 
Feynman rules 
based on two-component spinor notation have also  been given in 
\cite{DHM}.

Specific models for the soft 
terms can predict the masses and the mixing angles angles for the 
MSSM in terms of far fewer parameters. For example, 
in the MSUGRA models, the only free 
parameters not already measured by 
experiment are $m_0^2$, $m_{1/2}$, $A_0$, $\mu$, and $b$. In 
GMSB models, the free parameters include 
the scale $\Lambda$, the messenger mass scale $M_{\rm mess}$, 
the integer number $\nmess$ of copies of the minimal messengers, 
the goldstino decay constant $\langle F \rangle $, and the Higgs mass 
parameters $\mu$ and $b$. 

After RG evolving the soft terms down to the 
electroweak scale, one can demand that the scalar potential gives correct 
electroweak symmetry breaking. This allows us to trade $|\mu|$ and $b$ 
for one parameter $\tan\beta$, as in 
eqs.~(\ref{mubsub1})-(\ref{mubsub2}). So, to a reasonable approximation, 
the entire mass spectrum in MSUGRA models is determined by 
only five unknown parameters: $m_0^2$, $m_{1/2}$, $A_0$, $\tan\beta$, and 
Arg($\mu$), while in the simplest gauge-mediated supersymmetry breaking 
models one can pick parameters $\Lambda$, $M_{\rm mess}$, $\nmess$, 
$\langle F \rangle $, $\tan\beta$, and Arg($\mu$). Both frameworks are 
highly predictive. Of course, it is quite likely that the essential 
physics of supersymmetry breaking is not captured by either of these two 
scenarios in their minimal forms. 

\begin{figure} 
\vspace{-0.2cm}
\centerline{\psfig{figure=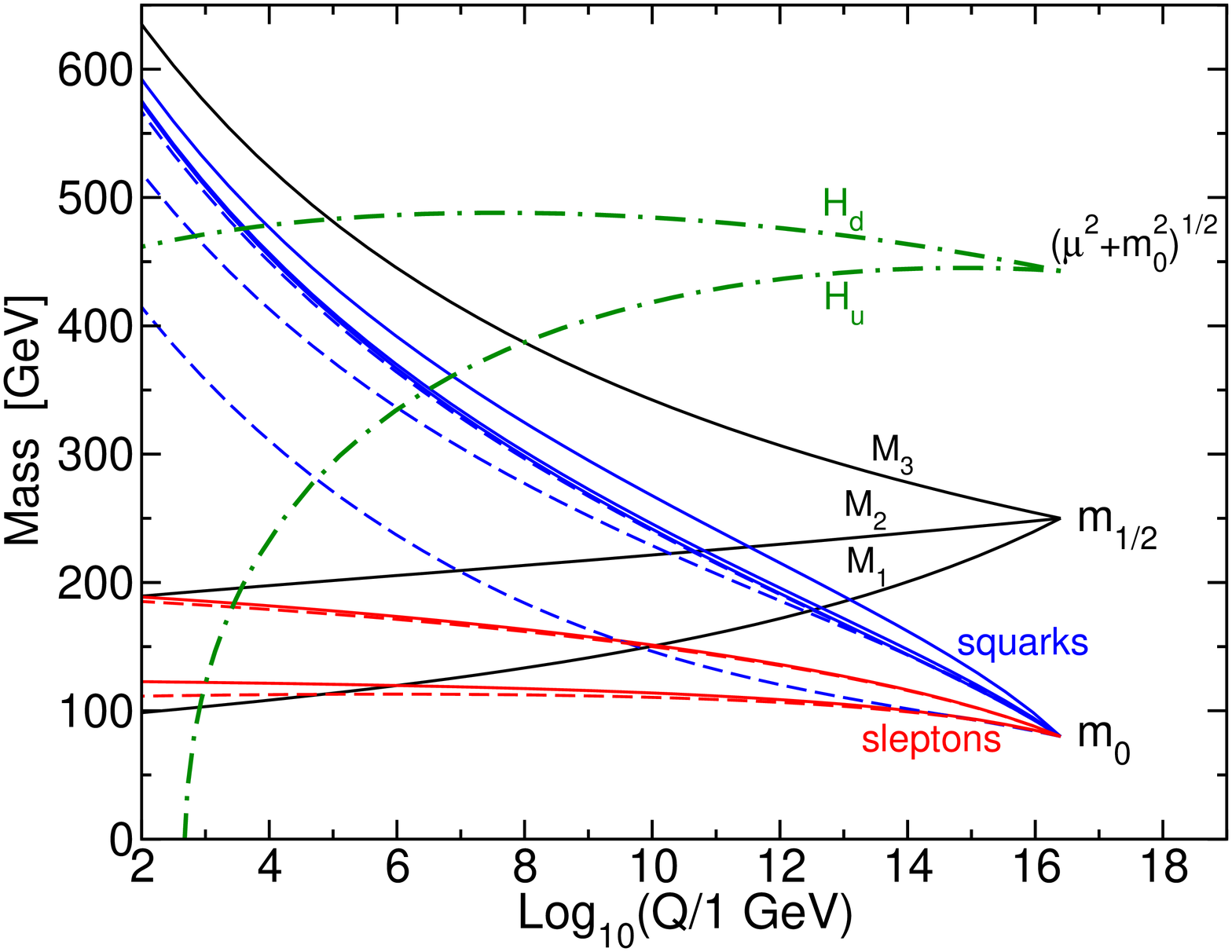,height=3.65in}} 
\vspace{-0.24cm}
\caption{RG evolution of scalar and gaugino mass parameters 
in the MSSM with MSUGRA boundary conditions imposed at $Q_0 = 1.5\times 
10^{16}$ GeV. The parameter 
$\mu^2 + m^2_{H_u}$ runs negative, provoking electroweak
symmetry breaking. 
\label{fig:running}}
\end{figure}

Figure \ref{fig:running} shows the RG running of scalar and gaugino 
masses in a sample model based on the MSUGRA boundary 
conditions imposed at $Q_0 = 1.5\times 10^{16}$ GeV.
[The parameter values used for this illustration were $m_0 = 300$ GeV, 
$m_{1/2} = -A_0 = 1000$ GeV, $\tan\beta = 15$, and 
sign($\mu$)$=+$, but these values were chosen more for their
artistic value in Figure \ref{fig:running}, and not as an attempt at realism.
The goal here is to understand the qualitative trends, 
rather than guess the correct numerical values.] 
The running gaugino masses are solid lines labeled by 
$M_1$, $M_2$, and $M_3$. The dot-dashed lines labeled $H_u$ and $H_d$ are 
the running values of the quantities $(\mu^2 + m_{H_u}^2)^{1/2}$ and 
$({\mu^2 + m_{H_d}^2})^{1/2}$, which appear in the Higgs potential. The 
other lines are the running squark and slepton masses, with dashed lines 
for the square roots of the third family parameters $m^2_{\sbar d_3}$, 
$m^2_{Q_3}$, $m^2_{\sbar u_3}$, $m^2_{L_3}$, and $m^2_{\sbar e_3}$ (from 
top to bottom), and solid lines for the first and second family sfermions. 
Note that $\mu^2 + m_{H_u}^2$ runs negative because of the effects of the 
large top Yukawa coupling as discussed above, providing for electroweak 
symmetry breaking. At the electroweak scale, the values of the Lagrangian 
soft parameters can be used to extract the physical masses, 
cross-sections, and decay widths of the particles, and other observables 
such as dark matter abundances and rare process rates. There are a variety 
of publicly available programs that do these tasks, including radiative 
corrections; see for example
\cite{ISAJET}-\cite{micrOMEGAs},\cite{FeynHiggs}.

Figure \ref{fig:sample} shows deliberately qualitative sketches of sample 
MSSM mass spectrum obtained from four different types of models 
assumptions.
\begin{figure}[p]
\begin{flushleft}
\mbox{
\hspace{-0.4cm}
\scalebox{1.25}{
\begin{picture}(190,180.)(0,0)
\Line(0,0)(190,0)\Line(190,0)(190,180.)
\Line(190,180.)(0,180.)\Line(0,180.)(0,0)
\Line(0,0)(190,0)
\Line(5,14.6336)(25,14.6336)
\Line(5,118.175)(25,118.175)
\Line(5,118.188)(25,118.188)
\Line(5,118.623)(25,118.623)
\Line(37,30.871)(57,30.871)
\Line(37,58.9073)(57,58.9073)
\Line(37,106.662)(57,106.662)
\Line(37,107.334)(57,107.334)
\Line(69,59.1933)(89,59.1933)
\Line(69,106.724)(89,106.724)
\Line(101,169.2)(121,169.2)
\Line(133,150.712)(153,150.712)
\Line(133,155.2)(153,155.2)
\Line(133,149.208)(153,149.208)
\Line(133,155.884)(153,155.884)
\Line(133,37.9827)(153,37.9827)
\Line(133,56.0104)(153,56.0104)
\Line(133,55.2272)(153,55.2272)
\Line(165,111.872)(185,111.872)
\Line(165,145.739)(185,145.739)
\Line(165,139.933)(185,139.933)
\Line(165,147.828)(185,147.828)
\Line(165,36.2047)(185,36.2047)
\Line(165,55.8985)(185,55.8985)
\Line(165,54.7796)(185,54.7796)
\rText(12.,20.)[][]{\textBlue$\scriptstyle h^0$}
\rText(9.8,113.5)[][]{$\scriptstyle H^0\, A^0$}
\rText(12.,126.)[][]{$\scriptstyle H^\pm$}
\rText(44.,36.5)[][]{$\scriptstyle\tilde N_1$}
\rText(44.,65.)[][]{$\scriptstyle\tilde N_2$}
\rText(44.,101.)[][]{$\scriptstyle\tilde N_3$}
\rText(44.,113.3)[][]{$\scriptstyle\tilde N_4$}
\rText(76,65.)[][]{$\scriptstyle\tilde C_1$}
\rText(76,113.3)[][]{$\scriptstyle\tilde C_2$}
\rText(106.,165.2)[c][]{$\scriptstyle\tilde g$}
\rText(138.6,163.5)[][]{$\scriptstyle\tilde d_L\,\tilde u_L$}
\rText(138.6,144.)[][]{$\scriptstyle\tilde u_R\,\tilde d_R$}
\rText(139.,61.)[][]{$\scriptstyle\tilde e_L$}
\rText(139.2,33.)[][]{$\scriptstyle\tilde e_R$}
\rText(139.,50.5)[][]{$\scriptstyle\tilde \nu_e$}
\rText(171.,106.6)[][]{$\scriptstyle\tilde t_1$}
\rText(169.8,154.5)[][]{$\scriptstyle\tilde t_2\,\tilde b_2$}
\rText(171.,135.)[][]{$\scriptstyle\tilde b_1$}
\rText(171.,31.8)[][]{$\scriptstyle\tilde \tau_1$}
\rText(171.,61.)[][]{$\scriptstyle\tilde \tau_2$}
\rText(171.,50.6)[][]{$\scriptstyle\tilde \nu_{\tau}$}
\Line(0,0)(190,0)
\rText(95,-10)[][]{\textBlack\small (a)\textBlue}
\end{picture}
}
\scalebox{1.25}{
\begin{picture}(190,180.)(0,0)
\Line(0,0)(190,0)\Line(190,0)(190,180.)
\Line(190,180.)(0,180.)\Line(0,180.)(0,0)
\Line(0,0)(190,0)
\Line(5,6.29337)(25,6.29337)
\Line(5,166.373)(25,166.373)
\Line(5,166.37)(25,166.37)
\Line(5,166.444)(25,166.444)
\Line(37,6.22496)(57,6.22496)
\Line(37,9.4611)(57,9.4611)
\Line(37,10.5135)(57,10.5135)
\Line(37,16.2859)(57,16.2859)
\Line(69,8.66653)(89,8.66653)
\Line(69,16.2544)(89,16.2544)
\Line(101,48.058)(121,48.058)
\Line(133,169.6)(153,169.6)
\Line(133,169.153)(153,169.153)
\Line(133,169.684)(153,169.684)
\Line(133,169.2)(153,169.2)
\Line(133,168.163)(153,168.163)
\Line(133,167.869)(153,167.869)
\Line(133,167.821)(153,167.821)
\Line(165,101.057)(185,101.057)
\Line(165,139.285)(185,139.285)
\Line(165,139.159)(185,139.159)
\Line(165,168.263)(185,168.263)
\Line(165,166.795)(185,166.795)
\Line(165,167.195)(185,167.195)
\Line(165,167.143)(185,167.143)
\rText(12.,11.7)[][]{\textBlue$\scriptstyle h^0$}
\rText(16.,162.)[][]{$\scriptstyle H^0\, A^0\, H^\pm$}
\rText(43.,22.7)[][]{$\scriptstyle\tilde N_i$}
\rText(75.5,22.9)[][]{$\scriptstyle\tilde C_i$}
\rText(106.7,43.5)[c][]{$\scriptstyle\tilde g$}
\rText(135.,175.6)[][]{$\scriptstyle\tilde u_R\, \tilde d_R\, \tilde u_L\, \tilde d_L$}
\rText(138.,164.)[][]{$\scriptstyle\tilde e_R\,\tilde\nu_e\,\tilde e_L$}
\rText(172.,162.2)[][]{$\scriptstyle\tilde \tau_1\, \tilde \tau_2\,\tilde\nu_\tau$}
\rText(172.,174.5)[][]{$\scriptstyle\tilde b_{2}$}
\rText(172.,95.)[][]{$\scriptstyle\tilde t_1$}
\rText(172.,133.9)[][]{$\scriptstyle\tilde b_1$}
\rText(172.,145.4)[][]{$\scriptstyle\tilde t_2$}
\Line(0,0)(190,0)
\rText(95,-10)[][]{\textBlack\small (b)\textBlue}
\end{picture}
}
}
\vspace{1cm}

\mbox{
\hspace{-0.4cm}
\scalebox{1.25}{
\begin{picture}(190,180.)(0,0)
\Line(0,0)(190,0)\Line(190,0)(190,180.)
\Line(190,180.)(0,180.)\Line(0,180.)(0,0)
\Line(0,0)(190,0)
\Line(5,12.6108)(25,12.6108)
\Line(5,80.472)(25,80.472)
\Line(5,80.4032)(25,80.4032)
\Line(5,80.9312)(25,80.9312)
\Line(37,22.471)(57,22.471)
\Line(37,41.7882)(57,41.7882)
\Line(37,61.4763)(57,61.4763)
\Line(37,63.5032)(57,63.5032)
\Line(69,41.7828)(89,41.7828)
\Line(69,63.4957)(89,63.4957)
\Line(101,127.258)(121,127.258)
\Line(133,161.261)(153,161.261)
\Line(133,168.984)(153,168.984)
\Line(133,160.522)(153,160.522)
\Line(133,169.2)(153,169.2)
\Line(133,28.0516)(153,28.0516)
\Line(133,56.9097)(153,56.9097)
\Line(133,56.2624)(153,56.2624)
\Line(165,149.034)(185,149.034)
\Line(165,164.142)(185,164.142)
\Line(165,159.525)(185,159.525)
\Line(165,163.401)(185,163.401)
\Line(165,27.3366)(185,27.3366)
\Line(165,56.9161)(185,56.9161)
\Line(165,56.1559)(185,56.1559)
\rText(12.,18.6)[][]{$\scriptstyle h^0$}
\rText(10.,76.)[][]{$\scriptstyle H^0\, A^0$}
\rText(12.,86.8)[][]{$\scriptstyle H^\pm$}
\rText(44,16.3)[][]{$\scriptstyle\tilde N_1$}
\rText(44.,35.7)[][]{$\scriptstyle\tilde N_2$}
\rText(44.,55.3)[][]{$\scriptstyle\tilde N_3$}
\rText(44.,70.)[][]{$\scriptstyle\tilde N_4$}
\rText(75,35.6)[][]{$\scriptstyle\tilde C_1$}
\rText(74.5,70.4)[][]{$\scriptstyle\tilde C_2$}
\rText(107.5,121.3)[c][]{$\scriptstyle\tilde g$}
\rText(138.4,175.6)[][]{$\scriptstyle\tilde d_L\,\tilde u_L$}
\rText(138.4,155.2)[][]{$\scriptstyle\tilde u_R\,\tilde d_R$}
\rText(140.,62.)[][]{$\scriptstyle\tilde e_L$}
\rText(140.,23.)[][]{$\scriptstyle\tilde e_R$}
\rText(140.,52.)[][]{$\scriptstyle\tilde \nu_e$}
\rText(172.,144.)[][]{$\scriptstyle\tilde t_1$}
\rText(170.,170.7)[][]{$\scriptstyle\tilde t_2, \scriptstyle \tilde b_2$}
\rText(172.,155.3)[][]{$\scriptstyle\tilde b_1$}
\rText(172.,22.9)[][]{$\scriptstyle\tilde \tau_1$}
\rText(172.,62.35)[][]{$\scriptstyle\tilde \tau_2$}
\rText(172.,52)[][]{$\scriptstyle\tilde \nu_{\tau}$}
\Line(0,0)(190,0)
\rText(95,-10)[][]{\textBlack\small (c)\textBlue}
\end{picture}
}
\scalebox{1.25}{
\begin{picture}(190,180.)(0,0)
\Line(0,0)(190,0)\Line(190,0)(190,180.)
\Line(190,180.)(0,180.)\Line(0,180.)(0,0)
\Line(0,0)(190,0)
\Line(5,14.7241)(25,14.7241)
\Line(5,69.829)(25,69.829)
\Line(5,69.7619)(25,69.7619)
\Line(5,70.561)(25,70.561)
\Line(37,31.5365)(57,31.5365)
\Line(37,50.2952)(57,50.2952)
\Line(37,54.8238)(57,54.8238)
\Line(37,66.9138)(57,66.9138)
\Line(69,49.9293)(89,49.9293)
\Line(69,66.867)(89,66.867)
\Line(101,169.2)(121,169.2)
\Line(133,152.032)(153,152.032)
\Line(133,157.647)(153,157.647)
\Line(133,151.58)(153,151.58)
\Line(133,157.969)(153,157.969)
\Line(133,23.5495)(153,23.5495)
\Line(133,48.939)(153,48.939)
\Line(133,47.8891)(153,47.8891)
\Line(165,142.476)(185,142.476)
\Line(165,155.514)(185,155.514)
\Line(165,150.57)(185,150.57)
\Line(165,153.814)(185,153.814)
\Line(165,22.7656)(185,22.7656)
\Line(165,49.0504)(185,49.0504)
\Line(165,47.8094)(185,47.8094)
\rText(12.,20.)[][]{$\scriptstyle h^0$}
\rText(10.,65.8)[][]{$\scriptstyle H^0\, A^0$}
\rText(12.,75.3)[][]{$\scriptstyle H^\pm$}
\rText(43.5,26.2)[][]{$\scriptstyle\tilde N_1$}
\rText(43.5,44.7)[][]{$\scriptstyle\tilde N_2$}
\rText(43.5,60.4)[][]{$\scriptstyle\tilde N_3$}
\rText(43.5,72.8)[][]{$\scriptstyle\tilde N_4$}
\rText(74,44.7)[][]{$\scriptstyle\tilde C_1$}
\rText(74,73.)[][]{$\scriptstyle\tilde C_2$}
\rText(107.2,165.)[c][]{$\scriptstyle\tilde g$}
\rText(138.5,164.)[][]{$\scriptstyle\tilde d_L\,\tilde u_L$}
\rText(138.5,146.7)[][]{$\scriptstyle\tilde u_R\,\tilde d_R$}
\rText(140.,53.7)[][]{$\scriptstyle\tilde e_L$}
\rText(140.,19.1)[][]{$\scriptstyle\tilde e_R$}
\rText(140.,43.8)[][]{$\scriptstyle\tilde \nu_e$}
\rText(172.,138.)[][]{$\scriptstyle\tilde t_1$}
\rText(170.,162.)[][]{$\scriptstyle\tilde t_2, \scriptstyle\tilde b_2$}
\rText(172.,148.7)[][]{$\scriptstyle\tilde b_1$}
\rText(172.,18.5)[][]{$\scriptstyle\tilde \tau_1$}
\rText(172.,54.)[][]{$\scriptstyle\tilde \tau_2$}
\rText(172.,43.5)[][]{$\scriptstyle\tilde \nu_{\tau}$\textBlack}
\rText(95,-10)[][]{\small (d)}
\end{picture}
}
}
\end{flushleft}
\caption{Four sample mass spectra for the undiscovered 
particles in the MSSM, for 
(a) MSUGRA with $m^2_0 \ll m_{1/2}^2$, 
(b) MSUGRA with $m^2_0 \gg m_{1/2}^2$,
(c) GMSB with $N_5=1$, and 
(d) GMSB with $N_5=3$.
Mass scales are not equal for the four cases, and are deliberately omitted.
These spectra are presented for entertainment purposes only! No warranty, expressed or implied, guarantees that they look anything like the real world. \label{fig:sample}}
\end{figure}
The first, in Figure \ref{fig:sample}(a), is the output from an 
MSUGRA model with
relatively low $m_0^2$ compared to $m^2_{1/2}$ (similar to fig.~\ref{fig:running}).
This model features a near-decoupling limit for the Higgs sector, and
a bino-like $\stilde N_1$ LSP, nearly degenerate wino-like 
$\stilde N_2, \stilde C_1$, and higgsino-like $\stilde N_3, \stilde N_4,
\stilde C_2$. The gluino is the heaviest superpartner. The squarks are
all much heavier than the sleptons, and the lightest sfermion is a stau.
(The second-family squarks and sleptons are nearly degenerate with those of the first family, and so are not shown separately.) 
Variations in the model 
parameters have important and predictable effects. For example, 
taking larger values of $\tan\beta$ with other model parameters 
held fixed will usually tend to lower $\stilde b_1$ and $\stilde \tau_1$ 
masses compared to those of the other sparticles. 
Taking
larger $m_0^2$ will tend to  
squeeze together the spectrum of squarks and sleptons and move 
them all higher compared to the 
neutralinos, charginos and 
gluino. This is illustrated in Figure \ref{fig:sample}(b), which instead has
$m_0^2 \gg m_{1/2}^2$. 
In this model, the heaviest chargino and neutralino are wino-like.

The third sample sketch, in fig.~\ref{fig:sample}(c), is obtained from a 
typical minimal GMSB model, with $N_5 = 1$ 
Here we see that the hierarchy between 
strongly interacting sparticles and weakly interacting ones is quite 
large. Changing the messenger scale or $\Lambda$ does not reduce the 
relative splitting between squark and slepton masses, because there is no 
analog of the universal $m_0^2$ contribution here. Increasing the number 
of messenger fields tends to decrease the squark and slepton masses 
relative to the gaugino masses, but still keeps the hierarchy between 
squark and slepton masses intact. In the model shown, the LSP is the 
nearly massless gravitino and the NLSP is a 
bino-like neutralino, but for larger number of messenger fields it could 
be either a stau, or else co-NLSPs $\tilde \tau_1$, $\tilde e_L$, $\tilde 
\mu_L$, depending on the choice of $\tan\beta$.

The fourth sample sketch, in fig.~\ref{fig:sample}(d), 
is of a typical GMSB model with a non-minimal messenger sector, $N_5=3$
Again the LSP is the nearly massless gravitino, but this 
time the NLSP is the lightest stau. The heaviest superpartner is the gluino, and the 
heaviest chargino and neutralino
are wino-like.

It would be a mistake to rely too heavily on specific scenarios for the 
MSSM mass and mixing spectrum, and the above illustrations are only
a tiny fraction of the available possibilities. However, it is also 
useful to keep in mind some general trends that often recur in various 
different models. Indeed, there has emerged a sort of folklore 
concerning likely features of the MSSM spectrum, partly based on 
theoretical bias and partly on the constraints inherent in many known viable 
softly-broken supersymmetric theories. We remark on these features mainly 
because they represent the prevailing prejudices among many supersymmetry 
theorists, which is certainly a useful thing to know even if one wisely 
decides to remain skeptical. For example, it is perhaps not unlikely that:
\begin{itemize}
\item[$\bullet$] The LSP is the lightest neutralino $\stilde N_1$, unless 
the gravitino is lighter or $R$-parity is not conserved. If $M_1 <  
M_2,|\mu|$, then $\stilde N_1$ is likely to be bino-like, with a mass 
roughly 
0.5 times the masses of $\stilde N_2$ and $\stilde C_1$ in many
well-motivated models. If, instead, 
$|\mu| < M_1,M_2$, then the LSP $\stilde N_1$ 
has a large higgsino content and 
$\stilde N_2$ and $\stilde C_1$ are not much heavier.
And, if $M_2 \ll M_1, |\mu|$, then the LSP will be a wino-like
neutralino, with a chargino only very slightly heavier.
\item[$\bullet$] The gluino will be much heavier than the lighter 
neutralinos and charginos. This is certainly true in the case of the 
``standard" gaugino mass relation eq.~(\ref{gauginomassunification}); more 
generally, the running gluino mass parameter grows relatively quickly as 
it is RG-evolved into the infrared because the QCD coupling is larger than 
the electroweak gauge couplings. So even if there are big corrections to 
the gaugino mass boundary conditions eqs.~(\ref{gauginounificationsugra}) 
or (\ref{gauginogmsb}), the gluino mass parameter $M_3$ is likely to come 
out larger than $M_1$ and $M_2$.
\item[$\bullet$] The squarks of the first and second families are nearly 
degenerate and much heavier than the sleptons. This is because each squark 
mass gets the same large positive-definite radiative corrections from 
loops involving the gluino. The left-handed squarks $\stilde u_L$, 
$\stilde d_L$, $\stilde s_L$ and $\stilde c_L$ are likely to be heavier 
than their right-handed counterparts $\stilde u_R$, $\stilde d_R$, 
$\stilde s_R$ and $\stilde c_R$, because of the effect parameterized
by $K_2$ in eqs.~(\ref{msdlform})-(\ref{mserform}).
\item[$\bullet$] The squarks of the first two families cannot be lighter 
than about 0.8 times the mass of the gluino in MSUGRA 
models, and about 0.6 times the mass of the gluino in the simplest 
gauge-mediated models as discussed in section \ref{subsec:origins.gmsb} if 
the number of messenger squark pairs is $\nmess \leq 4$.
In the MSUGRA case this is because the gluino mass feeds
into the squark masses through RG evolution; in the gauge-mediated case it
is because the gluino and squark masses are tied together by
eqs.~(\ref{gauginogmsbgen}) and (\ref{scalargmsbgen}). 
\item[$\bullet$] The lighter stop $\stilde t_1$ and the lighter sbottom 
$\stilde b_1$ are probably the lightest squarks. This is because stop and 
sbottom mixing effects and the effects of $X_t$ and $X_b$ in 
eqs.~(\ref{mq3rge})-(\ref{md3rge}) both tend to decrease the lighter stop 
and sbottom masses.
\item[$\bullet$] The lightest charged slepton is probably a stau $\stilde 
\tau_1$. The mass difference $m_{\tilde e_R}-m_{\tilde \tau_1}$ is 
likely to be significant if $\tan\beta$ is large, because of the effects 
of a large tau Yukawa coupling. For smaller $\tan\beta$, $\stilde \tau_1$ 
is predominantly $\stilde \tau_R$ and it is not so much lighter than 
$\stilde e_R$, $\stilde \mu_R$.
\item[$\bullet$] The left-handed charged sleptons $\stilde e_L$ and 
$\stilde \mu_L$ are likely to be heavier than their right-handed 
counterparts $\stilde e_R$ and $\stilde \mu_R$. This is because of the 
effect of $K_2$ in eq.~(\ref{mselform}). (Note also that $\Delta_{\tilde 
e_L} - \Delta_{\tilde e_R}$ is positive but very small because of the 
numerical accident $\sin^2\theta_W \approx 1/4$.)
\end{itemize}
It should be kept in mind that each of these prejudices 
might be defied by the real world.
The most important point is that by measuring the masses and mixing angles 
of the MSSM particles we will be able to gain a great deal of information 
that differentiate between competing proposals for the 
origin and mediation of supersymmetry breaking.


\section{Sparticle decays}\label{sec:decays}
\setcounter{equation}{0}
\setcounter{figure}{0}
\setcounter{table}{0}
\setcounter{footnote}{1}

This section contains a brief qualitative overview of the decay patterns
of sparticles in the MSSM, assuming that $R$-parity is conserved. We will 
consider in turn the possible decays of neutralinos, charginos, sleptons, 
squarks, and the gluino. If, as is most often assumed, the lightest 
neutralino $\NI$ is the LSP, then all decay chains will end up with it in 
the final state. Section \ref{subsec:decays.gravitino} discusses the 
alternative possibility that the gravitino/goldstino $\G$ is the LSP. 
For the sake of simplicity of notation, we will often not distinguish
between particle and antiparticle names and labels in this section, with
context and consistency (dictated by charge and color conservation)
resolving any ambiguities. 

\subsection{Decays of neutralinos and
charginos}\label{subsec:decays.inos}
\setcounter{equation}{0}
\setcounter{footnote}{1}

Let us first consider the possible two-body decays. Each neutralino and
chargino contains at least a small admixture of the electroweak gauginos
$\stilde B$, $\stilde W^0$ or $\stilde W^\pm$, as we saw in section
\ref{subsec:MSSMspectrum.inos}. So, $\stilde N_i$ and $\stilde C_i$ inherit
couplings of weak interaction strength to (scalar, fermion) pairs, as
shown in Figure~\ref{fig:gaugino}b,c. If sleptons or squarks are
sufficiently light, a neutralino or chargino can therefore decay into
lepton+slepton or quark+squark.  To the extent that sleptons are probably
lighter than squarks, the lepton+slepton final states are favored. A
neutralino or chargino may also decay into any lighter neutralino or
chargino plus a Higgs scalar or an electroweak gauge boson, because they
inherit the gaugino-higgsino-Higgs (see Figure~\ref{fig:gaugino}b,c) and
$SU(2)_L$ gaugino-gaugino-vector boson (see Figure~\ref{fig:gauge}c)
couplings of their components. So, the possible two-body decay modes for
neutralinos and charginos in the MSSM are: 
\beq
\stilde N_i \rightarrow
Z\stilde N_j,\>\>\, W\stilde C_j,\>\>\, h^0\stilde N_j,\>\>\, \ell \stilde
\ell,\>\>\,
\nu \stilde \nu,\>\>\,
[A^0 \stilde N_j,\>\>\, H^0 \stilde N_j,\>\>\, H^\pm
\stilde C_j^\mp,\>\>\,
q\stilde q],
\qquad\>\>\>{}
\label{nino2body}
\\
\stilde C_i \rightarrow
W\stilde N_j,\>\>\, Z\stilde C_1,\>\>\, h^0\stilde C_1,\>\>\, \ell \stilde
\nu,\>\>\,
\nu \stilde \ell,\>\>\,
[A^0 \stilde C_1,\>\>\, H^0 \stilde C_1,\>\>\, H^\pm \stilde N_j,\>\>\,
q\stilde q^\prime],
\qquad\>\>\>{}
\label{cino2body}
\eeq
using a generic notation $\nu$, $\ell$, $q$ for neutrinos, charged
leptons, and quarks. The final states in brackets are the more
kinematically implausible ones. (Since $m_{h^0} = 125$ GeV, it
is the most likely of the Higgs scalars to appear in these decays.) For
the heavier neutralinos and chargino ($\stilde N_3$, $\stilde N_4$ and
$\stilde C_2$), one or more of the two-body decays in
eqs.~(\ref{nino2body}) and (\ref{cino2body}) is likely to be kinematically
allowed. Also, if the decays of neutralinos and charginos with a
significant higgsino content into third-family quark-squark pairs are
open, they can be greatly enhanced by the top-quark Yukawa coupling,
following from the interactions shown in fig.~\ref{fig:topYukawa}b,c. 

It may be that all of these two-body modes are kinematically forbidden for
a given chargino or neutralino, especially for $\stilde C_1$ and $\stilde
N_2$ decays. In that case, they have three-body decays
\beq
\stilde N_i \rightarrow f f \stilde N_j,\>\>\>\,
\stilde N_i \rightarrow f f^\prime \stilde C_j,\>\>\>\,
\stilde C_i \rightarrow f f^\prime \stilde N_j,\>\>\>\,{\rm and}\>\>\>\,
\stilde C_2 \rightarrow f f \stilde C_1,\qquad\>\>\>\>\>{}
\label{cino3body}
\eeq
through the same (but now off-shell) gauge bosons, Higgs scalars,
sleptons, and squarks that appeared in the two-body decays
eqs.~(\ref{nino2body}) and (\ref{cino2body}). Here $f$ is generic notation
for a lepton or quark, with $f$ and $f^\prime$ distinct members of the same
$SU(2)_L$ multiplet (and of course one of the $f$ or $f'$ in each of these
decays must actually be an antifermion).  The chargino and neutralino
decay widths into the various final states can be found in
refs.~\cite{inodecays}-\cite{neutralinoloopdecays}. 

\begin{figure}
\begin{center}
\scalebox{1.42}{
\begin{picture}(90,35)(10,0)
\Line(0,0)(30,0)
\Line(30,0)(45,27)
\DashLine(30,0)(60,0){3}
\Line(60,0)(75,27)
\Line(60,0)(90,0)
\rText(0,6)[][]{$\scriptstyle\tilde N_i$}
\rText(42,6)[][]{$\scriptstyle\tilde f$}
\rText(43.5,27)[][]{$\scriptstyle  f$}
\rText(73.5,27)[][]{$\scriptstyle  f$}
\rText(84,6)[][]{$\scriptstyle\tilde N_1$}
\end{picture}
}
~~~~
\scalebox{1.42}{
\begin{picture}(90,35)(10,0)
\Line(0,0)(30,0)
\Line(30,0)(45,27)
\Photon(30,0)(60,0){1.5}{5}
\Line(60,0)(75,27)
\Line(60,0)(90,0)
\rText(0,6)[][]{$\scriptstyle\tilde N_i$}
\rText(42,6)[][]{$\scriptstyle Z$}
\rText(46.5,27)[][]{$\scriptstyle\tilde N_1$}
\rText(74.5,27)[][]{$\scriptstyle f$}
\rText(82,6)[][]{$\scriptstyle f$}
\end{picture}
}
~~~~
\scalebox{1.42}{
\begin{picture}(90,35)(10,0)
\Line(0,0)(30,0)
\Line(30,0)(45,27)
\DashLine(30,0)(66,0){3}
\Line(66,0)(80,27)
\Line(66,0)(96,0)
\rText(0,6)[][]{$\scriptstyle\tilde N_i$}
\rText(46,6)[][]{$\scriptstyle h^0\!\!\!,\, H^0\!\!\!,\, A^0$}
\rText(47,30)[][]{$\scriptstyle\tilde N_1$}
\rText(83.5,32)[][]{$\scriptstyle b,\,\tau,\>\ldots$}
\rText(93,6)[][]{$\scriptstyle b,\,\tau,\>\ldots$}
\end{picture}
}
\end{center}
\vspace{0.03cm}
\begin{center}
\scalebox{1.42}{
\begin{picture}(90,35)(10,0)
\Line(0,0)(30,0)
\Line(30,0)(45,27)
\DashLine(30,0)(60,0){3}
\Line(60,0)(75,27)
\Line(60,0)(90,0)
\rText(0,6)[][]{$\scriptstyle\tilde C_i$}
\rText(42,6)[][]{$\scriptstyle\tilde f$}
\rText(44.5,27)[][]{$\scriptstyle f'$}
\rText(74.5,27)[][]{$\scriptstyle f$}
\rText(84,6)[][]{$\scriptstyle\tilde N_1$}
\end{picture}
}
~~~~
\scalebox{1.42}{
\begin{picture}(90,35)(10,0)
\Line(0,0)(30,0)
\Line(30,0)(45,27)
\Photon(30,0)(60,0){1.5}{5}
\Line(60,0)(75,27)
\Line(60,0)(90,0)
\rText(0,6)[][]{$\scriptstyle\tilde C_i$}
\rText(42,6)[][]{$\scriptstyle W$}
\rText(46,28)[][]{$\scriptstyle\tilde N_1$}
\rText(74.5,27)[][]{$\scriptstyle f'$}
\rText(84,6)[][]{$\scriptstyle f$}
\end{picture}
}
~~~~
\scalebox{1.42}{
\begin{picture}(90,35)(10,0)
\Line(0,0)(30,0)
\Line(30,0)(45,27)
\DashLine(30,0)(60,0){3}
\Line(60,0)(75,27)
\Line(60,0)(90,0)
\rText(0,6)[][]{$\scriptstyle\tilde C_i$}
\rText(42,6)[][]{$\scriptstyle H^\pm$}
\rText(46,30.3)[][]{$\scriptstyle \tilde N_1$}
\rText(75,32.2)[][]{$\scriptstyle b,\, \tau,\,\ldots$}
\rText(84,6)[][]{$\scriptstyle t,\,\nu_\tau,\,\ldots$}
\end{picture}
}
\end{center}
\vspace{-0.25cm}
\caption{Feynman diagrams for neutralino and chargino decays with $\tilde
N_1$ in the final state. The intermediate scalar or vector boson in each
case can be either on-shell (so that actually there is a sequence of
two-body decays) or off-shell, depending on the sparticle mass spectrum.
\label{fig:NCdecays}}
\end{figure}
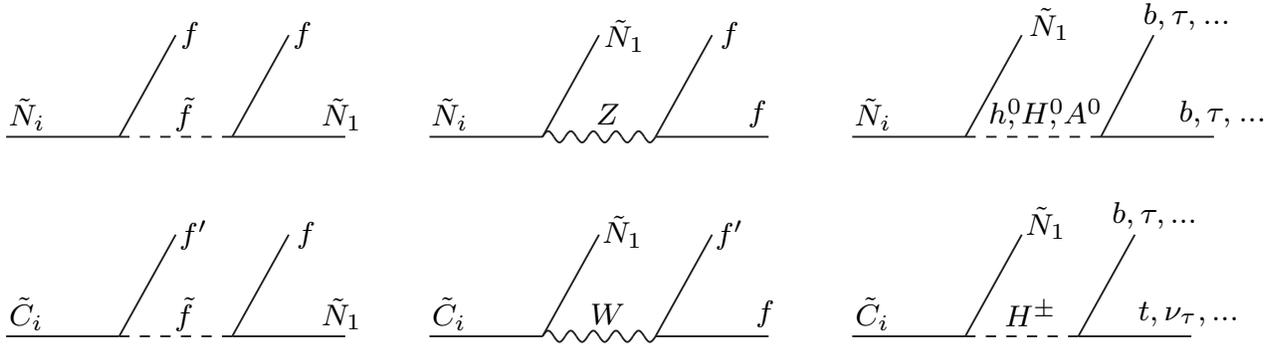

The Feynman diagrams for the neutralino and chargino decays with $\stilde
N_1$ in the final state that seem most likely to be important are shown in
figure~\ref{fig:NCdecays}. 
In many situations, the decays
\beq
\stilde C_1^\pm \rightarrow \ell^\pm \nu \stilde N_1,\qquad\quad
\stilde N_2 \rightarrow \ell^+\ell^- \stilde N_1
\label{eq:CNleptonic}
\eeq
can be particularly important for phenomenology, because the leptons in
the final state might result in clean signals. These decays are more
likely if the intermediate sleptons are relatively light, even if they
cannot be on-shell. Unfortunately, the enhanced mixing of staus, common in
models, may well result in larger branching fractions for both $\stilde
N_2$ and $\tilde C_1$ into final states with taus, rather than electrons
or muons. This is one reason why good tau identification may be very helpful 
in attempts to discover and study supersymmetry. 

In other situations, decays without isolated leptons in the final state
are more useful, so that one will not need to contend with
background events with missing energy coming from leptonic $W$ boson
decays in Standard Model processes. Then the decays of interest
are the ones with quark partons in the final state, leading to
\beq
\stilde C_1 \rightarrow jj \stilde N_1,\qquad\quad
\stilde N_2 \rightarrow jj \stilde N_1,
\label{eq:CNjetdecays}
\eeq
where $j$ means a jet. If the second of these decays goes through an
on-shell $h^0$, then these will usually be $b$-jets that reconstruct an invariant mass consistent with 125 GeV. 

\subsection{Slepton decays}\label{subsec:decays.sleptons}
\setcounter{equation}{0}
\setcounter{footnote}{1}

Sleptons can have two-body decays into a lepton and a chargino or 
neutralino, because of their gaugino admixture, as may be seen directly 
from the couplings in Figures~\ref{fig:gaugino}b,c. Therefore, the 
two-body decays
\beq
\stilde \ell \rightarrow \ell \stilde N_i,\>\>\>\>\>\,
\stilde \ell \rightarrow \nu \stilde C_i,\>\>\>\>\>\,
\stilde \nu \rightarrow \nu \stilde N_i,\>\>\>\>\>\,
\stilde \nu \rightarrow \ell \stilde C_i
\eeq
can be of weak interaction strength. In particular, the direct decays
\beq
\stilde \ell \rightarrow \ell \stilde N_1
\>\>\>\>\>{\rm and}\>\>\>\>\>
\stilde \nu \rightarrow \nu \stilde N_1
\label{sleptonrightdecay}
\eeq
are (almost\footnote{An exception occurs if the mass difference $m_{\tilde
\tau_1} - m_{\tilde N_1}$ is less than $m_{\tau}$.}) always kinematically
allowed if $\stilde N_1$ is the LSP. However, if the sleptons are
sufficiently heavy, then the two-body decays
\beq
\stilde \ell \rightarrow \nu \stilde C_{1}
,\>\>\>\>\>\,
\stilde \ell \rightarrow \ell \stilde N_{2}
,\>\>\>\>\>\,
\stilde \nu \rightarrow \nu \stilde N_{2}
,\>\>\>\>{\rm and}\>\>\>\>
\stilde \nu \rightarrow \ell \stilde C_{1}
\label{sleptonleftdecay}
\eeq
can be important. The right-handed sleptons do not have a coupling to the 
$SU(2)_L$ gauginos, so they typically prefer the direct decay $\stilde 
\ell_R \rightarrow \ell\NI$, if $\NI$ is bino-like. In contrast, the 
left-handed sleptons may prefer to decay as in 
eq.~(\ref{sleptonleftdecay}) rather than the direct decays to the LSP as 
in eq.~(\ref{sleptonrightdecay}), if the former is kinematically open and 
if $\stilde C_1$ and $\stilde N_2$ are mostly wino. This is because the 
slepton-lepton-wino interactions in Figure~\ref{fig:gaugino}b are 
proportional to the $SU(2)_L$ gauge coupling $g$, whereas the 
slepton-lepton-bino interactions in Figure~\ref{fig:gaugino}c are 
proportional to the much smaller $U(1)_Y$ coupling $g^\prime$. Formulas 
for these decay widths can be found in ref.~\cite{epprod}.

\subsection{Squark decays}\label{subsec:decays.squarks}
\setcounter{equation}{0}
\setcounter{footnote}{1}

If the decay $ \stilde q \rightarrow q\stilde g $ is kinematically 
allowed, it will usually dominate, because the quark-squark-gluino vertex 
in Figure~{\ref{fig:gaugino}}a has QCD strength. Otherwise, the squarks 
can decay into a quark plus neutralino or chargino: $ \stilde q 
\rightarrow q \stilde N_i$ or $ q^\prime \stilde C_i $. The direct decay 
to the LSP $\stilde q \rightarrow q \stilde N_1$ is always kinematically 
favored, and for right-handed squarks it can dominate if 
$\stilde N_1$ is mostly bino. However, the left-handed squarks may strongly prefer 
to decay into heavier charginos or neutralinos instead, for example 
$\stilde q \rightarrow q \stilde N_2$ or $q^\prime \stilde C_1$, because 
the relevant squark-quark-wino couplings are much bigger than the 
squark-quark-bino couplings. Squark decays to higgsino-like charginos and 
neutralinos are less important, except in the cases of stops and sbottoms, 
which have sizable Yukawa couplings. The gluino, chargino or neutralino 
resulting from the squark decay will in turn decay, and so on, until a 
final state containing $\stilde N_1$ is reached. This results in 
numerous and complicated decay chain possibilities called cascade decays 
\cite{cascades}.

It is possible that the decays $\stilde t_1 \rightarrow t\stilde g$ and 
$\stilde t_1 \rightarrow t \stilde N_1$ are both kinematically forbidden. 
If so, then the lighter top squark may decay only into charginos, by 
$\stilde t_1 \rightarrow b \stilde C_1$, or by a three-body decay
$\stilde t_1 \rightarrow b W \stilde N_1$. If even this decay is 
kinematically closed, then it has only the flavor-suppressed decay to a 
charm quark, $ \stilde t_1\rightarrow c \stilde N_1$, and the four-body 
decay $ \stilde t_1\rightarrow bff' \stilde N_1 $. These decays can be 
very slow \cite{stoptocharmdecay}, so that the lightest stop can be 
quasi-stable on the time scale relevant for collider physics, and can 
hadronize into bound states.

\subsection{Gluino decays}\label{subsec:decays.gluino}
\setcounter{equation}{0}
\setcounter{footnote}{1}

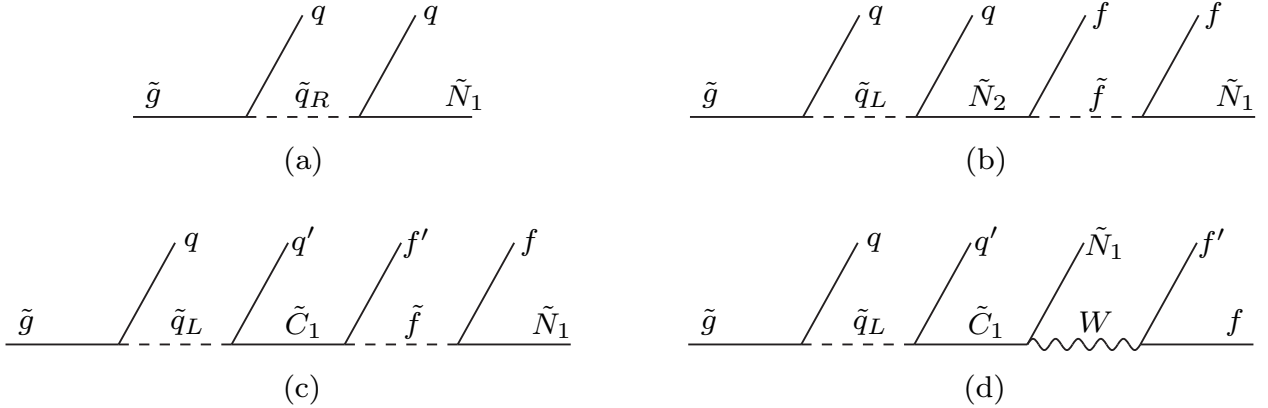
\begin{figure}
\begin{flushleft}
\scalebox{1.42}{
\begin{picture}(24,36)(0,0)
\end{picture}
}
\scalebox{1.42}{
\begin{picture}(138.5,36)(0,0)
\Line(0,0)(30,0)
\Line(30,0)(45,27)
\DashLine(30,0)(60,0){3}
\Line(60,0)(75,27)
\Line(60,0)(90,0)
\rText(0,6)[][]{$\scriptstyle\tilde g$}
\rText(42.5,6)[][]{$\scriptstyle\tilde q_R$}
\rText(44,27)[][]{$\scriptstyle q$}
\rText(74,27)[][]{$\scriptstyle q$}
\rText(83,6)[][]{$\scriptstyle\tilde N_1$}
\rText(40,-12)[][]{$\scriptstyle {\rm (a)}$}
\end{picture}
}
\scalebox{1.42}{
\begin{picture}(120,38)(0,0)
\Line(0,0)(30,0)
\Line(30,0)(45,27)
\DashLine(30,0)(60,0){3}
\Line(60,0)(75,27)
\Line(60,0)(90,0)
\Line(90,0)(105,27)
\DashLine(90,0)(120,0){3}
\Line(120,0)(135,27)
\Line(120,0)(150,0)
\rText(0,6)[][]{$\scriptstyle\tilde g$}
\rText(43,6)[][]{$\scriptstyle\tilde q_L$}
\rText(44,27)[][]{$\scriptstyle q$}
\rText(74,27)[][]{$\scriptstyle q$}
\rText(74,6)[][]{$\scriptstyle\tilde N_2$}
\rText(103,6)[][]{$\scriptstyle\tilde f$}
\rText(104,27)[][]{$\scriptstyle f$}
\rText(134,27)[][]{$\scriptstyle f$}
\rText(140,6)[][]{$\scriptstyle\tilde N_1$}
\rText(73.5,-12)[][]{$\scriptstyle {\rm (b)}$}
\end{picture}
}
\end{flushleft}
\vspace{0.25cm}
\begin{flushleft}
\scalebox{1.42}{
\begin{picture}(172,38)(0,0)
\Line(0,0)(30,0)
\Line(30,0)(45,27)
\DashLine(30,0)(60,0){3}
\Line(60,0)(75,27)
\Line(60,0)(90,0)
\Line(90,0)(105,27)
\DashLine(90,0)(120,0){3}
\Line(120,0)(135,27)
\Line(120,0)(150,0)
\rText(0,6)[][]{$\scriptstyle\tilde g$}
\rText(43,6)[][]{$\scriptstyle\tilde q_L$}
\rText(44,27)[][]{$\scriptstyle q$}
\rText(74,27.5)[][]{$\scriptstyle q'$}
\rText(74,6)[][]{$\scriptstyle\tilde C_1$}
\rText(103,6)[][]{$\scriptstyle\tilde f$}
\rText(104,27)[][]{$\scriptstyle f'$}
\rText(134,27)[][]{$\scriptstyle f$}
\rText(140,6)[][]{$\scriptstyle\tilde N_1$}
\rText(73.5,-12)[][]{$\scriptstyle {\rm (c)}$}
\end{picture}
}
\scalebox{1.42}{
\begin{picture}(150,38)(0,0)
\Line(0,0)(30,0)
\Line(30,0)(45,27)
\DashLine(30,0)(60,0){3}
\Line(60,0)(75,27)
\Line(60,0)(90,0)
\Line(90,0)(105,27)
\Photon(90,0)(120,0){1.5}{5}
\Line(120,0)(135,27)
\Line(120,0)(150,0)
\rText(0,6)[][]{$\scriptstyle\tilde g$}
\rText(43,6)[][]{$\scriptstyle\tilde q_L$}
\rText(44,27)[][]{$\scriptstyle q$}
\rText(74,27.5)[][]{$\scriptstyle q'$}
\rText(74,6)[][]{$\scriptstyle\tilde C_1$}
\rText(103,6)[][]{$\scriptstyle W$}
\rText(105.5,27)[][]{$\scriptstyle\tilde N_1$}
\rText(134,27)[][]{$\scriptstyle f'$}
\rText(140,6)[][]{$\scriptstyle f$}
\rText(73.5,-12)[][]{$\scriptstyle {\rm (d)}$}
\end{picture}
}
\end{flushleft}
\vspace{0.05cm}

\caption{Some of the many possible examples of gluino cascade decays 
ending with a neutralino LSP in the final state. The squarks appearing in 
these diagrams may be either on-shell or off-shell, depending on the mass 
spectrum of the theory.\label{fig:gluinocascades}}
\end{figure}

The decay of the gluino can only proceed through a squark, either on-shell 
or virtual.  If two-body decays $ \stilde g \rightarrow q\stilde q $ are 
open, they will dominate, again because the relevant gluino-quark-squark 
coupling in Figure~\ref{fig:gaugino}a has QCD strength.  Since the top and 
bottom squarks can easily be much lighter than all of the other squarks, 
it is quite possible that $ \stilde g \rightarrow t \stilde t_1$ and/or 
$\stilde g \rightarrow b \stilde b_1$ are the only available two-body 
decay mode(s) for the gluino, in which case they will dominate over all 
others. If instead all of the squarks are heavier than the gluino, the 
gluino will decay only through off-shell squarks, so $ \stilde g 
\rightarrow q q \stilde N_i$ and $ q q^\prime \stilde C_i $. The squarks, 
neutralinos and charginos in these final states will then decay as 
discussed above, so there can be many competing gluino decay chains. Some 
of the possibilities are shown in fig.~\ref{fig:gluinocascades}.
The cascade decays can have final-state branching fractions that are 
individually small and quite sensitive to the model parameters.

The simplest gluino decays, including the ones shown in 
fig.~\ref{fig:gluinocascades}, can have 0, 1, or 2 charged leptons (in 
addition to two or more hadronic jets) in the final state.  An important 
feature is that when there is exactly one charged lepton, it can have 
either charge with exactly equal probability.  This follows from the fact 
that the gluino is a Majorana fermion, and does not ``know" about electric 
charge; for each diagram with a given lepton charge, there is always an 
equal one with every particle replaced by its antiparticle.

\subsection{Decays to the gravitino/goldstino}\label{subsec:decays.gravitino}
\setcounter{equation}{0}
\setcounter{footnote}{1}

Most phenomenological studies of supersymmetry assume explicitly or 
implicitly that the lightest neutralino is the LSP. This is typically the 
case in gravity-mediated models for the soft terms. However, in 
gauge-mediated models (and in ``no-scale" models), the LSP is instead the 
gravitino. As we saw in section \ref{subsec:origins.gravitino}, a very 
light gravitino may be relevant for collider phenomenology, because it 
contains as its longitudinal component the goldstino, which has a 
non-gravitational coupling to all sparticle-particle pairs $(\stilde X, 
X$). The decay rate found in eq.~(\ref{generalgravdecay}) for $\stilde 
X\rightarrow X\G$ is usually not fast enough to compete with the other 
decays of sparticles $\stilde X$ as mentioned above, {\it except} in the 
case that $\stilde X$ is the next-to-lightest supersymmetric particle 
(NLSP). Since the NLSP has no competing decays, it should always decay 
into its superpartner and the LSP gravitino.

In principle, any of the MSSM superpartners could be the NLSP in models 
with a light goldstino, but most models with gauge mediation of 
supersymmetry breaking have either a neutralino or a charged lepton 
playing this role. The argument for this can be seen immediately from 
eqs.~(\ref{gauginogmsbgen}) and (\ref{scalargmsbgen}); since $\alpha_1 < 
\alpha_2,\alpha_3$, those superpartners with only $U(1)_Y$ interactions 
will tend to get the smallest masses. The gauge-eigenstate sparticles with 
this property are the bino and the right-handed sleptons $\stilde e_R$, 
$\stilde \mu_R$, $\stilde \tau_R$, so the appropriate corresponding mass 
eigenstates should be plausible candidates for the NLSP.

First suppose that $\stilde N_1$ is the NLSP in light goldstino models. 
Since $\stilde N_1$ contains an admixture of the photino (the linear 
combination of bino and neutral wino whose superpartner is the photon), 
from eq.~(\ref{generalgravdecay}) it decays into photon + 
goldstino/gravitino with a partial width
\beq
\Gamma (\NI \rightarrow \gamma \G ) \,=\,
2\times 10^{-3} \> \kappa_{1\gamma}\left ({m_{\NI}\over 100\>\rm{
GeV}}\right )^5
\left ( {\sqrt{\langle F \rangle}\over 100\>{\rm TeV}} \right )^{-4} \>
{\rm eV}.\qquad{}
\label{neutralinodecaywidth}
\eeq
Here $\kappa_{1\gamma} \equiv |{\bf N}_{11}\cos\theta_W + {\bf N}_{12}\sin 
\theta_W |^2$ is the ``photino content" of $\stilde N_1$, in terms of the 
neutralino mixing matrix ${\bf N}_{ij}$ defined by eq.~(\ref{diagmN}). We 
have normalized $m_{\NI}$ and $\sqrt{\langle F \rangle}$ to (very roughly) 
minimum expected values in gauge-mediated models. This width is much 
smaller than for a typical flavor-unsuppressed weak interaction decay, but 
it is still large enough to allow $\stilde N_1$ to decay before it has 
left a collider detector, if $\sqrt{\langle F\rangle}$ is less than a few 
thousand TeV in gauge-mediated models, or equivalently if $m_{3/2}$ is 
less than a keV or so when eq.~(\ref{gravitinomass}) holds. In fact, from 
eq.~(\ref{neutralinodecaywidth}), the mean decay length of an $\NI$ with 
energy $E$ in the lab frame is
\beq
d = 9.9 \times 10^{-3}\> {1\over \kappa_{1\gamma}}\,
({E^2/ m_{\NI}^2} - 1)^{1/2}
\left ({m_{\NI}\over 100\>\rm{ GeV}}\right )^{-5}
\left({\sqrt{\langle F \rangle}\over 100\>{\rm TeV}} \right )^{4}
\>{\rm cm},
\label{neutralinodecaylength}
\eeq
which could be anything from sub-micron to multi-kilometer, depending on 
the scale of supersymmetry breaking $\sqrt{\langle F \rangle}$. (In other 
models that have a gravitino LSP, including certain ``no-scale" models 
\cite{noscalephotons}, the same formulas apply with ${\langle F \rangle} 
\rightarrow \sqrt{3} m_{3/2} \MPlanck$.)

Of course, $\stilde N_1$ is not a pure photino, but contains also 
admixtures of the superpartner of the $Z$ boson and the neutral Higgs 
scalars. So, one can also have \cite{DDRT} $\NI\rightarrow Z\G$, $h^0\G$, 
$A^0\G$, or $H^0\G$, with decay widths given in ref.~\cite{AKKMM2}. Of 
these decays, the last two are unlikely to be kinematically allowed, and 
only the $\NI \rightarrow \gamma\G$ mode is guaranteed to be kinematically 
allowed for a gravitino LSP. Furthermore, even if they are open, the 
decays $\stilde N_1 \rightarrow Z\G$ and $\stilde N_1 \rightarrow h^0 \G$ 
are subject to strong kinematic suppressions proportional to 
$(1-m_Z^2/m_{\stilde N_1}^2)^4$ and $(1 - m_{h^0}^2/m_{\stilde N_1}^2)^4$, 
respectively, in view of eq.~(\ref{generalgravdecay}). Still, these decays 
may play an important role in phenomenology if ${\langle F\rangle }$ is 
not too large, $\stilde N_1$ has a sizable zino or higgsino content, and 
$m_{\stilde N_1}$ is significantly greater than $m_Z$ or $m_{h^0}$.

A charged slepton makes another likely candidate for the NLSP. Actually, 
more than one slepton can act effectively as 
the NLSP, even though one of them is slightly lighter, if they are 
sufficiently close in mass so that each has no kinematically allowed 
decays except to the goldstino. In GMSB models, the squared masses 
obtained by $\widetilde e_R$, $\widetilde \mu_R$ and $\widetilde \tau_R$ 
are equal because of the flavor-blindness of the gauge couplings. However, 
this is not the whole story, because one must take into account mixing 
with $\widetilde e_L$, $\widetilde \mu_L$, and $\widetilde \tau_L$ and 
renormalization group running.  These effects are very small for 
$\widetilde e_R$ and $\widetilde \mu_R$ because of the tiny electron and 
muon Yukawa couplings, so we can quite generally treat them as degenerate, 
unmixed mass eigenstates. In contrast, $\widetilde \tau_R$ usually has a 
quite significant mixing with $\widetilde \tau_L$, proportional to the tau 
Yukawa coupling. This means that the lighter stau mass eigenstate 
$\widetilde \tau_1$ is pushed lower in mass than $\widetilde e_R$ or 
$\widetilde \mu_R$, by an amount that depends most strongly on 
$\tan\beta$.  If $\tan\beta$ is not too large then the stau mixing effect 
leaves the slepton mass eigenstates $\widetilde e_R$, $\widetilde \mu_R$, 
and $\widetilde \tau_1$ degenerate to within less than $m_\tau \approx 1.8 
$ GeV, so they act effectively as co-NLSPs.  In particular, this means 
that even though the stau is slightly lighter, the three-body slepton 
decays $\widetilde e_R \rightarrow e\tau^\pm\widetilde \tau_1^\mp$ and 
$\widetilde \mu_R \rightarrow \mu\tau^\pm\widetilde \tau_1^\mp$ are not 
kinematically allowed; the only allowed decays for the three lightest 
sleptons are $\widetilde e_R\rightarrow e \G$ and $\widetilde \mu_R 
\rightarrow \mu\G$ and $\widetilde \tau_1 \rightarrow \tau \G$. This 
situation is called the ``slepton co-NLSP" scenario.

For larger values of $\tan\beta$, the lighter stau eigenstate $\stilde 
\tau_1$ is more than $1.8$ GeV lighter than $\widetilde e_R$ and 
$\widetilde \mu_R$ and $\NI$.  This means that the decays $\NI \rightarrow 
\tau\stilde \tau_1$ and $\widetilde e_R \rightarrow e \tau \stilde \tau_1$ 
and $\widetilde \mu_R \rightarrow \mu \tau \stilde\tau_1$ are open. Then 
$\widetilde \tau_1$ is the sole NLSP, with all other MSSM supersymmetric 
particles having kinematically allowed decays into it. This is called the 
``stau NLSP" scenario.

In any case, a slepton NLSP can decay like $\stilde \ell \rightarrow \ell 
\G$ according to eq.~(\ref{generalgravdecay}), with a width and decay 
length just given by eqs.~(\ref{neutralinodecaywidth}) and 
(\ref{neutralinodecaylength}) with the replacements $\kappa_{1\gamma} 
\rightarrow 1$ and $m_{\stilde N_1} \rightarrow m_{\stilde \ell}$. So, as 
for the neutralino NLSP case, the decay $\stilde \ell \rightarrow \ell\G$ 
can be either fast or very slow, depending on the scale of supersymmetry 
breaking.

If $\sqrt{\langle F \rangle}$ is larger than roughly $10^3$ TeV (or the 
gravitino is heavier than a keV or so), then the NLSP is so long-lived 
that it will usually escape a typical collider detector. If $\NI$ is the 
NLSP, then, it might as well be the LSP from the point of view of collider 
physics. However, the decay of $\NI$ into the gravitino is still important 
for cosmology, since an unstable $\NI$ is clearly not a good dark matter 
candidate while the gravitino LSP conceivably could be. On the other hand, 
if the NLSP is a long-lived charged slepton, then one can see its tracks 
(or possibly decay kinks) inside a collider detector \cite{DDRT}. The 
presence of a massive charged NLSP can be established by measuring an 
anomalously long time-of-flight or high ionization rate for a track in the 
detector.

\section{Experimental signals for supersymmetry}\label{sec:signals}
\setcounter{equation}{0}
\setcounter{figure}{0}
\setcounter{table}{0}
\setcounter{footnote}{1}

So far, the experimental study of supersymmetry has unfortunately been 
confined to setting limits. As we have already remarked in section 
\ref{subsec:mssm.hints}, there can be indirect signals for supersymmetry 
from processes that are rare or forbidden in the Standard Model but have 
contributions from sparticle loops. These include $\mu\rightarrow 
e\gamma$, $b\rightarrow s\gamma$, neutral meson mixing, electric dipole 
moments for the neutron and the electron, etc. There are also virtual 
sparticle effects on Standard Model predictions like $R_b$ (the fraction 
of hadronic $Z$ decays with $b\overline b$ pairs) \cite{Rb} and the 
anomalous magnetic moment of the muon \cite{muonmoment}, which 
exclude some models that would otherwise be viable. 
Extensions of the MSSM (including, but not limited, to GUTs) 
can quite easily predict proton decay and neutron-antineutron 
oscillations at potentially observable rates, even if $R$-parity is exactly 
conserved. However, it would be impossible to ascribe a positive result 
for any of these processes to supersymmetry in an unambiguous way. There 
is no substitute for the direct detection of sparticles and verification 
of their quantum numbers and interactions. In this section we will give an 
incomplete and qualitative review of some of the possible signals for 
direct detection of supersymmetry. LHC data and analyses 
are presently advancing this subject at a very high rate, so that any detailed and specific discussion would be obsolete on a time scale of weeks or months. The most recent
experimental results from the LHC are available at the 
ATLAS and CMS physics results web pages.

\subsection{Signals at hadron colliders}\label{subsec:signals.TEVLHC}
\setcounter{equation}{0}
\setcounter{footnote}{1}

At this writing, the CERN Large Hadron Collider (LHC) 
has already excluded significant chunks of supersymmetric parameter space,
based on proton-proton collisions amounting to about 5 fb$^{-1}$ at $\sqrt{s} = 7$ TeV,
20 fb$^{-1}$ at $\sqrt{s} = 8$ TeV, and 4 fb$^{-1}$ at $\sqrt{s} = 13$ TeV.
In many MSUGRA and similar models, gluinos and
squarks with masses well above 1 TeV are already excluded by LHC data, 
superseding the results
from the CDF and D$\emptyset$ detectors at the 
Fermilab Tevatron $p\overline p$ 
collider with $\sqrt{s} = 1.96$ TeV. 
Future planned increases in LHC integrated luminosity suggest that 
if supersymmetry is the solution to the hierarchy problem 
discussed in the Introduction, then the LHC 
has a good chance of finding direct evidence for it within the next few years.

At hadron colliders, sparticles can be produced in pairs from parton 
collisions of electroweak strength:
\beq
q \overline q^{\phantom '}\! &\rightarrow & \stilde  
C_i^+ \stilde C_j^-, 
\>\>
\stilde N_i \stilde N_j,
\qquad\quad
u \overline d \>\rightarrow\>  \stilde C_i^+ \stilde N_j,
\qquad\quad
d \overline u \>\rightarrow\>  \stilde C_i^- \stilde N_j,
\label{eq:qqbarinos}
\\
q \overline q^{\phantom '}\! &\rightarrow & \stilde \ell^+_i \stilde \ell^-_j,
\>\>\>
\stilde \nu_\ell \stilde \nu^*_\ell
\qquad\qquad\>\>
u \overline d \>\rightarrow\>  \stilde \ell^+_L \stilde \nu_\ell
\qquad\qquad\>
d \overline u \>\rightarrow\>  \stilde \ell^-_L \stilde \nu^*_\ell,
\label{eq:qqbarsleptons}
\eeq
as shown in fig.~\ref{fig:qqbarsusy}, and reactions of QCD strength:
\beq
gg &\rightarrow & \stilde g \stilde g, 
\>\>\,
\stilde q_i \stilde q_j^*,
\label{eq:gluegluegluinos}
\\
gq &\rightarrow & \stilde g \stilde q_i,
\label{eq:gluequarkgluinosquark}
\\
q \overline q &\rightarrow& \stilde g \stilde g, 
\>\>\,
\stilde q_i \stilde q_j^*,
\label{eq:qqbargluinosorsquarks}
\\
q q &\rightarrow& \stilde q_i \stilde q_j,
\label{eq:qqsquarks}
\eeq
as shown in figs.~\ref{fig:ggsusy} and \ref{fig:qqsusy}.
\begin{figure}[p]
\begin{center}
\begin{picture}(120,40)(0,13)
\SetWidth{0.85}
\Line(0,50)(33,25)
\Line(0,0)(33,25)
\Photon(33,25)(77,25){2.1}{4.5}
\Line(110,50)(77,25)
\Line(110,0)(77,25)
\rText(-8.5,43.2)[][]{$q$}
\rText(-8.5,8.5)[][]{$\overline q$}
\rText(50,36)[c][]{$\gamma,Z$}
\rText(113,44)[][]{$\stilde C_i^+$}
\rText(113,7)[][]{$\stilde C_j^-$}
\end{picture}
\hspace{1.5cm}
\begin{picture}(120,40)(0,13)
\SetWidth{0.85}
\Line(0,50)(55,50)
\Line(0,0)(55,0)
\Line(110,50)(55,50)
\Line(110,0)(55,0)
\DashLine(55,0)(55,50){5}
\rText(-6,43.2)[][]{$u$}
\rText(-6,6)[][]{$\overline u$}
\rText(53.2,27)[r][]{$\stilde d_L$}
\rText(113,45)[][]{$\stilde C_i^+$}
\rText(113,7)[][]{$\stilde C_j^-$}
\SetWidth{0.4}
\Photon(110,50)(55,50){-2.1}{5.5}
\Photon(110,0)(55,0){2.1}{5.5}
\end{picture}
\hspace{1.5cm}
\begin{picture}(120,40)(0,13)
\SetWidth{0.85}
\Line(0,50)(55,50)
\Line(0,0)(55,0)
\Line(110,0)(55,50)
\Line(79.75,22.5)(55,0)
\Line(110,50)(85.25,27.5)
\DashLine(55,0)(55,50){5}
\rText(-6,43)[][]{$d$}
\rText(-6,7.5)[][]{$\overline d$}
\rText(53.2,27)[r][]{$\stilde u_L$}
\rText(113,45)[][]{$\stilde C_i^+$}
\rText(113,7)[][]{$\stilde C_j^-$}
\SetWidth{0.4}
\Photon(110,0)(55,50){2.1}{7}
\Photon(79.75,22.5)(55,0){2.1}{3.5}
\Photon(110,50)(85.25,27.5){-2.1}{3.5}
\end{picture}
\end{center}
\vspace{0.02cm}
\begin{center}
\begin{picture}(120,52.5)(0,13)
\SetWidth{0.85}
\Line(0,50)(33,25)
\Line(0,0)(33,25)
\Photon(33,25)(77,25){2.1}{4.5}
\Line(110,50)(77,25)
\Line(110,0)(77,25)
\rText(-8.5,43.2)[][]{$q$}
\rText(-8.5,8.5)[][]{$\overline q$}
\rText(50,36)[c][]{$Z$}
\rText(113,44)[][]{$\stilde N_i$}
\rText(113,7)[][]{$\stilde N_j$}
\end{picture}
\hspace{1.5cm}
\begin{picture}(120,52.5)(0,13)
\SetWidth{0.85}
\Line(0,50)(55,50)
\Line(0,0)(55,0)
\Line(110,50)(55,50)
\Line(110,0)(55,0)
\DashLine(55,0)(55,50){5}
\rText(-6,43.2)[][]{$q$}
\rText(-6,7.2)[][]{$\overline q$}
\rText(50,27)[r][]{$\stilde q_{L,R}$}
\rText(113,45)[][]{$\stilde N_i$}
\rText(113,7)[][]{$\stilde N_j$}
\SetWidth{0.4}
\Photon(110,50)(55,50){-2.1}{5.5}
\Photon(110,0)(55,0){2.1}{5.5}
\end{picture}
\hspace{1.5cm}
\begin{picture}(120,52.5)(0,13)
\SetWidth{0.85}
\Line(0,50)(55,50)
\Line(0,0)(55,0)
\Line(110,0)(55,50)
\Line(79.75,22.5)(55,0)
\Line(110,50)(85.25,27.5)
\DashLine(55,0)(55,50){5}
\rText(-6,43)[][]{$q$}
\rText(-6,7.2)[][]{$\overline q$}
\rText(50,27)[r][]{$\stilde q_{L,R}$}
\rText(113,45)[][]{$\stilde N_i$}
\rText(113,7)[][]{$\stilde N_j$}
\SetWidth{0.4}
\Photon(110,0)(55,50){2.1}{7}
\Photon(79.75,22.5)(55,0){2.1}{3.5}
\Photon(110,50)(85.25,27.5){-2.1}{3.5}
\end{picture}
\end{center}
\vspace{0.02cm}
\begin{center}
\begin{picture}(120,52.5)(0,13)
\SetWidth{0.85}
\Line(0,50)(33,25)
\Line(0,0)(33,25)
\Photon(33,25)(77,25){2.1}{4.5}
\Line(110,50)(77,25)
\Line(110,0)(77,25)
\rText(-8.5,43.2)[][]{$u$}
\rText(-8.5,8.5)[][]{$\overline d$}
\rText(50,36)[c][]{$W^+$}
\rText(113,44)[][]{$\stilde C_i^+$}
\rText(113,7)[][]{$\stilde N_j$}
\end{picture}
\hspace{1.5cm}
\begin{picture}(120,52.5)(0,13)
\SetWidth{0.85}
\Line(0,50)(55,50)
\Line(0,0)(55,0)
\Line(110,50)(55,50)
\Line(110,0)(55,0)
\DashLine(55,0)(55,50){5}
\rText(-6,43.2)[][]{$u$}
\rText(-6,7.2)[][]{$\overline d$}
\rText(50,27)[r][]{$\stilde d_L$}
\rText(113,45)[][]{$\stilde C_i^+$}
\rText(113,7)[][]{$\stilde N_j$}
\SetWidth{0.4}
\Photon(110,50)(55,50){-2.1}{5.5}
\Photon(110,0)(55,0){2.1}{5.5}
\end{picture}
\hspace{1.5cm}
\begin{picture}(120,52.5)(0,13)
\SetWidth{0.85}
\Line(0,50)(55,50)
\Line(0,0)(55,0)
\Line(110,0)(55,50)
\Line(79.75,22.5)(55,0)
\Line(110,50)(85.25,27.5)
\DashLine(55,0)(55,50){5}
\rText(-6,43)[][]{$u$}
\rText(-6,7.2)[][]{$\overline d$}
\rText(50,27)[r][]{$\stilde u_L$}
\rText(113,45)[][]{$\stilde C_i^+$}
\rText(113,7)[][]{$\stilde N_j$}
\SetWidth{0.4}
\Photon(110,0)(55,50){2.1}{7}
\Photon(79.75,22.5)(55,0){2.1}{3.5}
\Photon(110,50)(85.25,27.5){-2.1}{3.5}
\end{picture}
\end{center}
\vspace{0.02cm}
\begin{center}
\begin{picture}(120,52.5)(0,13)
\SetWidth{0.85}
\Line(0,50)(33,25)
\Line(0,0)(33,25)
\Photon(33,25)(77,25){2.1}{4.5}
\Line(110,50)(77,25)
\Line(110,0)(77,25)
\rText(-8.5,43.2)[][]{$q$}
\rText(-8.5,8.5)[][]{$\overline q$}
\rText(50,36)[c][]{$\gamma,Z$}
\rText(113,44)[][]{$\stilde \ell_i^+$}
\rText(113,7)[][]{$\stilde \ell_j^-$}
\end{picture}
\hspace{1.5cm}
\begin{picture}(120,52.5)(0,13)
\SetWidth{0.85}
\Line(0,50)(33,25)
\Line(0,0)(33,25)
\Photon(33,25)(77,25){2.1}{4.5}
\Line(110,50)(77,25)
\Line(110,0)(77,25)
\rText(-8.5,43.2)[][]{$q$}
\rText(-8.5,8.5)[][]{$\overline q$}
\rText(50,36)[c][]{$Z$}
\rText(113,44)[][]{$\stilde \nu$}
\rText(113,7)[][]{$\stilde \nu^*$}
\end{picture}
\hspace{1.5cm}
\begin{picture}(120,52.5)(0,13)
\SetWidth{0.85}
\Line(0,50)(33,25)
\Line(0,0)(33,25)
\Photon(33,25)(77,25){2.1}{4.5}
\Line(110,50)(77,25)
\Line(110,0)(77,25)
\rText(-8.5,43.2)[][]{$u$}
\rText(-8.5,8.5)[][]{$\overline d$}
\rText(50,36)[c][]{$W^+$}
\rText(113,44)[][]{$\stilde \ell_i^+$}
\rText(113,7)[][]{$\stilde \nu_\ell$}
\end{picture}
\end{center}
\caption{Feynman diagrams for electroweak 
production of sparticles at hadron colliders
from quark-antiquark annihilation. The charginos and neutralinos
in the $t$-channel diagrams only couple because of their gaugino
content, for massless initial-state quarks, and so are drawn as
wavy lines superimposed on solid.
\label{fig:qqbarsusy}} 
\end{figure}
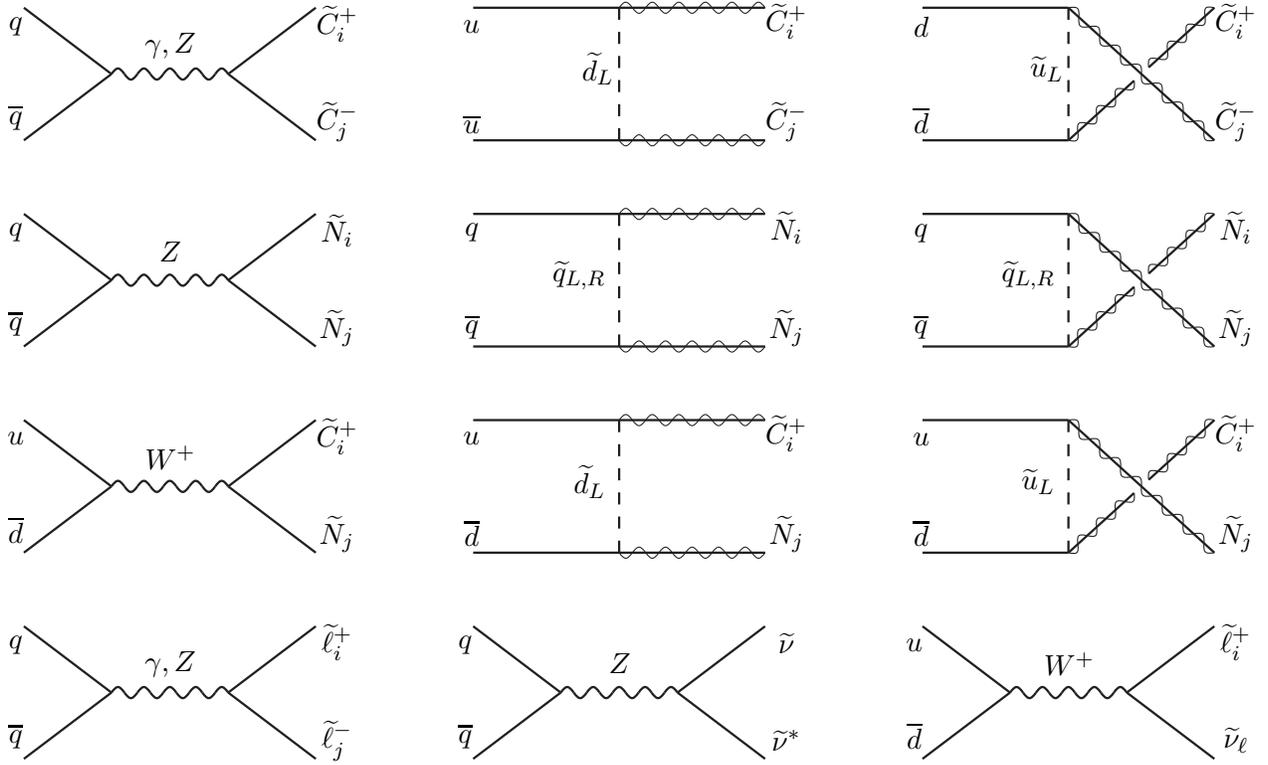
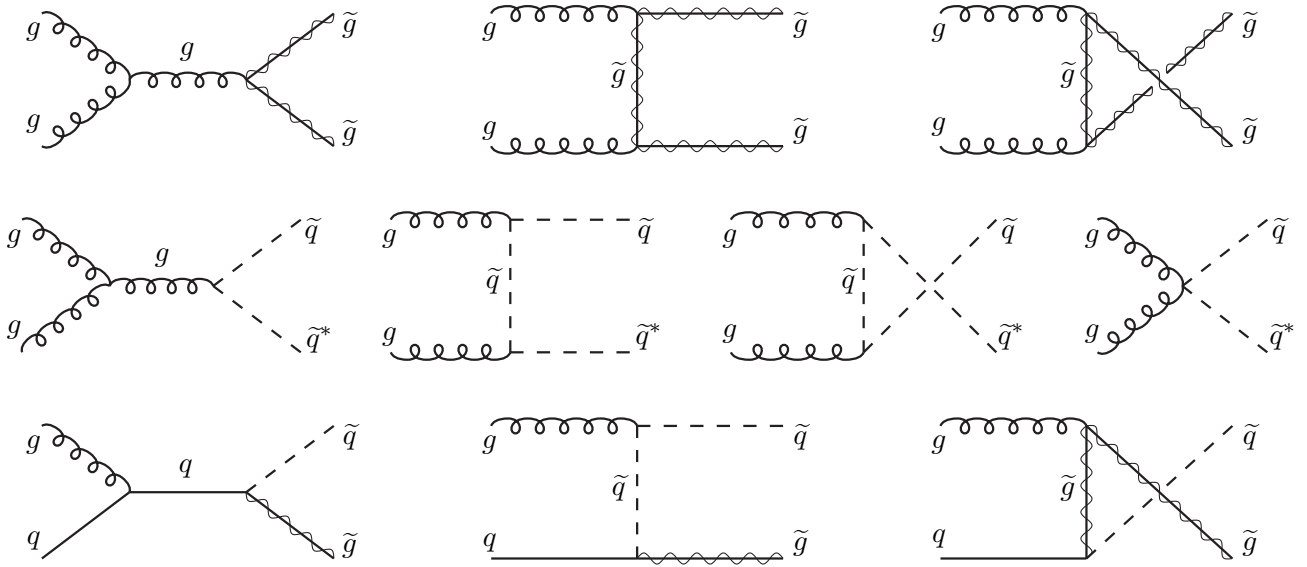
\begin{figure}
\begin{center}
\begin{picture}(120,52.5)(0,13)
\SetWidth{0.85}
\Gluon(0,50)(33,25){2.7}{4}
\Gluon(0,0)(33,25){-2.7}{4}
\Gluon(33,25)(77,25){2.7}{4}
\Line(110,50)(77,25)
\Line(110,0)(77,25)
\rText(-8.5,43.2)[][]{$g$}
\rText(-8.5,8.5)[][]{$g$}
\rText(49.5,36)[c][]{$g$}
\rText(111,46)[][]{$\stilde g$}
\rText(111,5)[][]{$\stilde g$}
\SetWidth{0.4}
\Photon(110,50)(77,25){2.1}{4}
\Photon(110,0)(77,25){-2.1}{4}
\end{picture}
\hspace{1.5cm}
\begin{picture}(120,52.5)(0,13)
\SetWidth{0.85}
\Gluon(0,50)(55,50){2.8}{5}
\Gluon(0,0)(55,0){-2.8}{5}
\Line(110,50)(55,50)
\Line(110,0)(55,0)
\Line(55,0)(55,50)
\rText(-6,43.2)[][]{$g$}
\rText(-6,6)[][]{$g$}
\rText(51,27)[r][]{$\stilde g$}
\rText(111.5,46)[][]{$\stilde g$}
\rText(111.5,6)[][]{$\stilde g$}
\SetWidth{0.4}
\Photon(110,50)(55,50){-2.1}{5.5}
\Photon(110,0)(55,0){2.1}{5.5}
\Photon(55,0)(55,50){2.1}{5}
\end{picture}
\hspace{1.5cm}
\begin{picture}(120,52.5)(0,13)
\SetWidth{0.85}
\Gluon(0,50)(55,50){2.8}{5}
\Gluon(0,0)(55,0){-2.8}{5}
\Line(110,0)(55,50)
\Line(79.75,22.5)(55,0)
\Line(110,50)(85.25,27.5)
\Line(55,0)(55,50)
\rText(-6,43)[][]{$g$}
\rText(-6,7.5)[][]{$g$}
\rText(51,27)[r][]{$\stilde g$}
\rText(111.5,46)[][]{$\stilde g$}
\rText(111.5,6)[][]{$\stilde g$}
\SetWidth{0.4}
\Photon(55,0)(55,50){-2.1}{5}
\Photon(110,0)(55,50){2.1}{7}
\Photon(79.75,22.5)(55,0){2.1}{3.5}
\Photon(110,50)(85.25,27.5){-2.1}{3.5}
\end{picture}
\end{center}
\vspace{0.02cm}
\begin{center}
\begin{picture}(112,52.5)(-4,13)
\SetWidth{0.85}
\Gluon(0,50)(33,25){2.7}{4}
\Gluon(0,0)(33,25){2.7}{4}
\Gluon(33,25)(72,25){2.7}{4}
\DashLine(105,50)(72,25){5}
\DashLine(105,0)(72,25){5}
\rText(-8.5,43.2)[][]{$g$}
\rText(-8.5,8.5)[][]{$g$}
\rText(47.5,36)[c][]{$g$}
\rText(104,46)[][]{$\stilde q$}
\rText(107,5)[][]{$\stilde q^*$}
\end{picture}
\hspace{0.85cm}
\begin{picture}(100,52.5)(0,13)
\SetWidth{0.85}
\Gluon(0,50)(45,50){2.8}{4}
\Gluon(0,0)(45,0){-2.8}{4}
\DashLine(90,50)(45,50){5}
\DashLine(90,0)(45,0){5}
\DashLine(45,0)(45,50){5}
\rText(-6,43.2)[][]{$g$}
\rText(-6,6)[][]{$g$}
\rText(41.2,27)[r][]{$\stilde q$}
\rText(90.4,46)[][]{$\stilde q$}
\rText(92,6)[][]{$\stilde q^*$}
\end{picture}
\hspace{0.75cm}
\begin{picture}(110,52.5)(0,13)
\SetWidth{0.85}
\Gluon(0,50)(50,50){2.8}{4}
\Gluon(0,0)(50,0){-2.8}{4}
\DashLine(100,0)(50,50){5}
\DashLine(100,50)(50,0){5}
\DashLine(50,0)(50,50){5}
\rText(-6,43)[][]{$g$}
\rText(-6,7)[][]{$g$}
\rText(47,27)[r][]{$\stilde q$}
\rText(99.3,46)[][]{$\stilde q$}
\rText(100.5,6)[][]{$\stilde q^*$}
\end{picture}
\hspace{0.75cm}
\begin{picture}(72,52.5)(0,13)
\SetWidth{0.85}
\Gluon(0,50)(32,25){2.8}{4}
\Gluon(0,0)(32,25){-2.8}{4}
\DashLine(64,0)(32,25){5}
\DashLine(64,50)(32,25){5}
\rText(-7.5,43)[][]{$g$}
\rText(-7.5,7.2)[][]{$g$}
\rText(63.2,46)[][]{$\stilde q$}
\rText(64.5,6)[][]{$\stilde q^*$}
\end{picture}
\end{center}
\vspace{0.02cm}
\begin{center}
\begin{picture}(120,52.5)(0,13)
\SetWidth{0.85}
\Gluon(0,50)(33,25){2.7}{4}
\Line(0,0)(33,25)
\Line(33,25)(77,25)
\DashLine(110,50)(77,25){5}
\Line(110,0)(77,25)
\rText(-8.5,43.2)[][]{$g$}
\rText(-8.5,7)[][]{$q$}
\rText(49,34)[c][]{$q$}
\rText(111,46)[][]{$\stilde q$}
\rText(111,5)[][]{$\stilde g$}
\SetWidth{0.4}
\Photon(110,0)(77,25){-2.1}{4}
\end{picture}
\hspace{1.5cm}
\begin{picture}(120,52.5)(0,13)
\SetWidth{0.85}
\Gluon(0,50)(55,50){2.8}{5}
\Line(0,0)(55,0)
\Line(110,0)(55,0)
\DashLine(110,50)(55,50){5}
\DashLine(55,0)(55,50){5}
\rText(-6,43.2)[][]{$g$}
\rText(-6,6)[][]{$q$}
\rText(51,27)[r][]{$\stilde q$}
\rText(111.5,46)[][]{$\stilde q$}
\rText(111.5,6)[][]{$\stilde g$}
\SetWidth{0.4}
\Photon(110,0)(55,0){-2.1}{5.5}
\end{picture}
\hspace{1.5cm}
\begin{picture}(120,52.5)(0,13)
\SetWidth{0.85}
\Gluon(0,50)(55,50){2.8}{5}
\Line(0,0)(55,0)
\Line(55,0)(55,50)
\Line(110,0)(55,50)
\DashLine(79.75,22.5)(55,0){5}
\DashLine(110,50)(85.25,27.5){5}
\rText(-6,43)[][]{$g$}
\rText(-6,7.5)[][]{$q$}
\rText(51,27)[r][]{$\stilde g$}
\rText(111.5,46)[][]{$\stilde q$}
\rText(111.5,6)[][]{$\stilde g$}
\SetWidth{0.4}
\Photon(55,0)(55,50){-2.1}{5}
\Photon(110,0)(55,50){2.1}{7}
\end{picture}
\end{center}
\caption{Feynman diagrams for gluino and squark production
at hadron colliders from gluon-gluon and gluon-quark 
fusion.\label{fig:ggsusy}} 
\end{figure}
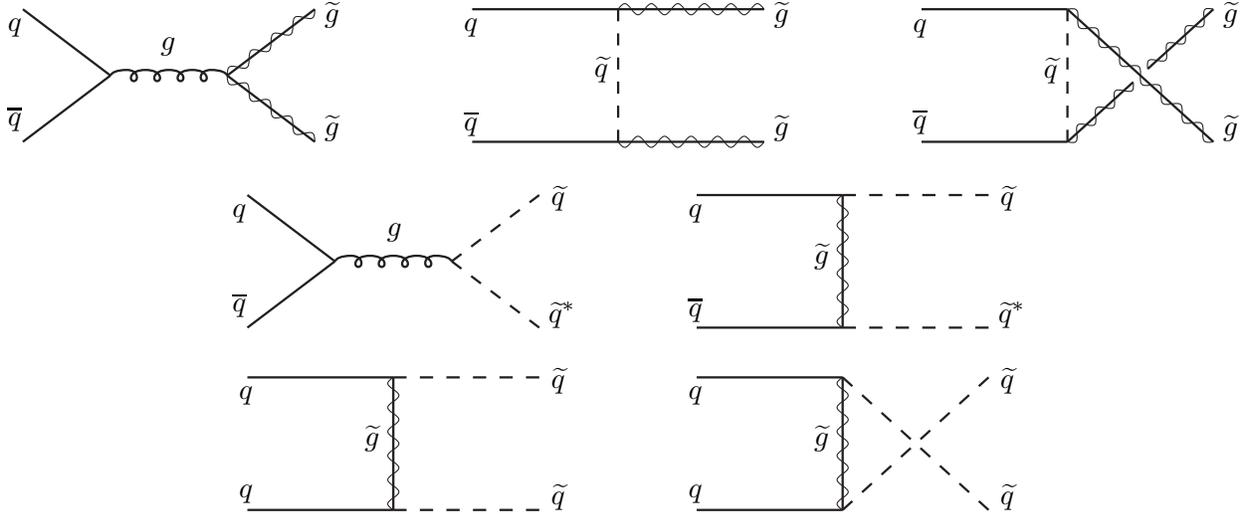
\begin{figure}[!t]
\begin{center}
\begin{picture}(120,42)(0,13)
\SetWidth{0.85}
\Line(0,50)(33,25)
\Line(0,0)(33,25)
\Gluon(33,25)(77,25){2.1}{4.5}
\Line(110,50)(77,25)
\Line(110,0)(77,25)
\rText(-8.5,43.2)[][]{$q$}
\rText(-8.5,8.5)[][]{$\overline q$}
\rText(49.5,36)[c][]{$g$}
\rText(111.5,48)[][]{$\stilde g$}
\rText(111.5,5)[][]{$\stilde g$}
\SetWidth{0.4}
\Photon(110,50)(77,25){2.1}{4}
\Photon(110,0)(77,25){-2.1}{4}
\end{picture}
\hspace{1.5cm}
\begin{picture}(120,42)(0,13)
\SetWidth{0.85}
\Line(0,50)(55,50)
\Line(0,0)(55,0)
\Line(110,50)(55,50)
\Line(110,0)(55,0)
\DashLine(55,0)(55,50){5}
\rText(-6,43.2)[][]{$q$}
\rText(-6,6.7)[][]{$\overline q$}
\rText(51.5,27.6)[r][]{$\stilde q$}
\rText(111.5,48)[][]{$\stilde g$}
\rText(111.5,5)[][]{$\stilde g$}
\SetWidth{0.4}
\Photon(110,50)(55,50){-2.1}{5.5}
\Photon(110,0)(55,0){2.1}{5.5}
\end{picture}
\hspace{1.5cm}
\begin{picture}(120,42)(0,13)
\SetWidth{0.85}
\Line(0,50)(55,50)
\Line(0,0)(55,0)
\Line(110,0)(55,50)
\Line(79.75,22.5)(55,0)
\Line(110,50)(85.25,27.5)
\DashLine(55,0)(55,50){5}
\rText(-6,43)[][]{$q$}
\rText(-6,7.5)[][]{$\overline q$}
\rText(51.5,27)[r][]{$\stilde q$}
\rText(111.5,48)[][]{$\stilde g$}
\rText(111.5,5)[][]{$\stilde g$}
\SetWidth{0.4}
\Photon(110,0)(55,50){2.1}{7}
\Photon(79.75,22.5)(55,0){2.1}{3.5}
\Photon(110,50)(85.25,27.5){-2.1}{3.5}
\end{picture}
\end{center}
\vspace{-0.25cm}
\begin{center}
\begin{picture}(120,52.5)(0,13)
\SetWidth{0.85}
\Line(0,50)(33,25)
\Line(0,0)(33,25)
\Gluon(33,25)(77,25){2.1}{4.5}
\DashLine(110,50)(77,25){5}
\DashLine(110,0)(77,25){5}
\rText(-8.5,43.2)[][]{$q$}
\rText(-8.5,8.5)[][]{$\overline q$}
\rText(50,36)[c][]{$g$}
\rText(112,49)[][]{$\stilde q$}
\rText(113.5,5)[][]{$\stilde q^*$}
\end{picture}
\hspace{1.5cm}
\begin{picture}(120,52.5)(0,13)
\SetWidth{0.85}
\Line(0,50)(55,50)
\Line(0,0)(55,0)
\DashLine(110,50)(55,50){5}
\DashLine(110,0)(55,0){5}
\Line(55,0)(55,50)
\rText(-6,43.2)[][]{$q$}
\rText(-6,6.6)[][]{$\overline q$}
\rText(50,27)[r][]{$\stilde g$}
\rText(112,49)[][]{$\stilde q$}
\rText(113.5,5)[][]{$\stilde q^*$}
\SetWidth{0.4}
\Photon(55,0)(55,50){2.1}{5.5}
\end{picture}
\end{center}
\vspace{-0.3cm}
\begin{center}
\begin{picture}(120,52.5)(0,13)
\SetWidth{0.85}
\Line(0,50)(55,50)
\Line(0,0)(55,0)
\DashLine(110,50)(55,50){5}
\DashLine(110,0)(55,0){5}
\Line(55,0)(55,50)
\rText(-6,43.2)[][]{$q$}
\rText(-6,6)[][]{$q$}
\rText(50,27)[r][]{$\stilde g$}
\rText(112,49)[][]{$\stilde q$}
\rText(112,5)[][]{$\stilde q$}
\SetWidth{0.4}
\Photon(55,0)(55,50){2.1}{5.5}
\end{picture}
\hspace{1.5cm}
\begin{picture}(120,52.5)(0,13)
\SetWidth{0.85}
\Line(0,50)(55,50)
\Line(0,0)(55,0)
\DashLine(110,0)(55,50){5}
\DashLine(110,50)(55,0){5}
\Line(55,0)(55,50)
\rText(-6,43.2)[][]{$q$}
\rText(-6,6)[][]{$q$}
\rText(50,27)[r][]{$\stilde g$}
\rText(112,49)[][]{$\stilde q$}
\rText(112,5)[][]{$\stilde q$}
\SetWidth{0.4}
\Photon(55,0)(55,50){2.1}{5.5}
\end{picture}
\end{center}
\caption{Feynman diagrams for gluino and squark production at hadron 
colliders from strong quark-antiquark annihilation and quark-quark scattering.
\label{fig:qqsusy}}
\end{figure}%
The reactions in (\ref{eq:qqbarinos}) and (\ref{eq:qqbarsleptons}) get 
contributions from electroweak vector bosons in the $s$-channel, and those 
in (\ref{eq:qqbarinos}) also have $t$-channel squark-exchange 
contributions that are of lesser importance in most models. The processes 
in (\ref{eq:gluegluegluinos})-(\ref{eq:qqsquarks}) get contributions from 
the $t$-channel exchange of an appropriate squark or gluino, and 
(\ref{eq:gluegluegluinos}) and (\ref{eq:qqbargluinosorsquarks}) also have 
gluon $s$-channel contributions. In a crude first approximation, for the 
hard parton collisions needed to make heavy particles, one may think of 
the Tevatron as a quark-antiquark collider, and the LHC as a gluon-gluon 
and gluon-quark collider. However, the signals are always an inclusive 
combination of the results of parton collisions of all types, and generally 
cannot be neatly separated.

At the Tevatron collider, the chargino and neutralino production processes 
(mediated primarily by valence quark annihilation into virtual weak 
bosons) tended to have the larger cross-sections, unless the squarks or 
gluino were rather light (less than 300 GeV or so, which is now clearly 
ruled out by the LHC). In a typical model 
where $\stilde C_1$ and $\stilde N_2$ are mostly $SU(2)_L$ gauginos and 
$\stilde N_1$ is mostly bino, the largest production cross-sections in 
(\ref{eq:qqbarinos}) belong to the $\stilde C_1^+\stilde C_1^-$ and 
$\stilde C_1\stilde N_2$ channels, because they have significant couplings 
to $\gamma,Z$ and $W$ bosons, respectively, and because of kinematics. At 
the LHC, the situation is typically reversed, with production of gluinos 
and squarks by gluon-gluon and gluon-quark fusion usually dominating. At both 
colliders, one can also have associated production of a chargino or 
neutralino together with a squark or gluino, but most models 
predict that the 
cross-sections (of mixed electroweak and QCD strength) are much lower than 
for the ones in (\ref{eq:qqbarinos})-(\ref{eq:qqsquarks}). Slepton pair 
production as in (\ref{eq:qqbarsleptons}) was quite small at the 
Tevatron, but might be observable eventually at the LHC \cite{sleptonLHC}. 
Cross-sections for sparticle production at hadron colliders can be found 
in refs.~\cite{gluinosquarkproduction}, and have been incorporated in 
computer programs including 
\cite{ISAJET},\cite{PYTHIA}-\cite{Alwall:2011uj}.

The decays of the produced sparticles result in final states with two 
neutralino LSPs, which escape the detector. The LSPs carry away at 
least $2 m_{\NI}$ of missing energy, but at hadron colliders only the 
component of the missing energy that is manifest as momenta transverse to 
the colliding beams, usually denoted $\Et$ or $E_T^{\rm miss}$ 
(although $\vec{\slashchar{p}}_T$ or $\vec{p}_T^{\hspace{1.5pt}\rm miss}$ 
might be more logical names) is observable. 
So, in general 
the observable signals for supersymmetry at hadron colliders are $n$ 
leptons + $m$ jets + $\Et$, where either $n$ or $m$ might be 0. There are 
important Standard Model backgrounds to these signals, especially 
from processes involving production of $W$ and $Z$ bosons that decay to 
neutrinos, which provide the $\Et$. Therefore it is important to identify 
specific signal region cuts for which the backgrounds can be reduced. Of course, 
the optimal choice of cuts  
depends on which sparticles are being produced and how they decay, facts that are not known in advance. Depending on the specific object of the search,
backgrounds can be further reduced by 
requiring at least some number $n$ of 
energetic jets, and imposing a cut on a variable $H_T$, 
typically defined to be the sum of the largest 
few (or all) of the $p_T$'s of the jets and leptons in each event. (Unfortunately, there is no standard definition of $H_T$.) Different signal regions can be defined by
how many jets are required in the event, the minimum $p_T$ cuts on those jets, how many 
jets are included in the definition of $H_T$, and other fine details. 
Alternatively, one can cut on $m_{\rm eff} \equiv H_T + \Et$ rather than $H_T$. 
Another cut that is often used in searches is to require a minimum value for the ratio of 
$\Et$ to either $H_T$ or $m_{\rm eff}$; the backgrounds tend to 
have smaller values of this ratio than a supersymmetric signal would.
LHC searches have also made use of more sophisticated kinematic observables,
such as $M_{T2}$ \cite{MT2}, $\alpha_T$ \cite{alphaT}, and razor variables
\cite{razor}.

The classic $\Et$ signal for supersymmetry at hadron colliders is events 
with jets and $\Et$ but no energetic isolated leptons. The latter 
requirement reduces backgrounds from Standard Model processes with 
leptonic $W$ decays, and is obviously most effective if the relevant 
sparticle decays have sizable branching fractions into channels with no 
leptons in the final state. The most important potential backgrounds are:
\begin{itemize}
\item detector mismeasurements of jet energies,
\item $W$+jets, with the $W$ decaying to $\ell\nu$, when the charged lepton is missed or absorbed into a jet,
\item $Z$+jets, with $Z \rightarrow \nu \bar \nu$,
\item $t\overline t$ production, with $W\rightarrow \ell\nu$, when the charged lepton is 
missed.
\end{itemize}
One must choose the $\Et$ cut high enough to reduce these backgrounds, and also to assist in efficient triggering. Requiring at least one very high-$p_T$ jet can also satisfy a trigger requirement. In addition, the first (QCD) 
background can be reduced by requiring that the transverse direction of the $\Et$ is not 
too close to the transverse direction of a jet. The jets$+\Et$ signature is a favorite possibility for the 
first evidence for supersymmetry to be found at the LHC. 
It can get important contributions from 
every type of sparticle pair production, except slepton pair production.

Another important possibility for the LHC is the single lepton plus jets plus $\Et$ signal 
\cite{LHCdiscovery}. It has a potentially large Standard Model 
background from production of 
$W\rightarrow\ell\nu$, either together with jets or from top decays. 
However, this background can 
be reduced by putting a cut on the transverse mass variable 
$m_T = \sqrt{2 p_T^\ell \Et [1 - \cos(\Delta \phi)]}$, 
where $\Delta \phi$ is the difference in azimuthal angle between the missing 
transverse momentum and the lepton. For $W$ decays, this is essentially always less than 
100 GeV even after detector resolution effects, 
so a cut requiring $m_T > 100$ GeV nearly eliminates those background 
contributions at the LHC. 
The single lepton plus jets signal can have an extremely large rate from various sparticle 
production modes, and may give a good discovery or confirmation signal at the LHC.

The same-charge dilepton signal \cite{likesigndilepton} has the advantage 
of relatively small
backgrounds. It can occur if the gluino decays with a 
significant branching fraction to hadrons 
plus a chargino, which can subsequently decay into a final state with a charged lepton, a 
neutrino, and $\stilde N_1$. Since the gluino doesn't know anything about 
electric charge, the 
charged lepton produced from each gluino decay can have either sign with equal 
probability, as discussed in section 
\ref{subsec:decays.gluino}. This means that gluino pair 
production or gluino-squark production will often lead 
to events with two leptons with the same 
charge (and uncorrelated flavors) plus jets and $\Et$. 
This signal can also arise from  squark 
pair production, for example if the squarks decay like 
$\stilde q \rightarrow q\stilde g$. The physics backgrounds at hadron 
colliders are very small, 
because the largest Standard Model sources for isolated lepton pairs, notably Drell-Yan,
$W^+W^-$, and $t\overline t$ production, can only yield opposite-charge dileptons. 
Despite the backgrounds just mentioned, 
opposite-charge dilepton signals, for example from 
slepton pair production, or slepton-rich decays 
of heavier superpartners, with subsequent decays 
$\stilde \ell \rightarrow \ell \NI$, may also eventually 
give an observable signal at the LHC.

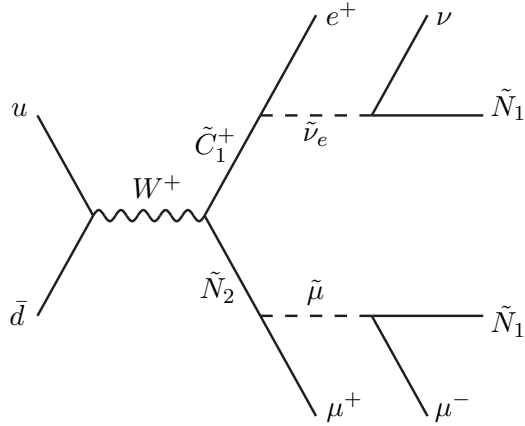
\begin{figure}
\begin{minipage}[]{0.45\linewidth}
\caption{A complete Feynman diagram for a clean (no high-$p_T$ hadronic 
jets) trilepton event at a hadron collider, from production of an on-shell 
neutralino and a chargino, with subsequent leptonic decays, leading in 
this case to $\mu^+\mu^-e^+ + \Et$.\label{fig:trilepton}}
\end{minipage}
\begin{minipage}[]{0.545\linewidth}
\begin{picture}(115,143)(-96,-68)
\SetScale{1.4}
\SetWidth{0.67}
\Line(0,0)(15,27)
\Line(0,0)(15,-27)
\Line(15,27)(30,54)
\Line(15,-27)(30,-54)
\DashLine(15,27)(45,27){3.5}
\DashLine(15,-27)(45,-27){3.5}
\Line(45,27)(75,27)
\Line(45,-27)(75,-27)
\Line(45,27)(60,54)
\Line(45,-27)(60,-54)
\Photon(-30,0)(0,0){1.5}{5}
\Line(-45,27)(-30,0)
\Line(-45,-27)(-30,0)
\Text(-18,11)[]{$ W^+$}
\Text(-70,40)[]{$ u$}
\Text(-71,-37)[]{$ \bar d$}
\Text(4,27)[]{$\tilde C_1^+$}
\Text(5,-27)[]{$\tilde N_2$}
\Text(42,31)[]{$ \tilde \nu_e$}
\Text(42,-29)[]{$ \tilde \mu$}
\Text(115,42)[]{$ \tilde N_1$}
\Text(115,-40)[]{$ \tilde N_1$}
\Text(94,-72.5)[]{$\mu^-$}
\Text(90.7,74.5)[]{$\nu$}
\Text(53,-72.5)[]{$\mu^+$}
\Text(52,76.5)[]{$e^+$}
\end{picture}
\end{minipage}
\end{figure}

The trilepton signal \cite{trilepton} is another possible discovery mode, 
featuring three leptons plus $\Et$, and possibly hadronic jets. At the 
Tevatron, this would most likely have come about from electroweak $\stilde 
C_1\stilde N_2$ production followed by the decays indicated in 
eq.~(\ref{eq:CNleptonic}), in which case high-$p_T$ hadronic activity 
should be absent in the event. A typical Feynman diagram for such an event 
is shown in fig.~\ref{fig:trilepton}.
It could also come from $\stilde g\stilde g$, $\stilde q\stilde g$, or 
$\stilde q \stilde q$ production, with one of the gluinos or squarks 
decaying through a $\stilde C_1$ and the other through a $\stilde N_2$
in a variety of different ways. 
This is the more likely origin at the LHC, at least in most benchmarks
based on MSUGRA or similar models. 
In that case, there will be 
very high-$p_T$ jets from the decays, in addition to the three leptons and 
$\Et$. These signatures rely on the $\stilde N_2$ having a significant 
branching fraction for the three-body decay to leptons in 
eq.~(\ref{eq:CNleptonic}). The competing two-body decay modes 
$\stilde N_2 \rightarrow h^0 \stilde N_1$ and $\stilde N_2 \rightarrow Z 
\stilde N_1$ are sometimes called ``spoiler" modes, since if they are 
kinematically allowed they can dominate, spoiling the trilepton signal. 
This is because if the $\stilde N_2$ decay is 
through an on-shell $h^0$ or $Z^0$, then the final state will likely include jets 
(especially bottom-quark jets in the case of $h^0$) rather than isolated leptons.
Although the trilepton signal is lost, supersymmetric events with 
$h^0 \rightarrow b \bar b$ following from $\stilde N_2 \rightarrow h^0 \stilde N_1$ 
could eventually be useful at the LHC, especially since we now know that $M_{h^0} = 125$
GeV.

One should also be aware of interesting signals that can appear for particular ranges of 
parameters. Final state leptons appearing in the signals listed above might be 
predominantly tau, and so a significant fraction could be realized as hadronic $\tau$ jets. 
This is because most models based on lepton universality at the input scale 
predict that $\stilde \tau_1$ is lighter than the selectrons 
and smuons. Similarly, supersymmetric events may have a preference for bottom jets, 
sometimes through decays involving top quarks because $\stilde t_1$ is relatively light, 
and sometimes because $\stilde b_1$ is expected to be lighter than the squarks of the first 
two families, and sometimes for both reasons. In such cases, there will be at least four 
potentially $b$-taggable jets in each event. Other things being equal, the larger 
$\tan\beta$ is, the stronger the preference for hadronic $\tau$ and $b$ jets will be in 
supersymmetric events.

\begin{figure}
\begin{minipage}[]{0.6\linewidth}
\caption{The theoretical shape of the dilepton invariant mass distribution 
from events with $\stilde N_2 \rightarrow \ell \stilde \ell \rightarrow
\ell^+\ell^-\stilde N_1$. No cuts or detector effects are included.
The endpoint is at $M_{\ell\ell}^{\rm max} = m_{\tilde N_2}
(1 - m^2_{\tilde \ell}/m^2_{\tilde N_2})^{1/2}
(1 - m^2_{\tilde N_1}/m^2_{\tilde \ell})^{1/2}.$
\label{fig:LHCendpoint}}
\end{minipage}
\begin{minipage}[]{0.339\linewidth} 
\begin{picture}(165,95)(-30,0)
\SetScale{1.35}
\LongArrow(0,0)(110,0)  
\LongArrow(0,0)(0,50)
\Text(9,77)[]{Events/GeV}
\Text(148,-10)[]{$M_{\ell\ell}$}
\Text(90,-10)[]{$M_{\ell\ell}^{\rm max}$}
\SetWidth{1.3}
\Line(0,0)(65,40)
\Line(65,40)(65,0)
\Line(65,0)(110,0)
\end{picture}
\end{minipage}
\end{figure}
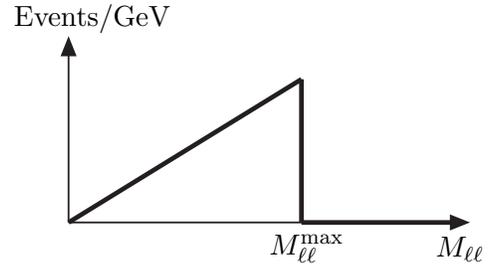

After evidence for the existence of supersymmetry is acquired, the LHC 
data can be used to extract sparticle masses by analyzing the kinematics 
of the decays. With a neutralino LSP always escaping the detector, there 
are no true invariant mass peaks possible. However, various combinations 
of masses can be measured using kinematic edges and other reconstruction 
techniques. For a particularly favorable possibility, suppose the decay of the second-lightest neutralino 
occurs in two stages through a real slepton, $\stilde N_2 \rightarrow \ell 
\stilde \ell \rightarrow \ell^+\ell^-\stilde N_1$. Then the resulting 
dilepton invariant mass distribution is as shown in 
fig.~\ref{fig:LHCendpoint}.%
It features a sharp edge, allowing a precision measurement of the 
corresponding combination of $\stilde N_2$, $\stilde \ell$, and 
$\stilde N_1$ masses \cite{LHCN2edge,LHCdileptonedge,ATLASTDR},
cuts will distort the shape, especially on the low end. There are 
significant backgrounds to this analysis, for example coming from 
$t\overline t$ production. However, the signal from $\stilde N_2$ 
has same-flavor leptons, while the background has contributions from 
different flavors. Therefore the edge can be enhanced by plotting the 
combination $[e^+e^-] + [\mu^+\mu^-] - [e^+\mu^-] - [\mu^+e^-]$, 
subtracting the background.

Heavier sparticle mass combinations can also be reconstructed at the LHC 
\cite{ATLASTDR}-\cite{Kawagoe}
using other kinematic 
distributions. For example, consider the gluino decay chain $\stilde g 
\rightarrow q \stilde q^* \rightarrow q \bar q \stilde N_2$ with $\stilde 
N_2 \rightarrow \ell \stilde \ell^* \rightarrow \ell^+ \ell^- \stilde N_1$ 
as above. By selecting events close to the dilepton mass edge as 
determined in the previous paragraph, one can reconstruct a peak in the 
invariant mass of the $jj\ell^+\ell^-$ system, which correlates well with 
the gluino mass. As another example, the decay $\stilde q_L \rightarrow q 
\stilde N_2$ with $\stilde N_2 \rightarrow h^0 \stilde N_1$ can be 
analyzed by selecting events near the peak from $h^0 \rightarrow b 
\overline b$. There will then be a broad $jb\bar b$ invariant mass 
distribution, with a maximum value that can be related to $m_{\tilde 
N_2}$, $m_{\tilde N_1}$ and $m_{\tilde q_L}$, since $m_{h^0} = 125$ GeV is known. 
There are many other similar opportunities, depending on the specific sparticle 
spectrum. These techniques may determine the sparticle mass 
differences much more accurately than the individual masses, so that the mass of 
the unobserved LSP will be constrained but not precisely 
measured.\footnote{A possible exception occurs if the lighter 
top squark has no 
kinematically allowed flavor-preserving 2-body decays, which requires
$m_{\tilde t_1} < m_{\tilde N_1} + m_t$ and 
$m_{\tilde t_1} < m_{\tilde C_1} + m_b$. Then the $\tilde t_1$ will 
live long enough to form
hadronic bound states. Scalar stoponium
might then be observable at the LHC via its rare $\gamma\gamma$ decay,
allowing a uniquely precise measurement of the mass through
a narrow peak (limited by detector resolution) in the diphoton
invariant mass spectrum \cite{Drees:1993yr,stoponium2}.}

Following the 2012 discovery of the 125 GeV Higgs boson, presumably $h^0$, the remaining
Higgs scalar bosons of the MSSM are also targets of searches at the 
the LHC. The heavier neutral Higgs scalars can be searched for in decays
\beq
&&
A^0/H^0 \>\rightarrow\> \tau^+\tau^-,\> \mu^+\mu^-,\> b\overline 
b,\>t\overline t,
\\
&&
H^0 \>\rightarrow\> h^0 h^0,
\\
&&
A^0 \>\rightarrow\> Z h^0 \>\rightarrow\> \ell^+ \ell^- b \overline b,
\eeq
with prospects that vary considerably depending on the parameters of the 
model. The charged Higgs boson may also appear at the LHC in 
top-quark decays, if $m_{H^+} < m_t$. 
If instead $m_{H^+} > m_t$, then one can look for
\beq
bg \rightarrow t H^-
\qquad\mbox{or}\qquad
gg \rightarrow t \overline b H^-,
\eeq
followed by the decay $H^- \rightarrow \tau^- \bar \nu_\tau$ or
$H^- \rightarrow \bar t b$ in each case, or the charge conjugates of these processes. 
More details on Higgs search projections and experimental results are 
available at the ATLAS and CMS physics results web pages.

The remainder of this subsection briefly considers the possibility that 
the LSP is the goldstino/gravitino, in which case the sparticle discovery 
signals discussed above can be significantly improved. If the NLSP is a 
neutralino with a prompt decay, then $\NI\rightarrow \gamma\G$ will yield 
events with two energetic, isolated photons plus $\Et$ from the escaping 
gravitinos, rather than just $\Et$. So at a hadron collider the signal is 
$\gamma\gamma+X+\Et$ where $X$ is any collection of leptons plus jets. The 
Standard Model backgrounds relevant for such events are quite small. If 
the $\NI$ decay length is long enough, then it may be measurable because 
the photons will not point back to the event vertex. This would be 
particularly useful, as it would give an indication of the 
supersymmetry-breaking scale $\sqrt{\langle F \rangle}$; see 
eq.~(\ref{generalgravdecay}) and the discussion in section 
\ref{subsec:decays.gravitino}. If the $\NI$ decay is outside of the 
detector, then one just has the usual leptons + jets + $\Et$ signals as 
discussed above in the neutralino LSP scenario.

In the case that the NLSP is a charged slepton, then the decay $\stilde 
\ell \rightarrow \ell\G$ can provide two extra leptons in each event, 
compared to the signals with a neutralino LSP. If the $\stilde \tau_1$ is 
sufficiently lighter than the other charged sleptons $\stilde e_R$, 
$\stilde \mu_R$ and so is effectively the sole NLSP, then events will 
always have a pair of taus. If the slepton NLSP is long-lived, one can 
look for events with a pair of very heavy charged particle tracks or a 
long time-of-flight in the detector. Since slepton pair production usually 
has a much smaller cross-section than the other processes in 
(\ref{eq:qqbarinos})-(\ref{eq:qqsquarks}), this will typically be 
accompanied by leptons and/or jets from the same event vertex, which may 
be of crucial help in identifying candidate events. It is also quite 
possible that the decay length of $\stilde \ell \rightarrow \ell\G$ is 
measurable within the detector, seen as a macroscopic kink in the charged 
particle track. This would again be a way to measure the scale
of supersymmetry breaking through eq.~(\ref{generalgravdecay}).

\subsection{Signals at $e^+e^-$ colliders}\label{subsec:signals.LEPNLC}
\setcounter{equation}{0}
\setcounter{footnote}{1}

At $e^+e^-$ colliders, all sparticles (except the gluino) can be produced 
in tree-level reactions:
\begin{figure}[p]
\begin{center}
\begin{picture}(140,45)(0,15)
\SetScale{0.9}\SetWidth{0.85}
\Line(0,0)(40,30)
\Line(0,60)(40,30)
\Photon(40,30)(90,30){2.4}{5}
\Line(90,30)(130,60)
\Line(90,30)(130,0)
\Text(-7,5)[]{$e^+$}
\Text(-7,58)[]{$e^-$}
\Text(61,39)[]{$\gamma, Z$}
\Text(129,56)[]{$\stilde C_i^-$}
\Text(129,4)[]{$\stilde C_j^+$}
\SetWidth{0.4}
\end{picture}
\hspace{1.75cm}
\begin{picture}(140,45)(0,15)
\SetScale{0.9}\SetWidth{0.85}
\Line(0,0)(60,0)
\Line(0,60)(60,60)
\Line(60,0)(120,0)
\Line(60,60)(120,60)
\DashLine(60,0)(60,60){5}
\Text(-2,6)[c]{$e^+$}
\Text(-2,60)[c]{$e^-$}
\Text(46,28)[c]{$\stilde \nu_e$}
\Text(122,56)[c]{$\stilde C_i^-$}
\Text(122,5)[c]{$\stilde C_j^+$}
\SetWidth{0.45}
\Photon(60,0)(120,0){2.1}{6}
\Photon(60,60)(120,60){2.1}{6}
\end{picture}
\end{center}
\caption{Diagrams for chargino pair production at $e^+e^-$ colliders.
\label{fig:eecharginoprod}}
\end{figure}
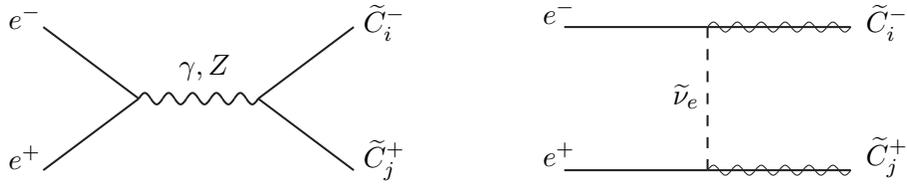
\begin{figure}[p]
\begin{center}
\begin{picture}(130,48)(0,15)
\SetScale{0.9}\SetWidth{0.85}
\Line(0,0)(40,30)
\Line(0,60)(40,30)
\Photon(40,30)(90,30){2.4}{5}
\Line(90,30)(130,60)
\Line(90,30)(130,0)
\Text(-5.5,5)[]{$e^+$}
\Text(-5.5,58)[]{$e^-$}
\Text(59,38)[]{$Z$}
\Text(125,57)[]{$\stilde N_i$}
\Text(125,3)[]{$\stilde N_j$}
\end{picture}
\hspace{0.93cm}
\begin{picture}(120,48)(0,15)
\SetScale{0.9}\SetWidth{0.85}
\Line(3,0)(60,0)
\Line(3,60)(60,60)
\Line(60,0)(117,0)
\Line(60,60)(117,60)
\DashLine(60,0)(60,60){5}
\Text(-3,5)[c]{$e^+$}
\Text(-3,59)[c]{$e^-$}
\Text(36,28)[c]{$\stilde e_L$, $\stilde e_R$}
\Text(115,58)[c]{$\stilde N_i$}
\Text(115,4)[c]{$\stilde N_j$}
\SetWidth{0.45}
\Photon(60,0)(117,0){2.1}{6}
\Photon(60,60)(117,60){2.1}{6}
\end{picture}
\hspace{0.9cm}
\begin{picture}(115,48)(0,15)
\SetScale{0.9}\SetWidth{0.85}
\Line(0,0)(60,0)
\Line(0,60)(60,60)
\Line(60,60)(120,0)
\Line(60,0)(85,25)
\Line(120,60)(95,35)
\DashLine(60,0)(60,60){5}
\Text(-5,5)[c]{$e^+$}
\Text(-5,59)[c]{$e^-$}
\Text(36,28)[c]{$\stilde e_L$, $\stilde e_R$}
\Text(117,56)[c]{$\stilde N_i$}
\Text(117,3)[c]{$\stilde N_j$}
\SetWidth{0.45}
\Photon(60,60)(120,0){2.1}{8}
\Photon(60,0)(85,25){2.1}{3}
\Photon(120,60)(95,35){2.1}{3}
\end{picture}
\end{center}
\caption{Diagrams for neutralino pair production at $e^+e^-$ colliders.
\label{fig:eeneutralinoprod}}
\end{figure}
\begin{figure}[p]
\begin{center}
\begin{picture}(140,48)(0,15)
\SetScale{0.9}\SetWidth{0.85}
\Line(0,0)(40,30)
\Line(0,60)(40,30)
\Photon(40,30)(90,30){2.4}{5}
\DashLine(90,30)(130,60){5}
\DashLine(90,30)(130,0){5}
\Text(-4,5)[]{$e^+$}
\Text(-4,59)[]{$e^-$}
\Text(61,38)[]{$\gamma, Z$}
\Text(128,57)[]{$\stilde \ell^-$}
\Text(128,5)[]{$\stilde \ell^+$}
\end{picture}
\hspace{1.75cm}
\begin{picture}(140,48)(0,15)
\SetScale{0.9}\SetWidth{0.85}
\Line(0,0)(60,0)
\Line(0,60)(60,60)
\DashLine(60,0)(120,0){5}
\DashLine(60,60)(120,60){5}
\Line(60,0)(60,60)
\Text(-4,8)[c]{$e^+$}
\Text(-4,59)[c]{$e^-$}
\Text(45,28)[c]{$\stilde N_i$}
\Text(118,59)[c]{$\stilde e^-$}
\Text(118,5)[c]{$\stilde e^+$}
\SetWidth{0.45}
\Photon(60,0)(60,60){2.1}{6}
\end{picture}
\end{center}
\caption{Diagrams for charged slepton pair production at $e^+e^-$ 
colliders.
\label{fig:eesleptonprod}}
\end{figure}
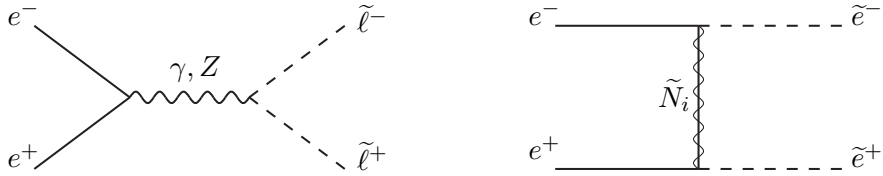
\begin{figure}[p]
\begin{center}
\begin{picture}(140,48)(0,15)
\SetScale{0.9}\SetWidth{0.85}
\Line(0,0)(40,30)
\Line(0,60)(40,30)
\Photon(40,30)(90,30){2.4}{5}
\DashLine(90,30)(130,60){5}
\DashLine(90,30)(130,0){5}
\Text(-5,5)[]{$e^+$}
\Text(-5,59)[]{$e^-$}
\Text(61,38)[]{$Z$}
\Text(126.2,57)[]{$\tilde \nu_\ell$}
\Text(126.6,2)[]{$\tilde \nu_\ell^*$}
\end{picture}
\hspace{1.75cm}
\begin{picture}(140,48)(0,15)
\SetScale{0.9}\SetWidth{0.85}
\Line(0,0)(60,0)
\Line(0,60)(60,60)
\DashLine(60,0)(120,0){5}
\DashLine(60,60)(120,60){5}
\Line(60,0)(60,60)
\Text(-4,8)[c]{$e^+$}
\Text(-4,59)[c]{$e^-$}
\Text(45,27)[c]{$\tilde C_i$}
\Text(118.5,57)[c]{$\tilde \nu_e$}
\Text(118.5,4)[c]{$\tilde \nu_e^*$}
\SetWidth{0.45}
\Photon(60,0)(60,60){2.1}{6}
\end{picture}
\end{center}
\caption{Diagrams for sneutrino pair production at $e^+e^-$ colliders.
\label{fig:eesneutrinoprod}}
\end{figure}
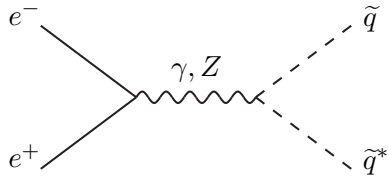
\begin{figure}[p]
\begin{center}
\begin{picture}(140,44)(0,15)
\SetScale{0.9}\SetWidth{0.85}
\Line(0,0)(40,30)
\Line(0,60)(40,30)
\Photon(40,30)(90,30){2.4}{5}
\DashLine(90,30)(130,60){5}
\DashLine(90,30)(130,0){5}
\Text(-6,5)[]{$e^+$}
\Text(-6,59)[]{$e^-$}
\Text(59,38)[]{$\gamma, Z$}
\Text(124.6,57)[]{$\stilde q$}
\Text(126.8,3)[]{$\stilde q^*$}
\end{picture}
\end{center}
\caption{Diagram for squark production at $e^+e^-$ colliders.
\label{fig:eesquarkprod}}
\end{figure}
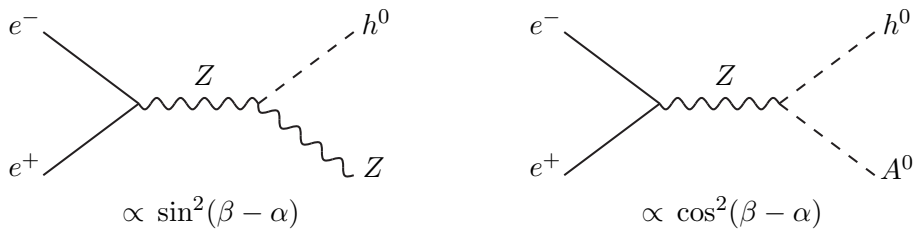
\begin{figure}[p]
\begin{center}
\begin{picture}(140,61)(0,-2)
\SetScale{0.9}\SetWidth{0.85}
\Line(0,0)(40,30)
\Line(0,60)(40,30)
\Photon(40,30)(90,30){-2.4}{5}
\DashLine(90,30)(130,60){5}
\Photon(90,30)(130,0){-2.4}{4.5}
\Text(-7,4)[]{$e^+$}
\Text(-7,58)[]{$e^-$}
\Text(61,38)[]{$Z$}
\Text(126,58)[]{$h^0$}
\Text(125,3)[]{$Z$}
\Text(63.5,-14)[c]{$\propto \,\sin^2(\beta - \alpha)$}
\end{picture}
\hspace{1.75cm}
\begin{picture}(140,61)(0,-2)
\SetScale{0.9}\SetWidth{0.85}
\Line(0,0)(40,30)
\Line(0,60)(40,30)
\Photon(40,30)(90,30){-2.4}{5}
\DashLine(90,30)(130,60){5}
\DashLine(90,30)(130,0){5}
\Text(-7,4)[]{$e^+$}
\Text(-7,58)[]{$e^-$}
\Text(61,38)[]{$Z$}
\Text(126,58)[]{$h^0$}
\Text(126,3)[]{$A^0$}
\Text(63.5,-14)[c]{$\propto\, \cos^2(\beta - \alpha)$}
\end{picture}
\end{center}
\caption{Diagrams for neutral Higgs scalar boson 
production at $e^+e^-$ colliders.\label{fig:eehiggs}}
\end{figure}
\beq
e^+e^- \rightarrow
\stilde C_i^+ \stilde C_j^-,\>\>\, \stilde N_i \stilde N_j,
\>\>\, \stilde \ell^+ \stilde \ell^-,\>\>\, \stilde \nu \stilde \nu^*,
\>\>\, \stilde q \stilde q^* ,
\label{eesignals}
\eeq
as shown in figs.~\ref{fig:eecharginoprod}-\ref{fig:eesquarkprod}. The 
important interactions for sparticle production are the 
gaugino-fermion-scalar couplings shown in Figures~\ref{fig:gaugino}b,c and 
the ordinary vector boson interactions. The cross-sections are therefore 
determined just by the electroweak gauge couplings and the sparticle 
mixings. They were calculated in ref.~\cite{epprod}, and are available
in computer programs 
\cite{ISAJET}, \cite{PYTHIA}-\cite{Herwig}, \cite{SUSYGEN}.

All of the processes in eq.~(\ref{eesignals}) get contributions from the 
$s$-channel exchange of the $Z$ boson and, for charged sparticle pairs, 
the photon. In the cases of $\stilde C_i^+ \stilde C_j^-$, $\stilde N_i 
\stilde N_j$, $\stilde e_R^+ \stilde e_R^-$, $\stilde e_L^+ \stilde 
e_L^-$, $\stilde e_L^\pm \stilde e_R^\mp$, and $\stilde \nu_e \stilde 
\nu_e^*$ production, there are also $t$-channel diagrams exchanging a 
virtual sneutrino, selectron, neutralino, neutralino, neutralino, and 
chargino, respectively. The $t$-channel contributions are significant if 
the exchanged sparticle is not too heavy. For example, the production of 
wino-like $\stilde C_1^+ \stilde C_1^-$ pairs typically suffers a 
destructive interference between the $s$-channel graphs with $\gamma,Z$ 
exchange and the $t$-channel graphs with $\stilde \nu_e$ exchange, if the 
sneutrino is not too heavy. In the case of sleptons, the pair production 
of smuons and staus proceeds only through $s$-channel diagrams, while 
selectron production also has a contribution from the $t$-channel 
exchanges of the neutralinos, as shown in Figure~\ref{fig:eesleptonprod}. 
For this reason, the selectron production cross-section
may be significantly larger than that of
smuons or staus at $e^+e^-$ colliders.

The pair-produced sparticles decay as discussed in section 
\ref{sec:decays}. If the LSP is the lightest neutralino, it will always 
escape the detector because it has no strong or electromagnetic 
interactions. Every event will have two LSPs leaving the detector, so 
there should be at least $2m_{\NI}$ of missing energy ($\Etot$). For 
example, in the case of $\stilde C_1^+ \stilde C_1^-$ production, the 
possible signals include a pair of acollinear leptons and $\Etot$, or one 
lepton and a pair of jets plus $\Etot$, or multiple jets plus $\Etot$. The 
relative importance of these signals depends on the branching fraction of 
the chargino into the competing final states, $\stilde C_1 \rightarrow 
\ell\nu \NI$ and $qq^\prime\NI$. In the case of slepton pair production, 
the signal should be two energetic, acollinear, same-flavor leptons plus 
$\Etot$. There is a potentially large Standard Model background for the 
acollinear leptons plus $\Etot$ and the lepton plus jets plus $\Etot$ 
signals, coming from $W^+W^-$ production with one or both of the $W$ 
bosons decaying leptonically. However, these and other Standard Model 
backgrounds can be kept under control with angular cuts, and beam 
polarization if available. It is not difficult to construct the other 
possible signatures for sparticle pairs, which can become quite 
complicated for the heavier charginos, neutralinos and squarks.

The MSSM neutral Higgs bosons can also be produced at $e^+e^-$ colliders,
with the principal processes of interest at low energies
\beq
e^+e^- \rightarrow h^0 Z,
\qquad\qquad
e^+e^-\rightarrow h^0A^0,
\eeq
shown in fig.~\ref{fig:eehiggs}. At tree-level, the first of these has a 
cross-section given by the corresponding Standard Model cross-section 
multiplied by a factor of $\sin^2(\beta - \alpha)$, which approaches 1 in 
the decoupling limit of $m_{A^0} \gg m_Z$ discussed in section 
\ref{subsec:MSSMspectrum.Higgs}. The other process is complementary, since 
(up to kinematic factors) its cross-section is the same but multiplied 
by $\cos^2(\beta - \alpha)$, which is significant if $m_{A^0}$ is not 
large. If $\sqrt{s}$ is high enough [note the mass relation 
eq.~(\ref{eq:m2Hpm})], one can also have
\beq
e^+e^-\rightarrow H^+ H^-,
\eeq
with a cross-section that is fixed, at tree-level, in terms of 
$m_{H^\pm}$, and also
\beq
e^+e^- \rightarrow H^0Z,
\qquad\qquad
e^+e^-\rightarrow H^0A^0,
\eeq
with cross-sections proportional to $\cos^2(\beta - \alpha)$ and 
$\sin^2(\beta - \alpha)$ respectively. Also, at sufficiently high 
$\sqrt{s}$, the process
\beq
e^+ e^- \rightarrow \nu_e \bar \nu_e h^0 
\eeq
following from $W^+W^-$ fusion provides the best way to study the Higgs 
boson decays, which can differ \cite{GunionHaber,HHG,Haber:1997dt} 
from those in the Standard Model.

The CERN LEP $e^+e^-$ collider conducted searches until November 2000, 
with various center of mass energies up to 209 GeV, yielding no firm 
evidence for superpartner production. The resulting limits 
\cite{LEPSUSYWG} on the charged sparticle masses are of order roughly half 
of the beam energy, minus taxes paid for detection and identification 
efficiencies, backgrounds, and the suppression of cross-sections near 
threshold. The bounds become weaker if the mass difference between the 
sparticle in question and the LSP (or another sparticle that the produced 
one decays into) is less than a few GeV, because then the available 
visible energy can be too small for efficient detection and identification.
Despite the strong limits coming from the LHC, some of the limits from LEP are still relevant, especially when the mass differences between supersymmetric particle
are small.

For example, LEP established limits $m_{\tilde e_R} > 99$ GeV and 
$m_{\tilde \mu_R} > 95$ GeV at 95\% CL, provided that $m_{\tilde \ell_R} - 
m_{\tilde N_1} > 10$ GeV, and that the branching fraction for $\ell_R 
\rightarrow \ell \stilde N_1$ is 100\% in each case. The limit for staus 
is weaker, and depends somewhat more strongly on the neutralino LSP mass.
The LEP chargino mass bound is approximately $m_{\tilde C_1} > 103$ GeV 
for mass differences $m_{\tilde C_1} - m_{\tilde N_1} > 3$ GeV, assuming 
that the chargino decays predominantly through a virtual $W$, or with 
similar branching fractions. However, this bound reduces to about 
$m_{\tilde C_1} > 92$ GeV for $100$ MeV $< m_{\tilde C_1} - m_{\tilde N_1} 
< 3$ GeV. For small positive mass differences 0 $< m_{\tilde C_1} - 
m_{\tilde N_1} <$ 100 MeV, the limit is again about $m_{\tilde C_1} > 103$ 
GeV, because the chargino is long-lived enough to have a displaced decay 
vertex or leave a track as it moves through the detector. These limits 
assume that the sneutrino is heavier than about 200 GeV, so that it does 
not significantly reduce the production cross-section by interference of 
the $s$- and $t$-channel diagrams in fig.~\ref{fig:eecharginoprod}. If the 
sneutrino is lighter, then the bound reduces, especially if $m_{\tilde C_1} - 
m_{\tilde \nu}$ is positive but small, so that the decay $\stilde C_1 
\rightarrow \tilde \nu \ell$ dominates but releases very little visible 
energy. More details on these and many other legacy limits from the LEP runs can 
be found at \cite{LEPSUSYWG} and \cite{RPP}.

If supersymmetry is the solution to the hierarchy problem, then the LHC 
may be able to establish strong evidence for it, and measure 
some of the sparticle mass differences, as discussed in the previous 
subsection. However, many important questions will remain. 
Competing theories can also produce missing energy signatures. The overall 
mass scale of sparticles may not be known as well as one might like. 
Sparticle production will be inclusive and overlapping and might be 
difficult to disentangle. A future $e^+e^-$ collider 
with sufficiently large $\sqrt{s}$
should be able to resolve these issues, and establish more firmly that 
supersymmetry is indeed responsible, to the exclusion of other 
candidate theories. In 
particular, the couplings, spins, gauge quantum numbers, and absolute 
masses of the sparticles will all be measurable.

At an $e^+e^-$ collider, the processes in eq.~(\ref{eesignals}) can all be 
probed close to their kinematic limits, given sufficient integrated 
luminosity. (In the case of sneutrino pair production, this assumes that 
some of the decays are visible, rather than just 
$\stilde\nu\rightarrow\nu\NI$.) Establishing the properties of the 
particles can be done by making use of polarized beams and the relatively 
clean $e^+e^-$ collider environment. For example, consider the production 
and decay of sleptons in $e^+e^- \rightarrow \stilde\ell^+\stilde\ell^-$ 
with $\stilde\ell \rightarrow \ell\NI$. The resulting leptons will have 
(up to significant but calculable effects of initial-state radiation, 
beamstrahlung, cuts, and detector efficiencies and resolutions) a flat 
energy distribution as shown in fig.~\ref{fig:ILCendpoints}.%
\begin{figure}
\begin{minipage}[]{0.6\linewidth}
\caption{The theoretical shape of the lepton energy distribution from 
events with $e^+e^- \rightarrow \stilde \ell^+\stilde \ell^- \rightarrow 
\ell^+\ell^-\stilde N_1\stilde N_1$ at an $e^+e^-$ collider. No cuts 
or initial state radiation or beamstrahlung or detector effects are 
included. The endpoints are $E_{{\rm max,min}} 
= \frac{\sqrt{s}}{4} (1 - m^2_{\tilde 
N_1}/m^2_{\tilde \ell})[1 \pm (1 - 4 m^2_{\tilde \ell}/s)^{1/2}]$, 
allowing precision reconstruction of both $\stilde \ell$ and $\stilde N_1$ 
masses.\label{fig:ILCendpoints}}
\end{minipage}
\begin{minipage}[]{0.339\linewidth} 
\begin{picture}(165,80)(-30,0)
\SetScale{1.35}
\LongArrow(0,0)(100,0)  
\LongArrow(0,0)(0,50)
\Text(9,77)[]{Events/GeV}
\Text(138,-10)[]{$E_{\ell}$}
\Text(93,-10)[]{$E_{\rm max}$}
\Text(21,-10)[]{$E_{\rm min}$}
\SetWidth{1.3}
\Line(0,0)(12,0)
\Line(12,0)(12,40)
\Line(12,40)(65,40)
\Line(65,40)(65,0)
\Line(65,0)(100,0)
\end{picture}
\end{minipage}
\end{figure}
By measuring the endpoints of this distribution, one can precisely and 
uniquely determine both $m_{\stilde\ell_R}$ and $m_{\stilde N_1}$. There 
is a large $W^+W^- \rightarrow \ell^+ \ell^{\prime -} \nu_\ell \bar 
\nu_{\ell'}$ background, but this can be brought under control using 
angular cuts, since the positively (negatively) charged leptons from the 
background tend to go preferentially along the same direction as the 
positron (electron) beam. Also, since the background has uncorrelated 
lepton flavors, it can be subtracted. Changing the polarization of the 
electron beam will even further reduce the background, and will also allow 
controlled variation of the production of right-handed and left-handed 
sleptons, to get at the electroweak quantum numbers.

More generally, inclusive sparticle production at a given fixed $e^+e^-$ 
collision energy will result in a superposition of various kinematic 
edges in lepton and jet energies, and distinctive distributions in 
dilepton and dijet energies and invariant masses. By varying the beam 
polarization and changing the beam energy, these observables give 
information about the couplings and masses of the sparticles. For example, 
in the ideal limit of a right-handed polarized electron beam, the reaction
\beq
e^-_R e^+ \rightarrow \stilde C_1^+ \stilde C_1^-
\eeq
is suppressed if $\stilde C_1$ is pure wino, because in the first diagram 
of fig.~\ref{fig:eecharginoprod} the right-handed electron only couples to 
the $U(1)_Y$ gauge boson linear combination of $\gamma,Z$ while the wino 
only couples to the orthogonal $SU(2)_L$ gauge boson linear combination, 
and in the second diagram the electron-sneutrino-chargino coupling 
involves purely left-handed electrons. Therefore, the polarized beam 
cross-section can be used to determine the charged wino mixing with the 
charged higgsino. Even more precise information about the sparticle masses 
can be obtained by varying the beam energy in small discrete steps very 
close to thresholds, an option unavailable at hadron colliders. The rise 
of the production cross-section above threshold provides information about 
the spin and ``handedness", because the production cross-sections for 
$\tilde \ell_R^+ \tilde \ell_R^-$ and $\tilde \ell_L^+ \tilde \ell_L^-$ 
are $p$-wave and therefore rise like $\beta^3$ above threshold, where 
$\beta$ is the velocity of one of the produced sparticles. In contrast, 
the rates for $\tilde e_L^\pm \tilde e_R^\mp$ and for chargino and 
neutralino pair production are $s$-wave, and therefore should rise like 
$\beta$ just above threshold.  By measuring the angular distributions of 
the final state leptons and jets with respect to the beam axis, the spins 
of the sparticles can be inferred. These will provide crucial tests that 
the new physics that has been discovered is indeed supersymmetry.

A sample of the many detailed studies along these lines can be found in 
refs.~\cite{ilcmassdetrefs}-\cite{ALC}. 
In general, a future
$e^+ e^-$ collider will provide an excellent way of testing softly-broken 
supersymmetry and measuring the model parameters, if it has enough energy. 
Furthermore, the processes $e^+e^- \rightarrow$ $h^0Z$, $h^0A^0$, $H^0Z$, 
$H^0A^0$, $H^+H^-$, and $h^0 \nu_e \bar \nu_e$ should be able to 
test the Higgs sector of supersymmetry at an $e^+e^-$ collider.

The situation may be qualitatively better if the 
gravitino is the LSP as in gauge-mediated models, because of the decays 
mentioned in section \ref{subsec:decays.gravitino}. If the lightest 
neutralino is the NLSP and the decay $\NI\rightarrow\gamma\G$ occurs 
within the detector, then even the process $e^+e^-\rightarrow \NI\NI$ 
leads to a dramatic signal of two energetic photons plus missing energy 
\cite{eeGMSBsignal}-\cite{AKKMM2}. 
There are significant backgrounds to the 
$\gamma\gamma\Etot$ signal, but they are easily removed by cuts. Each of 
the other sparticle pair-production modes eq.~(\ref{eesignals}) will lead 
to the same signals as in the neutralino LSP case, but now with two 
additional energetic photons, which should make the experimentalists' 
tasks quite easy. If the decay length for $\NI\rightarrow\gamma\G$ is much 
larger than the size of a detector, then the signals revert back to those 
found in the neutralino LSP scenario. In an intermediate regime for the 
$\NI\rightarrow\gamma\G$ decay length, one may see events with one or both 
photons displaced from the event vertex by a macroscopic distance.

If the NLSP is a charged slepton $\stilde \ell$, then $e^+e^-\rightarrow 
\stilde \ell^+\stilde \ell^-$ followed by prompt decays $\stilde 
\ell\rightarrow \ell \G$ will yield two energetic same-flavor leptons in 
every event, and with a different energy distribution than the acollinear 
leptons that would follow from either $\stilde C_1^+\stilde C_1^-$ or 
$\stilde \ell^+\stilde \ell^-$ production in the neutralino LSP scenario. 
Pair production of non-NLSP sparticles will yield unmistakable signals, 
which are the same as those found in the neutralino NLSP case but with two 
additional energetic leptons (not necessarily of the same flavor). An even 
more striking possibility is that the NLSP is a slepton that decays very 
slowly \cite{DDRT}. If the slepton NLSP is so long-lived that it decays 
outside the detector, then slepton pair production will lead to events 
featuring a pair of charged particle tracks with high ionization rates 
that betray their very large mass. If the sleptons decay within the 
detector, then one can look for large-angle
kinks in the charged particle tracks, or a 
macroscopic impact parameter. The pair production of any of the other 
heavy charged sparticles will also yield heavy charged particle tracks or 
decay kinks, plus leptons and/or jets, but no $\Etot$ unless the decay 
chains happen to include neutrinos. It may also be possible to identify 
the presence of a heavy charged NLSP by measuring its anomalously long 
time-of-flight through the detector.

In both the neutralino and slepton NLSP scenarios, a measurement of the 
decay length to $\stilde G$ would provide a great opportunity to measure 
the supersymmetry-breaking scale $\sqrt{\langle F \rangle}$, as discussed 
in section \ref{subsec:decays.gravitino}.

\subsection{Dark matter and its detection}\label{subsec:signals.darkmatter}
\setcounter{equation}{0}
\setcounter{footnote}{1}

Evidence from experimental cosmology has now solidified to the point that, 
with some plausible assumptions, the cold dark matter density is known to 
be \cite{cosmokramer,RPP}
\beq
\Omega_{\rm DM} h^2 \approx 0.120,
\label{eq:OmegaDM}
\eeq
with statistical errors of about 2\%, and systematic errors that are less 
clear. Here $\Omega_{\rm DM}$ is the average energy density in 
non-baryonic dark matter divided by the total critical density that would 
lead to a spatially flat homogeneous universe, and $h$ is the Hubble 
constant in units of 100 km sec$^{-1}$ Mpc$^{-1}$, observed to be $h^2 
\approx 0.46$ with an error of order 3\%. This translates into a cold dark 
matter density
\beq
\rho_{\rm DM} \approx 1.2 \times 10^{-6} \> {\rm GeV}/{\rm cm}^3 ,
\label{eq:rhodm}
\eeq 
averaged over very large distance scales.

One of the nice features of supersymmetry with exact $R$-parity 
conservation is that a stable electrically neutral LSP might be this cold 
dark matter.  There are three obvious candidates:  the lightest sneutrino, 
the gravitino, and the lightest neutralino. The possibility of a sneutrino 
LSP making up the dark matter with a cosmologically interesting density 
has been largely ruled out by direct searches \cite{sneutrinonotLSP} (see 
however \cite{sneutrinoLSP}). If the gravitino is the LSP, as in many 
gauge-mediated supersymmetry breaking models, then gravitinos from 
reheating after inflation \cite{gravitinoDMfromreheating} or from other 
sparticle decays \cite{gravitinoDMfromdecays} might be the dark matter, 
but they would be impossible to detect directly even if they have the 
right cosmological density today. They interact too weakly. The most 
attractive prospects for direct detection of supersymmetric dark matter, 
therefore, are based on the idea that the lightest neutralino $\NI$ is the 
LSP \cite{neutralinodarkmatter,darkmatterreviews}.

In the early universe, sparticles existed in thermal equilibrium with the 
ordinary Standard Model particles. As the universe cooled and expanded, 
the heavier sparticles could no longer be produced, and they 
eventually annihilated or 
decayed into neutralino LSPs.  Some of the LSPs 
pair-annihilated into final states not containing sparticles. If there are 
other sparticles that are only slightly heavier, then they existed in 
thermal equilibrium in comparable numbers to the LSP, and their 
co-annihilations are also important in determining the resulting dark 
matter density \cite{GriestSeckel,Gondolo:1990dk}. Eventually, as the 
density decreased, the annihilation rate became small compared to the 
cosmological expansion, and the $\NI$ experienced ``freeze out", with a 
density today determined by this small rate and the subsequent dilution 
due to the expansion of the universe.

\begin{figure}[!t]
\begin{center}
\begin{picture}(120,50)(0,17)  
\SetWidth{0.85}
\Line(0,50)(33,25)
\Line(77,25)(110,50)
\Line(33,25)(77,25)
\DashLine(0,0)(33,25){5}
\Photon(77,25)(110,0){-2.1}{4.5}
\rText(-8,9)[][]{$\stilde f$}
\rText(-6,58.5)[][]{$\stilde N_1$}
\rText(51.5,34)[][]{$f$}
\rText(109.5,56)[][]{$f$}
\rText(112,8)[][]{$\gamma, Z$}
\end{picture}
\hspace{1.5cm}
\begin{picture}(120,50)(0,16)
\SetWidth{0.85}
\Line(0,50)(110,50)
\DashLine(0,0)(55,0){5}
\DashLine(55,0)(55,50){5}
\Photon(55,0)(110,0){2}{5}
\rText(-2,9)[][]{$\stilde f$}
\rText(0,60)[][]{$\stilde N_1$} 
\rText(55.7,27)[][]{$\stilde f$}
\rText(105,59)[][]{$ f$}
\rText(107,9)[][]{$\gamma, Z$}
\end{picture}
\hspace{1.5cm}
\begin{picture}(120,50)(0,17)
\SetWidth{0.85}
\DashLine(0,50)(55,50){5}
\Line(55,50)(110,50)
\DashLine(0,0)(55,0){5}
\Line(55,0)(55,50)
\Line(55,0)(110,0)
\rText(0,9)[][]{$\stilde f$}
\rText(0,60)[][]{$\stilde f$}
\rText(57.5,27)[][]{$\stilde N_i$}
\rText(105,59)[][]{$f$}
\rText(105,8)[][]{$f$}  
\end{picture}
\end{center} 
\caption{Some contributions to the co-annihilation of dark matter
$\stilde N_1$ LSPs with slightly heavier sfermions, which in popular 
models are most plausibly staus (or perhaps top 
squarks).\label{fig:sfermioncoannihilation}}
\end{figure}
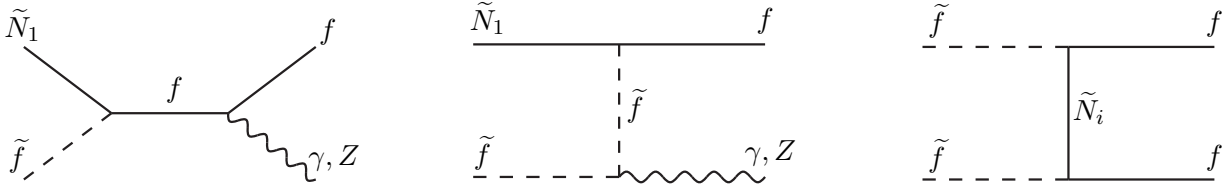

In order to get the observed dark matter density today, the thermal-averaged 
effective annihilation cross-section times the relative speed $v$ of the 
LSPs should be about \cite{darkmatterreviews}
\beq 
\langle \sigma v \rangle \>\sim\> 1\>{\rm pb} \>\sim\>
\alpha^2/(150\>{\rm GeV})^2, 
\eeq 
so a neutralino LSP naturally has, very roughly, the correct (electroweak) 
interaction strength and mass. More detailed and precise estimates can be 
obtained with publicly available computer programs 
\cite{DarkSUSY,micrOMEGAs}, so that the predictions of specific candidate 
models of supersymmetry breaking can be compared to 
eq.~(\ref{eq:OmegaDM}). Some of the diagrams that are typically important 
for neutralino LSP pair annihilation are shown in 
fig.~\ref{fig:darkmatterannihilation}. Depending on the mass of $\NI$, 
various other processes including $\NI\NI\rightarrow$$ZZ$, $Zh^0$, 
$h^0h^0$ or even $W^\pm H^\mp$, $Z A^0$, $h^0 A^0$, $h^0 H^0$, $H^0 A^0$, 
$H^0H^0$, $A^0 A^0$, or $H^+H^-$ may also have been important. Some of the 
diagrams that can lead to co-annihilation of the LSPs with slightly 
heavier sparticles are shown in figs.~\ref{fig:coannihilation} and 
\ref{fig:sfermioncoannihilation}.%
\begin{figure}
\begin{picture}(120,72)(0,-12)
\SetWidth{0.85}
\Line(0,0)(110,0)
\Line(0,50)(110,50)
\DashLine(55,0)(55,50){5}
\rText(0,9)[][]{$\stilde N_1$}
\rText(0,60)[][]{$\stilde N_1$}
\rText(55.5,27)[][]{$\tilde f$}
\rText(105,59)[][]{$ f$}
\rText(105,8)[][]{$\bar f$}
\Text(55,-16)[c]{(a)}
\end{picture}
\hspace{0.75cm}
\begin{picture}(180,72)(-10,-12)
\SetWidth{0.85}
\Line(0,0)(30,25)
\Line(0,50)(30,25)
\DashLine(30,25)(93,25){5}
\Line(123,0)(93,25)
\Line(123,50)(93,25)
\rText(-10,10)[][]{$\stilde N_1$}
\rText(-10,59)[][]{$\stilde N_1$}
\rText(56.8,32.8)[][]{$A^0$ ($h^0,H^0$)}
\rText(146,53)[][]{$b,t,\tau^-,\ldots$}
\rText(146,4)[][]{$\bar b,\bar t,\tau^+,\ldots$}
\Text(66.5,-16)[c]{(b)}
\end{picture}
\hspace{1.0cm}
\begin{picture}(120,72)(0,-12)
\SetWidth{0.85}
\Line(0,0)(55,0)
\Line(0,50)(55,50)
\Line(55,0)(55,50)
\Photon(55,0)(110,0){2}{5}
\Photon(55,50)(110,50){2}{5}
\rText(0,9)[][]{$\stilde N_1$}
\rText(0,60)[][]{$\stilde N_1$}
\rText(57.5,27)[][]{$\tilde C_i$}
\rText(105.5,60.2)[][]{$W^+$}
\rText(105.5,9.2)[][]{$W^-$}
\Text(55,-16)[c]{(c)}
\end{picture}
\caption{Contributions to the annihilation cross-section for neutralino
dark matter LSPs from (a) $t$-channel slepton and squark exchange,
(b) near-resonant annihilation through a Higgs boson
($s$-wave for $A^0$, and $p$-wave for $h^0$, $H^0$),
and (c) $t$-channel chargino exchange. 
\label{fig:darkmatterannihilation}}
\end{figure}
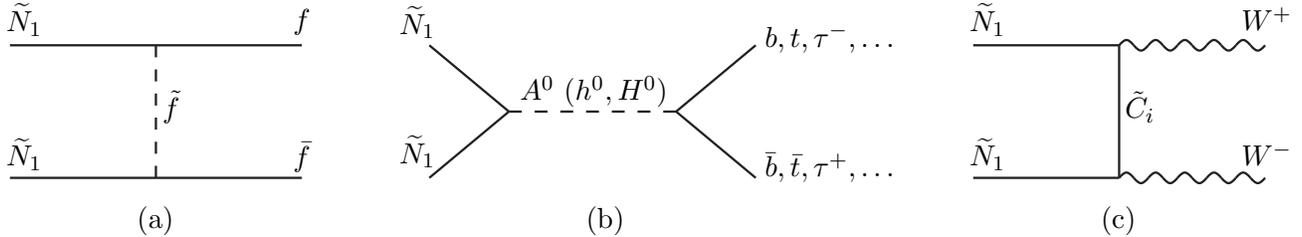
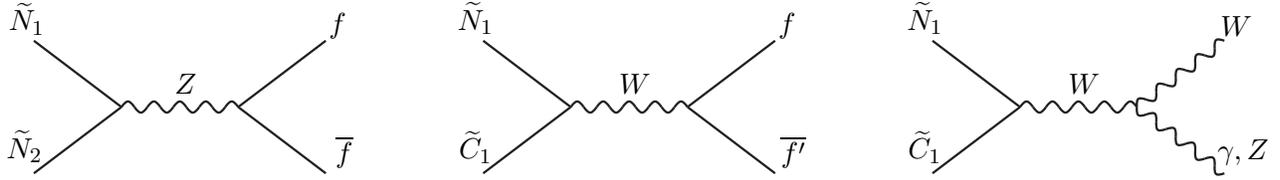
\begin{figure}[!t]
\begin{center}
\begin{picture}(120,50)(0,17)  
\SetWidth{0.85}
\Line(0,50)(33,25)
\Line(77,25)(110,50)
\Photon(33,25)(77,25){2.1}{4.5}
\Line(0,0)(33,25)
\Line(77,25)(110,0)
\rText(-9,10.2)[][]{$\stilde N_2$}
\rText(-8,58.5)[][]{$\stilde N_1$}
\rText(51.9,34)[][]{$Z$}
\rText(109.5,56)[][]{$f$}
\rText(112,8)[][]{$\overline{f}$}
\end{picture}
\hspace{1.5cm}
\begin{picture}(120,50)(0,17)  
\SetWidth{0.85}
\Line(0,50)(33,25)
\Line(77,25)(110,50)
\Photon(33,25)(77,25){2.1}{4.5}
\Line(0,0)(33,25)
\Line(77,25)(110,0)
\rText(-8.3,10)[][]{$\stilde C_1$}
\rText(-8.3,58.9)[][]{$\stilde N_1$}
\rText(51.9,34)[][]{$W$}
\rText(109.5,56)[][]{$f$}
\rText(112,8)[][]{$\overline{f'}$}
\end{picture}
\hspace{1.5cm}
\begin{picture}(120,50)(0,17)  
\SetWidth{0.85}
\Line(0,50)(33,25)
\Photon(77,25)(110,50){2.1}{4.5}
\Photon(33,25)(77,25){2.1}{4.5}
\Line(0,0)(33,25)
\Photon(77,25)(110,0){-2.1}{4.5}
\rText(-8.3,10)[][]{$\stilde C_1$}
\rText(-8.3,58.9)[][]{$\stilde N_1$}
\rText(51.9,34)[][]{$W$}
\rText(109.5,56)[][]{$W$}
\rText(112,8)[][]{$\gamma, Z$}
\end{picture}
\end{center} 
\caption{Some contributions to the co-annihilation of dark matter 
$\stilde N_1$ LSPs with slightly heavier $\stilde N_2$ and $\stilde C_1$.
All three diagrams are particularly important if the LSP is higgsino-like,
and the last two diagrams are important if the LSP is wino-like.
\label{fig:coannihilation}}
\end{figure}

If $\stilde N_1$ is mostly higgsino or mostly wino, then the the 
annihilation diagram fig.~\ref{fig:darkmatterannihilation}c and the 
co-annihilation mechanisms provided by fig.~\ref{fig:coannihilation} are 
typically much too efficient \cite{Mizuta:1992qp,EdsjoGondolo,DNRY} to 
allow the full required cold dark matter density, unless the LSP is very 
heavy, of order 1 TeV or more. This is often considered to be somewhat at 
odds with the idea that supersymmetry is the solution to the hierarchy 
problem; on the other hand, it is consistent with 
the lower bounds set on sparticle masses by the LHC. 
However, for lighter higgsino-like or wino-like LSPs, non-thermal 
mechanisms can be invoked to provide the right dark matter abundance 
\cite{AMSBphenotwo,winoDM}.

A recurring feature of many models of supersymmetry breaking is that the 
lightest neutralino is mostly bino. It turns out that in much of the 
parameter space not already ruled out by LEP with a bino-like $\stilde 
N_1$, the predicted relic density is too high, either because the LSP 
couplings are too small, or the sparticles are too heavy, or both, leading 
to an annihilation cross-section that is too low. To avoid this, there 
must be significant contributions to $\langle \sigma v \rangle$. The 
possibilities can be classified qualitatively in terms of the diagrams 
that contribute most strongly to the annihilation.

First, if at least one sfermion is not too heavy, the diagram of 
fig.~\ref{fig:darkmatterannihilation}a is effective in reducing the dark 
matter density. In models with a bino-like $\stilde N_1$, the most 
important such contribution usually comes from $\stilde e_R$, $\stilde 
\mu_R$, and $\stilde \tau_1$ slepton exchange. The region of parameter 
space where this works out right is often referred to by the jargon ``bulk 
region", because it corresponded to the main allowed region with dark 
matter density less than the critical density, before $\Omega_{\rm DM} 
h^2$ was accurately known and before the highest energy LEP searches had 
happened. However, the diagram of fig.~\ref{fig:darkmatterannihilation}a 
is subject to a $p$-wave suppression, and so sleptons that are light 
enough to reduce the relic density sufficiently are, in many models, also 
light enough to be excluded by LEP or LHC searches, 
or have difficulties with other indirect constraints. 
In the MSUGRA framework described in 
section \ref{subsec:origins.sugra}, the viable bulk region  
takes $m_0$ and $m_{1/2}$ less than about 100 GeV and 250 GeV 
respectively, depending on other parameters. Within MSUGRA, this part of
parameter space has now been thoroughly excluded by the LHC.
If the final state of neutralino pair annihilation 
is instead $t \overline t$, then there is no $p$-wave
suppression. This typically requires a top squark that is less than
about 150 GeV heavier than the LSP, which in turn has $m_{\tilde N_1}$ 
between about
$m_t$ and $m_t + 100$ GeV. This situation
does not occur in the MSUGRA framework,
but can be natural if the ratio of gluino and wino mass parameters, 
$M_3/M_2$, is 
smaller than the unification prediction of eq.~(\ref{eq:TiaEla}) by a 
factor of a few \cite{Compressed}. 

A second way of annihilating excess bino-like LSPs to the correct density 
is obtained if $2 m_{\tilde N_1} \approx m_{A^0}$, or $m_{h^0}$, or 
$m_{H^0}$, as shown in fig.~\ref{fig:darkmatterannihilation}b, so that the 
cross-section is near a resonance pole. An $A^0$ resonance annihilation 
will be $s$-wave, and so more efficient than a $p$-wave $h^0$ or $H^0$ 
resonance. Therefore, the most commonly found realization involves 
annihilation through $A^0$. Because the $A^0 b \overline b$ coupling is 
proportional to $m_b \tan\beta$, this usually entails large values of 
$\tan\beta$ \cite{DNDM}. (Annihilation through $h^0$ is also possible 
\cite{hfunnel}, if the LSP mass is close to $m_{h^0}/2 = 62.5$ GeV.) 
The region of parameter space where this happens is often 
called the ``$A$-funnel" or ``Higgs funnel" or ``Higgs resonance region".

A third effective annihilation mechanism is obtained if $\stilde N_1$ 
mixes to obtains a significant higgsino or wino admixture. Then both 
fig.~\ref{fig:darkmatterannihilation}c and the co-annihilation diagrams of 
fig.~\ref{fig:coannihilation} can be important 
\cite{EdsjoGondolo}. In the ``focus point" region of parameter space, 
where $|\mu|$ is not too large, an LSP with a significant higgsino content 
can yield the correct relic abundance even for very heavy squarks and 
sleptons \cite{focuspointDM}. This is motivated by focusing properties of 
the renormalization group equations, which allow $|\mu|^2 \ll m_0^2$ in 
MSUGRA models \cite{hyperbolic,focuspoint}. In fact, within MSUGRA, squarks are required to be 
very heavy, typically several TeV. This possibility is attractive, given the 
LHC results that exclude most models with squarks lighter than 1 TeV.
It is also possible to arrange for just enough wino content in the LSP to do the 
job \cite{winocontentDM}, by choosing $M_1/M_2$ appropriately.

A fourth possibility, the ``sfermion co-annihilation region" of parameter 
space, is obtained if there is a sfermion that happens to be less than a 
few GeV heavier than the LSP \cite{GriestSeckel}. In many model 
frameworks, this is most naturally the lightest stau 
\cite{staucoannihilation}, but it could also be the lightest top squark 
\cite{stopcoannihilation}. A significant density of this sfermion will 
then coexist with the LSP around the freeze-out time, and so annihilations 
involving the sfermion with itself or with the LSP, including those of the 
type shown in fig.~\ref{fig:sfermioncoannihilation}, will further dilute 
the number of sparticles and so the eventual dark matter density.

It is important to keep in mind that a set of MSSM Lagrangian 
parameters that ``fails" to predict the correct relic dark matter abundance 
by the standard thermal mechanisms is {\em not} ruled out as a model for 
collider physics. This is because simple extensions can completely change 
the dark matter relic abundance prediction without changing the predictions for 
colliders much or at all. For example, if the model predicts a neutralino 
dark matter abundance that is too small, one need only assume another 
sector (even a completely disconnected one) with a stable neutral 
particle, or that the dark matter is supplied by some non-thermal 
mechanism such as out-of-equilibrium decays of heavy particles. If the 
predicted neutralino dark matter abundance appears to be too large, one 
can assume that $R$-parity is slightly broken, so that the offending LSP 
decays before nucleosynthesis; this would require some other unspecified 
dark matter candidate. Or, the dark matter LSP might be some particle that 
the lightest neutralino decays into. One possibility is a gravitino LSP 
\cite{gravitinoDMfromdecays}.  Another example is obtained by extending 
the model to solve the strong CP problem with an invisible axion, which 
can allow the LSP to be a very weakly-interacting axino \cite{axinoDM} 
(the fermionic supersymmetric partner of the axion). In such cases, the 
dark matter density after the lightest neutralino decays would be reduced 
compared to its naively predicted value by a factor of $m_{\rm 
LSP}/m_{\tilde N_1}$, provided that other sources for the LSP relic 
density are absent. A correct density for neutralino LSPs can also be 
obtained by assuming that they are produced non-thermally in reheating of 
the universe after neutralino freeze-out but before nucleosynthesis 
\cite{gelgon}. Finally, in the absence of a compelling explanation for the 
apparent cosmological constant, it seems possible that the standard model 
of cosmology will still need to be modified in ways not yet imagined.

If neutralino LSPs really make up the cold dark matter, then their local 
mass density in our neighborhood ought to be of order 0.3 GeV/cm$^3$ [much 
larger than the density averaged over the largest scales, 
eq.~(\ref{eq:rhodm})] in order to explain the dynamics of our own galaxy. 
LSP neutralinos could be detectable directly through their weak 
interactions with ordinary matter, or indirectly by their ongoing 
annihilations. However, the dark matter halo is subject to significant uncertainties 
in density, velocity, and clumpiness, so even if the Lagrangian 
parameters were known exactly, the signal rates would be quite indefinite, 
possibly even by orders of magnitude.

The direct detection of $\NI$ depends on their elastic scattering off of 
heavy nuclei in a detector. At the parton level, $\NI$ can interact with a 
quark by virtual exchange of squarks in the $s$-channel, or Higgs scalars 
or a $Z$ boson in the $t$-channel. It can also scatter off of gluons 
through one-loop diagrams. The scattering mediated by neutral Higgs 
scalars is suppressed by tiny Yukawa couplings, but is coherent for the 
quarks and so can actually be the dominant contribution for nuclei with 
larger atomic weights, if the squarks are heavy. The energy transferred to 
the nucleus in these elastic collisions is typically of order tens of keV 
per event. There are important backgrounds from natural radioactivity and 
cosmic rays, which can be reduced by shielding and pulse-shape analysis. A 
wide variety of current or future experiments are sensitive to some, but 
not all, of the parameter space of the MSSM that predicts a dark matter 
abundance in the range of eq.~(\ref{eq:OmegaDM}).

Another, more indirect, way to detect neutralino LSPs is through ongoing 
annihilations. This can occur in regions of space where the density is 
greatly enhanced. If the LSPs lose energy by repeated elastic scattering 
with ordinary matter, they can eventually become concentrated inside 
massive astronomical bodies like the Sun or the Earth. In that case, the 
annihilation of neutralino pairs into final states leading to neutrinos is 
the most important process, since no other particles can escape from the 
center of the object where the annihilation is going on. In particular, 
muon neutrinos and antineutrinos from $\NI\NI \rightarrow W^+ W^-$ or 
$ZZ$, (or possibly $\NI\NI \rightarrow \tau^+ \tau^-$ or 
$\nu\overline\nu$, although these are $p$-wave suppressed) will travel 
large distances, and can be detected in neutrino telescopes. The neutrinos 
undergo a charged-current weak interaction in the earth, water, or ice 
under or within the detector, leading to energetic upward-going muons 
pointing back to the center of the Sun or Earth.

Another possibility is that neutralino LSP annihilation in the galactic 
center (or the halo) could result in high-energy photons from cascade 
decays of the heavy Standard Model particles that are produced.  These 
photons could be detected in air Cerenkov telescopes or in space-based 
detectors. There are also interesting possible signatures from neutralino 
LSP annihilation in the galactic halo producing detectable quantities of 
high-energy positrons or antiprotons.

More information on these possibilities, and the various experiments that 
can exploit them, can be found from refs.~\cite{darkmatterreviews} and 
papers referred to in them.

\section{Beyond minimal supersymmetry}\label{sec:variations}
\setcounter{equation}{0}
\setcounter{figure}{0}
\setcounter{table}{0}
\setcounter{footnote}{1}

In this section I will briefly outline a few of my favorite variations 
on the basic picture of the MSSM discussed above. First, the possibility 
of $R$-parity violation is considered in section 
\ref{subsec:variations.RPV}. Another obvious way to extend the MSSM is 
to introduce new chiral supermultiplets, corresponding to scalars and 
fermions that are all sufficiently heavy to have avoided discovery so 
far. This requires that the new chiral supermultiplets must form a real 
representation of the Standard Model gauge group; they can then have a 
significant positive effect on the Higgs boson mass through loop 
corrections, as described in section \ref{subsec:variations.vectorlike}. 
However, the simplest possibility for adding particles is to put them in 
just one gauge-singlet chiral supermultiplet; this can raise the Higgs 
boson mass at tree level, as discussed in section 
\ref{subsec:variations.NMSSM}. The resulting model is also attractive 
because it can solve the $\mu$ problem that was described in sections 
\ref{subsec:mssm.superpotential} and \ref{subsec:MSSMspectrum.Higgs}. 
Two other solutions to the $\mu$ problem, based on including 
non-renormalizable superpotential terms or K\"ahler potential terms, are 
discussed in section \ref{subsec:variations.munonrenorm}. The MSSM could 
also be extended by introducing new gauge interactions that are 
spontaneously broken at high energies. The possibilities here include 
GUT models like $SU(5)$ or $SO(10)$ or $E_6$, which unify the Standard 
Model gauge interactions, with important implications for rare processes 
like proton decay and $\mu \rightarrow e \gamma$.  Superstring models 
also usually enlarge the Standard Model gauge group at high energies.  
One or more Abelian subgroups could survive to the TeV scale, leading to 
a $Z'$ massive vector boson. There is a vast literature on these 
possibilities, but I will concentrate instead on the implications of 
just adding a single $U(1)$ factor that is assumed to be spontaneously 
broken at energies beyond the reach of any foreseeable collider. As 
described in section \ref{subsec:variations.Dterms}, the broken gauge 
symmetry can still leave an imprint on the soft supersymmetry-breaking 
Lagrangian at low energies.

\subsection{$R$-parity violation}\label{subsec:variations.RPV}
\setcounter{equation}{0}

In the preceding, it has been assumed that $R$-parity (or equivalently matter parity) is
an exact symmetry of the MSSM. This assumption precludes renormalizable
proton decay and predicts that the LSP should be stable, but despite these
virtues $R$-parity is not inevitable. Because of the threat of proton
decay, one expects that if $R$-parity is violated, then in the
renormalizable Lagrangian either B-violating or L-violating couplings are
allowed, but not both, as explained in section \ref{subsec:mssm.rparity}. 
There are also upper bounds on the individual $R$-parity violating
couplings \cite{RPVreviews}. 

One proposal is that matter parity can be replaced by an alternative
discrete symmetry that still manages to forbid proton decay at the level
of the renormalizable Lagrangian. The $Z_2$ and $Z_3$ possibilities have
been cataloged in ref.~\cite{baryontriality}, where it was found that
provided no new particles are added to the MSSM, that the discrete
symmetry is family-independent, and that it can be defined at the level of
the superpotential, there is only one other candidate besides matter
parity. That other possibility is a $Z_3$ discrete symmetry
\cite{baryontriality}, which was originally called ``baryon parity" but is
more appropriately referred to as ``baryon triality". The baryon triality
of any particle with baryon number B and weak hypercharge $Y$ is defined
to be
\beq
Z_3^{\rm B} = {\rm exp}\left ({2\pi i} [{\rm B}-2Y]/3 \right ).
\eeq
This is always a cube root of unity, since B$-2Y$ is an integer for every
MSSM particle. The symmetry principle to be enforced is that the product
of the baryon trialities of the particles in any term in the Lagrangian
(or superpotential) must be 1. This symmetry conserves baryon number at
the renormalizable level while allowing lepton number violation; in other
words, it allows the superpotential terms in eq.~(\ref{WLviol}) but
forbids those in eq.~(\ref{WBviol}). In fact, baryon triality conservation
has the remarkable property that it absolutely forbids proton
decay \cite{noprotondecay}. The reason for this is simply that baryon
triality requires that B can only be violated in multiples of 3 units
(even in non-renormalizable interactions), while any kind of proton decay
would have to violate B by 1 unit. So it is eminently falsifiable.
Similarly, baryon triality conservation predicts that experimental
searches for neutron-antineutron oscillations will be negative, since they
would violate baryon number by 2 units. However, baryon triality
conservation does allow the LSP to decay. If one adds some new chiral
supermultiplets to the MSSM (corresponding to particles that are
presumably very heavy), one can concoct a variety of new candidate
discrete symmetries besides matter parity and baryon triality. Some of
these will allow B violation in the superpotential, while forbidding the
lepton number violating superpotential terms in eq.~(\ref{WLviol}). 

Another idea is that matter parity is an exact symmetry of the underlying 
superpotential, but it is spontaneously broken by the VEV of a scalar with 
$P_R=-1$.  One possibility is that an MSSM sneutrino gets a VEV 
\cite{sneutvevRPV}, since sneutrinos are scalars carrying L=1. However, 
there are strong bounds \cite{nonsneutvevRPV} on $SU(2)_L$-doublet 
sneutrino VEVs $\langle \stilde \nu \rangle \ll m_Z$ coming from the 
requirement that the corresponding neutrinos do not have large masses. It 
is somewhat difficult to understand why such a small VEV should occur, 
since the scalar potential that produces it must include soft sneutrino 
squared-mass terms of order $m^2_{\rm soft}$. One can get around this by 
instead introducing a new gauge-singlet chiral supermultiplet with L=$-1$. 
The scalar component can get a large VEV, which can induce L-violating 
terms (and in general B-violating terms also) in the low-energy effective 
superpotential of the MSSM \cite{nonsneutvevRPV}.

In any case, if $R$-parity is violated, then the collider searches for
supersymmetry can be completely altered. The new couplings imply
single-sparticle production mechanisms at colliders, besides the usual
sparticle pair production processes. First, one can have $s$-channel
single sfermion production. At electron-positron colliders, the $\lambda$
couplings in eq.~(\ref{WLviol}) give rise to $e^+e^-\rightarrow \stilde
\nu$. At the LHC, single sneutrino or charged slepton
production, $q \bar q \rightarrow \stilde \nu$ or $\stilde \ell$ are
mediated by $\lambda'$ couplings, and single squark production $qq
\rightarrow \stilde {\bar q}$ is mediated by $\lambda''$ couplings in
eq.~(\ref{WBviol}). 

Second, one can have $t$-channel exchange of sfermions, providing for
gaugino production in association with a standard model fermion. At
electron-positron colliders, one has $e^+ e^- \rightarrow \stilde C_i
\ell$ mediated by $\stilde \nu_e$ in the $t$-channel, and $e^+ e^-
\rightarrow \stilde N_i \nu$ mediated by selectrons in the $t$-channel, if
the appropriate $\lambda$ couplings are present. At the 
LHC, one can look for the partonic processes $q\overline q \rightarrow
(\stilde N_i\>{\rm or}\>\stilde C_i\>{\rm or}\>\tilde g)+(\ell\>{\rm
or}\>\nu)$, mediated by $t$-channel squark exchange if $\lambda'$
couplings are present. If instead $\lambda''$ couplings are present, then
$qq \rightarrow (\stilde N_i\>{\rm or}\>\stilde C_i\>{\rm or}\>\tilde g)+
q$, again with squarks exchanged in the $t$-channel, provides a possible
production mechanism. 

Next consider sparticle decays. In many cases, the $R$-parity violating
couplings are already constrained by experiment, or expected from more
particular theoretical models, to be smaller than electroweak gauge
couplings \cite{RPVreviews}. If so, then the heavier sparticles will
usually decay to final states containing the LSP, as in section
\ref{sec:decays}. However, now the LSP can also decay; if it is a
neutralino, as most often assumed, then it will decay into three Standard
Model fermions. The collider signals to be found depend on the type of
$R$-parity violation. 

\begin{figure}
\begin{center}
\begin{picture}(135,61)(0,-18)
\SetWidth{0.85}
\Line(0,0)(45,0)
\Line(45,0)(67.5,40.5)
\DashLine(45,0)(90,0){4.5}
\Line(90,0)(112.5,40.5)
\Line(90,0)(135,0)
\Text(7,10)[c]{$\stilde N_1$}
\Text(70,9)[c]{$\stilde \ell$}
\Text(73,41)[c]{$\ell$}
\Text(118.5,41)[c]{$\ell'$}
\Text(135,7)[c]{$\nu''$}
\Text(90.5,-6.5)[c]{$\lambda$}
\Text(67.5,-18.3)[c]{(a)}
\end{picture}
\hspace{1.15cm}
\begin{picture}(135,61)(0,-18)
\SetWidth{0.85}
\Line(0,0)(45,0)
\Line(45,0)(67.5,40.5)
\DashLine(45,0)(90,0){4.5}
\Line(90,0)(112.5,40.5)
\Line(90,0)(135,0)
\Text(7,10)[c]{$\stilde N_1$}
\Text(70,9)[c]{$\stilde \nu''$}
\Text(75,42)[c]{$\nu''$}
\Text(118,41)[c]{$\ell$}
\Text(135,7)[c]{$\ell'$}
\Text(90.5,-6.5)[c]{$\lambda$}
\Text(67.5,-18.3)[c]{(b)}
\end{picture}
\hspace{1.15cm}
\begin{picture}(135,61)(0,-18)
\SetWidth{0.85}
\Line(0,0)(45,0)
\Line(45,0)(67.5,40.5)
\DashLine(45,0)(90,0){4.5}
\Line(90,0)(112.5,40.5)
\Line(90,0)(135,0)
\Text(7,10)[c]{$\stilde N_1$}
\Text(70,9)[c]{$\stilde \ell$}
\Text(73,41)[c]{$\ell$}
\Text(118,41)[c]{$q$}
\Text(133,8)[c]{$q'$}
\Text(90.5,-6.5)[c]{$\lambda'$}
\Text(67.5,-18.3)[c]{(c)}
\end{picture}

\vspace{1cm}
\begin{picture}(135,51)(0,-8)
\SetWidth{0.85}
\Line(0,0)(45,0)
\Line(45,0)(67.5,40.5)
\DashLine(45,0)(90,0){4.5}
\Line(90,0)(112.5,40.5)
\Line(90,0)(135,0)
\Text(7,10)[c]{$\stilde N_1$}
\Text(70,8.2)[c]{$\stilde \nu$}
\Text(73,41)[c]{$\nu$}
\Text(118,41)[c]{$q$}
\Text(135,9)[c]{$q'$}
\Text(90.5,-6.5)[c]{$\lambda'$}
\Text(67.5,-18.3)[c]{(d)}
\end{picture}
\hspace{1.15cm}
\begin{picture}(135,51)(0,-8)
\SetWidth{0.85}
\Line(0,0)(45,0)
\Line(45,0)(67.5,40.5)
\DashLine(45,0)(90,0){4.5}
\Line(90,0)(112.5,40.5)
\Line(90,0)(135,0)
\Text(7,10)[c]{$\stilde N_1$}
\Text(70,9)[c]{$\stilde q$}
\Text(73,41)[c]{$q$}
\Text(117,41)[l]{$\ell$ or $\nu$}
\Text(135,9)[c]{$q'$}
\Text(90.5,-6.5)[c]{$\lambda'$}
\Text(67.5,-18.3)[c]{(e)}
\end{picture}
\hspace{1.15cm}
\begin{picture}(135,51)(0,-8)
\SetWidth{0.85}
\Line(0,0)(45,0)
\Line(45,0)(67.5,40.5)
\DashLine(45,0)(90,0){4.5}
\Line(90,0)(112.5,40.5)
\Line(90,0)(135,0)
\Text(7,10)[c]{$\stilde N_1$}
\Text(70,9)[c]{$\stilde q$}
\Text(73,41)[c]{$q$}
\Text(119.5,41)[c]{$q'$}
\Text(133,9)[c]{$q''$}
\Text(90.5,-6.5)[c]{$\lambda''$}
\Text(67.5,-18.3)[c]{(f)}
\end{picture}
\end{center}
\caption{Decays of the $\NI$ LSP in models with $R$-parity violation, 
with lepton number not conserved (a)-(e) [see eq.~(\ref{WLviol})], and 
baryon number not conserved (f) [see eq.~(\ref{WBviol})].
\label{fig:rparityviolation}}
\end{figure}
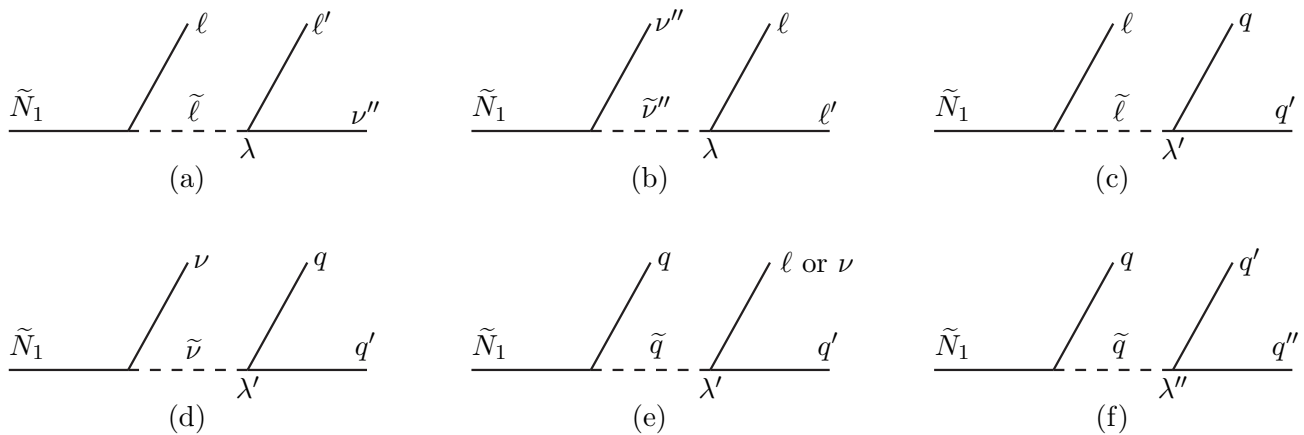

Lepton number violating terms of the type $\lambda$ as in
eq.~(\ref{WLviol}) will lead to final states from $\stilde N_1$ decay with
two oppositely charged, and possibly different flavor, leptons and a
neutrino, as in Figure~\ref{fig:rparityviolation}a,b. 
Couplings of the $\lambda'$ type will cause $\stilde N_1$ to decay to a
pair of jets and either a charged lepton or a neutrino, as shown in
Figure~\ref{fig:rparityviolation}c,d,e.  Signals with L-violating LSP
decays will therefore always include charged leptons or large missing
energy, or both. 

On the other hand, if terms of the form $\lambda^{\prime\prime}$ in
eq.~(\ref{WBviol}) are present instead, then there are B-violating decays
$\stilde N_1 \rightarrow q q^\prime q^{\prime\prime}$ from diagrams like
the one shown in Figure~\ref{fig:rparityviolation}f. In that case,
supersymmetric events will always have lots of hadronic activity, and will
only have physics missing energy signatures when the other parts of the
decay chains happen to include neutrinos. 

There are other possibilities, too.  The decaying LSP need not be $\NI$.
Sparticles that are not the LSP can, in principle, decay directly to
Standard Models quarks and leptons, if the $R$-parity violating couplings
are large enough. The $t$-channel exchange of sfermions can produce a pair
of Standard Model fermions, leading to indirect sparticle signatures.
Or, if the $R$-parity violating couplings are sufficiently small, then the
LSP will usually decay outside of collider detectors, and the model will
be difficult or impossible to distinguish from the $R$-parity conserving
case. Surveys of experimental constraints and future
prospects can be found in \cite{RPVreviews}. 

\subsection{Extra vectorlike chiral 
supermultiplets}\label{subsec:variations.vectorlike}
\setcounter{equation}{0}

An interesting way to extend the MSSM is by adding extra particles 
in chiral supermultiplets. 
It has now become clear that together 
the new fields must form a vectorlike (self-conjugate) representation 
of the Standard Model gauge group. Otherwise, the only way the new 
fermions could have masses large enough to have avoided discovery would be through
extremely large Yukawa couplings to the Higgs VEVs. 
These couplings would in turn lead to very large 
corrections to the 125 GeV Higgs boson production cross-section at the LHC 
through loop effects, as well as corrections to electroweak precision observables, 
both in contradiction with the observations. In contrast, the addition of
chiral supermultiplets with vectorlike quantum numbers to the MSSM does not lead to such problems, and can help to raise the lightest Higgs boson mass up to 125 GeV
in models where it would otherwise be too light 
\cite{Moroi:1991mg}-\cite{Graham:2009gy}.

If the new vectorlike chiral supermultiplets live in the fundamental representation of 
$SU(2)_L$ or $SU(3)_c$, or are charged under $U(1)_Y$, then they must come in pairs with opposite gauge quantum numbers. 
If we call such a pair 
$\Phi_i$ and $\overline \Phi_i$, then there is an allowed superpotential 
mass term of the form
\beq
W = M_i \Phi_i \overline \Phi_i ,
\label{eq:Wvectorlikemass}
\eeq
which does not involve any interactions with the Higgs boson. Note that 
such electroweak singlet mass terms 
can arise from whatever mechanism also gives rise to the $\mu$ term of the MSSM. Three 
such possible mechanisms are described below in sections \ref{subsec:variations.NMSSM} and
\ref{subsec:variations.munonrenorm}. Whatever that mechanism is, it is 
reasonable to suppose that it operates the same way to produce the masses $M_i$ 
with the same order of magnitude as $\mu$, i.e.~at the TeV scale. 

Because the new vectorlike particle have mostly electroweak singlet 
masses, they do not impact Higgs boson production and decay, and 
decouple from precision electroweak observables involving the $Z$ and 
$W$ self-energies and the Standard Model fermions.  In order for the 
lightest of the new particles to not cause problems as stable relics 
from thermal production in the early universe, one may suppose that 
either $\Phi_i$ or $\overline \Phi_i$ has the same gauge quantum numbers 
as one of the MSSM quark and lepton chiral superfields, allowing small 
mixing Yukawa couplings to the Higgs boson. This small mixing allows the 
new vectorlike fermions to decay to Standard Model fermions.

If they are indeed at the TeV scale, the new particles can be 
pair-produced at the LHC, either through gluon fusion or through 
$s$-channel $W$ or $Z$ boson diagrams. Thus one can look for heavy 
cousins of the top quark, bottom quark, and/or tau lepton; call them 
$t'$, $b'$, and $\tau'$. These fermions will have decays that depend on 
the choice of mixing terms between them and the Standard Model fermions. 
The easiest way to minimize possible flavor problems in low-energy 
experiments is to assume that the mixing is primarily with the third 
family. Then the relevant decays will be:
\beq
&&t' \rightarrow Zt,\>\>\>h^0 t,\>\>\> W^+b,
\\
&&b' \rightarrow Zb,\>\>\>h^0 b,\>\>\> W^-t,
\\
&&\tau' \rightarrow Z\tau,\>\>\>h^0 \tau,\>\>\> W^-\nu,
\eeq
with branching ratios that depend on the type of mixing Yukawa coupling. 
The possibilities and the resulting branching ratio predictions
are discussed in detail in \cite{Martin:2009bg}. 
If the Yukawa couplings that mix the
new fermions to the Standard Model fermions are larger than about $10^{-6}$, these
2-body decays will occur promptly within collider detectors. The scalar partners
of these fermionic states are likely to be much heavier, because they have soft 
supersymmetry-breaking contributions to their masses. In addition, for a given
mass, production 
cross-sections for scalars tend to be lower than for fermions, so it is 
most likely that the new vectorlike fermions will be discovered first.

In order to raise the Higgs boson mass, one can also introduce a Yukawa 
coupling between the new chiral supermultiplets 
and the MSSM Higgs fields. As an example, suppose 
there are extra vectorlike chiral supermultiplets in the following 
representations of $SU(3)_c \times SU(2)_L \times U(1)_Y$:
\beq
{\cal Q} &=& ({\bf 3}, {\bf 2}, +1/6),\qquad\quad
\overline{\cal Q} \>=\> ({\bf \overline{3}}, {\bf 2}, -1/6),
\\
{\cal U} &=& ({\bf 3}, {\bf 1}, +2/3),\qquad\quad
\overline{\cal U} \>=\> ({\bf \overline{3}}, {\bf 1}, -2/3).
\eeq
Then the allowed superpotential terms include:
\beq
W &=& M_{\cal Q} {\cal Q}\overline{\cal Q} + 
M_{\cal U} {\cal U}\overline{\cal U}
+ k H_u {\cal Q}\overline{\cal U}
\eeq
where $M_{\cal Q}$ and $M_{\cal U}$ are electroweak singlet masses 
as in eq.~(\ref{eq:Wvectorlikemass}), and $k$ is a Yukawa 
coupling, which can be large and yet provide only a subdominant contribution to the
masses of the vectorlike states. There is an infrared-stable 
quasi-fixed point at $k\approx 1.05$, giving a natural expectation for its magnitude
\cite{Martin:2009bg}. 
This coupling mediates a positive 1-loop contribution to 
lightest Higgs scalar boson mass, provided that 
the masses of the new scalars are larger than the masses of the new fermions.
(This is similar to the 1-loop contribution from the top/stop sector.)   
An approximate formula for this contribution, 
with several simplifying assumptions, is \cite{Babu:2008ge}:
\beq
\Delta (m_{h^0}^2) = \frac{3}{4 \pi^2} k^4 v^2 \sin^4\beta \left [
\ln(x) - \frac{1}{6} (5-1/x)(1-1/x) \right ]
\label{eq:deltamhvectorlike}
\eeq
Here $x = M_S^2/M_F^2$, and it is assumed 
that the scalars in ${\cal Q}, \overline{\cal Q},
{\cal U}$, and $\overline{\cal U}$ are approximately degenerate with 
each other with average mass $M_S$, and likewise for the new fermions with 
average mass $M_F \approx M_{\cal Q} \approx M_{\cal U} $, that 
$k v_u$ is a small perturbation on these masses,
and that the mixing in the new scalar sector is small. 
It is also assumed that the Higgs bosons are in 
the decoupling limit described at the end of section
\ref{subsec:MSSMspectrum.Higgs}. For $x>1$, 
eq.~(\ref{eq:deltamhvectorlike}) is positive definite 
and monotonically increasing with $x$. For example, with $x=4$, the correction 
to the Higgs boson mass can be about 10 GeV. (Results for the Higgs mass 
correction with these assumptions relaxed can be found in \cite{Martin:2009bg}; mixing in the scalar sector increases the Higgs mass correction.)
Note that even in the limit of very large $M_F$, the contribution to $m_{h^0}^2$
does not decouple, provided only that the hierarchy $x>1$ is maintained. Despite 
this non-decoupling contribution to $m_{h^0}^2$, the contributions to 
precision electroweak observables do decouple quadratically (like $m_W^2/M_F^2$), and so are quite benign \cite{Martin:2009bg}.

The positive contribution to the Higgs mass from extra vectorlike quarks 
is a plausible way to rescue supersymmetric theories that 
would otherwise have difficulty in accommodating the 125 GeV Higgs boson. For example,
GMSB models typically predict much lower $m_{h^0}$, unless all of the superpartners are 
well out of reach of the LHC, because they imply small top-squark mixing. 
However, including extra vectorlike quarks with a large Yukawa coupling
allows the MSSM superpartners to be as light as their direct experimental 
limits in GMSB models, while still allowing
$m_{h^0} = 125$ GeV 
\cite{Endo:2011mc}-\cite{Martin:2012dg}.

\subsection{The next-to-minimal supersymmetric standard
model}\label{subsec:variations.NMSSM}
\setcounter{equation}{0}

The simplest possible extension of the particle content of the MSSM is 
obtained by adding a new gauge-singlet chiral supermultiplet that is 
even under matter parity. The resulting model 
\cite{NMSSM}-\cite{nMSSM} 
is often 
called the next-to-minimal supersymmetric standard model or NMSSM or 
(M+1)SSM. The most general renormalizable superpotential for this field 
content is
\beq
W_{\rm NMSSM} \>=\> W_{\rm MSSM}
+ \lambda S H_u H_d + \frac{1}{3} \kappa S^3 + \frac{1}{2} \mu_S S^2  ,
\label{NMSSMwww}
\eeq
where $S$ stands for both the new chiral supermultiplet and its scalar 
component. There could also be a term linear in $S$ in $W_{\rm NMSSM}$, 
but in global supersymmetry it can always be removed by redefining $S$ by 
a constant shift. The soft supersymmetry-breaking Lagrangian is
\beq
{\cal L}_{\rm soft}^{\rm NMSSM} \>=\> {\cal L}_{\rm soft}^{\rm MSSM}
-\bigl (
a_{\lambda} S H_u H_d - {1\over 3} a_\kappa S^3 + {1\over 2} b_S S^2
+ t S
+ \conj \bigr ) -m_S^2 |S|^2
.
\label{eq:NMSSMsoft}
\eeq
The tadpole coupling $t$ could be subject to dangerous quadratic 
divergences in supergravity \cite{NMSSMtadpole} unless it is highly 
suppressed or forbidden by some additional symmetry at very high energies.

One of the virtues of the NMSSM is that it can provide a solution to the
$\mu$ problem mentioned in sections \ref{subsec:mssm.superpotential} and
\ref{subsec:MSSMspectrum.Higgs}. To understand this, suppose we
set\footnote{The even more economical case with only $t \sim m_{\rm
soft}^3$ and $\lambda$ and $a_\lambda$ nonzero is also viable and
interesting \cite{nMSSM}.} $\mu_S = \mu = 0$ so that there are no mass
terms or dimensionful parameters in the superpotential at all, and also
set the corresponding terms $b_S = b = 0$ and $t=0$ in the
supersymmetry-breaking Lagrangian.  If 
$\lambda$, $\kappa$, $a_\lambda$, and $a_{\kappa}$ are chosen
auspiciously, then phenomenologically acceptable VEVs will be induced for
$S$, $H_u^0$, and $H_d^0$.  By doing phase rotations on these fields, all
three of $s \equiv \langle S \rangle$ and $v_u = v \sin\beta = \langle
H_u^0 \rangle$ and $v_d = v \cos\beta = \langle H_d^0 \rangle$ can be made
real and positive. In this convention, $a_\lambda + \lambda \kappa^* s$
and $a_{\kappa} + 3 \lambda^* \kappa v_u v_d/s$ will also be real and
positive. 

However, in general, this theory could have unacceptably large CP 
violation. This can be avoided by assuming that $\lambda$, $\kappa$, 
$a_\lambda$ and $a_\kappa$ are all real in the same convention that makes 
$s$, $v_u$, and $v_d$ real and positive; this is natural if the mediation 
mechanism for supersymmetry breaking does not introduce new CP violating 
phases, and is assumed in the following. To have a stable minimum with 
respect to variations in the scalar 
field phases, it is required that $a_\lambda + 
\lambda \kappa s > 0$ and $a_\kappa (a_\lambda + \lambda \kappa s) + 3 
\lambda \kappa a_\lambda v_u v_d/s > 0$. (An obvious sufficient, but not
necessary, way to achieve these two conditions is to assume that 
$\lambda \kappa > 0$ and $a_\kappa>0$ and $a_\lambda > 0$.)

An effective $\mu$-term for $H_u H_d$ will arise from
eq.~(\ref{NMSSMwww}), with \beq \mu_{\rm eff} \>=\> \lambda s. \eeq It is
determined by the dimensionless couplings and the soft terms of order
$m_{\rm soft}$, instead of being a free parameter conceptually independent
of supersymmetry breaking. With the conventions chosen here, the
sign of $\mu_{\rm eff}$ (or more generally its phase) is the same as that
of $\lambda$. Instead of eqs.~(\ref{mubsub2}), (\ref{mubsub1}), the
minimization conditions for the Higgs potential are now: 
\beq
m^2_{H_u} + \lambda^2 (s^2 + v^2 \cos^2\beta) 
- (a_\lambda + \lambda \kappa s) s \cot\beta
- (m_Z^2/2) \cos(2\beta) &=& 0,
\\
m^2_{H_d} + \lambda^2 (s^2 + v^2 \sin^2\beta) 
-(a_\lambda + \lambda \kappa s) s \tan\beta
+ (m_Z^2/2) \cos(2\beta) &=& 0,
\\
m^2_{S} + \lambda^2 v^2 + 2 \kappa^2 s^2 - a_\kappa s
-(\kappa\lambda  + a_\lambda/2 s) v^2 \sin(2\beta) &=& 0.
\eeq
The effects of radiative corrections $\Delta V(v_u,v_d,s)$ to the
effective potential are included by replacing $m_S^2 \rightarrow m_S^2 +
[\partial (\Delta V)/\partial s]/2s$, in addition to
eq.~(\ref{eq:Vradcor}). 

The absence of dimensionful terms in $W_{\rm NMSSM}$, and the
corresponding terms in $V_{\rm soft}^{\rm NMSSM}$, can be enforced by
introducing a new symmetry. The simplest way is to notice that the new
superpotential and Lagrangian will be invariant under a $Z_3$ discrete
symmetry, under which every field in a chiral supermultiplet transforms as
$\Phi \rightarrow e^{2 \pi i/3} \Phi$, and all gauge and gaugino fields
are inert. Imposing this symmetry indeed eliminates $\mu$, $\mu_S$, $b$,
$b_S$, and $t$. However, if this symmetry were exact, then because it must
be spontaneously broken by the VEVs of $S$, $H_u$ and $H_d$, domain walls
are expected to be produced in the electroweak symmetry breaking phase
transition in the early universe \cite{NMSSMdomainwalls}. These would
dominate the cosmological energy density, and would cause unobserved
anisotropies in the microwave background radiation. Several ways of
avoiding this problem have been proposed, including late inflation after
the domain walls are formed, embedding the discrete symmetry into a
continuous gauged symmetry at very high energies, or allowing either
higher-dimensional terms in the Lagrangian or a very small $\mu$ term to
explicitly break the discrete symmetry. 

The NMSSM contains, besides the particles of the MSSM, a real $P_R=+1$ 
scalar, a real $P_R=+1$ pseudo-scalar, and a $P_R=-1$ Weyl fermion 
``singlino". These fields have no gauge couplings of their own, so they 
can only interact with Standard Model particles by mixing with the neutral 
MSSM fields with the same spin and charge. The real scalar mixes with the 
MSSM particles $h^0$ and $H^0$, and the pseudo-scalar mixes with $A^0$. 
One of the effects of replacing the $\mu$ term by the dynamical field $S$ 
is that the lightest Higgs boson squared mass is raised, by an amount bounded 
at tree-level by:
\beq
\Delta(m_{h^0}^2) &\leq& \lambda^2 v^2 \sin^2 (2 \beta).
\eeq
This extra contribution comes from the $|F_S|^2$ contribution to 
the scalar potential. Its effect is limited, because there is an upper bound 
$\lambda \lsim 0.8$ if one 
requires that $\lambda$ not have a Landau pole in its RG running below the 
GUT mass scale.  
Also, the neutral Higgs scalars have reduced couplings to the electroweak 
gauge bosons, compared to those in the Standard Model, because of the 
mixing with the singlets. Because the 125 GeV Higgs boson discovered by the LHC
appears to have properties like those of a Standard Model Higgs boson, it seems unlikely 
to have a large admixture of the single field $S$. This means that there could be
a yet-undiscovered neutral Higgs scalar that is mostly electroweak singlet and
even lighter than 125 GeV.

The odd $R$-parity singlino $\stilde S$ mixes with the four MSSM 
neutralinos, so there are really five neutralinos now. The singlino could 
be the LSP, depending on the parameters of the model, and so could be the
dark matter \cite{NMSSMdarkmatter}. The neutralino mass matrix in the 
$\psi^0 = (\stilde B, \stilde W^0, \stilde H_d^0, \stilde H_u^0, \stilde S)$ 
gauge-eigenstate basis is: 
\beq
{\bf M}_{\stilde N} \,=\, \pmatrix{
  M_1 & 0 & -g' v_d/\sqrt{2} & g' v_u/\sqrt{2} & 0\cr
  0 & M_2 & g v_d/\sqrt{2} & -g v_u/\sqrt{2} & 0\cr
  -g' v_d/\sqrt{2} & g v_d/\sqrt{2} & 0 & -\lambda s & -\lambda v_u\cr
  g' v_u/\sqrt{2} & -g v_u/\sqrt{2}& -\lambda s & 0 & -\lambda v_d\cr 
  0 & 0 & -\lambda v_u & -\lambda v_d & 2 \kappa s}.
\label{NMSSMNinomassmatrix}
\eeq
[Compare eq.~(\ref{preneutralinomassmatrix}).] For small $v/s$ and
$\lambda v/\kappa s$, mixing effects of the singlet Higgs scalar and the singlino are
small, and they nearly decouple. In that case, the phenomenology of the
NMSSM is almost indistinguishable from that of the MSSM. For larger
$\lambda$, the mixing is important and the experimental signals for
sparticles and the Higgs scalars can be altered in important ways
\cite{NMSSMpheno}-\cite{nMSSM}, \cite{NMHDECAY}.

\subsection{The $\mu$-term from non-renormalizable 
Lagrangian terms}\label{subsec:variations.munonrenorm}
\setcounter{equation}{0}

The previous subsection described how the NMSSM can provide a solution to the
$\mu$ problem. Another possible solution involves generating $\mu$ from 
non-renormalizable Lagrangian terms. If the non-renormalizable terms are in the
superpotential,
this is called the Kim-Nilles mechanism\cite{KimNilles}, 
and if they are in the K\"ahler potential
it is called the Giudice-Masiero mechanism\cite{GiudiceMasiero}.

It is useful to note that when the $\mu$ term is set to zero,
the MSSM superpotential 
has a global $U(1)$ Peccei-Quinn symmetry, with charges listed in Table 
\ref{table:PecceiQuinn}. 
This symmetry cannot be an exact symmetry of the Lagrangian, since it has
an $SU(3)_c$ anomaly. However, if all other sources of Peccei-Quinn breaking
are small, then there must result a pseudo-Nambu-Goldstone boson, the axion.
If the scale of the breaking is too low, then the axion would be ruled
out by astrophysical observations, so one must introduce an additional
explicit breaking of the Peccei-Quinn symmetry. This is what happens in the
NMSSM of the previous section. 
On the other hand, if the scale of Peccei-Quinn breaking is
such that the axion decay constant
is in the range
\beq
10^{9}\>\mbox{GeV} \> \lsim \,f\, \lsim\> 10^{12}\>\mbox{GeV},
\label{eq:axionfrange}
\eeq
then the resulting axion is of the invisible DFSZ type \cite{DFSZ}
that is consistent with present astrophysical constraints. This is an
enticing possibility, since it links the solution to the strong CP problem to
supersymmetry breaking. 
\renewcommand{\arraystretch}{1.4}
\begin{table}[tb]
\begin{center}
\begin{tabular}{|c|c|c|c|c|c|c|c|}
\hline
& $H_u$ & $H_d$ & $Q$ & $L$ & $\overline u$ & $\overline d$ &
$\overline e$ \\
\hline Peccei-Quinn charge & $+1$ & $+1$ & $-1$ & $-1$ & $0$ & $0$ &
$0$ \\ \hline
\end{tabular}
\caption{Peccei-Quinn charges of MSSM chiral superfields.
These charges are not unique,
as one can add to them any multiple of the weak hypercharge or 
B$-$L.\label{table:PecceiQuinn}} 
\end{center}
\end{table}

To illustrate the Kim-Nilles mechanism, consider 
the non-renormalizable superpotential
\beq
W = \frac{\lambda_\mu}{2\MPlanck} S^2 H_u H_d,
\eeq
where $S$ is an $SU(3)_c \times SU(2)_L \times U(1)_Y$ singlet chiral
superfield, and
$\lambda_\mu$
is a dimensionless coupling normalized by the reduced Planck mass $\MPlanck$. From Table 
\ref{table:PecceiQuinn}, $S$ has Peccei-Quinn charge $-1$.
If $S$ obtains a VEV that is parametrically of order
\beq
\langle S \rangle \sim
\sqrt{m_{\rm soft} M_P},
\label{eq:KimNillesSVEV}
\eeq
then the spontaneous breaking of the
Peccei-Quinn symmetry gives rise to an invisible 
axion of the DFSZ type \cite{DFSZ}, with a decay constant
$f \sim \langle S \rangle$ that will 
automatically be in the range eq.~(\ref{eq:axionfrange}).
The low-energy effective theory will then contain the usual $\mu$ term, with
\beq
\mu = \frac{\lambda_\mu}{2 M_P} \langle S^2 \rangle  \sim m_{\rm soft},
\eeq
simultaneously solving 
the $\mu$ problem and the strong
CP problem. It is natural to also
have a dimensionless, holomorphic soft supersymmetry-breaking term in the
Lagrangian of the form:
\beq
-{\cal L}_{\rm soft} = \frac{a_b}{M_P} S^2 H_u H_d + {\rm c.c.},
\eeq
where $a_b$ is of order $m_{\rm soft}$.
The $b$ term in the MSSM will then arise as \beq
b = \frac{a_b}{M_P} \langle S^2
\rangle,
\eeq
and will be of order $m_{\rm soft}^2$, as required for
electroweak symmetry breaking.

To ensure the required spontaneous breaking with a stable vacuum, 
one can introduce an additional
non-renormalizable superpotential term, in several different possible ways
\cite{MurayamaSuzukiYanagida}-\cite{Martinaxinos}.
For example, one could take \cite{Martinaxinos}:
\beq
W = \frac{\lambda_S}{4 M_{P}} S^2 S^{\prime 2},
\eeq
where $S'$ is a chiral superfield with Peccei-Quinn charge $+1$.
This implies a scalar potential that stabilizes $S$ and $S'$ at
large field strength:
\beq
V_S \,=\, |F_S|^2 + |F_{S'}|^2 \,=\, \frac{|\lambda_S|^2}{4 M_P^2}
|S S'|^2 (|S|^2 + |S'|^2) .
\eeq
There is also a soft supersymmetry-breaking Lagrangian:
\beq
-{\cal L}_{\rm soft} = V_{\rm soft} = m^2_S |S|^2  + m^2_{S'}
|S'|^2 - \left ( \frac{a_S}{4 M_P} S^2 S^{\prime 2}  + {\rm c.c.} \right ),
\eeq
where $m^2_S$ and $m^2_{S'}$ are of order $m_{\rm soft}^2$
and $a_S$ is of order $m_{\rm soft}$. The total scalar potential
$V_S + V_{\rm soft}$
will have
an appropriate VEV of order eq.~(\ref{eq:KimNillesSVEV}) provided that
$m^2_S$, $m^2_{S'}$ are
negative or if $a_S$ is sufficiently large. For example, with $m^2_S =
m^2_{S'}$ for simplicity, there will be a non-trivial minimum of
the potential if $|a_S|^2 - 12 m_S^2 |\lambda_S|^2 > 0$,
and it will be a global minimum of the potential if $|a_S|^2 - 16 m_S^2
|\lambda_S|^2 > 0$.

One pseudo-scalar degree of freedom, a mixture of $S$ and $S'$,
is the axion, with a very small mass. The rest of the chiral
supermultiplet from which the axion came will have masses of order
$m_{\rm soft}$, but couplings to the MSSM that are highly suppressed.
However, if one of the fermionic members of this chiral supermultiplet
(a singlino that can be properly called an ``axino'' $\tilde a$, and
which has tiny mixing with the MSSM neutralinos $\tilde N_i$) is lighter
than all of the MSSM odd $R$-parity particles,
then it could be the LSP dark matter. If its relic density 
arises predominantly from decays of the would-be LSP $\tilde N_1$, then today 
$\Omega_{\rm DM} h^2$ today can be obtained from that one would have obtained for $\tilde 
N_1$ if it were stable, but just
suppressed by a factor of $m_{\tilde a}/m_{\tilde N_1}$. It is
also possible that the decay of $\tilde N_1$ to $\tilde a$ could occur
within a collider detector, rarely and with a macroscopic decay length but
just often enough to provide a signal in a sufficiently large sample of
superpartner pair production events \cite{Martinaxinos}.

There are several variations on the theme given above.
The non-renormalizable superpotential could instead
have the schematic form
$S^3 S' + S S' H_u H_d$
as in the original explicit model of this type
\cite{MurayamaSuzukiYanagida}, or
$S^3 S' + S^2 H_u H_d$
as in \cite{ChoiChunKim},
or $S S^{\prime 3} + S^2 H_u H_d$ as in \cite{Martinaxinos},
each entailing a different assignment
of Peccei-Quinn charges for the gauge singlet fields, 
but with qualitatively similar behavior.
One can also introduce more than two new fields that break the
Peccei-Quinn symmetry at the intermediate scale.

The Giudice-Masiero mechanism instead relies on 
a non-renormalizable contribution to the
K\"ahler potential in addition to the usual 
canonical terms for the MSSM Higgs fields:
\beq
K = H_u H_u^* + H_d H_d^* + \Bigl (\frac{\lambda_\mu}{M_P} H_u H_d X^* +
{\rm c.c.} \Bigr ) + \ldots. \label{eq:KahlerpotentialforGiudiceMasiero}
\eeq
Here $\lambda_\mu$ is a dimensionless coupling parameter and 
$X$ has Peccei-Quinn charge $+2$, and is a chiral superfield responsible for
spontaneous breaking of supersymmetry through its auxiliary $F$ field.
Giudice and Masiero showed
\cite{GiudiceMasiero} that in supergravity, the presence of such
couplings in the K\"ahler potential
will always give rise to a non-zero $\mu$ with a natural
order-of-magnitude of $m_{\rm soft}$. The $b$ term arises
similarly with order-of-magnitude $m_{\rm soft}^2$. The actual values of
$\mu$ and $b$ depend on contributions to the full superpotential and K\"ahler potential
involving the hidden-sector fields including $X$; see
\cite{GiudiceMasiero} for details. These terms do
not have any other direct effect on phenomenology, so without faith in a
complete underlying theory it will be difficult to correlate them with future
experimental results.

One way of understanding the origin of the $\mu$ term in the Giudice-Masiero
class of models is to consider the low-energy effective theory below $M_P$
involving a non-renormalizable K\"ahler potential term of the form in
eq.~(\ref{eq:KahlerpotentialforGiudiceMasiero}). Even if not present in
the fundamental theory, this term could arise from radiative corrections
\cite{HallLykkenWeinberg}. If
the auxiliary field for $X$ obtains a VEV, then one obtains \beq
\mu = \frac{\lambda_\mu}{M_P} \langle F_X^* \rangle .
\eeq
This will be of the correct order of magnitude if parametrically
$\langle F_X^* \rangle \sim m_{\rm soft} M_P$, which is indeed the typical
size
assigned to the $F$-terms of the hidden sector in Planck-scale mediated
models of supersymmetry breaking. The
$b$ term in the soft supersymmetry breaking sector at low energies could
arise in this effective field theory picture from K\"ahler potential terms of
the form $K = \frac{\lambda_b}{M_P^2} Y^* Z H_u H_d$, where $\langle F_Y^*
\rangle \sim \langle F_Z \rangle \sim m_{\rm soft} M_P$. 
However, this is not necessary, because with $\mu
\not= 0$, the low-energy non-zero value of $b$ will arise from
threshold effects and renormalization group running.
One
could also identify both of the fields $Y,Z$ with $X$, at the cost of explicitly violating
the Peccei-Quinn symmetry.

\subsection{Extra $D$-term contributions to scalar 
masses}\label{subsec:variations.Dterms}
\setcounter{equation}{0}

Another way to generalize the MSSM is to include additional gauge
interactions. The simplest possible gauge extension introduces just one
new Abelian gauge symmetry; call it $U(1)_X$. If it is broken at a very
high mass scale, then the corresponding vector gauge boson and gaugino
fermion will both be heavy and will decouple from physics at the TeV scale
and below. However, as long as the MSSM fields carry $U(1)_X$ charges, the
breaking of $U(1)_X$ at an arbitrarily high energy scale can still leave a
telltale imprint on the soft terms of the MSSM \cite{Dterms}. 

To see how this works, let us consider the scalar potential for a model in
which $U(1)_X$ is broken. Suppose that the MSSM scalar fields, denoted
generically by $\phi_i$, carry $U(1)_X$ charges $x_i$. We also introduce a
pair of chiral supermultiplets $\Splus$ and $\Sminus$ with $U(1)_X$
charges normalized to $+1$ and $-1$ respectively. These fields are
singlets under the Standard Model gauge group $SU(3)_C \times SU(2)_L
\times U(1)_Y$, so that when they get VEVs, they will just accomplish the
breaking of $U(1)_X$. An obvious guess for the superpotential containing
$\Splus$ and $\Sminus$ is $W = M \Splus \Sminus$, where $M$ is a
supersymmetric mass. However, unless $M$ vanishes or is very small, it
will yield positive-semidefinite quadratic terms in the scalar potential
of the form $V = |M|^2 (|\Splus|^2 + |\Sminus|^2)$, which will force the
minimum to be at $\Splus = \Sminus = 0$. Since we want $\Splus$ and
$\Sminus$ to obtain VEVs, this is unacceptable. Therefore we assume that
$M$ is 0 (or very small) and that the leading contribution to the
superpotential comes instead from a non-renormalizable term, say: 
\beq
W \,=\, {\lambda\over 2 \MPlanck} \Splus^2 \Sminus^2.
\label{wfordterms}
\eeq
The equations 
of motion for the auxiliary fields are then $F^*_{\Splus} = -\partial 
W/\partial \Splus = -(\lambda/\MPlanck)\Splus \Sminus^{2}$ and 
$F^*_{\Sminus} = -\partial W/\partial \Sminus = -(\lambda/\MPlanck)\Sminus 
\Splus^{2}$, and the corresponding contribution to the scalar potential is
\beq
V_F \>=\> |F_{\Splus}|^2 + |F_{\Sminus}|^2  \>=\>
{|\lambda|^2\over \MPlanck^2}
\Bigl ( |\Splus|^4 |\Sminus|^2 + |\Splus|^2 |\Sminus|^4 \Bigr )  .
\eeq
In addition, there are supersymmetry-breaking terms that must be taken 
into account:
\beq
V_{\rm soft} \,=\, m_+^2 |\Splus|^2 + m_-^2 |\Sminus|^2 -
\left ({a\over 2\MPlanck} \Splus^2 \Sminus^2 + \conj\right ).
\eeq
The terms with $m_+^2$ and $m_-^2$ are soft squared masses for
$\Splus$ and $\Sminus$. They could come from a minimal supergravity
framework at the Planck scale, but in general they will be renormalized
differently, due to different interactions for $\Splus$ and $\Sminus$,
which we have not bothered to write down in eq.~(\ref{wfordterms}) because
they involve fields that will not get VEVs. The last term is a ``soft" 
term analogous to the $a$ terms in 
eq.~(\ref{lagrsoft}), with $a$ of order $m_{\rm soft}$. The coupling
$a/2\MPlanck$ is actually dimensionless, but should be treated as soft
because of its origin and its tiny magnitude. Such terms arise from the
supergravity Lagrangian in an exactly analogous way to the usual soft
terms. Usually one can just ignore them, but this one plays a crucial role
in the gauge symmetry breaking mechanism. The scalar potential for terms
containing $\Splus$ and $\Sminus$ is: 
\beq
V =
{1\over 2} g_X^2 \Bigl ( |\Splus|^2 - |\Sminus|^2 + \sum_i x_i |\phi_i|^2
\Bigr )^2 + V_F + V_{\rm soft}.
\label{xpotential}
\eeq
The first term involves the square of the $U(1)_X$ $D$-term [see
eqs.~(\ref{solveforD}) and (\ref{fdpot})], and $g_X$ is the $U(1)_X$ gauge
coupling. The scalar potential eq.~(\ref{xpotential}) has a nearly
$D$-flat direction, because the $D$-term part vanishes for $\phi_i=0$ and
any $|\Splus| = |\Sminus|$. Without loss of generality, we can take $a$
and $\lambda$ to both be real and positive for purposes of minimizing the
scalar potential. As long as $a^2 - 6 \lambda^2 (m_+^2 + m_-^2) >
0$, there is a minimum of the potential very near the flat direction: 
\beq
\langle \Splus \rangle^2 \,\approx\, \langle \Sminus \rangle^2
\, \approx \, 
\Bigl [a + \sqrt{ a^2 - 6 \lambda^2 (m_+^2 + m_-^2) } \Bigr ]
\MPlanck/6 \lambda^2
\eeq
(with $\langle \phi_i\rangle = 0$), so $\langle \Splus \rangle \approx
\langle \Sminus \rangle \sim {\cal O}(\sqrt{m_{\rm soft} \MPlanck})$. This
is also a global minimum of the potential if $a^2 - 8 \lambda^2 (m_+^2
+ m_-^2) > 0$. Note that $m_+^2 + m_-^2 < 0$ is a sufficient,
but not necessary, condition. The $V_F$ contribution is what stabilizes
the scalar potential at very large field strengths. The VEVs of $\Splus$
and $\Sminus$ will typically be of order $10^{10}$ GeV or so. Therefore
the $U(1)_X$ gauge boson and gaugino, with masses of order $g_X \langle
S_\pm\rangle$, will play no role in collider physics. 

However, there is also necessarily a small deviation from $\langle 
\Splus\rangle = \langle \Sminus \rangle$, as long as $m_+^2 \not= 
m_-^2$. At the minimum of the potential with $\partial V/\partial 
\Splus = \partial V/\partial \Sminus = 0$, the leading order difference in 
the VEVs is given by
\beq
 \langle \Splus \rangle^2 - \langle \Sminus \rangle^2 
\,=\, -\langle D_X \rangle/g_X 
\,\approx\,  (m_-^2 - m_+^2)/2 g_X^2,
\eeq
assuming that $\langle \Splus \rangle$ and $\langle \Sminus \rangle$ are 
much larger than their difference. After integrating out $\Splus$ and 
$\Sminus$ by replacing them using their equations of motion expanded 
around the minimum of the potential, one finds that the MSSM scalars 
$\phi_i$ each receive a squared-mass correction
\beq
\Delta m_i^2 \,=\,  -x_i g_X \langle D_X \rangle\, ,
\label{dxtermcorrections}
\eeq
in addition to the usual soft terms from other sources. The $D$-term
corrections eq.~(\ref{dxtermcorrections}) can be roughly of the order of
$m_{\rm soft}^2$ at most, since they are all proportional to $m_-^2-
m_+^2$. The result eq.~(\ref{dxtermcorrections}) does not
actually depend on the choice of the non-renormalizable superpotential, as
long as it produces the required symmetry breaking with large VEVs; this
is a general feature. The most important feature of
eq.~(\ref{dxtermcorrections}) is that each MSSM scalar squared mass
obtains a correction just proportional to its charge $x_i$ under the
spontaneously broken gauge group, with a universal factor $g_X \langle D_X
\rangle$. In a sense, the soft supersymmetry-breaking terms $m_+^2$ and
$m_-^2$ have been recycled into a non-zero $D$-term for $U(1)_X$,
which then leaves its ``fingerprint" on the spectrum of MSSM scalar
masses. From the point of view of TeV scale physics, the quantity $g_X
\langle D_X \rangle$ can simply be taken to parameterize our ignorance of
how $U(1)_X$ got broken. Typically, the charges $x_i$ are rational numbers
and do not all have the same sign, so that a particular candidate $U(1)_X$
can leave a quite distinctive pattern of mass splittings on the squark and
slepton spectrum. As long as the charges are family-independent, the
squarks and sleptons with the same electroweak quantum numbers remain
degenerate, maintaining the natural suppression of flavor-mixing effects. 

The additional gauge symmetry $U(1)_X$ in the above discussion can stand
alone, or may perhaps be embedded in a larger non-Abelian gauge group. If
the gauge group for the underlying theory at the Planck scale contains
more than one new $U(1)$ factor, then each can make a contribution like
eq.~(\ref{dxtermcorrections}). Additional $U(1)$ gauge group 
factors are quite
common in superstring models, and in the GUT groups $SO(10)$ and $E_6$, 
suggesting optimism about the existence of
corresponding $D$-term corrections. Once one merely assumes the existence
of additional $U(1)$ gauge groups at very high energies, it is unnatural
to assume that such $D$-term contributions to the MSSM scalar masses
should vanish, unless there is an exact symmetry that enforces $m_+^2 =
m_-^2$. The only question is whether or not the magnitude of the
$D$-term contributions is significant compared to the usual minimal
supergravity and RG contributions. Therefore, efforts to
understand the sparticle spectrum of the MSSM may need to take into
account the possibility of $D$-terms from additional gauge groups. 

\section{Concluding remarks}\label{sec:outlook}
\setcounter{equation}{0}
\setcounter{figure}{0}
\setcounter{table}{0}
\setcounter{footnote}{1}

In this primer, I have tried to convey some of the more essential 
features of supersymmetry as a theory of physics beyond the Standard Model. 
One of the nicest qualities of supersymmetry is that so much is known 
about its implications already, despite the present lack of direct experimental evidence. 
The interactions of the Standard Model particles and their superpartners are 
fixed by supersymmetry, up to mass mixing effects due to supersymmetry 
breaking. Even the terms and stakes of many of the important outstanding 
questions, especially the paramount issue ``How is supersymmetry 
broken?", are already rather clear. That this can be so is a testament 
to the unreasonably predictive quality of the symmetry itself.

At this writing, LHC searches have been performed
based on 5 fb$^{-1}$ at $\sqrt{s} = 7$ TeV,
20 fb$^{-1}$ at $\sqrt{s} = 8$ TeV, and 4 fb$^{-1}$ at $\sqrt{s} = 13$ TeV.
These searches have not found any evidence for superpartners, and 
have put strong lower bounds on the masses of squarks and the gluino in 
large classes of models. Even for the weakly interacting 
superpartners, the mass limits have begun to exceed those from LEP, in some 
cases greatly so. The earliest search strategies used by ATLAS and CMS were 
tuned to simple and optimistic templates, including the the MSUGRA 
scenario with new parameters $m^2_0$, $m_{1/2}$, $A_0$, $\tan\beta$ and 
Arg$(\mu )$, and the GMSB scenario with new parameters $\Lambda$, 
$M_{\rm mess}$, $\nmess$, $\langle F \rangle$, $\tan\beta$, and ${\rm 
Arg}(\mu )$. However, the only indispensable idea of supersymmetry is 
simply that of a symmetry between fermions and bosons. Nature may or may 
not be kind enough to realize this beautiful idea within one of the 
specific frameworks that have already been explored well by theorists. 
More recent searches reported by the LHC experimental collaborations 
probe the more general supersymmetric parameter space, including 
$R$-parity violating models, and models in which small mass differences 
or decay modes with softened visible energies  
make the detection of supersymmetry more difficult.

While the present lack of direct evidence for sparticles is 
disappointing, it is at least consistent with the observation of 
$m_{h^0} = 125$ GeV. As noted above, this value of the lightest Higgs 
boson mass points to top squarks that are quite heavy, at least within 
the MSSM with small or moderate stop mixing. In many model frameworks, 
the top-squark masses are correlated, through radiative corrections, 
with the masses of the other squarks and the gluino. Therefore, based only on 
the information that $m_{h^0} = 125$ GeV, one could have surmised 
that supersymmetry probably would not be discovered early at the LHC, and that 
perhaps even with $\sqrt{s} =13$ or 14 TeV the discovery of sparticles is 
not favored, contrary to earlier expectations. A more 
optimistic inference one could draw is that the MSSM is likely to be augmented 
with additional particles or interactions that raise the $h^0$ mass, as 
discussed for example in sections \ref{subsec:variations.vectorlike}
and \ref{subsec:variations.NMSSM}.

It is also worth nothing that most of the {\em other} theories that had been 
put forward as solutions to the hierarchy problem are in no better 
shape than supersymmetry is, given the discovery of the 
125 GeV Higgs boson as well as the lack of other evidence for 
exotic physics at the LHC in the runs at 7 and 8 TeV. 
In fact, many of the competitors to supersymmetry 
in this regard have now been eliminated. Therefore, based on a belief that 
the hierarchy problem needs a solution at the TeV scale, and the alternatives are less than compelling,
I personally maintain a guarded optimism that supersymmetry 
will be discovered at the LHC in the higher energy runs that have just begun. 

If supersymmetry is experimentally verified, the discovery will not be 
an end, but rather a beginning in high energy physics. It seems likely 
to present us with questions and challenges that we can only guess at 
presently. The measurement of sparticle masses, production 
cross-sections, and decay modes will rule out some models for 
supersymmetry breaking and lend credence to others. We will be able to 
test the principle of $R$-parity conservation, the idea that 
supersymmetry has something to do with the dark matter, and possibly 
make connections to other aspects of cosmology including baryogenesis 
and inflation. Other fundamental questions, like the origin of the $\mu$ 
parameter and the rather peculiar hierarchical structure of the Yukawa 
couplings may be brought into sharper focus with the discovery of 
superpartners. Understanding the precise connection of supersymmetry to 
the electroweak scale will surely open the window to even deeper levels 
of fundamental physics.


\addcontentsline{toc}{section}{Acknowledgments}
\section*{Acknowledgments} This is an extended and revised version 
of a chapter in the volume {\it Perspectives on Supersymmetry} 
(World Scientific, 1998) at the kind invitation of Gordy Kane. I am 
thankful to him and to James Wells for many helpful comments and 
suggestions on this primer. I am also indebted to my other collaborators 
on supersymmetry and related matters:
Ben Allanach,
Sandro Ambrosanio,
Nima Arkani-Hamed,
Diego Casta\~no,
Ray Culbertson,
Yanou Cui,
Michael Dine,
Manuel Drees,
Herbi Dreiner,
Tony Gherghetta,
Howie Haber,
Ian Jack, 
Tim Jones,
Chris Kolda,
Graham Kribs,
Nilanjana Kumar,
Tom LeCompte,
Stefano Moretti,
David Morrissey,
Steve Mrenna,
Jian-ming Qian,
Dave Robertson,
Roberto Ruiz de Austri,
Scott Thomas,
Kazuhiro Tobe,
Mike Vaughn, 
Graham Wilson,
Youichi Yamada,
James Younkin,
2890 members of ATLAS,
and especially Pierre Ramond, for many illuminating and inspiring 
conversations. 
Corrections to 
previous versions have been provided by
Daniel Arnold,
Howie Baer,
Jorge de Blas,
Meike de With,
Herbi Dreiner,
Paddy Fox,
Hajime Fukuda,
Peter Graf,
Gudrun Hiller, 
Graham Kribs, 
Bob McElrath,
Matt Reece,
Ver\'onica Sanz,
Frank Daniel Steffen,
Shufang Su, 
John Terning, 
Keith Thomas, 
Scott Thomas, 
Sean Tulin,
and
Robert Ziegler. 
I will be grateful to receive further corrections at 
{\tt spmartin@niu.edu}, and a list of them is maintained at 
{\tt http://www.niu.edu/spmartin/primer}.
I thank the Aspen Center for Physics, Fermilab, the Kavli Institute 
for Theoretical Physics in Santa Barbara, and SLAC for their hospitality, 
and the students of 
the 2013 and 2005 ICTP Summer Schools on Particle Physics,
the 2011 TASI Summer School,
the 2010 PreSUSY Summer School in Bonn,
the 2008 CERN/Fermilab Hadron Collider Physics Summer School,
and PHYS 686 at NIU in Spring 2004,
for asking interesting questions. This work was supported in part by the U.S. 
Department of Energy, and by National Science Foundation 
grants PHY-9970691, PHY-0140129, PHY-0456635, PHY-0757325, PHY-1068369,
and PHY-1417028.


\addcontentsline{toc}{section}{References}

\end{document}